\documentclass{article}
\usepackage[utf8]{inputenc}
\usepackage{amsmath,amssymb,amsfonts,stmaryrd}
\usepackage{physics}
\usepackage{dsfont}
\usepackage{graphicx}
\usepackage{xcolor}
\usepackage{cancel}
\usepackage{hyperref}
\hypersetup{
    colorlinks=true,
    linkcolor=black,
    filecolor=red,
    urlcolor=blue,
    citecolor=red
    }
\usepackage{bm} 
\usepackage{subfigure} 
\usepackage{environ}
\usepackage{amsthm}
\usepackage{url}
\usepackage[margin=0.75in]{geometry}
\usepackage[mathscr]{euscript}
\usepackage[scr=boondox]{mathalpha}
\usepackage{booktabs}
\usepackage{array}
\usepackage{adjustbox}
\usepackage{tikz}
\usetikzlibrary{arrows.meta,positioning,calc}
\usepackage{pgfplots}
\usepackage{tikz-cd}
\usepackage{tabularx}
\usepackage{enumitem}
\usepackage[most]{tcolorbox}
\pgfplotsset{compat=1.18}

\newcommand{\RR}{{\mathbb R}}
\newcommand{\CC}{{\mathbb C}}

\newcommand{\calP}{{\mathcal{P}}}

\newcommand{\Z}{{\mathbb Z}}

\newtheorem{theorem}{Theorem}[section]
\newtheorem{lemma}[theorem]{Lemma}

\newtheorem{proposition}[theorem]{Proposition}
\newtheorem{claim}[theorem]{Proposition}
\newtheorem{corollary}[theorem]{Corollary}
\newtheorem{remark}[theorem]{Remark}

\newtheorem{example}[theorem]{Example}

\newtcolorbox{mainbox}{breakable,colback=blue!3,colframe=blue!45!black,boxrule=0.5pt,arc=2pt,left=5pt,right=5pt,top=3pt,bottom=3pt}

\newtheorem{definition}[theorem]{Definition}
\newtheorem*{exampleu}{Example}

\newcommand{\idea}{\par\smallskip\noindent\textit{Idea.}\ }
\newcommand{\movedproof}[1]{ The proof is in Appendix~\hyperref[pf:#1]{\ref*{pf:#1}}.}
\newcommand{\code}[1]{[[#1]]}

\DeclareMathOperator{\coker}{coker}
\DeclareMathOperator{\Ima}{Im}

\newcommand{\FF}{\mathbb F}
\newcommand{\TrW}{\mathrm{Tr}}

\NewEnviron{eqs}{%
\begin{align}\begin{split}
    \BODY
\end{split}\end{align}
}

\newcommand{\Cay}{\mathrm{Cay}}
\newcommand{\hCay}{\widehat{\mathrm{Cay}}}
\newcommand{\Hom}{\mathrm{Hom}}
\newcommand{\supp}{\mathrm{supp}}
\newcommand{\rk}{\mathrm{rk}}

\newtcolorbox{sketchbox}[1][Sketch: the proof below is outlined, not complete]{enhanced,colback=gray!6,colframe=gray!55,boxrule=0.4pt,arc=2pt,left=5pt,right=5pt,top=3pt,bottom=3pt,title={#1},fonttitle=\small\sffamily,coltitle=black,colbacktitle=gray!18,toptitle=1pt,bottomtitle=1pt}
\newtcolorbox{statusbox}{colback=orange!4,colframe=orange!60!black,boxrule=0.5pt,arc=2pt,left=5pt,right=5pt,top=3pt,bottom=3pt}

\title{Twisted Compactification, Quotient Quantum Codes and Logical Gates}
\author{Po-Shen Hsin$^{1}$}

\begin{document}

\maketitle

\begin{center}
${}^1$ Department of Mathematics, King's College London, Strand, London WC2R 2LS, UK
\end{center}

\bigskip
\begin{abstract}
Many quantum codes, including balanced-product, bivariate-bicycle and genon codes and the five-qubit code, are quotient codes: a larger
code folded onto itself by identifying each check with its image under a deck transformation composed with a symmetry. We show that this
is the lattice form of twisted compactification, in which a theory on $X\times F$ carries a symmetry background along $F$, and we use the
correspondence in both directions. For monomial quotients of Pauli stabilizer codes, a transfer theorem gives the logical qudits, logical
operators and distance bounds from those of the parent, with a Tate cohomology group counting the logical operators that are new. For
higher-form and cup-product symmetries the twist is an entangling diagonal circuit; when a finite cover trivializes the background, the
code is a dressed quotient, so abelian Pauli codes give non-abelian topological order such as $D(D_4)$. The dressings have Clifford level
at most one less than the degree of the twist, and for cubic twists the code dimension is a count of solutions of quadratic equations in
cup products, which certifies that the code is not a Pauli stabilizer code when it is not a power of two; on tori and slabs the distance
is linear in the size. Twists also create logical gates: three toric codes on $S^3$ have no logical qubits, but on $\mathbb{RP}^3$ the
same circuit is a logical $CCZ$. A quarter-turn twist of the 3D fermionic toric code gives codes of linear distance, from $[[768,2,9]]$
up, whose $p+ip$ pump is an explicit finite-time quasi-local evolution that preserves the code space exactly; using the spin-topological
response formula of Barkeshli, Hsin and Kobayashi, we identify its logical action as $T$ up to Clifford factors. Finally, twisting
abelian gauge theories reach only the quantum doubles $D(G)$ with $[G,G]$ central.
\end{abstract}

\bigskip
\eject
\tableofcontents
\setcounter{tocdepth}{2}
\bigskip

\section{Introduction}
\label{sec:intro}

Many quantum error-correcting codes are made by folding a large code onto itself: one identifies each check of a code with its images
under a symmetry. Balanced-product, bivariate-bicycle and genon codes, the first good quantum LDPC codes and the five-qubit code are all of
this form. Physicists know the same operation as twisted compactification, in which a theory is placed on a small space with a symmetry
applied as one goes around it. This paper shows that the two are one construction and uses the correspondence in both directions. Its main
outputs, stated precisely in Section~\ref{sec:mainres}, are the following.
\begin{itemize}[leftmargin=1.5em,itemsep=1pt,topsep=3pt]
\item \emph{One theorem for many families.} The logical qudits, logical operators and distance bounds of a quotient follow from those of
its parent, and a Tate cohomology group counts the new logical operators (Theorem~\ref{thm:general}).
\item \emph{Codes glued by entangling circuits.} Cubic cup-product twists turn abelian toric codes into codes beyond the Pauli stabilizer
framework, such as a code with non-abelian order $D(D_4)$, dimension 22, which no qubit Pauli stabilizer code has, and linear distance
(Corollaries~\ref{cor:dresseddist} and~\ref{cor:cert}).
\item \emph{Gates created by the twist.} Three toric codes on $S^3$ have no logical qubits, but on $\mathbb{RP}^3=S^3/\Z_2$ the same
circuit is a logical $CCZ$ (Proposition~\ref{prop:twisted}).
\item \emph{A route to a logical $T$ in three dimensions.} A quarter-turn twist of the 3D fermionic toric code gives codes of linear
distance, from $\code{768,2,9}$ up, with an explicit quasi-local pump that preserves the code space exactly; by the response formula of
Barkeshli, Hsin and Kobayashi~\cite{Barkeshli:2023bta} it acts as a logical $T$ up to Clifford factors (Proposition~\ref{prop:quarter},
Table~\ref{tab:quarter}).
\item \emph{A limit.} Twisting abelian gauge theories reaches only the quantum doubles $D(G)$ with $[G,G]$ central
(Propositions~\ref{claim:reach} and~\ref{prop:nilp}).
\end{itemize}

\subsection{Twisted compactification and quotient codes}
\label{sec:introtwo}

\paragraph{Twisted compactification.} To compactify a theory on a closed space $F$ is to take space to be $X\times F$ and to view the
result as a theory on the remaining space $X$, as a thin cylinder looks like a line from far away. On a circle, every field must return to
itself when it goes once around. A symmetry of the theory can twist this condition: the field returns only up to a symmetry transformation
$U$, as an antiperiodic fermion returns with a sign. This is the twisted boundary condition of Scherk and Schwarz~\cite{ScherkSchwarz1979}.
Equivalently, one cuts the circle at a point and inserts there a wall that acts by $U$ on everything that crosses it, a codimension-one
symmetry defect. On a general $F$ one inserts a network of such walls, or, equivalently, turns on a flat background gauge field for the
symmetry along $F$. The data of the twist depends on the symmetry. 
Throughout, we focus on invertible symmetries. An ordinary symmetry with group $G$ gives a holonomy $\rho:\pi_1(F)\to G$, the symmetry
picked up around each loop of $F$. A $q$-form symmetry $A$, which acts on $q$-dimensional objects such as the Wilson lines of a gauge theory for
$q=1$, gives a class $B\in H^{q+1}(F;A)$. A wall can also carry a gauged SPT phase, a phase that depends on the fields on the wall, exchange
electric and magnetic objects as a duality does, or be non-invertible. Twisted compactification is a
standard tool in the study of generalized symmetries~\cite{Kaidi:2022uux,Gukov:2020btk,GukovHsinPei2025}: the compactified theory keeps the
symmetries that commute with the background, and gains new ones from the geometry and the higher-form symmetries of $F$. In topological
phases, symmetry defects give twist defects and genons~\cite{Bombin2010,BarkeshliJianQi2013,Barkeshli:2014cna}. The closest physics
precedent for this paper is the reduction of a three-dimensional topological order on a circle threaded by a flux, which resolves it into
sectors labeled by the flux and determines loop braiding and the modular data of the 3D phase~\cite{WangLevin2014,JiangMesarosRan2014,Moradi:2014cfa}. Section~\ref{sec:ft} reviews what we need.

\paragraph{A first example.} Take two repetition codes, each on a ring of $N$ qubits with checks $Z_iZ_{i+1}$, $i\in\Z_N$. Each ring carries
the Hilbert space $(\mathbb C^2)^{\otimes N}$, and exactly two computational basis states satisfy all its checks: $|0\cdots0\rangle$ and $|1\cdots1\rangle$, in
which all qubits agree. They are the two ground states of the Ising ferromagnet on the ring, or, in field theory, the two vacua of the phase
in which the $\Z_2$ symmetry $\prod_iX_i$ is spontaneously broken, and each ring encodes one qubit. The two rings carry $(\mathbb C^2)^{\otimes2N}$,
and their code space is spanned by the four product states
\[
|a\cdots a\rangle\otimes|b\cdots b\rangle,\qquad a,b\in\{0,1\},
\]
where $a$ is the common value of the qubits of the first ring and $b$ that of the second. Now twist the pair by the symmetry that exchanges
the two rings, qubit $i$ of one with qubit $i$ of the other. Put the wall on the link between qubits $N$ and $1$: the check across it now
joins qubit $N$ of the first ring to qubit $1$ of the second, and qubit $N$ of the second to qubit $1$ of the first. The two rings become one
ring of $2N$ qubits, and only $|0\cdots0\rangle$ and $|1\cdots1\rangle$ on all $2N$ qubits satisfy its checks. These are the two states above with $a=b$,
so the twist removes one logical qubit. Field theory gives this count without solving the checks. The twisted theory has as many states
as the trace of the symmetry on the states of the untwisted theory (Proposition~\ref{prop:trace}). The exchange acts on the four states by
$|a\cdots a\rangle\otimes|b\cdots b\rangle\mapsto|b\cdots b\rangle\otimes|a\cdots a\rangle$: it fixes the two with $a=b$ and swaps the other two, so its trace is
$2$. The twisted pair is also a quotient code: start from two
rings of $2N$ qubits and identify qubit $i$ of each ring with qubit $i+N$ of the other. This paper carries out the same identification for
codes in every dimension and for symmetries well beyond permutations, such as the exchange of $X$ and $Z$ and entangling circuits.

\paragraph{Quotient codes.} A quotient code, like the twisted pair above, is built from a larger parent code and a group $\Gamma$ of its symmetries that acts freely on
the qudits. One identifies each qudit with its images, and each stabilizer with its images, and keeps one representative of each orbit
(Definition~\ref{def:quot}). Many code families are built this way, under different names. Toric codes on twisted
tori~\cite{KovalevPryadko2012,BombinMartinDelgado2007,Liang2025} and hyperbolic surface codes~\cite{BreuckmannTerhal2016} are quotients of codes on
the plane or on a tiling by a lattice or a Fuchsian group. Balanced products~\cite{BreuckmannEberhardt2021},
lifted products~\cite{PanteleevKalachev2022LP,PanteleevKalachev2022good}, fibre-bundle codes~\cite{HastingsHaahODonnell2021}, two-block
group-algebra and bivariate-bicycle codes~\cite{LinPryadko2024,Bravyi2024,WangMueller2024} and quantum Tanner codes~\cite{LeverrierZemor2022} are
quotients of product codes by a diagonal group action. The first asymptotically good quantum LDPC codes are among
them~\cite{PanteleevKalachev2022good,LeverrierZemor2022}. A symmetry that exchanges $X$ and $Z$ gives codes that are not CSS, such as the
five-qubit code on a twisted torus~\cite{Wen2003,BonillaAtaides2021}, and the quotients of bivariate-bicycle codes by a ZX-duality, which we
call genon codes (Section~\ref{sec:genon}) and which are related to the genon double covers of~\cite{BurtonGenons2024}. Clifford-deformed surface
codes~\cite{DuaCliffordDeformed2024} change the letters of the checks by single-qubit Cliffords without a quotient.

\paragraph{This paper.} We show that the two constructions are the same and turn this into a framework for building codes and their
logical gates, with one principle and a four-step method.

\begin{mainbox}
\textbf{Principle.} A twisted boundary condition identifies each operator on a cover with its image under the symmetry. For a code, this
identification is the quotient by the deck group acting together with the symmetry. A \emph{deck transformation} of a cover $\tilde F\to F$ permutes the
sheets: it is a homeomorphism of $\tilde F$ that maps every point to a point over the same point of $F$, such as the translation by $L$ of a
circle of length $2L$ covering a circle of length $L$. For a finite regular cover the deck transformations form a finite group $\Gamma$, the
\emph{deck group}, which acts freely on $\tilde F$ with $\tilde F/\Gamma=F$.

\smallskip
\textbf{Method} (Section~\ref{sec:lattice}).
\begin{enumerate}[leftmargin=1.5em,label=\arabic*.]
\item \emph{Lift.} Pass to a finite cover $\tilde F\to F$ on which the background becomes trivial. The parent is the code on
$X\times\tilde F$ without the background.
\item \emph{Identify.} Divide by the deck group, acting by deck transformations on $\tilde F$ and by the symmetry on the fields. For a monomial
symmetry, a permutation of qudits with single-qudit Cliffords, this is an ordinary quotient. For a cup-product defect the symmetry is an entangling diagonal circuit and the result is a dressed
quotient.
\item \emph{Compute.} Read off rate, logical operators, distance bounds and the Clifford level of the dressings of the checks from the cohomology of the deck
group and of $F$, with coefficients in the logical data of the parent.
\item \emph{Descend.} Parent symmetries that commute with the deck group, symmetries of $(F,\rho)$, and the higher-form symmetries of the
compactification become logical gates of the quotient, with the locality listed in Section~\ref{sec:method}.
\end{enumerate}
\end{mainbox}

The framework applies to every background in this paper, a holonomy, a $q$-form class or a cup-product defect, and to parents that are
Pauli stabilizer codes, commuting-projector codes, or codes whose checks are already non-Clifford. Steps 1 and 2 need a finite cover that
trivializes the background; without one the dressed Gauss laws still define the code, but not as a quotient (Remark~\ref{rem:cover}).

\subsection{The correspondence}
\label{sec:introcorr}

\begin{figure}[tbp]
\centering
\begin{tikzpicture}[scale=0.62,>=Stealth]
\foreach \psx/\pname/\pmap/\pylo/\pyhi/\pya/\pyb/\plet in {0/{(a) translation}/{t}/2/3/2/3/X,7.2/{(b) reflection}/{t\circ r}/1/2/2/1/X,14.4/{(c) duality}/{t\circ H}/2/3/2/3/Z}{
\begin{scope}[xshift=\psx cm]
\fill[blue!4] (-0.5,-0.5) rectangle (3.5,4.5);
\draw[densely dashed,red!70!black,thick] (-0.5,-0.5) -- (-0.5,4.5);
\draw[densely dashed,red!70!black,thick] (3.5,-0.5) -- (3.5,4.5);
\foreach \i in {0,...,3} \foreach \j in {0,...,4} \fill[gray!70] (\i,\j) circle (1.6pt);
\fill[blue!20] (3,2) rectangle (3.5,3);
\draw[blue!70!black,thick] (3.5,2) -- (3,2) -- (3,3) -- (3.5,3);
\fill[blue!20] (-0.5,\pylo) rectangle (0,\pyhi);
\draw[blue!70!black,thick] (-0.5,\pylo) -- (0,\pylo) -- (0,\pyhi) -- (-0.5,\pyhi);
\node[circle,draw=blue!70!black,fill=white,inner sep=0.4pt,font=\tiny] at (3,2) {$X$};
\node[circle,draw=blue!70!black,fill=white,inner sep=0.4pt,font=\tiny] at (3,3) {$X$};
\node[circle,draw=blue!70!black,fill=white,inner sep=0.4pt,font=\tiny] at (0,\pya) {$\plet$};
\node[circle,draw=blue!70!black,fill=white,inner sep=0.4pt,font=\tiny] at (0,\pyb) {$\plet$};
\draw[->,red!70!black] (3.5,4.75) to[out=150,in=30] node[above,font=\scriptsize]{$\pmap$} (-0.5,4.75);
\node[font=\small] at (1.5,-1.1) {\pname};
\end{scope}}
\end{tikzpicture}
\caption{The principle on the simplest example (schematic). The shaded strip $0\le x<L$ is a fundamental domain, the dashed lines are the two
sides of the seam, and the arrow is the deck transformation $t$ acting together with the symmetry. A check of the parent that crosses the seam
is glued to its image: (a) at the same height, (b) at the mirrored height, (c) with its letters exchanged by the duality. The quotients are the
toric code on a smaller torus, the toric code on a Klein bottle, and a code that is not CSS, such as the five-qubit code. The example after
Definition~\ref{def:quot} and Example~\ref{ex:run1} give the details.}
\label{fig:principle}
\end{figure}
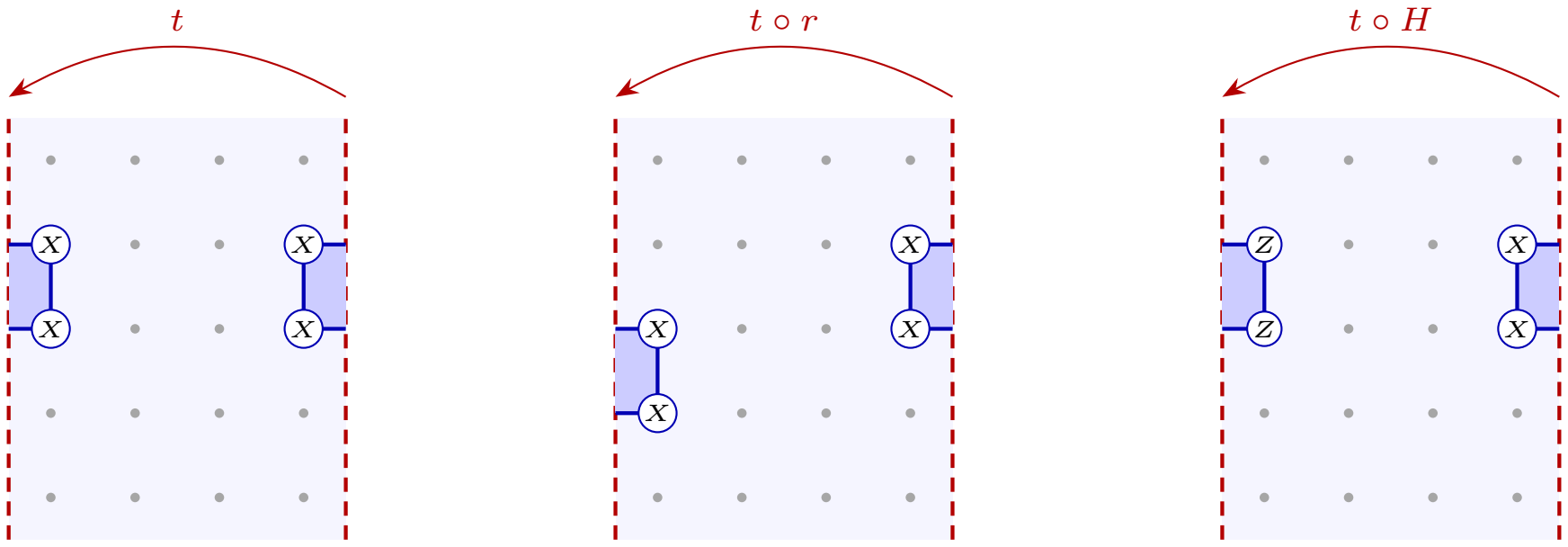

\paragraph{Why twisted compactification is a quotient.} Unfold the compact direction to a cover $\tilde F$ on which the background is
trivial. A field that goes once around a cycle $\gamma$ of $F$ returns acted on by $\rho(\gamma)$. So the fields on $X\times F$ are the fields on
$X\times\tilde F$ that are invariant under the deck group acting together with the symmetry. On the lattice, a stabilizer that crosses the
defect is glued to its image under the symmetry. This is exactly the identification that defines a quotient code. The toric code on a
$2L\times L'$ torus shows the three basic cases (Figure~\ref{fig:principle}, and the example after Definition~\ref{def:quot}).
\begin{itemize}[leftmargin=1.5em]
\item A translation twist gives the toric code on the $L\times L'$ torus.
\item A reflection twist glues each check to its mirror image and gives the toric code on a Klein bottle.
\item An electromagnetic-duality twist glues $X$-checks to $Z$-checks. For the rotated toric code on a twisted torus it gives the five-qubit
code (Example~\ref{ex:run1}). For the 4D toric code with 2-form fields it gives the 3D fermionic toric code (\hyperref[ex:emfour]{example at the end of Section~\ref*{sec:general}}).
\end{itemize}

\paragraph{Entangling twists.} For a higher-form or cup-product symmetry the twist is not a permutation but an entangling diagonal circuit,
for instance a line of $CZ$ gates between two toric codes. The identification of the principle then conjugates each stabilizer by that
circuit where it crosses the wall, and the result is still a quotient, in a dressed sense (Section~\ref{sec:dressedquot}). This is the
sharpest form of the correspondence, and the main results below are organized around it.

\paragraph{Quotient, not restriction.} The quotient code is not the deck-invariant subspace of the parent Hilbert space. It has one
qudit for each orbit, and the configurations constant on orbits span a smaller space: $|00\rangle$ and $|11\rangle$ for two qubits
exchanged by $\Z_2$, against a three-dimensional invariant subspace. The words quotient and descent cover four operations, two on codes
and two on circuits, which Section~\ref{sec:method} separates.

\paragraph{Using the correspondence.} The correspondence is used in both directions. From field theory to codes, it says which twist gives
a code with given properties, and it gives the answer: the code dimension is the partition function on $X\times F\times S^1$, the logical
operators are the twisted sectors, and the logical gates are the symmetries of the compactified theory. From codes to field theory, a quotient code
realizes the compactified theory on the lattice, with explicit checks, logical operators and distance. Appendix~\ref{app:dict} lists the dictionary entry by entry
(Table~\ref{tab:dict}) and writes known code families in this form (Table~\ref{tab:known}).

\subsection{Main results}
\label{sec:mainres}

The results fall into three groups: the construction of codes from a twist, the computation of what a quotient keeps, and the logical
gates. The central result is the correspondence for cup-product symmetries, stated in the box below. Each item says what question the result
answers, what we find and why it matters, and names the precise statement; the longer proofs are in Appendix~\ref{app:proofs}.

\begin{mainbox}
\textbf{The correspondence for invertible symmetry} (Sections~\ref{sec:higher} and~\ref{sec:dressedquot}). Take a parent made of $\Z_p$ gauge
fields, each a form of any degree, with one qudit on each cell of that degree; 1-form fields are toric codes. A cup-product symmetry
multiplies each state by a phase given by a cup product of the fields, such as $a_1\cup a_2$ for two toric codes or $a_1\cup a_2\cup a_3$
for three. A background $B$ says where the twist acts; for a wall, $B$ is dual to the wall. For every such symmetry and background, local
Gauss laws dressed by the background define a commuting-projector code, which depends only on the cohomology class of $B$
(Definition~\ref{def:gauss}, Proposition~\ref{prop:twisted-code}). If a finite cover makes $B$ trivial, this code is the dressed quotient of the
parent without background by the deck transformations of the cover, each composed with the diagonal circuit of the symmetry on a region
determined by $B$ (Theorem~\ref{thm:dressedquot}). So an entangling circuit, and not only a permutation, can serve as the gluing. If no
finite cover makes $B$ trivial, as for a non-zero class in $H^2(S^2;\Z_p)$, the dressed Gauss laws still define a code, but the construction
of Theorem~\ref{thm:dressedquot} does not produce it (Remark~\ref{rem:cover}).
\end{mainbox}

\paragraph{Construction.}
\begin{enumerate}[label=(\roman*),leftmargin=2em]
\item \emph{Non-abelian order from abelian Pauli codes.} Pauli stabilizer codes are the best understood quantum codes, but in two
dimensions their anyons are all abelian~\cite{BombinDuclosCianciPoulin2012,Bombin2014,Haah2021TI} and their locality-preserving logical gates
are Clifford~\cite{BravyiKoenig2013}. Codes with non-abelian topological order need checks that are not Pauli operators, and in two
dimensions they can go past this limit, as the logical $T$ of Example~\ref{ex:kzhgate} does. Such codes have so far been built one model at
a time. Earlier quotient codes were glued by permutations and other monomial maps, which keep a Pauli code Pauli. Since an entangling
circuit can also glue the sheets, codes with non-abelian order are dressed quotients of abelian Pauli codes whenever a finite cover
trivializes the background (Theorem~\ref{thm:dressedquot}). For example, three 3D toric codes compactified on a circle with a cubic wall carry
$D(D_4)$ order. So non-abelian codes can be built systematically from well-understood parents.
\item \emph{Twisting has a ceiling that gauging does not.} Twisting and gauging are the two standard ways to build new phases from old
ones, and one might hope that twisting copies of abelian codes reaches every non-abelian order. We show
(Proposition~\ref{prop:nilp}) that twisted abelian doubles, and what stacking and condensation make from them, have nilpotency class at most
two, that is, the fusion product $x\otimes\bar x$ of every anyon with its antiparticle is a sum of abelian anyons; among quantum doubles these
are the $D(G)$ with $[G,G]$ central, such as $D(D_4)$ and $D(Q_8)$. Compactifying abelian gauge theories with
holonomies by automorphisms of the gauge group, higher-form classes or cup-polynomial defects gives twisted abelian doubles
(Proposition~\ref{claim:reach}), so these twists stay at class two, while one gauging step already gives $D(D_8)$, which is not of this kind. This marks the
reach of the construction and says what gauging adds; for example, the higher-level gates of~\cite{WSN2025} on the doubles of larger
dihedral groups need gauging.
\end{enumerate}

\paragraph{Computation.}
\begin{enumerate}[label=(\roman*),leftmargin=2em,resume]
\item \emph{The parameters of a quotient from those of its parent.} Each quotient code family so far has come with its own computation of
rate and distance, and this computation is usually the hardest part of the work. The transfer theorem gives one description for all of
them. For a monomial quotient of a Pauli stabilizer code by a free action, it expresses the rate, the logical operators and bounds on the
distance through those of the parent (Theorem~\ref{thm:general}). The only logical operators with no parent counterpart are winding
sectors, which wrap around the new space and which a Tate cohomology group counts. The work for a given family is reduced to computing
invariant representatives, norm images and winding sectors; these can still be hard to compute, and the resulting bound can be weak, as in
the running example. Three further tools complete the computation. The number of logical qudits follows from a spectral sequence for flat
backgrounds (Theorem~\ref{thm:ss}) and from a cohomology group for quadratic twists (Theorem~\ref{thm:hfcount}). The distance of a quotient is
at least the injectivity radius of the cover, the radius up to which the quotient looks like the parent, unless the parent has a lighter
logical operator inside a ball of that radius (Proposition~\ref{prop:injdist}). For the dressed codes, which need not be Pauli stabilizer
codes, a diagonal dressing never creates a light logical operator, so the distance is at least a constant times the size of the region on
which the undressed toric codes look like those of flat space (Theorem~\ref{thm:dresseddist}). For example, the code of $D(D_4)$ on the slab
is $\code{15L^2,\log_222,\Theta(L)}$, a code of dimension 22 on $15L^2$ qubits, or $((15L^2,22,\Theta(L)))$ in the notation of general
codes (Corollary~\ref{cor:dresseddist}). Together these reduce the design of monomial quotients to the choice of a parent and a symmetry,
followed by a computation on the parent; for cubic dressed quotients the dimension and a distance bound are known, but not yet a formula
for the logical operators.
\item \emph{The degree of the twist decides the physics.} Cup-product twists are graded by the degree of their phase polynomial in the qudit
variables: $a_1\cup a_2$ is quadratic and $a_1\cup a_2\cup a_3$ is cubic. A twist of degree $m$ dresses the checks at Clifford level at most
$m-1$ (Proposition~\ref{prop:clifflevel}). So a quadratic twist, which is a Clifford circuit, gives Pauli checks and a new Pauli code, whose
logical qudits a cohomology group counts, and non-abelian order needs a cubic twist. A cubic twist can produce non-Pauli checks, and the resulting
code need not admit a Pauli stabilizer description, so its code space is hard to compute directly; we show that its dimension is the number of solutions of a system of quadratic equations in cup
products (Theorem~\ref{thm:pointcount}). So the degree tells a designer which twists to try, and a count of solutions replaces the hardest
part of the computation. Three properties are kept apart: checks that are not Pauli, a code that is not a Pauli stabilizer code, and
non-abelian order. The degree bounds only the level of the dressings, the count certifies the second property when its value is not a power
of two, and the third comes from the field theory, as for type III.
\item \emph{A cheap certificate that a code is not Pauli.} A commuting-projector code may be a Pauli stabilizer code in disguise, after a
change of basis, and this is not easy to rule out. The point count gives the code dimension exactly, and a dimension that is neither 0 nor
a power of 2 rules out every qubit Pauli stabilizer description, local or not (Corollary~\ref{cor:cert}). For the code of $D(D_4)$ in (iii), three 3D toric codes
with a cubic defect on the slab $T^2\times[0,1]$, the count is 22. The test needs one number and no knowledge of the logical
operators, so it applies across the growing range of non-Pauli codes; when the dimension is a power of 2, it says nothing.
\end{enumerate}

\paragraph{Gates.}
\begin{enumerate}[label=(\roman*),leftmargin=2em,resume]
\item \emph{A twist creates gates by creating the logical states they act on.} A logical gate is usually a symmetry of a code that acts on
the logical qubits the code already has, so a parent with no logical qubits seems to offer nothing. A twist changes this. The quotient has
new logical states, the twisted sectors, which record how the fields wind around the non-contractible cycles of the new space. On these
states a cup-product symmetry of the parent acts by a phase that depends only on the topology of the quotient, a bordism invariant of the
deck group, and not on the details of the parent (Proposition~\ref{prop:twisted}). For example, three toric codes on $S^3$ have no logical
qubits, but on $\mathbb{RP}^3=S^3/\Z_2$ the same circuit is a logical $CCZ$. So the circuit comes from the parent and the gate it performs
comes from the topology, and finding a gate becomes a computation in group homology. Propositions~\ref{prop:descent} and~\ref{prop:lift} say
which parent symmetries descend.
\item \emph{A twist can do what independent copies cannot.} One might think that a twist which permutes several copies of a code around a
cycle only relabels them. It does more. On independent copies of a toric code, cup-product circuits give phases that are products of the
fields, such as $CCZ$. A twist that permutes the copies identifies them, and the phase becomes a power of one field, which can be lifted to
finer roots of unity. This gives the qutrit $T$ on one 3D qutrit toric code on $T^3$ of~\cite{BDEHT2026cupsII} (Proposition~\ref{prop:tori}),
a gate that no circuit on the independent copies induces. In a different way, a non-invertible parent symmetry, which is not a unitary
operator, becomes a logical unitary gate on every quotient on which its fusion outcomes act as scalars (Lemma~\ref{lem:noninv}), the lattice
form of an observation of~\cite{GukovHsinPei2025}; an example in which this gate is not also induced by an invertible parent symmetry is
open (Section~\ref{sec:discussion}).
\item \emph{A logical $T$ with linear distance in three dimensions.} Non-Clifford gates are the expensive part of fault-tolerant quantum
computation, so a constant-time logical $T$ on codes of growing distance is valuable. The 3D fermionic toric code has a $\Z_8$ pump, a
finite-time evolution that gives a logical $CCZ$ on the three-torus~\cite{FidkowskiHastings2024}. A quarter-turn twist changes the space.
On the resulting flat 3-manifolds the code has distance linear in their size, and we construct the same pump explicitly and prove that it
preserves the code space exactly and is quasi-local; using the spin-topological response formula of~\cite{Barkeshli:2023bta}, we identify its
logical action as $T$ up to Clifford factors (Proposition~\ref{prop:quarter}). Table~\ref{tab:quarter} gives exact distances of
small members, from $\code{768,2,9}$ to $\code{6144,2,17}$. So the twist selects which non-Clifford gate a fixed parent symmetry performs. The pump is a finite-time
evolution of a quasi-local Hamiltonian, so the gate is approximated to any accuracy $\varepsilon$ by a circuit of constant depth whose
gates have range $O(\log(n/\varepsilon))$ in the number $n$ of qubits. No exact circuit of bounded depth and bounded range can pump a chiral layer,
and the known $CCZ$ and $T$ gates of the fermionic toric code~\cite{FidkowskiHastings2024,Barkeshli:2023bta} are constant depth in the same
sense. The distance bound uses that the fermionic toric code has no finitely supported logical operator on $\mathbb R^3$, which we prove
(Lemma~\ref{lem:ftclocal}).
\end{enumerate}
Design rules and no-go results that follow from these statements are collected at the start of Section~\ref{sec:discussion}, and the
reading of known code families and gates as twisted compactifications in Section~\ref{sec:intronew} and Appendix~\ref{app:dict}.

\paragraph{The running example.} One small code is followed through the paper, in Examples~\ref{ex:run1}--\ref{ex:run4}, with a
certificate in Appendix~\ref{app:runcert}. The parent is the rotated toric code $\code{20,2,4}$: one qubit on each of the 20 sites of the
torus $\Z^2/2\Lambda$, where $\Lambda=(1+2i)\Z[i]$ is the lattice spanned by $(1,2)$ and $(-2,1)$, and one check on each plaquette, $X^{\otimes4}$
and $Z^{\otimes4}$ in a checkerboard pattern. The group is $\Gamma=\Lambda/2\Lambda\cong\Z_2^2$, the translations by $\Lambda$ modulo $2\Lambda$; a
translation that exchanges the two colours of plaquettes is composed with the transversal Hadamard, which exchanges $X$ and $Z$, so every
element maps checks to checks. Identifying the qubits along the orbits gives the five-qubit code $\code{5,1,3}$, with checks $XZZXI$ and
its cyclic shifts, which is the Wen plaquette model on the twisted torus $\Z^2/\Lambda$ (Example~\ref{ex:run1}). In field theory it is the
toric code with a wall of its electromagnetic duality, which exchanges the charges $e$ and $m$, and the trace formula gives its one
logical qubit (Example~\ref{ex:run3}). The transfer theorem explains the counts. The Tate group is one-dimensional, so one of the two
logical Paulis is a winding sector that no parent logical explains, and the factorization theorem splits the quotient into a translation,
which gives the toric code $\code{10,2,3}$, and a duality, which fixes the distance 3 (Example~\ref{ex:run2}). Finally, only 20 of the 360
monomial-Clifford automorphisms of the five-qubit code descend from the parent, and the familiar order-3 transversal gate is not among
them (Example~\ref{ex:run4}).

\subsection{Relation to previous work}
\label{sec:intronew}

We see four gaps in the literature. Each paragraph below says what was known and what this paper adds, and
Appendix~\ref{app:dict} gives the detailed comparison for individual codes and gates. For known codes, phases and gates we claim only the
general statement and its reading as a twisted compactification. The results collected in Section~\ref{sec:mainres} are new in the
generality stated.

\paragraph{1. From examples to one construction.} Quotient codes were developed family by family, for lattices, group algebras, products
and fibre bundles, each with its own proof of the parameters. Breuckmann and Eberhardt identified balanced products as
quotients~\cite{BreuckmannEberhardt2021}, and Hastings, Haah and O'Donnell introduced fibre-bundle codes~\cite{HastingsHaahODonnell2021}; our
spectral sequence (Theorem~\ref{thm:ss}) is Leray--Serre in form, as for their codes. Here all these families are one construction:
balanced-product, fibre-bundle, lifted-product, quantum Tanner, bivariate-bicycle and two-block codes, Haah's code, the twisted toric codes
including the five-qubit code, genon codes and Clifford-hierarchy stabilizer codes are all twisted compactifications (Table~\ref{tab:known}),
and so are the known logical $CZ$, $CCZ$ and $T$ gates of 2D and 3D topological codes (Table~\ref{tab:create}). Each is a choice of a
parent, a symmetry and an internal space, so a search for new codes can range over these choices instead of over check matrices. The field
theory says what to twist, and the transfer theorem and the spectral sequence say what comes out, for any finite group, abelian or not,
including dualities. Appendix~\ref{app:dict} works out five cases that show the reach: the optimal twisted toric
codes of Kovalev and Pryadko and of Bomb\'in and Martin-Delgado~\cite{KovalevPryadko2012,BombinMartinDelgado2007} as duality twists, a
colour-code quotient by the triality $X\to Y\to Z$ that uses all three Pauli types~\cite{KesselringColorTwist2018}, Haah's
code~\cite{PhysRevA.83.042330} as a balanced product, the non-abelian two-block codes of~\cite{LinPryadko2024}, and the good quantum LDPC
codes~\cite{PanteleevKalachev2022good,LeverrierZemor2022}, whose expanding non-abelian twist makes the distance linear.

\paragraph{2. From permutations to entangling twists.} Field theory twists by any symmetry, but lattice quotients used permutations
and, in special cases, dualities. For higher-form and cup-product symmetries the twist operator is an entangling circuit, and the quotient
must be dressed by it. We give the general lattice construction with all curvature terms, and say when it is a quotient (a trivializing cover
exists) and when the construction does not apply (a class on $S^2$ survives every finite cover). Some outputs are not Pauli stabilizer codes
(Corollary~\ref{cor:cert}).

\paragraph{3. Two meanings of ``non-abelian''.} The word is used for two different phenomena, which we label (N1) and (N2), and the
framework separates them. (N1), a \emph{non-abelian twist}: the twist group is non-abelian but the code is a Pauli stabilizer code, as for
lifted products and two-block codes over non-abelian groups. (N2), \emph{non-abelian order}: the code need not be a Pauli stabilizer code; it is a commuting-projector code whose phase is a non-abelian topological order, such as
$D(D_4)$. By the level rule (Proposition~\ref{prop:clifflevel}), (N2) needs a twist operator of degree at least 3, and the (N2) codes obtained
this way are Clifford-hierarchy stabilizer codes in the sense of~\cite{KZH2025chs}. Non-abelian qLDPC codes of type (N2), in which a
triple cup product dresses the Gauss laws of abelian codes on CW and sheaf complexes, are constructed in~\cite{ZKH2026nonabelian,ZTKH2026sheaf},
where they are called twisted codes; here they are dressed codes (Definition~\ref{def:gauss}).

\paragraph{4. From gates on a code to gates from a twist.} The logical gates of a fixed code come from parent symmetries, such as transversal,
fold-transversal~\cite{BreuckmannBurton2024} and automorphism gates~\cite{EberhardtSteffan2024}; from isometries and mapping classes of the
internal space; and from higher-form and cup-product symmetries, whose operators on cycles are constant-depth diagonal
circuits~\cite{Barkeshli:2022edm,Barkeshli:2023bta,ZhuSikanderPortnoyCrossBrown2023,HKZ2024classifying,BDET2024cups,Lin2024sheaves,KZH2025chs}.
What was missing is how a twist changes them: which parent gates survive, which new ones appear, and when a symmetry of the
quotient comes from the parent. Results (vi) and (vii) of Section~\ref{sec:mainres}, with the design rules of Section~\ref{sec:discussion},
answer these questions. The example gates are known, and Appendix~\ref{app:dict} gives the source of each. Two evaluations are
new as far as we know: the formula~\cite[eq.~(26)]{Barkeshli:2023bta} of Barkeshli, Hsin and Kobayashi gives a logical $T$ on the
quarter-turn space and a logical $CS$ on $S^3/Q_8$. For the quarter turn we also construct the pump and prove that it descends exactly, that
it is quasi-local and that the distance is linear in the size (Proposition~\ref{prop:quarter}); for $S^3/Q_8$ no pump is known.

\paragraph{Organization.} Section~\ref{sec:ft} reviews the field theory of twisted compactification. Section~\ref{sec:lattice} carries out
steps 1--3 of the method on the lattice: monomial quotients and the transfer theorem, flat backgrounds, and higher-form and cup-polynomial
backgrounds with their quotient description, counts and distances. Section~\ref{sec:sym} carries out step 4, from descent and the Clifford level to
fermionic pumps under rotation twists (Section~\ref{sec:fermionic}), and Section~\ref{sec:discussion} collects design rules and open problems.
The appendices are ordered from the conceptual to the computational. Appendix~\ref{app:dict} collects the dictionary between twisted
compactification and quotient codes and writes known codes and gates in this form. Appendix~\ref{app:tech} collects the lattice
conventions, auxiliary lemmas, side remarks and worked cochain calculations, among them three statements the main results rest on: the
local cochain lemma, the equivariant $p+ip$ pump and the absence of finitely supported logical operators in the fermionic toric code. Appendix~\ref{app:proofs} contains the longer proofs, Appendix~\ref{sec:cert} the certificates that a code is not
local-Clifford CSS or not a Pauli stabilizer code, and Appendix~\ref{app:checks} documents the computations, with inputs, methods and
outputs.
\section{Field theory of twisted compactification}
\label{sec:ft}

This section describes twisted compactification in field theory, and Sections~\ref{sec:lattice}--\ref{sec:sym} carry out the same
construction on the lattice. The results all come from one picture. A twist is a codimension-one wall
wrapped on the internal space. Cutting space open along the wall shows that the fields on its two sides are identified by the symmetry,
which is the twisted boundary condition. Evaluating one partition function in two ways (Figure~\ref{fig:trace}) then counts the states of the
compactified theory as the trace of the wall operator. Sections~\ref{sec:walls}--\ref{sec:cupft} apply this picture to one wall, to a network
of walls (a flat bundle), to defects of higher codimension (higher-form backgrounds) and to walls that carry a gauged SPT phase (cup-polynomial
defects). Section~\ref{sec:gauging} contrasts twisting with gauging and argues that twisting abelian gauge theories reaches only anyon
theories of nilpotency class at most two, and Section~\ref{sec:symft} turns to the symmetries of the compactified theory.

\subsection{Walls and the compactification rule}
\label{sec:walls}

\begin{figure}[tbp]
\centering
\begin{tikzpicture}[>=Stealth]
\begin{scope}
\fill[blue!8] (0,0) rectangle (2.2,2.4);
\fill[orange!12] (4.0,0) rectangle (6.2,2.4);
\fill[gray!30] (2.2,0) rectangle (4.0,2.4);
\draw[thick] (2.2,0) -- (2.2,2.4); \draw[thick] (4.0,0) -- (4.0,2.4);
\node at (1.1,1.5) {$a_L$};
\node[font=\small] at (1.1,0.95) {$G_L$ field};
\node at (5.1,1.5) {$a_R$};
\node[font=\small] at (5.1,0.95) {$G_R$ field};
\node[font=\small] at (3.1,1.55) {$\omega(a_L,a_R)$};
\node[font=\small,align=center] at (3.1,0.8) {$(a_L,a_R)$\\$\in K$};
\node[font=\small] at (3.1,2.65) {thickened wall};
\draw[->] (0,-0.3) -- (6.2,-0.3) node[midway,below] {\small transverse direction};
\node[font=\small,align=center] at (3.1,-1.35) {(a) a wall between two gauge theories};
\end{scope}
\begin{scope}[xshift=9.6cm,yshift=1.2cm]
\fill[violet!8, even odd rule] (0,0) circle (1.6) (0,0) circle (0.85);
\fill[gray!30] (70:0.85) arc (70:110:0.85) -- (110:1.6) arc (110:70:1.6) -- cycle;
\draw (0,0) circle (1.6); \draw (0,0) circle (0.85);
\draw[thick] (70:0.85) -- (70:1.6); \draw[thick] (110:0.85) -- (110:1.6);
\node at (90:1.22) {\small$\omega$};
\node at (132:1.22) {\small$a_L$};
\node at (48:1.22) {\small$a_R$};
\node at (0,0) {\small$a_L=a_R$};
\draw[->] (40:1.85) arc (40:-220:1.85);
\node[font=\small,align=center] at (0,-2.55) {(b) periodic transverse direction:\\$K'=K\cap G^{\rm diag}$, action $\iota^*\omega$};
\end{scope}
\end{tikzpicture}
\caption{The compactification rule. (a) A thickened wall between two gauge theories: the fields $a_L$ and $a_R$ extend into the wall, where
they are constrained to $(a_L,a_R)\in K$ and weighted by the topological action $\omega$. (b) When the transverse direction is a circle, the
region to the right of the wall continues around the circle (arrow) to the region on its left, so $a_L$ and $a_R$ are the same field. The modes
constant along the circle obey $a_L=a_R$, and the compactified theory is the $K'=K\cap G^{\rm diag}$ gauge theory with action $\iota^*\omega$, together
with the fields that wrap the circle. For the electromagnetic wall, $K=G\times G$ and $\omega=\pi\int a_L\cup a_R$ give $K'=G$ with action
$\pi\int a\cup a$ (\hyperref[ex:emfour]{example at the end of Section~\ref*{sec:general}}).}
\label{fig:wall}
\end{figure}
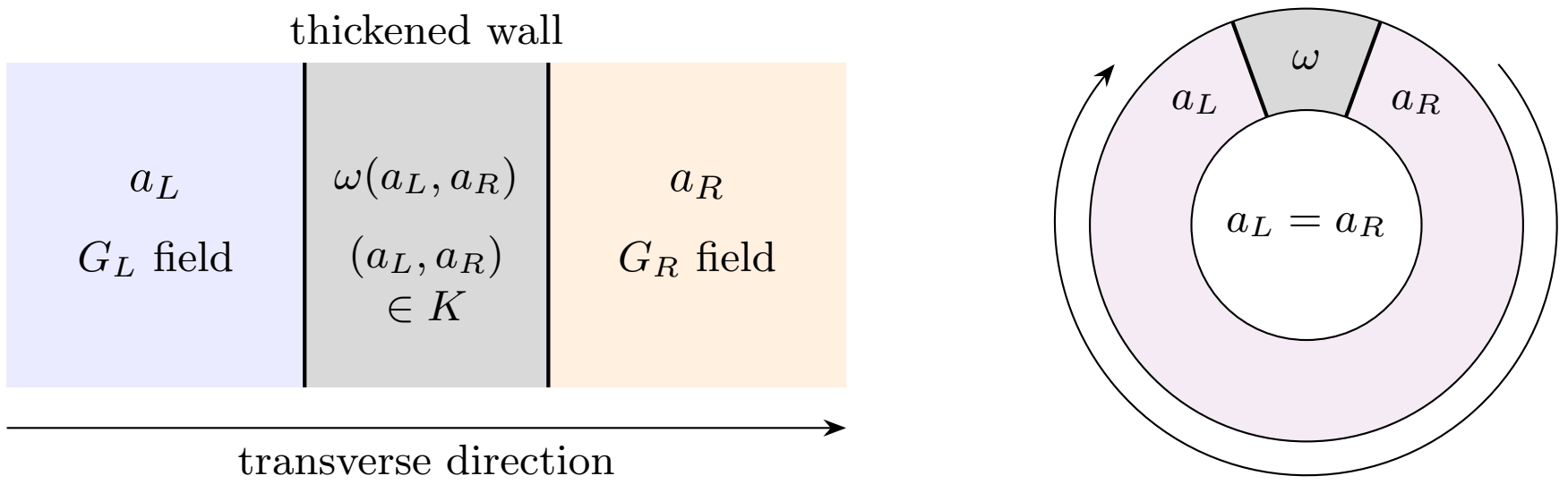

The basic twist is a single wall: a codimension-one defect across which every field is acted on by a symmetry. In a gauge theory with a
finite gauge group every gapped wall has a simple description in terms of the gauge fields on its two sides, and compactifying on a circle
with one wall identifies the two sides (Figure~\ref{fig:wall}). This gives the rule for the compactified theory, and evaluating one
partition function in two ways gives its number of states.

Let $G$ be a finite group and consider $G$ gauge theory in $D$ space dimensions, possibly with a Dijkgraaf--Witten
action~\cite{Dijkgraaf:1989hb}. A gapped codimension-one defect between two copies of the theory is labeled by a subgroup $K\subset
G_L\times G_R$, which constrains the gauge fields restricted to the wall to $(a_L,a_R)\in K$, and a topological action $\omega\in
H^D(K,U(1))$ on the wall~\cite{Cordova:2024jlk} (Figure~\ref{fig:wall}(a)). The construction behind this is gauging on the wall~\cite{Roumpedakis:2022aik}. For an
$n$-form gauge theory with finite abelian $G$ the constraint becomes $(a_L,a_R)\in Z^n(W;K)$ and $\omega\in H^D(B^nK;U(1))$. Three families
are used repeatedly:
\begin{itemize}[leftmargin=1.5em]
\item \emph{automorphism walls}, $K=\{(a,\alpha(a))\}$ for $\alpha\in\mathrm{Aut}(G)$ and $\omega=0$, such as charge conjugation; for $\omega\neq0$ an automorphism with $\alpha^*\omega=\omega+\delta\lambda$ is a symmetry only after it is dressed by the gauged SPT operator $e^{i\int\lambda}$, and one that does not preserve $[\omega]$ becomes non-invertible~\cite{HsinKobayashi2025aut};
\item \emph{SPT walls}, $K=G^{\rm diag}$ and $\omega\neq0$;
\item \emph{condensation walls}, for example $K=H_L\times H_R$, on which charges trivial on $H$ condense.
\end{itemize}
Invertibility cannot be read off $K$ alone: the electromagnetic (EM) wall of the $\Z_2$ $n$-form theory in $2n$ dimensions has $K=G\times
G$ and is invertible because $\omega=\pi\int a_L\cup a_R$ is non-degenerate. For $n=2$, compactifying on a circle with this wall gives the 3D
fermionic toric code (\hyperref[ex:emfour]{example at the end of Section~\ref*{sec:general}}).

Twisted compactification with the wall means that $a_L=a_R$ after going around the circle (Figure~\ref{fig:wall}(b)), so the gauge fields
along the non-compact directions are projected to $K'=K\cap G^{\rm diag}$. With $\iota:K'\to K$ the inclusion, the compactified theory is the $K'$ gauge theory with
action $\iota^*\omega$, together with a sector of fields wrapping the circle. For an automorphism wall $u$ of $\Z_N$ the two sectors are
\begin{equation}
H_1(S^1;\Z_N^u)=\ker(u-1),\qquad H_0(S^1;\Z_N^u)=\coker(u-1),
\label{eq:circlecoeff}
\end{equation}
so the compactified theory is the $\ker(u-1)$ $n$-form gauge theory plus the $\coker(u-1)$ $(n-1)$-form gauge theory. We state the rule for
abelian $G$. For non-abelian $G$ and a fixed holonomy the results of Section~\ref{sec:flatlat} hold for any $G$; the sum
over holonomy sectors labeled by conjugacy classes is the gauged version (Section~\ref{sec:gauging}), and we do not develop it.

Let $W$ be a wall and write $\mathcal H_W(Y)$ for the Hilbert space of the parent on $Y\times S^1$ with $W$ inserted along $Y\times\{pt\}$: this is
the Hilbert space of the compactified theory on $Y$. Write $\hat W$ for the operator of $W$ when it fills a whole spatial slice, here
$Y\times S^1$, at one instant of time.

\begin{claim}[Trace formula]\label{prop:trace}
Suppose the parent realizes a bosonic unitary TQFT and $W$ is invertible, and let $\hat W$ be the operator that the TQFT assigns to $W$ on the
slice $Y\times S^1$, with the counterterms of the wall and the framing induced from $Y\times S^1\times S^1$. Then
\[
\dim\mathcal H_W(Y)=\TrW_{\mathcal H(Y\times S^1)}\hat W .
\]
If an operator $U=e^{i\theta}\hat W$ is known only up to a phase, for instance the logical operator that a lattice symmetry of the code on
$Y\times S^1$ induces on its code space, then $\dim\mathcal H_W(Y)=|\TrW U|$, and this does not depend on the orientation of $W$.
\end{claim}

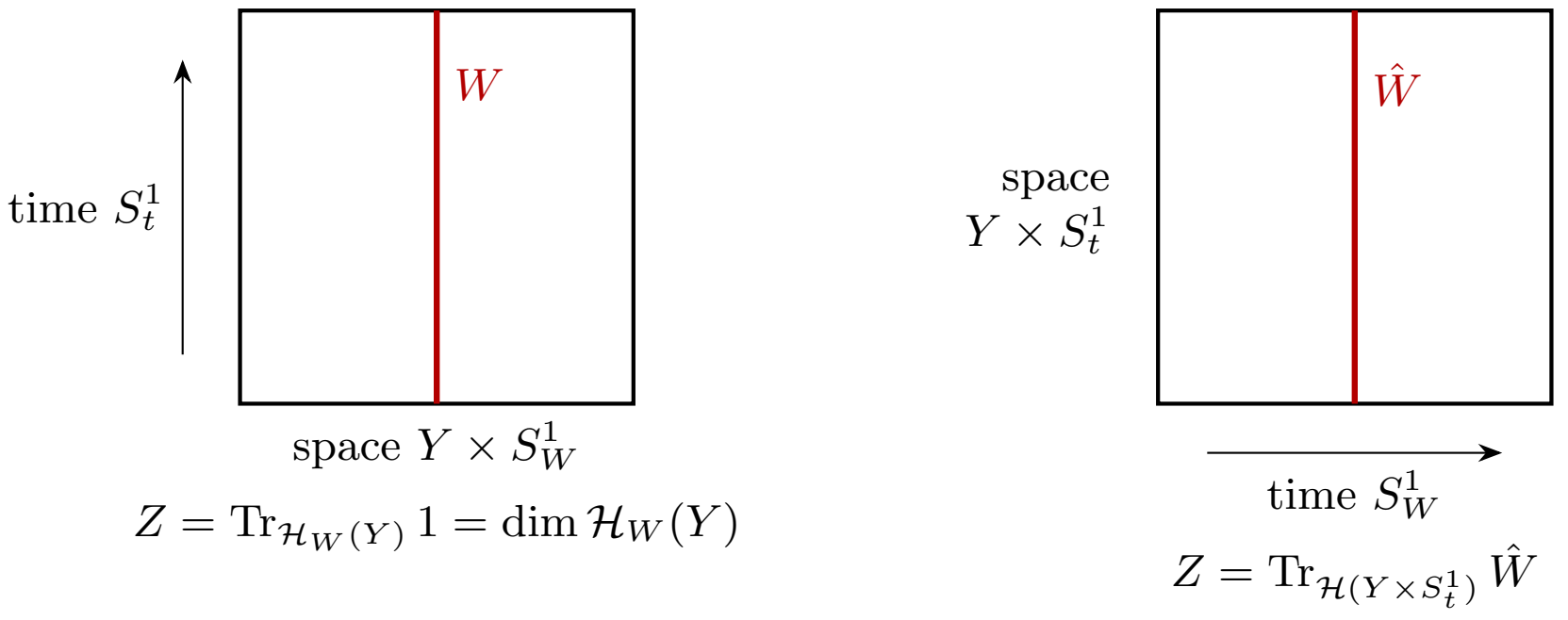
\begin{figure}[ht]
\centering
\begin{tikzpicture}[scale=1.15,>=Stealth]
\begin{scope}
\draw[thick] (0,0) rectangle (2.4,2.4);
\draw[very thick,red!70!black] (1.2,0) -- (1.2,2.4);
\node[red!70!black,right] at (1.2,1.95) {\small $W$};
\draw[->] (-0.35,0.3) -- (-0.35,2.1) node[midway,left] {\small time $S^1_t$};
\node[below] at (1.2,0) {\small space $Y\times S^1_W$};
\node at (1.2,-0.75) {\small $Z=\TrW_{\mathcal H_W(Y)}\,1=\dim\mathcal H_W(Y)$};
\end{scope}
\begin{scope}[xshift=5.6cm]
\draw[thick] (0,0) rectangle (2.4,2.4);
\draw[very thick,red!70!black] (1.2,0) -- (1.2,2.4);
\node[red!70!black,right] at (1.2,1.95) {\small $\hat W$};
\draw[->] (0.3,-0.3) -- (2.1,-0.3) node[midway,below] {\small time $S^1_W$};
\node[left] at (0,1.2) {\small \begin{tabular}{r}space\\ $Y\times S^1_t$\end{tabular}};
\node at (1.2,-1.05) {\small $Z=\TrW_{\mathcal H(Y\times S^1_t)}\,\hat W$};
\end{scope}
\end{tikzpicture}
\caption{The two channels of the partition function on $Y\times S^1_W\times S^1_t$ (the factor $Y$ is suppressed). The wall (red) sits at a
point of $S^1_W$ and extends along $Y\times S^1_t$. Left: with $S^1_t$ as time it is a defect in space, and the path integral counts states of
$\mathcal H_W(Y)$. Right: with $S^1_W$ as time it fills the spatial slice $Y\times S^1_t$ at one instant, and the path integral is the trace of
$\hat W$ on the untwisted Hilbert space.}
\label{fig:trace}
\end{figure}

\begin{proof}
Evaluate the partition function on $Y\times S^1_W\times S^1_t$ with the wall along $Y\times\{pt\}\times S^1_t$ in the two ways of
Figure~\ref{fig:trace}. Taking $S^1_t$ as time, space is $Y\times S^1_W$ with the defect at a point of $S^1_W$, so the partition function is
$\TrW_{\mathcal H_W(Y)}1=\dim\mathcal H_W(Y)$, the Hamiltonian being zero. Taking $S^1_W$ as time, space is $Y\times S^1_t$ and the wall is an
operator inserted at one time, so the partition function is $\TrW_{\mathcal H(Y\times S^1_t)}\hat W$. Both are evaluations of the same
partition function by the gluing axioms, so they agree exactly when $\hat W$ is the operator of the TQFT, which fixes its phase; the trace is
then a positive integer. An operator fixed by another convention, such as a lattice symmetry, can differ from it by a phase, from local
counterterms on the wall or, in a theory with framing dependence, from the framing, and the absolute value removes that phase. Reversing the
orientation of $W$, or of $S^1_W$, replaces $\hat W$ by $\hat W^{-1}=\hat W^\dagger$, and $\TrW\hat W^\dagger=\overline{\TrW\hat W}$ has the same absolute value.
\end{proof}

\paragraph{Variants.} The trace formula has analogues for spin theories and for non-invertible walls, and for a Clifford wall it
becomes a character of the Weil representation (Appendix~\ref{app:tech-ft}). The codes below use only invertible walls.

\begin{exampleu}[A charge-conjugation wall]
Let the parent be the $\Z_3$ toric code in two dimensions, the $\Z_3$ gauge theory with $n=1$, and let $W$ be the charge-conjugation wall
$u=-1$. Since 2 is invertible mod 3, both sectors in~\eqref{eq:circlecoeff} vanish, $\ker(u-1)=\coker(u-1)=0$, so the compactified theory on a
circle $Y$ is trivial: $\dim\mathcal H_W(S^1)=1$, and no logical qutrit survives. The trace formula gives the same answer. The nine ground
states of the toric code on the torus $Y\times S^1$ are labeled by the anyons $e^am^b$, $a,b\in\Z_3$, the wall operator $\hat W$ sends the label
$(a,b)$ to $(-a,-b)$, and only $(0,0)$ is fixed, so $|\TrW\hat W|=1$. On the lattice the twist is the quotient of the toric code on
$\Z_{2L}\times\Z_{L'}$ by the translation by $L$ composed with the charge conjugation $X\mapsto X^{-1}$, $Z\mapsto Z^{-1}$ on every qutrit, and
Corollary~\ref{cor:ss}(iii) gives $k'=0$, because $H_*(S^1;\FF_3^-)=0$. For $\Z_2$ the charge conjugation is trivial, and the simplest
non-trivial wall is the electromagnetic duality of the running example, with $|\TrW\hat W|=2$ (Example~\ref{ex:run3}).
\end{exampleu}

\subsection{Flat bundles and mapping classes}
\label{sec:flat}

Instead of a circle one can compactify on a complex $F$ with walls along a network of codimension-one loci. A topological wall can be
moved without changing the theory, so only the symmetry picked up around each closed loop of $F$ matters. Such a configuration is a flat
bundle with fibre the theory: a homomorphism $\rho:\pi_1(F)\to\mathcal S$ into the symmetries, internal or geometric, of the theory on
the remaining space $X$. For permutation-type twists with field coefficients the compactified count is homology with local coefficients,
\begin{equation}
\dim\mathcal H_\rho(X)=\prod_j\big|H_{n-j}(X)\otimes H_j(F;\Z_N^\rho)\big|,
\label{eq:localcoeff}
\end{equation}
which for $F=S^1$ is~\eqref{eq:circlecoeff}. The mapping class group of the slice gives a second kind of twist: the gluing relates $a_L$
and $a_R$ by a diffeomorphism, $a_R=f^*a_L$, and the compactified space is the mapping torus $T_f$. Its count follows from the Wang
sequence,
\begin{equation}
k'=\dim\ker(f_*-1|_{H_n(Y;\FF_p)})+\dim\ker(f_*-1|_{H_{n-1}(Y;\FF_p)}).
\label{eq:wang}
\end{equation}
For example, the rotation of the square torus $Y=T^2$ by $90^\circ$ acts on $H_1(Y;\FF_p)$ by $\left(\begin{smallmatrix}0&-1\\1&0\end{smallmatrix}\right)$
and trivially on $H_0$. So the 3D toric code ($n=1$) on its mapping torus, the quarter-turn space $T^3/\Z_4$, has $k'=1+1=2$ for $p=2$ and
$k'=0+1=1$ for odd $p$; these are the dimensions of $H_1(T^3/\Z_4;\FF_p)$, since $H_1(T^3/\Z_4;\Z)=\Z\oplus\Z_2$.
Section~\ref{sec:method} explains which mapping classes give quotient codes.

\subsection{Higher-form backgrounds}
\label{sec:hfft}

For a $q$-form symmetry with abelian group $A$, defects have codimension $q+1$, and the twist datum is a class $B\in H^{q+1}(F;A)$,
Poincar\'e dual to the defect locus. For $q=0$, $H^1(F;A)=\Hom(\pi_1F,A)$ recovers abelian flat bundles. The native higher-form symmetries
of an $n$-form toric code are its Wilson and 't~Hooft operators, which are Pauli; their backgrounds only change signs
(Corollary~\ref{cor:paulinogo}). The useful ones are cup-product symmetries
\begin{equation}
U_Q(\Sigma)=\exp\Big(\tfrac{2\pi i}{p}\int_\Sigma Q(a)\Big),\qquad \dim\Sigma=\deg Q=D-q,
\end{equation}
which are diagonal Clifford circuits of $CZ$ or $S$ type. The coupling to a background is $S_B=\frac{2\pi}{p}\int Q(a)\cup B$. Expanding
$a=\sum_ra^{(r)}\otimes\omega_r$ in a cohomology basis of $F$ gives the reduced action
\begin{equation}
S_B=\tfrac{2\pi}{p}\sum_{r,s}M^B_{rs}\int_XQ(a^{(r)},a^{(s)}),\qquad M^B_{rs}=\int_F\omega_r\cup\omega_s\cup B .
\label{eq:MB}
\end{equation}
The monodromy of the circle case is replaced by a bilinear pairing on $H^*(F)$. On $T^{q+1}$ with $B$ the fundamental class only the
unwrapped sector couples; on a ball $D^{q+1}$ with relative $B$ only $\int_XQ(a)$ survives, for every $q$.

The simplest case is two toric codes in two dimensions with $Q=a_1a_2$, a circuit of $CZ$ gates, and $F=S^1$ with $B$ its generator, so $q=0$
and the defect is a line of $CZ$ gates. Only the unwrapped sector couples, and~\eqref{eq:MB} is the Dijkgraaf--Witten twist $\pi\int a_1a_2$
of the $\Z_2\times\Z_2$ gauge theory on $X$ and time. With $X$ a circle, its Hilbert space is one-dimensional instead of four-dimensional, while the
wrapped sector keeps its four states, so the twist removes two of the four logical qubits. The lattice code is the code with a $CZ$ wall of
Section~\ref{sec:gausslaw}, which has $k'=2$.

\subsection{Cup-polynomial defects}
\label{sec:cupft}

Every gauged SPT of a stack of $\Z_2$ gauge fields gives a topological operator $U_\omega(\Sigma)=(-1)^{\int_\Sigma\omega(a_1,\dots,a_r)}$. Its Clifford
level is at most the degree of $\omega$: the cluster-state term $a_1a_2$ is at level at most 2, the type-I--III terms $a^3$, $a_1^2a_2$,
$a_1a_2a_3$ at level at most 3, and the type-IV term $a_1a_2a_3a_4$ at level at most 4. The level can be lower when the phase polynomial
collapses on the lattice; for a $\Z_2$ 0-form field, for instance, $a\cup a\cup a=a$ as functions.

\begin{claim}[The gauge theory of a cup defect]\label{prop:ext}
For $\omega=\sum_kb_k\cup Q_k(a)$ with $a_i$ 1-form gauge fields and $b_k$ further gauge fields, the reduced theory is the gauge theory of the
central extension $1\to\Z_2^s\to G\to\Z_2^r\to1$ with extension classes $[Q_k]$.
\end{claim}
\begin{proof}
Write $G$ gauge fields as pairs $(a,c)$ with $\delta c_k=Q_k(a)$, impose the constraint with Lagrange multipliers $b_k$, and integrate out $c_k$.
\end{proof}

For example $a_1a_2\mapsto D_4$, $a_1^2+a_1a_2+a_2^2\mapsto Q_8$ and $a^2\mapsto\Z_4$; the type-III theory is $D^{\omega_{\rm III}}(\Z_2^3)\simeq
D(D_4)$~\cite{deWildPropitius1995}. Here we need only that a cubic defect can produce (N2) order, that is, non-abelian topological order (gap~3 of Section~\ref{sec:intronew}), as type III does; Section~\ref{sec:hfcountsec} counts the
resulting codes on the lattice.

\subsection{Twisting versus gauging}
\label{sec:gauging}

In twisted compactification the background is fixed. In gauging it is summed over, which makes it a dynamical gauge field. Summing over
non-abelian $G$-bundles on $X$ gives quantum-double codes with ground space $\CC[\Hom(\pi_1X,G)/G]$. This is the other route to (N2) order,
and it is not a twist: the background is summed over, not fixed.

The two routes are separated by an invariant of the anyon theory. For a fusion category $\mathcal B$ let $\mathcal B_{\rm ad}$ be the fusion subcategory
generated by the products $x\otimes\bar x$ of its simple objects. $\mathcal B$ is \emph{nilpotent of class at most $c$} if $c$ steps of
$\mathcal B\supseteq\mathcal B_{\rm ad}\supseteq(\mathcal B_{\rm ad})_{\rm ad}\supseteq\cdots$ reach $\mathrm{Vec}$~\cite{GelakiNikshych2008}. Class 1 means that every anyon is
abelian, and class at most 2 that every $x\otimes\bar x$ is a sum of abelian anyons.

\begin{proposition}[Twisted abelian doubles have class at most two]\label{prop:nilp}
\begin{enumerate}[label=(\alph*)]
\item Every twisted abelian double $D^\omega(A)$ is nilpotent of class at most 2.
\item Class at most 2 is kept under stacking, under passing to a fusion subcategory, and under condensation, both for the category of modules
and for its local modules.
\item For a finite group $G$ the following are equivalent: $D(G)$ has class at most 2; $[G,G]\subseteq Z(G)$; $D(G)$ is braided equivalent to a
twisted abelian double.
\end{enumerate}
\end{proposition}
\idea A twisted abelian double is graded by flux, and $x\otimes\bar x$ has trivial flux, where the anyons are pure charges, all abelian. Stacking,
restriction and condensation are tensor functors or subcategories, so they cannot make $x\otimes\bar x$ non-abelian. In $D(G)$ the charges alone
detect whether $G/Z(G)$ is abelian, since $\mathrm{Rep}(G)_{\rm ad}=\mathrm{Rep}(G/Z(G))$. Conversely, when $G/Z(G)$ is abelian, the anyons with central flux and a
charge pulled back from $G/Z(G)$ are abelian bosons that form a Lagrangian subcategory, and that makes $D(G)$ a twisted abelian double.
\movedproof{prop:nilp}

Proposition~\ref{prop:nilp} is about anyon theories. Its consequence for twisted compactification is the following, for the twists it
covers.

\begin{claim}[The reach of twisting]\label{claim:reach}
Let the parent be a stack of abelian gauge theories, $\Z_p$ gauge fields of any form degrees with any Dijkgraaf--Witten twist, such as
copies of toric codes, on $X\times F$ with $\dim X=2$. Compactify it on a closed $F$ with a background for any combination of: a holonomy
acting by automorphisms of the gauge group, such as permutations of copies and charge conjugation; a class for a higher-form symmetry; a
cup-polynomial defect. Then in each sector of the wrapped fields the anyon theory on $X$ is a twisted abelian double. So by
Proposition~\ref{prop:nilp} it has nilpotency class at most 2, as does everything obtained from it by stacking and condensation, and a quantum
double $D(G)$ arises in this way only when $[G,G]\subseteq Z(G)$.
\end{claim}
\begin{proof}
Fields wrapped on $F$ become fields of lower degree on $X$, which either label sectors or are again abelian gauge fields. A holonomy by
automorphisms keeps the gauge group abelian, replacing it by invariants and coinvariants as in~\eqref{eq:circlecoeff}. A higher-form
background couples to the reduced fields through the pairing~\eqref{eq:MB}, and a cup-polynomial wall adds a Dijkgraaf--Witten twist
(Proposition~\ref{prop:ext}). So each sector is an abelian gauge theory with a twist, whose anyon theory is a twisted abelian double, and
Proposition~\ref{prop:nilp}(b) covers stacking and condensation.
\end{proof}

Proposition~\ref{claim:reach} does not cover holonomies by dualities, which act on the anyon theory but not on the gauge group, nor non-invertible
backgrounds; we expect the same conclusion for the former but have not shown it. Proposition~\ref{prop:ext} produces $D(G)$ for every central
extension $G$ of $\Z_2^r$ by $\Z_2^s$, such as $D_4$ and $Q_8$. Warman and
Sch\"afer-Nameki realize the logical gate $T^{1/N}=\mathrm{diag}(1,e^{i\pi/(4N)})$ on $D(D_{4N})$, with $D_{4N}$ the dihedral group of order $8N$, and for
$8N=2^n$ a gate at level $n$ of the Clifford hierarchy with $n$ qubits per edge~\cite{WSN2025}. For $N=1$ this is the gate of
Example~\ref{ex:kzhbdry}. For $N\ge2$ the commutator subgroup $\langle r^2\rangle$ has order $2N$ and is not central, so these codes lie beyond
the twists of Proposition~\ref{claim:reach}. They come from gauging: $D(D_{4N})$ is the $\Z_{4N}$ toric code with its charge conjugation gauged, and
already this one step goes past class 2. So class 2 is where the two routes separate.

\subsection{Symmetries of the compactified theory}
\label{sec:symft}

Symmetries of the theory on $X$ come from isometries of $F$, mapping classes of $F$ and internal symmetries, provided they preserve the
background~\cite{Gukov:2020btk,GukovHsinPei2025}. Isometries are symmetries of the full theory; a mapping class with no isometric
representative, such as a Dehn twist on $T^2$, is a symmetry of a subsector only. $\mathrm{MCG}(F)$ acts on the operators wrapped on $F$
through $\mathrm{Aut}(H^*(F;D),\text{pairing})$, projectively, with the phase fixed by a quadratic refinement. The identity component of
$\mathrm{Iso}(F)$ acts trivially on $H^*(F;D)$ and combines with the wrapped symmetries into higher groups. The wrapped operators'
pairing is a mixed anomaly of the compactified symmetries, and a polarization (a maximal isotropic subgroup $L$) gives an absolute theory
with symmetry $L^\vee$. Section~\ref{sec:sym} builds these symmetries on the codes: the quotient codes are the relative theory, the higher groups
give only Clifford gates, and in the constructions studied here non-Clifford gates come from a symmetry that pumps a cubic response, the
SPT-decorated part of the higher group~\cite{Barkeshli:2022edm}.

\section{Quotient code construction as twisted compactification on the lattice}
\label{sec:lattice}

This section carries out steps 1--3 of the method of Section~\ref{sec:introtwo} on the lattice, for every background in the paper.
Section~\ref{sec:method} states the principle in lattice terms, explains each step, and lists in Table~\ref{tab:method} which construction applies
to which background. The remaining subsections prove the entries of the table, from the simplest identification to the most general one.
Each subsection starts with the idea and the simplest example and then states the result. Lattice conventions, auxiliary lemmas and
side remarks are collected in Appendix~\ref{app:tech-lattice}, and the longer proofs in Appendix~\ref{app:proofs}.

\subsection{The principle and the method}
\label{sec:method}

\paragraph{The principle on the lattice.} Let $\tilde F\to F$ be a finite regular cover with deck group $\Gamma$ on which the background becomes
trivial; $\Gamma$ consists of the deck transformations, the maps of $\tilde F$ that permute the points over each point of $F$ (Section~\ref{sec:introtwo}), and $F=\tilde F/\Gamma$. A field configuration on $X\times F$ is then a configuration on $X\times\tilde F$ that obeys the twisted boundary condition: moving from a
cell to its image under a deck transformation $g$ acts on the fields by the symmetry $U_g$ that the background assigns to $g$. On operators,
the boundary condition identifies an operator $O$ near a cell $\tilde\sigma$ with $U_gOU_g^\dagger$ near $g\tilde\sigma$. For the parent code on $X\times\tilde F$,
imposing this identification on every stabilizer is the quotient by the group $\Gamma'=\{g\circ U_g\}$. Every construction in this section is this one
identification, made precise for one class of symmetries $U_g$.
\begin{itemize}[leftmargin=1.5em]
\item For a holonomy $\rho:\pi_1(F)\to G$, $U_g=\rho(g)$ acts on the fibre (Definition~\ref{def:twisted}).
\item For a $q$-form class or a cup-product defect with $\pi^*B=\delta\tilde\eta$ on the cover, $U_g=U(\zeta_g)$ is the diagonal circuit of the cup-product
symmetry on the cocycle $\zeta_g=\tilde\eta-g^*\tilde\eta$ (Theorem~\ref{thm:dressedquot}(c)).
\end{itemize}

\paragraph{Step 1: lift.} The method needs a finite cover on which the background is trivial. For a holonomy this is the cover with deck
group $\rho(\pi_1F)$, which is finite exactly when the holonomy has finite image. Finite-order mapping classes of surfaces are of this kind
by Nielsen realization. Infinite-order ones, such as the Anosov map of $T^2$, give codes on mapping tori that the method does not produce
from a finite product code, since no finite cover trivializes the holonomy. For a non-zero class $[B]\in H^{q+1}(F;\Z_p)$ a trivializing cover has degree divisible by $p$, and it need not exist: a non-zero class on $S^2$
survives every finite cover (Remark~\ref{rem:cover}). In that case the dressed Gauss laws define the code on $X\times F$ itself, with no
cover.

\paragraph{Step 2: identify.} The form of the identification depends on how $U_g$ acts.
\begin{enumerate}[leftmargin=2em,label=(\alph*)]
\item If $U_g$ is a monomial Clifford, that is, a permutation of qudits composed with single-qudit Cliffords, the identification is the monomial
quotient of Definition~\ref{def:quot} (Section~\ref{sec:setup}). A flat background acting on a CSS code by chain maps gives a balanced product,
which is such a quotient when the action is monomial on the chosen bases (Section~\ref{sec:flatlat}).
\item If $U_g$ is an entangling diagonal circuit, as for higher-form and cup-product backgrounds, the identification conjugates stabilizers by
a circuit, and the result is the dressed quotient of Definition~\ref{def:descent} (Section~\ref{sec:dressedquot}). For quadratic defects it is again a
monomial quotient, of the dressed parent $\mathcal C_{\tilde B}$ on the cover.
\item If no trivializing cover exists, the dressed Gauss laws of Definition~\ref{def:gauss} define the code directly (Section~\ref{sec:higher}).
\end{enumerate}
Condensation walls are not of this kind: they add stabilizers along the wall rather than identify them, and we do not treat them.
In this paper ``quotient'' always means one of the identifications (a) and (b).

\paragraph{Step 3: compute.} The identification turns data of the parent into cohomology.
\begin{itemize}[leftmargin=1.5em]
\item \emph{Logical operators.} For a monomial quotient they are the parent logical classes with invariant representatives, extended by the
Tate group $\hat H^0(\Gamma;S)$ of winding sectors (Theorem~\ref{thm:general}). So the design rate is kept when $\Gamma$ acts freely on the checks, and winding sectors can appear only
when $p$ divides $|\Gamma|$.
\item \emph{Flat backgrounds.} The logicals are the homology of $F$ with local coefficients in the logicals of the fibre, up to the
differentials of a spectral sequence (Theorem~\ref{thm:ss}).
\item \emph{Higher-form backgrounds.} The twist removes the logical classes that pair with $B$. The count is $\dim\ker\cup[B]$ in the quadratic case
(Theorem~\ref{thm:hfcount}, Proposition~\ref{prop:cone}), and the number of zeros of a polynomial map in the cubic case (Theorem~\ref{thm:pointcount}).
\item \emph{Distance and level.} Distances are bounded by lifting and descending logical operators (Theorem~\ref{thm:general}(iii)--(iv)). The
Clifford level of the dressings of the checks is bounded by the degree of the twist operator (Propositions~\ref{prop:clifflevel} and~\ref{prop:chslevel}).
\end{itemize}
Step 4, the descent of symmetries to logical gates, is Section~\ref{sec:sym}.

\paragraph{Four operations.} The words quotient and descent cover four operations, which the box separates. The first two are step 2
of the method, and the last two are used in Section~\ref{sec:sym}.
\begin{mainbox}
\begin{enumerate}[leftmargin=1.5em,label=\arabic*.]
\item \emph{Monomial quotient} (Definition~\ref{def:quot}): a Pauli stabilizer parent and a free action by qudit permutations with
single-qudit Cliffords give the Pauli stabilizer code $\pi(S)$, with one qudit for each orbit; $\pi$ is a linear map of Pauli vectors.
\item \emph{Dressed quotient} (Definition~\ref{def:descent}): generators of the parent, each an $X$-string times a diagonal phase, permuted by a
free action, are replaced by their descents. This is not a homomorphism of operator algebras, so it is defined for a code with a generating
family; for the dressed Gauss laws it gives the dressed Gauss-law code (Theorem~\ref{thm:dressedquot}).
\item \emph{Descent of a circuit} (Propositions~\ref{prop:descent} and~\ref{prop:cupgate}): a parent symmetry given by a local rule that commutes
with the action is built on the quotient by the same rule, with one local factor for each orbit, not evaluated on invariant
configurations. Monomial symmetries give one layer of wire relabeling and single-qudit Cliffords, not always geometrically local (Section~\ref{sec:descent});
cup-product symmetries give circuits of bounded depth and range.
\item \emph{Descent of a pump} (Proposition~\ref{prop:gaussdescent}): an equivariant single-particle evolution on the cover, for each spin
structure of the cover that descends, is restricted to the sectors of the compatible spin structures of the quotient. The result is a
finite-time quasi-local evolution, within $\varepsilon$ of constant-depth circuits with gates of range $O(\log(n/\varepsilon))$.
\end{enumerate}
In none of them is the quotient code the deck-invariant subspace of the parent Hilbert space (Section~\ref{sec:introcorr}).
\end{mainbox}

\begin{table}[ht]
\centering\small
\begin{tabularx}{\textwidth}{@{}>{\raggedright}p{3.0cm}>{\raggedright}p{3.0cm}>{\raggedright}p{3.0cm}>{\raggedright}X>{\raggedright\arraybackslash}p{2.6cm}@{}}
\toprule
Background on $F$ & Symmetry $U_g$ & Identification & Output & Where \\
\midrule
holonomy, monomial on the bases & permutation with single-qudit Cliffords & monomial quotient & Pauli stabilizer code & Def.~\ref{def:quot}, Thms.~\ref{thm:general}, \ref{thm:balanced} \\
holonomy by chain maps & chain automorphism & balanced product & CSS code & Thms.~\ref{thm:balanced}(a), \ref{thm:ss} \\
duality holonomy & $ZX$-duality with a permutation & monomial quotient that changes letters & Pauli stabilizer code, need not be CSS & Thms.~\ref{thm:factor}, \ref{thm:transfer} \\
$q$-form class, quadratic defect & $CZ$-type circuit $U(\zeta_g)$ & monomial quotient of the dressed parent & Pauli stabilizer code & Thms.~\ref{thm:dressedquot}(d), \ref{thm:hfcount} \\
cubic or higher cup defect & $CCZ$-type circuit $U(\zeta_g)$ & dressed quotient & commuting-projector code, dressings of level at most $\deg\omega-1$ & Thms.~\ref{thm:dressedquot}, \ref{thm:pointcount}; Prop.~\ref{prop:clifflevel} \\
class with no trivializing cover & none & dressed Gauss laws on $X\times F$ & as in the two rows above & Def.~\ref{def:gauss}, Prop.~\ref{prop:twisted-code} \\
any background, Clifford-hierarchy parent of level $\ell$ & diagonal gate of level $m$ with a $CNOT$ circuit & descent of $X$-monomial checks & dressings of level at most $\max(\ell,m-1)$ & Prop.~\ref{prop:chslevel} \\
\bottomrule
\end{tabularx}
\caption{Step 2 of the method: which identification applies to which background, and what it produces.}
\label{tab:method}
\end{table}

\subsection{Monomial quotients}
\label{sec:setup}

This subsection gives step 2 in its simplest form, in which the symmetry $U_g$ acts on qudits by a permutation composed with
single-qudit Cliffords. We fix conventions for Pauli stabilizer codes, define the quotient, and show the identification of stabilizers on the
toric code.

\paragraph{Pauli stabilizer codes.} Let $p$ be prime and consider $n$ qudits of dimension $p$, with $X|j\rangle=|j+1\rangle$ and
$Z|j\rangle=\varpi^j|j\rangle$, $\varpi=e^{2\pi i/p}$. Up to phases, the Pauli operators $X(x)Z(z)$ form the $\FF_p$-vector space $\calP_n\cong\FF_p^{2n}$,
and two of them commute up to the phase $\varpi^{\langle P,P'\rangle}$ for the symplectic form $\langle(x,z),(x',z')\rangle=x\cdot z'-z\cdot x'$. A
Pauli stabilizer group is an abelian group of Pauli operators not containing non-trivial scalars; its image $S\subset\calP_n$ is an isotropic
subspace, the code space has dimension $p^k$ with $k=n-\dim S$, the logical group is $\mathcal L=S^\perp/S$, and the distance $d$ is the minimum
weight of an element of $S^\perp\setminus S$. We use the same letter for the group and its image when no phase is involved. Every chain
complex of $\FF_p$-vector spaces $C_{n+1}\xrightarrow{\partial}C_n\xrightarrow{\partial}C_{n-1}$ defines a CSS code with qudits on a basis of
$C_n$, $X$-checks $X(\delta c)$ for $c\in C^{n-1}$ and $Z$-checks $Z(\partial f)$ for $f\in C_{n+1}$; its logical group is $H^n\oplus H_n$. A
\emph{monomial Clifford} is a qudit permutation composed with single-qudit Cliffords: translations, Hadamard-type dualities $X\leftrightarrow Z$
composed with a permutation, charge conjugation, and for $p=2$ the triality $X\to Y\to Z$.

\paragraph{Composite dimensions.} All results are stated for prime $p$, and most examples have $p=2$; Appendix~\ref{app:tech-lattice} says
which of them extend to composite dimensions.

\paragraph{Parents and symmetries.} A \emph{parent} is a Pauli stabilizer group $S$, or more generally a commuting-projector code, together with a
geometry: a cell complex for topological codes, or a chain complex of free $\FF_p[\Gamma]$-modules for algebraic codes. We use three kinds
of symmetry.
\begin{enumerate}[leftmargin=1.5em]
\item \emph{0-form, monomial.} A group acting by $q\mapsto gq$ on qudits followed by an on-site unitary $c_g(q)$; a monomial-Clifford action
when the $c_g(q)$ are Clifford. Permutations preserve CSS structure, and Hadamard-type $c_g$ are ZX-dualities.
\item \emph{$q$-form.} Operators $U(\Sigma)$ on cycles $\Sigma$ of codimension $q$ that are topological on the code space. The native ones are Pauli;
the cup-product ones $U_Q(\Sigma)=\varpi^{\int_\Sigma Q(a)}$ are diagonal circuits (Section~\ref{sec:higher}).
\item \emph{Non-invertible.} Condensation walls, which add stabilizers along the wall rather than reglue space; we do not count them here.
\end{enumerate}

\begin{definition}[Quotient code]\label{def:quot}
A \emph{monomial-Clifford action} of a finite group $\Gamma$ on $S$ sends the Pauli $P$ on qudit $q$ to $c_g(q)\,P\,c_g(q)^{\dagger}$ on qudit
$gq$. It is a \emph{symmetry} if $g(S)=S$ for all $g$, and \emph{free} if no qudit is fixed by a non-identity element. Choosing orbit
representatives $r$, the quotient map $\pi:\calP_{n}\to\calP_{n/|\Gamma|}$ sends the Pauli $P$ on $q=hr$ to $c_h^{-1}P$ on $r$, and the
\emph{quotient code} is $S/\Gamma:=\pi(S)$.
\end{definition}

If an orbit returned to a qudit with a non-trivial Clifford the identification would be inconsistent; freeness excludes this. A different
choice of representatives changes $\pi$ by a product of single-qudit Cliffords on the quotient, so $S/\Gamma$ is defined up to local Clifford
equivalence.

\begin{exampleu}[Translation, reflection and duality quotients of the toric code]
Let $S$ be the toric code on the square lattice of the torus $\Z_{2L}\times\Z_{L'}$, $L,L'\ge3$, with qubits on edges, $X$-checks on stars and
$Z$-checks on plaquettes, so $n=4LL'$ and $k=2$. Take orbit representatives in the strip $0\le x<L$ (Figure~\ref{fig:principle}).
\begin{enumerate}[label=(\alph*),leftmargin=2em]
\item \emph{Translation.} $\Gamma=\Z_2$ acts by $t(x,y)=(x+L,y)$ with every $c_g=1$; this is a free symmetry. The map $\pi$ fixes every check
inside the strip and sends a check that crosses the cut $x=L$ to the check of the same shape at $x=0$: each stabilizer is identified with its
image under $t$. The quotient is the toric code on $\Z_L\times\Z_{L'}$, again with $k=2$.
\item \emph{Reflection.} $\Gamma=\Z_2$ acts by the glide reflection $g(x,y)=(x+L,-y)$, which is free and maps stars to stars and plaquettes to
plaquettes. Now a check that crosses the cut at height $y$ is identified with its reflected image at height $-y$; for instance the vertical
edge from $(L,y)$ to $(L,y+1)$ is identified with the edge from $(0,-y-1)$ to $(0,-y)$. The quotient is the toric code on a Klein bottle, with
$n=2LL'$ and $k=\dim H_1(K;\FF_2)=2$. In field theory this is the $\Z_2$ gauge theory with a reflection defect along the cut: the twisted boundary
condition $a(x+L,y)=a(x,-y)$ is exactly this identification of stabilizers (Theorem~\ref{thm:balanced}, and the example after
Definition~\ref{def:twisted}).
\item \emph{Duality.} If some $c_g$ is a non-trivial Clifford, the identification also changes the letters. In Example~\ref{ex:run1} a check
that crosses the cut is identified with its image under a translation composed with transversal $H$, so $X$-checks are glued to $Z$-checks and
the quotient, the five-qubit code, is not CSS.
\end{enumerate}
\end{exampleu}

\begin{proposition}[Validity]\label{prop:valid}
If $\Gamma$ is a free symmetry, $\pi(S)$ is isotropic and defines a Pauli stabilizer code on $n'=n/|\Gamma|$ qudits.
\end{proposition}
\begin{proof}
The induced symplectic form is $\langle\pi a,\pi b\rangle'=\sum_{g\in\Gamma}\langle a,gb\rangle$, and each term vanishes since $gb\in S$.
\end{proof}

\paragraph{Phases.} Here $\pi(S)\subset\calP_{n'}$ is a phase-free isotropic subspace: $\pi$ is a linear map of Pauli vectors, not a
multiplicative map of signed Pauli operators. A stabilizer code is obtained by choosing commuting Pauli lifts of order $p$ of an independent
generating set of $\pi(S)$; they generate a group with no non-trivial scalars. Such lifts exist because $\pi(S)$ is isotropic, and two
choices differ by a character of $\pi(S)$, which is conjugation by a Pauli operator. So the quotient code is defined up to Pauli
equivalence, and its parameters do not depend on the choice. For dressed quotients the dressed Gauss laws supply the phases
(Theorem~\ref{thm:dressedquot}).

\paragraph{Three maps.} The coinvariant projection $\pi$ sums the transported components over each orbit. The \emph{lift} of a quotient Pauli
$P$ is $P^\uparrow=\sum_{g\in\Gamma}g(\tilde P)$, where $\tilde P$ places $P$ on the orbit representatives; it is $\Gamma$-invariant and has weight
$|\Gamma|\,|P|$. The \emph{descent} $a^\downarrow$ of an invariant Pauli is its restriction to the representatives, so $(P^\uparrow)^\downarrow=P$. Descent
is not $\pi$: for $\Gamma=\Z_2$ in characteristic two, $\pi(a)=0$ for every invariant $a$. With the norm $\mathsf N=\sum_gg$, the lift of $\pi(s)$ is
$\mathsf N(s)$ and $\langle P,Q\rangle'=\langle\tilde P,Q^\uparrow\rangle$.

\begin{example}[Running example, part 1: the quotient]\label{ex:run1}
Let $\Lambda=\langle(1,2),(-2,1)\rangle\subset\Z^2$, the ideal $(1+2i)$ of the Gaussian integers, and put qubits on the 20 sites of the torus
$\Z^2/2\Lambda$. The rotated toric code has one check on each plaquette: the plaquette with lower-left corner $x$ acts on $x$, $x+e_1$, $x+e_2$,
$x+e_1+e_2$ by $X^{\otimes4}$ if $x_1+x_2$ is even and by $Z^{\otimes4}$ if it is odd. The colouring is well defined because $2\Lambda$ lies in the
even sublattice, and the parent is $\code{20,2,4}$. The group $\Gamma=\Lambda/2\Lambda\cong\Z_2^2$ acts by $g_u=T_u\circ H^{\otimes20}$ if $u$ has odd
coordinate sum and by $g_u=T_u$ otherwise, where $T_u$ is the translation by $u$. For odd $u$ the translation exchanges the two colours of
plaquettes and $H$ exchanges the letters, so every $g_u$ is a monomial-Clifford symmetry, and it is free because $u\notin2\Lambda$. The orbits
are the five classes $j=x_2-2x_1\bmod5$ of $\Z^2/\Lambda$, and a plaquette covers the four consecutive classes $j-2,\dots,j+1$. With the
representatives of odd coordinate sum, the transport is $H$ on the even sites, and the quotient checks are
\[
XZZXI,\quad IXZZX,\quad XIXZZ,\quad ZXIXZ,\quad ZZXIX,
\]
the five-qubit code $\code{5,1,3}$~\cite{Laflamme1996,Bennett1996}. It is the Wen plaquette model~\cite{Wen2003,BonillaAtaides2021} on the twisted
torus $\Z^2/\Lambda$, on which no checkerboard colouring exists.
\end{example}

\subsection{What a quotient keeps: transfer and factorization}
\label{sec:general}

This subsection carries out step 3 for monomial quotients. The principle is that every quantity of the quotient is a quantity of the parent
restricted to invariant operators, corrected by the cohomology of $\Gamma$: the norm map sends parent stabilizers to quotient stabilizers, and
its failure to be onto the invariants is the Tate group $\hat H^0(\Gamma;S)$. Concretely, an operator on the quotient is the same as a parent
operator that looks the same on every copy of the fundamental domain, its lift. A logical operator of the quotient therefore lifts either to a
parent logical, or to an invariant parent stabilizer that is not the sum of the images of a single stabilizer, such as two parallel strings
on a double cover; on the quotient the two strings become one non-contractible string. These are the winding sectors, and they form the
Tate group.

\begin{theorem}[Transfer]\label{thm:general}
Let a finite group $\Gamma$, possibly non-abelian and possibly containing dualities, act freely by monomial Cliffords preserving $S$, and put
$n'=n/|\Gamma|$.
\begin{enumerate}[label=(\roman*)]
\item $k'=n'-\dim\mathsf N(S)$. If $S$ is generated by $m'$ orbits of checks, then $k'\ge n'-m'$. If moreover $\Gamma$ acts freely on these
checks, so that the parent has $m=|\Gamma|\,m'$ of them, the design rate is preserved, $(n'-m')/n'=(n-m)/n$; a check whose support is
invariant under some $g\ne1$ has a smaller orbit, and then the design rate of the quotient is lower.
\item Quotient logicals correspond to $(S^\perp)^\Gamma/\mathsf N(S)$, and there is an exact sequence
\[
0\to\hat H^0(\Gamma;S)\to(S^\perp)^\Gamma/\mathsf N(S)\to(S^\perp)^\Gamma/S^\Gamma\to0,\qquad\hat H^0(\Gamma;S)=S^\Gamma/\mathsf N(S).
\]
The right-hand term is the space of parent logical classes that admit a $\Gamma$-invariant representative (Remark~\ref{rem:invclasses}).
\item $d'\ge\min(d/|\Gamma|,\,d_{\rm wind})$, where $d_{\rm wind}=\min\{|s|/|\Gamma|:s\in S^\Gamma\setminus\mathsf N(S)\}$. If $\hat H^0(\Gamma;S)=0$, then
$d'=d_\Gamma/|\Gamma|$, with $d_\Gamma$ the least weight of a $\Gamma$-invariant Pauli operator in $S^\perp\setminus S$, and $d'/n'\ge d/n$.
\item If some minimum-weight parent logical $\ell$ has $\mathsf N(\ell)\notin\mathsf N(S)$, then $d'\le d$. In that case the quotient raises the
relative distance by a factor of at most $|\Gamma|$. Without this hypothesis no upper bound on $d'$ in terms of $d$ is claimed.
\end{enumerate}
\end{theorem}
\idea An operator on the quotient is the same thing as a $\Gamma$-invariant operator on the parent, its lift, and a parent
stabilizer $s$ appears on the quotient as its norm $\mathsf N(s)=\sum_gg(s)$. So the quotient is computed by invariant parent operators, with
one new feature: an invariant stabilizer need not be a norm. Such a stabilizer, like two parallel strings on a double cover, is trivial on the
parent but descends to a non-trivial logical, and these form the Tate group. The distance bounds come from asking which of the two kinds a
lifted logical is.
\movedproof{thm:general}

\begin{proposition}\label{prop:count}
$k'=\frac12\big(\dim(S^\perp)^\Gamma-\dim\mathsf N(S)\big)$. If $p\nmid|\Gamma|$, then $\hat H^0(\Gamma;S)=0$ and $k'=\frac12\dim\mathcal L^\Gamma$. For
$\Gamma=\Z_2=\langle g\rangle$ and $p=2$, $k'=\frac12(k+\hat h(S))$ with $\hat h(S)=\dim\ker(1+g)|_S-\dim\Ima(1+g)|_S$.
\end{proposition}
\idea Count on the parent: the quotient stabilizers are $\mathsf N(S)$ and the quotient normalizer is $(S^\perp)^\Gamma$. When
$p\nmid|\Gamma|$, averaging over $\Gamma$ makes every invariant a norm. For $\Z_2$ in characteristic 2 it does not, and the correction is the Tate
dimension $\hat h$.
\movedproof{prop:count}

For groups of odd order $\hat H^0=0$ in characteristic 2, so winding sectors only appear for even-order groups.

\begin{exampleu}[Winding sectors of the translation quotient]
Take the translation quotient of the toric code, from $\Z_{2L}\times\Z_{L'}$ to $\Z_L\times\Z_{L'}$ by $t(x,y)=(x+L,y)$, with $p=2$. A $Z$-string of the
quotient in the $x$-direction, of length $L$, lifts to one $Z$-string of length $2L$ around the parent, a parent logical with an invariant
representative. A $Z$-string in the $y$-direction at $x=x_0$ lifts to the two parallel strings at $x_0$ and $x_0+L$. Their product is the product of
the plaquettes between them, an invariant stabilizer. It is not of the form $s\,t(s)$ for a stabilizer $s$: such a product is the product over a
$t$-invariant set of plaquettes, while neither the strip between the two strings nor its complement is $t$-invariant. So it is a winding sector. The same holds for $X$-strings, so $\hat H^0(\Gamma;S)\cong\FF_2^2$, $\hat h=2$, and
Proposition~\ref{prop:count} gives $k'=\frac12(2+2)=2$: the quotient keeps both logical qubits with half the qubits. Here $d_{\rm wind}=L'$, half
the weight of two parallel strings, so Theorem~\ref{thm:general}(iii) gives $d'\ge\min(L,L'/2)$, against the exact value $\min(L,L')$.
\end{exampleu}

The genon orbifold of the same parent has $\hat h=0$ and $k'=1$, and Haah's code~\cite{PhysRevA.83.042330} at $L=2$ goes from $k=6$ to $k'=3$.
For $D_3\times\Z_2$ acting on a product code, $\dim\hat H^0=k'=2$, so every quotient logical is a winding logical (Appendix~\ref{app:checks}).

\begin{theorem}[Factorization]\label{thm:factor}
Let $\tilde\Gamma$ act freely. Suppose it has an index-2 subgroup $\Gamma_0$ of CSS-preserving elements, and an element $\delta\notin\Gamma_0$ that is a
ZX-duality with $\delta^2\in\Gamma_0$. Then $S/\tilde\Gamma=(S/\Gamma_0)/\bar\delta$, where $\bar\delta$ is a free involutive ZX-duality of the CSS code $S/\Gamma_0$.
\end{theorem}
\idea Divide in two steps. The CSS-preserving half of the group gives a CSS code, and the remaining duality acts on it as a free
involution, so the second step is the involutive case.
\begin{proof}
Quotients by a normal subgroup compose. $\bar\delta^2$ is the image of $\delta^2\in\Gamma_0$, which acts trivially on $S/\Gamma_0$. The map $\bar\delta$ fixes the
$\Gamma_0$-orbit of $q$ iff $\gamma^{-1}\delta q=q$ for some $\gamma\in\Gamma_0$, which is impossible, since $\gamma^{-1}\delta\neq1$ lies in the other coset and $\tilde\Gamma$
acts freely.
\end{proof}

By Theorem~\ref{thm:factor}, the general duality-extended case reduces to the involutive one:

\begin{theorem}[Duality quotients of CSS codes]\label{thm:transfer}
Let $C$ be a CSS qubit code $\code{n,k,d_X,d_Z}$ with an involutive ZX-duality $g=\sigma\circ H^{\otimes n}$, where $\sigma$ is a fixed-point-free
qubit permutation mapping $X$-checks onto $Z$-checks. Then $C/g$ is a Pauli stabilizer code $\code{n/2,k/2,d'}$ with
\[
\tfrac12\,d_X\ \le\ d'=\min_{x}\tfrac12\,|x\cup\sigma(x)|\ \le\ d_X,
\]
the minimum over non-trivial $X$-logicals $x$, and no check of $C/g$ is heavier than in $C$. In particular a good CSS family with such
dualities gives a good family with at least the same relative distance.
\end{theorem}
\idea The duality swaps the $X$- and $Z$-checks, so every invariant stabilizer has the form $s\,g(s)$ and there are no winding
sectors. An invariant logical is an $X$-logical $x$ together with its mirror image $\sigma(x)$ written in $Z$, and the quotient sees it once, at
half its weight.
\begin{proof}
$S=S_X\oplus S_Z$ and $g$ swaps the summands, so every invariant stabilizer has the form $a+ga$ and $\hat h(S)=0$. By
Proposition~\ref{prop:count} and Theorem~\ref{thm:general}(iii), $k'=k/2$ and the quotient logicals are the descents of the invariant parent
logical operators, at half weight. An invariant logical operator has the form $x\cdot\sigma(x)^Z$ with $x$ an $X$-logical, non-trivial iff $x$ is, of weight $|x\cup\sigma x|$.
\end{proof}

\begin{example}[Running example, part 2: logicals, Tate term and distance]\label{ex:run2}
For the action of Example~\ref{ex:run1}, $\dim\mathsf N(S)=4$ and $\dim S^\Gamma=5$, so $\hat H^0(\Gamma;S)\cong\FF_2$ and Theorem~\ref{thm:general}(i) gives
$k'=5-4=1$. By (ii) the two-dimensional logical space of the quotient is an extension of the one-dimensional space of parent logical
classes with an invariant representative by the Tate group: one quotient logical lifts to a non-trivial parent logical, the other to a parent
stabilizer, a winding sector. The space $\mathcal L^\Gamma$ of invariant parent logical classes is two-dimensional, so one invariant class has no
invariant representative (Remark~\ref{rem:invclasses}). The winding distance is $d_{\rm wind}=3$, attained by an invariant stabilizer of weight 12, so (iii) gives only $d'\ge\min(4/4,3)=1$; some weight-4 parent $Z$-logical has a norm outside
$\mathsf N(S)$, so (iv) gives $d'\le4$. The exact value is $d'=3$, and the factorization sharpens the bounds. The translation subgroup
$\Gamma_0=\{1,T_{(-1,3)}\}$ gives the toric code $\code{10,2,3}$ on $\Z^2/\Lambda_0$, with $\Lambda_0$ the even-sum vectors of $\Lambda$. Here
$\hat H^0(\Gamma_0;S)\cong\FF_2^2$ and $k'=\frac12(k+\hat h)=2$: no logical qubit is lost although the qubit count halves, because two of the four
quotient logical Paulis lift to pairs of parallel parent strings, which are stabilizers. The remaining involution is a free ZX-duality of
this CSS code, and Theorem~\ref{thm:transfer} gives $d'=\min_x\frac12|x\cup\sigma(x)|=3$, the upper end of $\frac32\le d'\le3$. All numbers are exact;
Appendix~\ref{app:runcert} lists the matrices, logical representatives, distance witnesses and the complete searches behind them.
\end{example}

\begin{exampleu}[A higher-form electromagnetic duality]\phantomsection\label{ex:emfour}
The electromagnetic wall of the $\Z_2$ $n$-form theory in $2n$ dimensions (Section~\ref{sec:walls}) has the same lattice form for every $n$; for
$n=1$ it is the duality of the running example. Take $n=2$, the 4D toric code with qubits on the 2-cells of $\Z^4$, $X$-checks $X(\delta e)$ on
1-cells and $Z$-checks on the faces of 3-cells. The shift $t_v$ by $v=(\frac12,\frac12,\frac12,\frac12)$ maps $j$-cells to $(4-j)$-cells, so $t_v$
composed with $H$ on every qubit maps the $X$-checks onto the $Z$-checks: it is the electromagnetic duality, a higher-form ``Hadamard plus
half shift''. Its square is a translation, so on $\tilde M=\mathbb R^4/\langle Le_1,Le_2,Le_3,2v\rangle$, where the parent is $\code{6L^3,6}$, it is a
free involutive ZX-duality, and Theorem~\ref{thm:transfer} gives a quotient $\code{3L^3,3}$ on $\mathbb R^4/\langle Le_1,Le_2,Le_3,v\rangle$, which is $T^3_L$
times a circle along $v$ with one electromagnetic wall. Each orbit of 2-cells has one member in the hyperplane $x_4=0$, so the quotient has one
qubit on each plaquette of the cubic torus $T^3_L$, and its checks are
\[
W_c=Z(\partial c)\ \text{for each cube }c,\qquad G_e=X(\delta e)\,Z_{e+h}Z_{e-h}\ \text{for each edge }e,\qquad h=(\tfrac12,\tfrac12,\tfrac12),
\]
both of weight 6, where $e\pm h$ is the plaquette with that centre; the checks of the 1-cells along $e_4$ and of the 3-cells containing $e_4$ give the
same $W_c$ and $G_e$. The plaquettes $e\pm h$ are the $f$ with $\int(e\cup f+f\cup e)\ne0$ for the cup product~\eqref{eq:cubicalcup}, so $G_e$ is the
Gauss law of a 2-form field $b$ dressed by $\pi\int b\cup b$, and the checks have the form $W_c$, $G_e=X^{\delta e}Z^{\lambda_e}$ of the exact bosonization
of~\cite{ChenKapustin2019} (Section~\ref{sec:fermionic}). Field theory gives the same result. The compactification rule of Section~\ref{sec:walls} for the
electromagnetic wall, $K=G\times G$ with $\omega=\pi\int a_L\cup a_R$, sets $a_L=a_R=b$ and leaves the action $\pi\int b\cup b=\pi\int w_2\cup b$ (by the Wu
formula), the $\Z_2$ gauge theory in $3+1$ dimensions whose charge is a fermion. The trace formula gives $k'=3$: on the $2^6$ states of the 4D
toric code on $T^4$, labeled by $\alpha\in H^2(T^4;\Z_2)$, the duality acts, up to a phase, by $|\alpha\rangle\mapsto2^{-3}\sum_\beta(-1)^{\int\alpha\cup\beta}|\beta\rangle$,
whose trace is $2^{-3}\sum_\alpha(-1)^{\int\alpha\cup\alpha}=8$ because the intersection form of $T^4$ is even. Check T24 confirms on the lattice that $k'=3$,
that the checks are those above, and, by the T-junction test of Levin and Wen~\cite{LevinWen2003}, that the point excitation is a fermion, against
a boson for the same checks without the $Z$ dressings. So the quotient is the 3D fermionic toric code. Its distance is $2L$ for $2\le L\le7$, by
exhaustive search, and the product of $Y$ over a staircase of $2L$ plaquettes along a face diagonal is a logical operator for every $L$. With the
same number of physical and logical qubits, the 3D toric code on the same torus has distance $L$. We leave the distance for all $L$, and the
cases $n\ge3$, to separate work.
\end{exampleu}

\subsection{Flat backgrounds: twisted compactification as a balanced product}
\label{sec:flatlat}

This subsection shows that a flat background is an instance of the two previous subsections. With a finite-image holonomy, step 1 is the
cover with deck group $\rho(\pi_1F)$, step 2 is a monomial quotient of a product code, and step 3 computes the logicals as homology of $F$ with
local coefficients.

\paragraph{Twist data.} Let $F$ be a finite cell complex, the internal space. A \emph{flat background} on $F$ is a homomorphism
$\rho:\pi_1(F)\to G$ up to conjugation (a flat $G$-bundle, non-abelian allowed), or a class $B\in H^{q+1}(F;A)$ for a $q$-form symmetry with
abelian group $A$, or both at once (a map to the classifying space of a higher group, which we do not use). This subsection treats holonomies;
Section~\ref{sec:higher} treats classes $B$.

\begin{definition}[Twisted compactification, holonomy]\label{def:twisted}
Let the fibre $Y$ carry a code complex $C(Y)$ with an action of $G$, let $\Gamma=\rho(\pi_1F)$ be finite, and let $\tilde F\to F$ be the regular cover with
deck group $\Gamma$. If $G$ acts by chain automorphisms, the \emph{twisted compactification} is
\[
C(Y\times_\rho F)\;:=\;C(\tilde F)\otimes_{\FF_p[\Gamma]}C(Y)\;=\;\big(C(\tilde F)\otimes C(Y)\big)_\Gamma .
\]
If $G$ acts by monomial Cliffords that are not chain maps, such as dualities, it is the quotient code of Definition~\ref{def:quot} of the product
code on $\tilde F\times Y$ by the diagonal action of $\Gamma$.
\end{definition}

\begin{exampleu}[A reflection holonomy]
Let $Y$ be the circle with $L'$ edges, so that $C(Y)$ is the complex $C_1(Y)\to C_0(Y)$, let $G=\Z_2$ act by the reflection $r(y)=-y$, which sends
the edge $[y,y+1]$ to $-[-y-1,-y]$, and let $F$ be the circle with $L$ edges, with holonomy $r$ around it. Then $\tilde F$ is the circle with $2L$
edges, $C(Y\times_\rho F)$ is the cellular chain complex of the Klein bottle, and with qudits in degree 1 its CSS code is the toric code on the
Klein bottle. For $p=2$ it is the reflection quotient of the example after Definition~\ref{def:quot}; for odd $p$ the action is monomial with
$\lambda_r=-1$ on the edges of $Y$, so the transport is the gate $Z\mapsto Z^{-1}$ there (Theorem~\ref{thm:balanced}(b)). The Wang sequence of
Corollary~\ref{cor:ss}(i) gives the count: $r_*$ is the identity on $H_0(Y;\FF_p)$ and $-1$ on $H_1(Y;\FF_p)$, so
$k'=\dim\coker(r_*-1|_{H_1})+\dim\ker(r_*-1|_{H_0})$ is $1+1=2$ for $p=2$ and $0+1=1$ for odd $p$, the dimensions of $H_1(K;\FF_p)$.
\end{exampleu}

\begin{theorem}[Flat bundles are balanced products]\label{thm:balanced}
Let $C(Y)$ carry a $G$-action by chain automorphisms and let $\rho:\pi_1F\to G$ have finite image $\Gamma$.
\begin{enumerate}[label=(\alph*)]
\item (Any chain action.) $C(\tilde F)\otimes_{\FF_p[\Gamma]}C(Y)$ is the cellular chain complex of $F$ with local coefficients in $C(Y)$: as a vector
space it is $\bigoplus_{\sigma\in F}C(Y)$, with one copy of $C(Y)$ for each cell of $F$, and its differential is the local-coefficient differential. The
diagonal action of $\Gamma$ on the cells of $\tilde F\times Y$ is free.
\item (Monomial actions.) Suppose in addition that every $g\in G$ maps each basis vector of $C_j(Y)$, for $j=n-1,n,n+1$ (checks, qudits and
checks), to a non-zero multiple of a basis vector, $g\,e_c=\lambda_g(c)\,e_{gc}$. Then $g$ acts on the CSS code of $Y$ by the monomial Clifford that
moves qudit $c$ to $gc$ and applies the multiplication gate $|j\rangle\mapsto|\lambda_g(c)^{-1}j\rangle$, so $Z(z)\mapsto Z(gz)$ on chains and $X(x)\mapsto
X\big((g^{-1})^{\top}x\big)$ on cochains. This is a free monomial-Clifford symmetry of the product code on $\tilde F\times Y$, and its quotient in the sense of
Definition~\ref{def:quot} is the CSS code of the complex in (a).
\end{enumerate}
Without the hypothesis of (b), a chain automorphism can mix basis vectors, and the balanced product of (a) is defined but is not a quotient in the sense
of Definition~\ref{def:quot}.
\end{theorem}
\idea On the cover the code is an ordinary product, and the deck group acts freely on its cells. Taking coinvariants of a free
action keeps one copy of $C(Y)$ over each cell of $F$, and the only memory of the cover is the gluing: passing from a cell of $F$ to a face, the
fibre is identified with itself by the holonomy. That is the local-coefficient differential. When the action is monomial, coinvariants are
exactly the quotient of Definition~\ref{def:quot}.
\movedproof{thm:balanced}

\begin{theorem}[Spectral sequence]\label{thm:ss}
In the setting of Theorem~\ref{thm:balanced}, filter $C(Y\times_\rho F)$ by $F$-degree. There is a spectral sequence
\[
E^2_{pq}=H_p\big(F;H_q(Y)_\rho\big)\ \Longrightarrow\ H_{p+q}\big(C(Y\times_\rho F)\big),
\]
and $k'=\dim H_n$ when the qubits sit in total degree $n$.
\end{theorem}
\idea Compute the homology of the fibre first, over each cell of $F$, and then let the holonomy act on it. This is the Serre spectral
sequence of the bundle $Y\times_\rho F\to F$, written for cellular chains.
\begin{proof}
$E^0=\bigoplus_\sigma C(Y)$ with $d^0=\partial_Y$, so $E^1_{pq}=C_p(F)\otimes H_q(Y)$. The differential $d^1$ is the local-coefficient boundary for the
induced action of $\rho$ on $H_q(Y)$.
\end{proof}

The theorem concerns chain complexes, so it applies to CSS codes with $G$ acting by chain automorphisms. Duality twists are not chain maps;
their quotients are counted by Theorem~\ref{thm:general} and Proposition~\ref{prop:count}, and in field theory by the Weil
character of Section~\ref{sec:walls}. Higher-form and cup-polynomial twists are counted in Section~\ref{sec:higher}, not by this spectral
sequence.

\begin{corollary}\label{cor:ss}
\begin{enumerate}[label=(\roman*)]
\item ($F=S^1$.) The sequence has two columns and degenerates, giving the Wang sequence
$k'=\dim\coker(\rho_*-1|_{H_n})+\dim\ker(\rho_*-1|_{H_{n-1}})$.
\item ($F$ a rose $R$ with holonomies $a_1,\dots,a_\Delta$.) Also two columns, with
\[
H_0(R;M)=M\Big/\sum_i(a_i-1)M,\qquad H_1(R;M)=\ker\Big(M^\Delta\to M,\ (m_i)\mapsto\sum_i(a_i-1)m_i\Big).
\]
\item (Coefficient twists.) If $\rho$ acts on $C(Y)$ only through coefficients, the twisted product is $C(Y)\otimes C(F;\FF_p^\rho)$ as complexes and
$k'=\sum_j\dim H_{n-j}(Y)\cdot\dim H_j(F;\FF_p^\rho)$.
\end{enumerate}
\end{corollary}

Item (ii) applies to lifted-product and quantum Tanner codes, whose internal space is a rose or a dipole of circles. Item (iii) is the
twisted K\"unneth formula. For example, a charge-conjugation twist around one handle of a genus-$g$ surface kills $H_0$ and $H_2$ and, by
the Euler characteristic $\chi(\Sigma_g)=2-2g$, leaves $2g-2$ dimensions of $H_1(\Sigma_g;\FF_p^-)$.

\begin{example}[Running example, part 3: the field theory]\label{ex:run3}
In field-theory terms, Example~\ref{ex:run1} is the $\Z_2$ gauge theory on $T^2=\mathbb R^2/\Lambda$ with a flat background for its electromagnetic duality
symmetry. The holonomy is a homomorphism $\rho:\Lambda\to\Z_2$, factoring through $\Lambda/2\Lambda$, that sends $v=(1,2)$ and $(-2,1)$ to the duality and $(-1,3)$
to the identity, so there is one duality wall
along the cycle dual to $v$. On the lattice the duality is a half-period translation composed with transversal $H$, the ``Hadamard plus half
shift'' of Table~\ref{tab:dict}. The image of the holonomy is $\Z_2$, so the twofold cover $\Z^2/\Lambda_0$, with $\Lambda_0=\ker\rho$ the
vectors of $\Lambda$ with even coordinate sum, already trivializes it. The fourfold cover $\Z^2/2\Lambda$ of Example~\ref{ex:run1} is a choice:
its deck group $\Lambda/2\Lambda\cong\Z_2^2$ is larger than the image of the holonomy, and its subgroup $\Lambda_0/2\Lambda$ acts by translations. Example~\ref{ex:run2}
factors the quotient through the twofold cover, whose code is $\code{10,2,3}$, followed by the duality alone. The trace
formula (Proposition~\ref{prop:trace}) gives $\dim\mathcal H_W(S^1)=|\TrW\hat W|=2$, since the duality acting on the four ground states of the
toric code on $T^2$ fixes the sectors $1$ and $f$ and exchanges $e$ and $m$; this is $k'=1$, as found in Example~\ref{ex:run2}. The duality is not a
chain map, so Theorem~\ref{thm:ss} does not apply here and the count comes from Section~\ref{sec:general}.
\end{example}

\subsection{Higher-form and cup-polynomial backgrounds: dressed Gauss laws}
\label{sec:higher}
\label{sec:corr}

For higher-form and cup-product backgrounds the symmetry $U_g$ of the principle is an entangling diagonal circuit. This subsection and the
next three extend Sections~\ref{sec:setup}--\ref{sec:flatlat} to this case in two ways. The code is defined by local Gauss laws on the compactified space
itself, with no cover (Sections~\ref{sec:gausslaw} and~\ref{sec:twistedcode}). When a finite cover trivializes the background, the code is a
quotient of a parent in a generalized sense, by deck transformations composed with entangling diagonal circuits
(Section~\ref{sec:dressedquot}). The results of Section~\ref{sec:general} carry over only for quadratic backgrounds
(Remark~\ref{rem:carry}).

\paragraph{The idea.} Let $U_\omega=\varpi^{\int\omega(a)}$ be a cup-product symmetry of the parent, and apply it only on a region:
$U_\eta=\varpi^{\int\omega(a)\cup\eta}$, with $\eta$ the indicator cochain of the region. Conjugating a Gauss law $X_i(\delta e)$ by $U_\eta$ multiplies it by
a diagonal phase, and by Stokes this phase depends on $\eta$ only through $B=\delta\eta$, the wall at the edge of the region. Gauss laws away from
the wall are unchanged, and a Gauss law that touches the wall is dressed by the response $\omega$ of the charges it moves. Written with $B$
alone, the dressed Gauss laws make sense for any cocycle $B$, also one with no primitive, on $M$ itself and with no cover. These \emph{twisted
Gauss laws} define the code. For two toric codes and $\omega=a_1\cup a_2$, $U_\omega$ is a circuit of $CZ$ gates, and a Gauss law of one copy that
touches the wall picks up a $Z$ of the other copy (Figure~\ref{fig:czwall}). The rest of this subsection makes the idea precise: the setting
(cup products and integration), the cup polynomials, the dressed Gauss laws and their algebra (Section~\ref{sec:gausslaw}), and the code they
define (Section~\ref{sec:twistedcode}).

\paragraph{Setting.} Let $p$ be prime and $M$ a finite pure cubical complex of dimension $D$ with a \emph{cup structure}: each cell is
identified with a standard cube, with a base point $x$ and an ordered set of directions $I$, compatibly with taking faces, so that we write it
$(x,I)$. On a torus or a box this is the coordinate structure, and a finite cover inherits a cup structure by pulling back. We use $\Z_p$
cochains and the cubical cup product~\cite{Chen:2021ppt}
\begin{equation}
(\alpha\cup\beta)(x,I)=\sum_{J\subset I,\ |J|=|\alpha|}\epsilon(J,I\setminus J)\,\alpha(x,J)\,\beta(x+e_J,I\setminus J).
\label{eq:cubicalcup}
\end{equation}
It satisfies the Leibniz rule $\delta(\alpha\cup\beta)=\delta\alpha\cup\beta+(-1)^{|\alpha|}\alpha\cup\delta\beta$ exactly on cochains, and it is local:
$(\alpha\cup\beta)(\sigma)$ depends only on $\alpha$ and $\beta$ on faces of $\sigma$, since the cells $(x,J)$ and $(x+e_J,I\setminus J)$
in~\eqref{eq:cubicalcup} are faces of $(x,I)$.

\begin{exampleu}[Cup products on a square]
On the square $(x,\{1,2\})$ with vertices $0=x$, $1=x+e_1$, $2=x+e_2$, $3=x+e_1+e_2$, write $\alpha_{01}$ for the value of a 1-cochain on the
edge from $0$ to $1$. For 1-cochains $\alpha,\beta$ and a 0-cochain $h$,~\eqref{eq:cubicalcup} gives
\[
(\alpha\cup\beta)(x,\{1,2\})=\alpha_{01}\beta_{13}-\alpha_{02}\beta_{23},\qquad (h\cup\beta)(x,\{i\})=h(x)\,\beta(x,\{i\}),\qquad
(\beta\cup h)(x,\{i\})=\beta(x,\{i\})\,h(x+e_i).
\]
The first factor is read at the start of each path from $0$ to $3$ and the second at its end, so the product is not graded commutative on
cochains, only on cohomology. For a $\Z_2$ 1-cochain, $a\cup a$ on a square is $a_{01}a_{13}+a_{02}a_{23}$, the sum over the two paths from $0$ to $3$
of the product of their two edges.
\end{exampleu}

\paragraph{Integration.} Fix a $D$-chain $[M]$ and write $\int y=\langle y,[M]\rangle$. The only property of $\int$ that we use is Stokes' theorem,
$\int\delta x=0$. It holds when $[M]$ is the fundamental cycle of a closed cubical manifold, orientable if $p$ is odd, and also on a manifold
with boundary when the background is relative (Appendix~\ref{app:tech-lattice}). Since only the Leibniz rule and the locality of the cup product
are used, everything below also holds for the simplicial cup product of a simplicial complex with a branching.

\paragraph{Fields and checks.} There are fields $a_1,\dots,a_r$ with $a_i\in C^{n_i}(M;\Z_p)$: one qudit for each $n_i$-cell and each field, with
$Z$-basis $|a\rangle$. $X_i(c)$ shifts $a_i$ by the cochain $c$. For each field $i$ and $(n_i+1)$-cell $f$ the \emph{flux check} is the diagonal
operator $\varpi^{(\delta a_i)(f)}$, a product of $Z$'s, with $+1$ projector $P_{i,f}$; for a set $V$ of such pairs, $P_V=\prod_{(i,f)\in V}P_{i,f}$. For
a cell $e$, $\mathrm{st}(e)$ is the set of $D$-cells that have $e$ as a face, and $V_e$ the set of flux checks $(j,f)$ with $f$ a face of a cell of
$\mathrm{st}(e)$. The \emph{undressed code} $\mathcal C_0$ has the flux checks and the Gauss laws $X_i(\delta e)$.

\paragraph{Cup polynomials.} A \emph{cup polynomial} is $\omega(a)=\sum_mc_m\,m(a)$ with $c_m\in\Z_p$ and ordered monomials
$m(a)=a_{m_1}\cup\dots\cup a_{m_k}$, all of the same form degree $|\omega|=\sum_sn_{m_s}\le D$. The \emph{order} $o(\omega)$ of $\omega$ is the largest number
of slots, and $\omega$ is \emph{multilinear} if no field occurs twice in a monomial. We keep apart the form degrees $n_i$ and $|\omega|$,
the order $o(\omega)$, the degree of a phase polynomial in the qudit variables, the Clifford level, and the space dimension $D$; the degree of
a field or a cochain always means its form degree. A background is a cocycle $B\in Z^{D+1-|\omega|}(M;\Z_p)$,
so $q=D-|\omega|$. We also allow $|\omega|=D+1$ on a closed $M$, with $B=1$, so $q=-1$; this gives the codes of Example~\ref{ex:kzh}(i), and Appendix~\ref{app:tech-lattice}
lists the statements that hold in this case. For a monomial $m$ let $I_i(m)$ be the set of slots holding $a_i$ and $\mathrm{pos}(s)=\sum_{t<s}n_{m_t}$ the degree in front of slot
$s$. For $\emptyset\ne S\subseteq I_i(m)$ and $e\in C^{n_i-1}$, $m^e_S(a)$ denotes $m$ with $e$ in the slot $\min S$, $\delta e$ in the other slots of $S$,
and $a_{m_t}$ in the remaining slots.

\begin{exampleu}[Cup polynomials]
For two 1-form fields in $D=2$, $\omega=a_1\cup a_2$ is multilinear of order 2 with $|\omega|=2$, so $q=0$ and $B$ is a 1-cocycle, the dual of a
wall. For 1-form fields in $D=3$ the cubic polynomials of types I, II and III of Section~\ref{sec:cupft} are $a\cup a\cup a$, $a_1\cup a_1\cup a_2$ and
$a_1\cup a_2\cup a_3$, all of order 3 with $q=0$, and only type III is multilinear; type IV is $a_1\cup a_2\cup a_3\cup a_4$ in $D=4$. For one field of
degree $n$ and $\omega=a\cup a$, the monomial has $I(m)=\{1,2\}$, $\mathrm{pos}(1)=0$ and $\mathrm{pos}(2)=n$, and the three sets $S=\{1\},\{2\},\{1,2\}$
give $m^e_S(a)=e\cup a$, $a\cup e$ and $e\cup\delta e$.
\end{exampleu}

\subsubsection{Dressed Gauss laws}
\label{sec:gausslaw}

\begin{definition}[Dressed Gauss law]\label{def:gauss}
The dressed Gauss law of field $i$ at an $(n_i-1)$-cell $e$ is
\begin{equation}
G_{i,e}=X_i(\delta e)\,\varpi^{f_{i,e}(a)},\qquad f_{i,e}(a)=\sum_mc_m\sum_{\emptyset\ne S\subseteq I_i(m)}(-1)^{\mathrm{pos}(\min S)+|\omega|}\int m^e_S(a)\cup B .
\label{eq:gauss}
\end{equation}
The \emph{dressed code} $\mathcal C_B$ is the code of the flux checks and the operators $G_{i,e}$, made precise in
Proposition~\ref{prop:twisted-code}.
\end{definition}

For cubic $\omega$, dressed codes are of the kind called twisted codes in~\cite{ZKH2026nonabelian,ZTKH2026sheaf}, where $CZ$ dressings from a
triple cup product couple three copies of an abelian code on a CW or sheaf complex and give a non-abelian qLDPC code. We say
\emph{dressed} so that \emph{twisted} refers only to the boundary condition of twisted compactification.

The dressing involves $B$ but no primitive of it. It is the conjugation of $X_i(\delta e)$ by $\varpi^{\int\omega(a)\cup\eta}$ with the first $\delta e$
moved onto $\eta$ by Stokes, up to curvature terms (Lemma~\ref{lem:cochain}(i)). For the mutual quadratic case $\omega=a_1\cup a_2$, with
$n_1+n_2+q=D$,
\begin{equation}
G^{(1)}_e=X_1(\delta e)\,Z_2\Big(c\mapsto(-1)^{n_1+n_2}\textstyle\int e\cup c\cup B\Big),\qquad
G^{(2)}_{e'}=X_2(\delta e')\,Z_1\Big(c\mapsto(-1)^{n_2}\textstyle\int c\cup e'\cup B\Big);
\label{eq:mutualgauss}
\end{equation}
for one field with $\omega=\kappa\,a\cup a$ the term $S=\{1,2\}$ adds a constant phase,
$G_e=X(\delta e)\,Z\big(c\mapsto \kappa[\int e\cup c\cup B+(-1)^n\int c\cup e\cup B]\big)\,\varpi^{\kappa\int e\cup\delta e\cup B}$; and for type III at $p=2$,
$G_{1,e}=X_1(\delta e)(-1)^{\int e\cup a_2\cup a_3\cup B}$ and its two cyclic analogues, which are $CZ$-dressed.

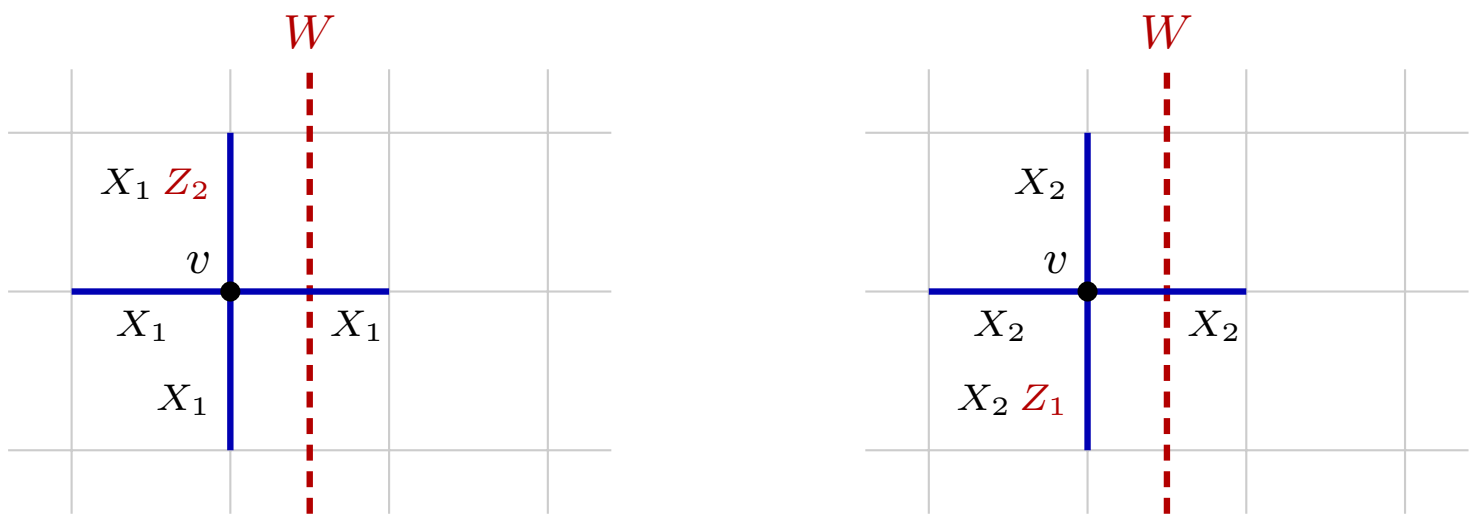
\begin{figure}[tbp]
\centering
\begin{tikzpicture}[scale=1.05]
\begin{scope}
\draw[gray!40] (-2.4,-1) -- (1.4,-1) (-2.4,0) -- (1.4,0) (-2.4,1) -- (1.4,1) (-2,-1.4) -- (-2,1.4) (-1,-1.4) -- (-1,1.4) (0,-1.4) -- (0,1.4) (1,-1.4) -- (1,1.4);
\draw[very thick,red!70!black,densely dashed] (-0.5,-1.4) -- (-0.5,1.4) node[above] {\small $W$};
\draw[very thick,blue!70!black] (-2,0) -- (0,0) (-1,-1) -- (-1,1);
\fill (-1,0) circle (1.8pt);
\node[above left,font=\small] at (-1,0) {$v$};
\node[below,font=\scriptsize] at (-1.55,0) {$X_1$};
\node[below,font=\scriptsize] at (-0.2,0) {$X_1$};
\node[left,font=\scriptsize] at (-1,-0.68) {$X_1$};
\node[left,font=\scriptsize] at (-1,0.68) {$X_1\,\textcolor{red!70!black}{Z_2}$};
\node[font=\small] at (-0.5,-1.9) {(a) Gauss law of copy 1 at $v$};
\end{scope}
\begin{scope}[xshift=5.4cm]
\draw[gray!40] (-2.4,-1) -- (1.4,-1) (-2.4,0) -- (1.4,0) (-2.4,1) -- (1.4,1) (-2,-1.4) -- (-2,1.4) (-1,-1.4) -- (-1,1.4) (0,-1.4) -- (0,1.4) (1,-1.4) -- (1,1.4);
\draw[very thick,red!70!black,densely dashed] (-0.5,-1.4) -- (-0.5,1.4) node[above] {\small $W$};
\draw[very thick,blue!70!black] (-2,0) -- (0,0) (-1,-1) -- (-1,1);
\fill (-1,0) circle (1.8pt);
\node[above left,font=\small] at (-1,0) {$v$};
\node[below,font=\scriptsize] at (-1.55,0) {$X_2$};
\node[below,font=\scriptsize] at (-0.2,0) {$X_2$};
\node[left,font=\scriptsize] at (-1,0.68) {$X_2$};
\node[left,font=\scriptsize] at (-1,-0.68) {$X_2\,\textcolor{red!70!black}{Z_1}$};
\node[font=\small] at (-0.5,-1.9) {(b) Gauss law of copy 2 at $v$};
\end{scope}
\end{tikzpicture}
\caption{The dressed Gauss laws of two toric codes with a $CZ$ wall $W$, the line dual to $B=\theta_x$, at a vertex $v$ next to the wall. Each
edge carries one qubit of each copy. (a) The star of copy 1 is $X_1(\delta v)$ times $Z_2$ on the vertical edge above $v$. (b) The star of copy 2
is $X_2(\delta v)$ times $Z_1$ on the vertical edge below $v$. Gauss laws away from the wall are unchanged. The two operators overlap in two
anticommuting pairs of letters, so they commute.}
\label{fig:czwall}
\end{figure}

\begin{exampleu}[Two toric codes with a $CZ$ wall]
Take $p=2$, $D=2$, two 1-form fields on the square torus $\Z_L\times\Z_{L'}$, $\omega=a_1\cup a_2$, so $q=0$, and $B=\theta_x$, the 1-cocycle equal to 1
on the horizontal edges from $x=-1$ to $x=0$; it is Poincar\'e dual to a vertical line $W$. By~\eqref{eq:mutualgauss} and the cubical cup
product~\eqref{eq:cubicalcup}, the star of copy 1 at a vertex $v$ with $v_1=-1$ becomes $X_1(\delta v)\,Z_2$ on the vertical edge from $v$ to
$v+e_2$, the star of copy 2 there becomes $X_2(\delta v)\,Z_1$ on the vertical edge from $v-e_2$ to $v$, and every other check is unchanged
(Figure~\ref{fig:czwall}). The
star of copy 1 at $v$ and the star of copy 2 at $v$ or at $v+e_2$ overlap in two anticommuting pairs, so they commute, as
Lemma~\ref{lem:cochain}(iii) requires. The line $W$ is
the wall of the symmetry $U_\omega=(-1)^{\int a_1\cup a_2}$, a circuit of $CZ$ gates that maps $m_1\mapsto m_1e_2$ and $m_2\mapsto m_2e_1$. The
product of all stars of copy 1 is now the Wilson loop of copy 2 along $W$, and vice versa, so the two electric loops along $W$, which pair
with $B$, become stabilizers, and Theorem~\ref{thm:hfcount}(i) gives $k'=1+1=2$, against $k=4$ without the wall. On the double cover
$\Z_{2L}\times\Z_{L'}$ the background is $\delta\tilde\eta$ with $\tilde\eta$ the indicator of a strip of width $L$, and
Theorem~\ref{thm:dressedquot} gives both quotient descriptions: the code is the quotient of the two-wall code by the translation $t$ by $L$ (a
permutation, part (d)), and the dressed quotient of the undressed parent by $t\circ U_\omega$ (part (c), since $\zeta_t=1$).
\end{exampleu}

\paragraph{Why the dressed Gauss laws define a code.} All the algebra of the dressed Gauss laws comes from Stokes' theorem applied to cup
products. Conjugating $X_i(\delta e)$ by the twist operator on a region multiplies it by a sum of cup products that contain $\delta e$. Moving the
$\delta$ onto the indicator of the region gives the dressing, and moving it onto a field gives a flux, which vanishes on flux-free states. The
same manipulation shows that the dressed Gauss laws commute with the flux checks, commute with each other on flux-free states and have
order $p$ there; for $\omega$ of order at most 2 they commute and have order $p$ exactly. Lemma~\ref{lem:cochain} (Appendix~\ref{app:tech-lattice})
states this precisely.

\subsubsection{The dressed code}
\label{sec:twistedcode}

\begin{proposition}[The dressed code]\label{prop:twisted-code}
\begin{enumerate}[label=(\alph*)]
\item The operators $\Pi_{i,e}=P_{V_e}\cdot\frac1p\sum_{t=0}^{p-1}G_{i,e}^t$ and the flux projectors $P_{j,f}$ are pairwise commuting orthogonal projectors, and
$\Pi_{i,e}$ acts only on qudits on faces of cells of $\mathrm{st}(e)$. Their common $+1$ eigenspace is
$\mathcal C_B=\{\psi:\ P_{j,f}\psi=\psi\ \text{and}\ G_{i,e}\psi=\psi\ \text{for all}\ (j,f),(i,e)\}$, the ground space of the commuting-projector Hamiltonian
$-\sum\Pi_{i,e}-\sum P_{j,f}$ when it is non-zero.
\item If $\omega$ has order at most 2, the $G_{i,e}$ are Pauli operators that commute with each other and with the flux checks on the whole
Hilbert space. If the group they generate contains no non-trivial scalar, it is a Pauli stabilizer group with code space $\mathcal C_B$; otherwise
$\mathcal C_B=0$.
\item (Gauge invariance.) For every cochain $\mu$ of degree $D-|\omega|$, $\Pi^{B+\delta\mu}_{i,e}=U_\mu\Pi^B_{i,e}U_\mu^\dagger$ with $U_\mu=\varpi^{\int\omega(a)\cup\mu}$, a
product of commuting diagonal gates, one on the qudits of each $D$-cell, of bounded depth under the bounded geometry of Appendix~\ref{rem:depth}. So $\mathcal C_{B+\delta\mu}=U_\mu\,\mathcal C_B$, and the code depends only on $[B]$ up to such circuits. For order at most 2,
$U_\mu G^B_{i,e}U_\mu^\dagger$ equals $G^{B+\delta\mu}_{i,e}$ times a product of flux checks.
\item (Naturality.) A cellular automorphism $\phi$ of $M$ that preserves the cup structure and $[M]$ acts on qudits by a permutation with
$\phi\,G^B_{i,e}\phi^{-1}=G^{\phi_*B}_{i,\phi(e)}$, where $\phi_*B=(\phi^{-1})^*B$. Translations of a torus and deck transformations of a cover with the
pulled-back cup structure are of this kind. A reflection or a permutation of coordinate directions in general is not: it satisfies the same
identity only with the cup product transported by $\phi$, which induces the same product on cohomology but can differ on cochains.
\end{enumerate}
\end{proposition}
\idea By the cochain lemma (Lemma~\ref{lem:cochain}), once the fluxes near $e$ vanish, the dressed Gauss laws behave like commuting unitaries of order $p$.
Averaging the powers of each one and multiplying by its local flux projector therefore gives commuting local projectors. Gauge invariance
is the lemma with $B$ replaced by $\delta\mu$, and naturality holds because the dressing is built from the cup product alone.
\movedproof{prop:twisted-code}

\begin{proposition}[Clifford-level rule]\label{prop:clifflevel}
Let $\omega$ have order $o$.
\begin{enumerate}[label=(\alph*)]
\item Every dressing $f_{i,e}$ is a polynomial of degree at most $o-1$ in the qudit variables. So $o\le2$ gives Pauli Gauss laws and a Pauli stabilizer
code (Proposition~\ref{prop:twisted-code}(b)), and $o=3$ gives Gauss laws dressed by diagonal phases of degree at most 2, which for $p=2$ are
products of $CZ$ and $Z$ gates.
\item If $B=\delta\eta$ on the faces of the cells of $\mathrm{st}(e)$, then on the range of $P_{V_e}$ the dressing is the discrete derivative of the phase
polynomial of the twist operator $U_\eta=\varpi^{\int\omega(a)\cup\eta}$, so its degree is at most $m-1$ when that polynomial has degree $m\le o$.
\end{enumerate}
For $p=2$, and for odd $p$ when the degree is less than $p$, the Clifford level of a diagonal gate equals the degree of its phase
polynomial~\cite{CuiGottesmanKrishna2017}. So the Gauss laws are $X$-strings dressed at level at most $o-1$, and at most $m-1$ under (b). The bound is for the dressing: a Gauss law
with a non-trivial $X$-string is itself at level at most $\max(1,o-1)$, and $\max(1,m-1)$ under (b), since multiplying a diagonal gate by a
Pauli operator keeps its level (Section~\ref{sec:chsparent}).
Proposition~\ref{prop:chslevel} extends (b) to parents whose checks are already dressed.
\end{proposition}
\idea A dressing is a discrete derivative: it is the change of the phase polynomial of the twist operator when the field is
shifted by $\delta e$. A discrete derivative lowers the degree by one, and the Clifford level of a diagonal gate is the degree of its phase
polynomial.
\movedproof{prop:clifflevel}

\begin{exampleu}[A wall on a circle]
Take $p=2$, $D=1$ and $M$ the circle with vertices $v\in\Z_L$, $L\ge3$, and edges $e_v=[v,v+1]$. Put a 0-form field $a$ on the vertices and a
1-form field $b$ on the edges, one qubit per cell for each. A 0-form field has no Gauss law, and its flux checks $Z_{a,v}Z_{a,v+1}$ make it a
repetition code; $b$ has the Gauss laws $X_{b,e_{v-1}}X_{b,e_v}$ and no flux checks, a repetition code in the $X$ basis. The undressed code
has $k=b_0+b_1=2$. Take $\omega=a\cup b$, so $|\omega|=1$ and $q=0$, and let $B$ be the 1-cocycle equal to 1 on $e_0$ only, dual to a point of the
circle. For the field $b$ the dressing~\eqref{eq:gauss} is $\int a\cup v\cup B$. The cubical cup evaluates 0-cochains at the first vertex of an
edge, so this is $a(0)v(0)$, non-zero only for $v=0$, and the only dressed check is
\[
G_{b,0}=X_{b,e_{-1}}X_{b,e_0}\,Z_{a,0}.
\]
It is a Pauli operator, as Proposition~\ref{prop:clifflevel}(a) predicts for order 2. With two 0-form fields $a_1,a_2$ and the cubic
$\omega=a_1\cup a_2\cup b$, the same computation gives $G_{b,0}=X_{b,e_{-1}}X_{b,e_0}\,CZ(a_{1,0},a_{2,0})$, dressed at Clifford level $2=o-1$. These two
codes, on $2L$ and $3L$ qubits, illustrate Theorems~\ref{thm:dressedquot}, \ref{thm:hfcount} and~\ref{thm:pointcount} below.
\end{exampleu}

By Proposition~\ref{prop:clifflevel}(a), a fixed background of a Clifford symmetry, whose phase polynomial has order at most 2, never produces
(N2) order, non-abelian topological order, for any $q$, any $F$ and any holonomy.

\begin{corollary}[Pauli backgrounds change only phases]\label{cor:paulinogo}
If every twist operator in the deck-times-symmetry group is a Pauli operator, as for the native Wilson and 't~Hooft symmetries of a
toric code, the group acts on $\FF_p^{2n}$ like the deck group alone. The code then has the binary stabilizer space of the plain quotient by the deck group
and differs from it only in the phases of its generators: with consistent phases it is Pauli-equivalent to that quotient,
and otherwise it is zero.
\end{corollary}
\begin{proof}
Conjugation by a Pauli operator fixes every Pauli operator up to a phase.
\end{proof}

\subsection{Quotient description on a trivializing cover}
\label{sec:dressedquot}

This subsection is step 2(b): when a finite cover trivializes $B$, the dressed code is the identification of the parent with its image under
deck transformations composed with the circuits $U(\zeta_g)$. The reason is the idea of Section~\ref{sec:higher} read backwards. On the cover,
$\pi^*B=\delta\tilde\eta$, so the wall is undone by the symmetry $U(\tilde\eta)$ applied on the region $\tilde\eta$. A deck transformation $g$ moves this
region, and $\zeta_g=\tilde\eta-g^*\tilde\eta$ measures the mismatch, which the symmetry must supply when a cell is identified with its image. For the
code with a $CZ$ wall on the double cover, $\tilde\eta$ is the indicator of a strip of width $L$, $\zeta_t=1$, and the identification is by $t\circ U_\omega$.

\begin{definition}[Descent of $X$-monomial operators; dressed quotient]\label{def:descent}
Let $\Gamma$ act freely by permutations on a set $Q$ of qudits, and let $\{G_\alpha\}_{\alpha\in A}$ be a family of operators of the form
$G_\alpha=X(c_\alpha)\,e^{2\pi ih_\alpha(a)}$ with $h_\alpha(a)\in\RR/\Z$, an $X$-string times a diagonal phase, permuted by $\Gamma$: $gG_\alpha g^{-1}=G_{g\alpha}$ for an action of $\Gamma$ on $A$. The
quotient Hilbert space has one qudit for each $\Gamma$-orbit $\bar q$ of qudits, with configurations $a\in\Z_p^{Q/\Gamma}$ and their pull-backs
$\pi^*a\in(\Z_p^Q)^\Gamma$. The \emph{descent} of $G_\alpha$ is
\[
\bar G_{[\alpha]}=X(\pi_*c_\alpha)\,e^{2\pi ih_\alpha(\pi^*a)},\qquad(\pi_*c)(\bar q)=\sum_{q\in\bar q}c(q).
\]
It depends only on the orbit $[\alpha]$, since $\pi_*c_{g\alpha}=\pi_*c_\alpha$ and $h_{g\alpha}(\pi^*a)=h_\alpha(g^{-1}\cdot\pi^*a)=h_\alpha(\pi^*a)$. For the dressed Gauss laws
$G_{i,e}=X_i(\delta e)\,\varpi^{f_{i,e}}$, with $\varpi=e^{2\pi i/p}$, the phase is $h=f_{i,e}/p$; general $h$ allows the
dyadic phases of Section~\ref{sec:chsparent}. $Z$-type checks descend in
the same way, $Z(z)\mapsto Z(\pi_*z)$. If the parent is a commuting-projector code generated by such a family and $Z$-type checks, its
\emph{dressed quotient} is the code generated in the same way by the descended operators, whenever these again satisfy the conclusions of
Lemma~\ref{lem:cochain}(ii)--(iv).
\end{definition}

Descent is defined on a generating family rather than on the code, because the descent of a product need not be the product of the
descents (Appendix~\ref{app:tech-lattice}). The next theorem fixes the family, the dressed Gauss laws and flux checks, and shows that for it the
descended operators themselves are the dressed Gauss laws and flux checks on $M$, a valid code that depends only on $[B]$.

\begin{exampleu}[Descent identifies variables]
The basic instance is the code with a $CZ$ wall above: its dressed Gauss laws on the double cover are permuted by the translation $t$, and their
descents are the dressed Gauss laws on $\Z_L\times\Z_{L'}$. Descent can also lower the degree of a dressing. If a generator contains a $CZ$
between two qubits $q$ and $gq$ of one orbit, the term $\frac12a_qa_{gq}$ of $h_\alpha$ becomes $\frac12\bar a^2=\frac12\bar a$, a single $Z$ on the orbit
qubit; similarly an $S$ gate on each of the two qubits, $\frac14(a_q+a_{gq})$, descends to $\frac12\bar a$. This is why
descent never raises the level (Proposition~\ref{prop:chslevel}(b)), and why it is defined on generators: the descent of a product need not be the
product of the descents.
\end{exampleu}

\begin{theorem}[Quotient description]\label{thm:dressedquot}
Let $M=X\times F$ with $B=\pi_F^*B_F$, let $\tilde F\to F$ be a finite regular cubical cover with deck group $\Gamma$, and $\tilde M=X\times\tilde F\to M$, with the
pulled-back cup structure.
Suppose that $\tilde B=\pi^*B$ is exact, $\tilde B=\delta\tilde\eta$, and that $\omega$ is multilinear or the covering is injective on every closed cube of
$\tilde M$.
\begin{enumerate}[label=(\alph*)]
\item On $\tilde M$, $\mathcal C_{\tilde B}=U(\tilde\eta)\,\mathcal C_0$ with $U(\tilde\eta)=\varpi^{\int\omega(\tilde a)\cup\tilde\eta}$, and $\Gamma$ acts freely on qudits and permutes the
dressed Gauss laws and flux checks of $\mathcal C_{\tilde B}$.
\item The descent along $\Gamma$ of the dressed Gauss laws and flux checks of $\mathcal C_{\tilde B}$ is the set of dressed Gauss laws and flux checks of
$\mathcal C_B$ on $M$. So $\mathcal C_B$ is the dressed quotient of $\mathcal C_{\tilde B}$ by $\Gamma$, independently of the trivialization $\tilde\eta$ and of the
orbit representatives.
\item In the frame of the undressed parent, $U(\tilde\eta)^\dagger\,g\,U(\tilde\eta)=g\circ U(\zeta_g)$ with $\zeta_g=\tilde\eta-g^*\tilde\eta$ a cocycle. We write
$\mathcal C_0/\Gamma'$, $\Gamma'=\{g\circ U(\zeta_g)\}$, for the dressed quotient of (b). For $q=0$ and connected $\tilde M$, $\zeta_g$ is a constant and $U(\zeta_g)$ is
a power of the global symmetry $U_\omega=\varpi^{\int\omega(a)}$; for type III this is the transversal $CCZ$, and $\mathcal C_B=\mathcal C_0/\langle t\circ CCZ\rangle$
for a deck translation $t$.
\item For $\omega$ of order at most 2, $S_B=S_{\tilde B}/\Gamma$ in the sense of Definition~\ref{def:quot}, a quotient by a free permutation action.
\end{enumerate}
\end{theorem}
\idea On the cover the background is a coboundary, so the dressed code there is the undressed code conjugated by $U(\tilde\eta)$,
and the deck group permutes its checks. Identifying each check with its translates gives the dressed checks downstairs, because a cup product
in which every factor but one is pulled back can be pushed forward cube by cube (the projection formula). Moving the conjugation by
$U(\tilde\eta)$ onto the deck transformations turns $g$ into $g\circ U(\zeta_g)$.
\movedproof{thm:dressedquot}

\begin{exampleu}[The circle on its double cover]
For the codes on the circle, let $\tilde M$ be the circle of length $2L$ and $t$ the translation by $L$. The pulled-back background is 1 on $e_0$
and $e_L$, and it is $\delta\tilde\eta$ for $\tilde\eta$ the indicator of the vertices $1,\dots,L$. Then $\zeta_t=\tilde\eta-t^*\tilde\eta$ is the constant 1, and
$U(\zeta_t)=U_\omega=(-1)^{\int a\cup b}=\prod_vCZ(a_v,b_{e_v})$, one $CZ$ between each vertex qubit and the edge qubit that follows it. By (c) the
code with one wall is the quotient of two repetition codes on the doubled circle by the translation composed with this circuit,
$\mathcal C_B=\mathcal C_0/\langle t\circ\prod_vCZ(a_v,b_{e_v})\rangle$, and by (d) it is also the plain translation quotient of the code with two walls, at
$e_0$ and $e_L$. On the code space of $\mathcal C_0$ the circuit acts as the logical $CZ$, $(-1)^{\alpha\beta}$ on $|\alpha,\beta\rangle$, by
Proposition~\ref{prop:cupgate}. For the cubic $\omega=a_1\cup a_2\cup b$ the circuit is the transversal $\prod_vCCZ(a_{1,v},a_{2,v},b_{e_v})$.
\end{exampleu}

\paragraph{When the method applies.} A trivializing cover need not exist: a non-zero class in $H^2(S^2;\Z_p)$ survives every finite cover,
and then the Gauss laws of Section~\ref{sec:gausslaw} define the code directly (Remark~\ref{rem:cover}). For $\omega$ of order at most 2 the
dressed code is an ordinary monomial quotient of the dressed parent, so Section~\ref{sec:general} applies to it; for higher order the
dimension and distance results below take its place (Remark~\ref{rem:carry}).

\subsection{Counting logical qudits for higher-form backgrounds}
\label{sec:hfcountsec}

This subsection is step 3 for higher-form and cup-product backgrounds. The mechanism is visible in the code with a $CZ$ wall. In the
undressed code the product of the stars of copy 1 over all vertices is the identity; with the wall it is the electric loop of copy 2 along the
wall, which therefore becomes a stabilizer. In general, the product of the Gauss laws of field $i$ over a cocycle $\lambda$ is a diagonal operator,
built from electric operators paired with $\lambda$ through $B$, and the logical states on which it is not trivial are removed. For
quadratic $\omega$ the removed classes form a subspace, and the count is the dimension of a kernel (Theorem~\ref{thm:hfcount}). For cubic $\omega$
the surviving classes are the zeros of a polynomial map, and the count is a number of points (Theorem~\ref{thm:pointcount}). In both cases
the count is a statement about cup products on $M$.

\begin{theorem}[Exact count for quadratic twists]\label{thm:hfcount}
Let $M$ be a closed cubical manifold of dimension $D$, orientable if $p$ is odd, with fundamental class $[M]$, and let $B\in Z^{q+1}(M;\FF_p)$.
Write $\cup[B]:H^n(M;\FF_p)\to H^{n+q+1}(M;\FF_p)$. Suppose $\mathcal C_B\ne0$; in case (i) this always holds.
\begin{enumerate}[label=(\roman*)]
\item For the mutual twist $\omega=a_1\cup a_2$: $k'=\dim\ker(\cup[B]|_{H^{n_1}})+\dim\ker(\cup[B]|_{H^{n_2}})$.
\item For one field, $\omega=\kappa\,a\cup a$ with $\kappa\not\equiv0$: $k'=\dim\ker(\cup[B]|_{H^n})$ if $p$ is odd and $n$ is even; otherwise $k'=b_n$, the
undressed count.
\end{enumerate}
Manifolds with boundary are treated separately in Corollary~\ref{cor:boundary}.
\end{theorem}
\idea The twist does not change the $X$-parts of the checks. What changes is that a product of Gauss laws whose $X$-parts cancel,
which was the identity, becomes a $Z$-string $Z(\Phi\lambda)$: an electric logical of the other field, paired with the relation $\lambda$ through $B$.
It becomes a stabilizer, and its conjugate logical is removed with it. The number of removed pairs is the rank of the pairing
$([\lambda],[c])\mapsto\int\lambda\cup c\cup B$, which Poincar\'e duality turns into the rank of $\cup[B]$.
\movedproof{thm:hfcount}

In words: the twist turns into stabilizers exactly the logicals that pair non-trivially with $B$. The formula agrees with direct
computation in about 50 lattice cases, for example $k'=7$ against 10 for one 2-form field on $T^5$ with $B=dy_0dy_1$ and $p=3$, and $k'=0$
for ball backgrounds, as Corollary~\ref{cor:boundary} predicts (Appendix~\ref{app:checks}).

\begin{exampleu}[Counting with Theorem~\ref{thm:hfcount}]
On the circle with $\omega=a\cup b$, case (i) has $n_1=0$ and $n_2=1$. The map $\cup[B]:H^0(S^1)\to H^1(S^1)$ sends 1 to $[B]\ne0$, so its kernel is
0, and $\cup[B]:H^1\to H^2=0$ has kernel $H^1$. So $k'=0+1=1$, against $k=2$. On the lattice, the product of all Gauss laws of $b$ is no longer
the identity but $Z_{a,0}$, the logical $Z$ of the repetition code of $a$; it becomes a stabilizer, the logical qubit of $a$ is lost, and $b$
keeps its own. In the proof (Appendix~\ref{app:proofs3}), $Z_{a,0}$ is $Z(\Phi\lambda)$ for the relation $\lambda=1\in Z^0$. For two toric codes on $T^2$ with a $CZ$ wall, $\cup[B]$ has
rank one on each $H^1(T^2)=\FF_2^2$, and $k'=1+1=2$ against 4, as in the example of Section~\ref{sec:gausslaw}. Case (ii) shows the role of
the sign. For the toric code with $\omega=a\cup a$, $n=1$ is odd, the pairing is multiplied by $1+(-1)^1=0$, and $k'=b_1=2$ for every $B$. For one
$\Z_3$ 2-form field on $T^4$ with $\omega=a\cup a$ and $B=dy_0$, $\cup[B]:H^2(T^4)\to H^3(T^4)$ kills the three classes $dy_0dy_j$ and is injective on the
span of the other three, so $k'=3$ against $b_2=6$.
\end{exampleu}

Proposition~\ref{prop:cone} (Appendix~\ref{app:tech-lattice}) recasts this count as a spectral sequence, in which the dressing is a perturbation
of the undressed code that removes logical operators in pairs.

\paragraph{Cubic defects: no linear spectral sequence.} For $\omega$ of order 3 the dressed Gauss laws are in general not Pauli, and then there is no Pauli stabilizer complex.
So the count cannot be a number of logical qubits computed linearly, such as the dimension $k$ of the abutment of a filtered complex of
$\FF_2$-vector spaces with $\dim\mathcal C_B=2^k$, because type III on the slab $T^2\times[0,1]$ has dimension 22 (Theorem~\ref{thm:pointcount} and Corollary~\ref{cor:boundary}(iii)). What survives is the electric half of Proposition~\ref{prop:cone}, with $d_1$ replaced by the
polynomial map
\[
Q_B:\ \bigoplus_iH^{n_i}(M;\Z_2)\to\bigoplus_i\big(H^{n_i-1}(M;\Z_2)\big)^*,\qquad Q_B(\alpha)_i(\lambda)=\int\lambda\cup\partial_i\omega(\alpha)\cup B,
\]
of degree $o(\omega)-1$: Theorem~\ref{thm:pointcount} says $\dim\mathcal C_B=|Q_B^{-1}(0)|$, the number of points of its zero fibre. For example, for type
III on the slab $T^2\times[0,1]$ with relative $B$ (Corollary~\ref{cor:boundary}(iii)), the classes are $\alpha\in H^1(T^2;\Z_2)^{\oplus3}$ and $\lambda\in
H^0=\Z_2$, and the conditions are $\alpha_2\cup\alpha_3=\alpha_3\cup\alpha_1=\alpha_1\cup\alpha_2=0$ in $H^2(T^2;\Z_2)$, three quadratic equations in six variables. Two classes of
$H^1(T^2;\Z_2)=\FF_2^2$ have zero cup product exactly when one of them is zero or they are equal, so the solutions are the zero triple and, for each
of the three non-zero classes $v$, the seven non-zero triples in $\{0,v\}^3$: $1+3\cdot7=22$ points. For quadratic $\omega$, $Q_B$ is
linear, its zero fibre is the kernel of $d_1$ on electric classes, and the count is $2^{k'}$. That a higher-form twist can only lower the count
follows from Theorem~\ref{thm:hfcount} for quadratic $\omega$ and from the upper bound of Theorem~\ref{thm:dressdim} in general.

\begin{theorem}[Cubic and higher defects: a point count]\label{thm:pointcount}
Let $p=2$, let $M$ be a closed cubical manifold, and let $\omega$ be a cup polynomial with no linear terms, that is, every monomial has at
least two slots, in fields of degrees $n_i$; a field may occupy several slots of a monomial. Then
\[
\dim\mathcal C_B=\#\Big\{\alpha\in\textstyle\bigoplus_iH^{n_i}(M;\Z_2):\ \int\lambda\cup\partial_i\omega(\alpha)\cup B=0\ \ \forall i,\ \forall\lambda\in H^{n_i-1}\Big\},
\]
where $\lambda\cup\partial_i\omega(\alpha)$ denotes the sum, over the monomials of $\omega$ and over the slots that hold $a_i$, of the monomial with that slot
replaced by $\lambda$ and the other slots evaluated at $\alpha$. In particular $\mathcal C_B\ne0$.
\end{theorem}
\idea The code is the flux-free subspace cut down by the dressed Gauss laws. A product of Gauss laws over a relation $\lambda$ is
diagonal, and Theorem~\ref{thm:dressdim} says that the dimension counts the configurations on which all these diagonal operators act as $+1$.
The operator for $\lambda$ is the phase $(-1)^{\int\lambda\cup\partial_i\omega(a)\cup B}$, and Stokes shows that it depends only on the classes of $\lambda$ and $a$,
so the count becomes a count of cohomology classes. For quadratic $\omega$ the conditions are linear, and for cubic $\omega$ they are quadratic
equations.
\movedproof{thm:pointcount}

For quadratic $\omega$ the conditions are linear and the count is a power of 2, recovering Theorem~\ref{thm:hfcount}. For cubic $\omega$ the
conditions are quadratic equations, the count is the number of points of a variety, and non-powers of 2 occur. We name such a count by
the space $X$ of the compactified code and take for $M$ the thin slab $X\times[0,1]$ with $B$ the relative generator
(Corollary~\ref{cor:boundary}(iii)), the setting of $D(D_4)$ on $X$. Then type III on $T^2$ gives 22 (the pairwise cup-orthogonal triples
counted above), $b\cup a_1\cup a_2$ on $T^3$, with $b$ a 2-form field (the $D_4$ theory of Proposition~\ref{prop:ext}), gives 92, and type IV on
$T^3$ gives 1576. On the closed manifold $T^2\times S^1$ with $B$ dual to $T^2\times\{pt\}$ the same formula gives $22\cdot2^3=176$ for type III,
because the components of the $\alpha_i$ along the circle are unconstrained. These agree with the group
theory, $|\Hom(\Z^3,D_4)|/|D_4|=22$, $|\Hom(\Z^4,D_4)|/8=92$ and $(65536-2\cdot20160)/16=1576$, and with the lattice in all 30 cases computed, on closed manifolds and on slabs with relative $B$ (Corollary~\ref{cor:boundary}). A count that is neither 0 nor a power of 2
certifies that the code is not a Pauli stabilizer code (Corollary~\ref{cor:cert}; the certificates are collected in Appendix~\ref{sec:cert}). 

\begin{exampleu}[The smallest cubic count]
On the circle with $\omega=a_1\cup a_2\cup b$ and $B$ dual to a point, the classes are $(\alpha_1,\alpha_2,\beta)\in H^0\oplus H^0\oplus H^1=\FF_2^3$. The 0-form
fields have no Gauss laws ($H^{-1}=0$), and for $b$ the condition with $\lambda=1\in H^0$ is $\int\alpha_1\cup\alpha_2\cup\lambda\cup B=\alpha_1\alpha_2=0$. So
Theorem~\ref{thm:pointcount} gives $\dim\mathcal C_B=3\cdot2=6$. On the lattice, the product of all Gauss laws of $b$ is $CZ(a_{1,0},a_{2,0})$, which
must act as $+1$: the two repetition codes keep the logical states $|00\rangle$, $|01\rangle$ and $|10\rangle$ but lose $|11\rangle$, while $b$ keeps its
two states. Since 6 is not a power of 2, Corollary~\ref{cor:cert} shows that this code is not a Pauli stabilizer code. Direct diagonalization at
$L=3,4$ gives 6, and 2 for the quadratic wall.
\end{exampleu}

\begin{corollary}[Manifolds with boundary]\label{cor:boundary}
Let $M$ be a compact cubical manifold with boundary, orientable if $p$ is odd, with relative fundamental chain $[M]$; for example
$M=X\times D^{q+1}$ with $X$ closed. Use the following conventions.
\begin{itemize}[leftmargin=1.5em]
\item \emph{Fields are absolute cochains.} Field $i$ has a qudit on every $n_i$-cell of $M$, including the cells of $\partial M$, a Gauss law at every
$(n_i-1)$-cell and a flux check at every $(n_i+1)$-cell (smooth boundary). The undressed code has $k=\sum_i\dim H^{n_i}(M;\FF_p)$.
\item \emph{The background is relative.} $B\in Z^{q+1}(M,\partial M;\FF_p)$ vanishes on every cell contained in $\partial M$, and so do $\eta$ and $\mu$.
\end{itemize}
Then Lemma~\ref{lem:cochain} and Proposition~\ref{prop:twisted-code} hold, and, with $\cup[B]:H^n(M;\FF_p)\to H^{n+q+1}(M,\partial M;\FF_p)$ and $\mathcal C_B\neq0$:
\begin{enumerate}[label=(\roman*)]
\item for $\omega=a_1\cup a_2$, $k'=\dim\ker(\cup[B]|_{H^{n_1}(M)})+\dim\ker(\cup[B]|_{H^{n_2}(M)})$;
\item for $\omega=\kappa\,a\cup a$, $k'=\dim\ker(\cup[B]|_{H^n(M)})$ if $p$ is odd and $n$ is even, and $k'=\dim H^n(M)$ otherwise;
\item for $p=2$ and any $\omega$ without linear terms, Theorem~\ref{thm:pointcount} holds as stated, with
$\alpha\in\bigoplus_iH^{n_i}(M;\Z_2)$, $\lambda\in H^{n_i-1}(M;\Z_2)$ and $\int$ the pairing with $[M,\partial M]$.
\end{enumerate}
For $M=X\times D^{q+1}$ and $B$ the pull-back of a generator of $H^{q+1}(D^{q+1},\partial D^{q+1})$, every map $\cup[B]$ is an isomorphism, so the mutual
twist has $k'=0$.
\end{corollary}
\idea With the background relative to the boundary, every cup product with $B$ vanishes on $\partial M$, so Stokes holds with no
boundary term and the proofs of Lemma~\ref{lem:cochain}, Proposition~\ref{prop:twisted-code} and Theorems~\ref{thm:hfcount}
and~\ref{thm:pointcount} go through. Poincar\'e duality is replaced by Lefschetz duality between absolute and relative
cohomology.
\movedproof{cor:boundary}

\subsection{Clifford-hierarchy parents}
\label{sec:chsparent}

The method does not require a Pauli parent.
The parents so far are Pauli stabilizer codes. The dressed codes of Section~\ref{sec:higher} with $\omega$ of order $o\ge3$ are of a more general kind: their
$Z$-type checks are Pauli operators (the flux checks), and their $X$-type checks are $X$-strings dressed by diagonal gates of level at most
$o-1$, which commute only on the common $+1$ eigenspace of the $Z$-type checks. Kobayashi, Zhu and Hsin call codes of this kind
\emph{Clifford-hierarchy stabilizer codes}~\cite{KZH2025chs}. Their main examples are the twisted $\Z_2^N$ Dijkgraaf--Witten codes in $N-1$
dimensions, whose $X$-type checks are dressed by $C^{N-2}Z$ gates. Such a code can itself be the parent of a quotient or of a twist, and this
subsection extends Proposition~\ref{prop:clifflevel} to that case. We take $p=2$ and write $\mathcal C^{(\ell)}$ for level $\ell$ of the Clifford
hierarchy, with $\mathcal C^{(1)}$ the Pauli group and $\mathcal C^{(0)}$ the scalars, all up to phases.

We use the classification of diagonal gates of Cui, Gottesman and Krishna~\cite{CuiGottesmanKrishna2017}. Write a diagonal gate as
$D|a\rangle=e^{2\pi i h(a)}|a\rangle$ with $h(a)=\sum_Sc_S\,a^S$ multilinear in the bits, $a^S=\prod_{j\in S}a_j$, and $c_S\in\Z[\frac12]/\Z$; this expansion is
unique. If $c_S$ has denominator $2^{j_S}$, then $D\in\mathcal C^{(\ell)}$ if and only if $|S|+j_S-1\le\ell$ for every $S\ne\emptyset$ with $c_S\neq0$. For example,
$CZ$ ($\frac12a_1a_2$) has level 2, $CS$ ($\frac14a_1a_2$) and $T$ ($\frac18a_1$) level 3, and $C^{N-2}Z$ level $N-1$. Two consequences: the diagonal
gates in $\mathcal C^{(\ell)}$ form a group, and multiplying a gate in $\mathcal C^{(\ell)}$ by a Pauli operator, or conjugating it by a Clifford, keeps it in
$\mathcal C^{(\ell)}$.

\begin{definition}[$X$-monomial codes of level $\ell$]\label{def:xmono}
A code on qubits is \emph{$X$-monomial of level $\ell$} if it is the common $+1$ eigenspace of Pauli $Z$-checks $Z(v)$, $v\in V$, and of operators
$G_c=X(c)\,D_c$, with $D_c$ diagonal in $\mathcal C^{(\ell)}$, such that every $G_c$ commutes with every $Z(v)$ and the $G_c$ commute pairwise on the
common $+1$ eigenspace $P_V$ of the $Z(v)$. Level 1 gives Pauli stabilizer codes, CSS when every $D_c$ is trivial. By
Proposition~\ref{prop:clifflevel}, every dressed code $\mathcal C_B$ with $\omega$ of
order $o$ is $X$-monomial of level at most $o-1$, with the flux checks as $Z$-checks. The level is that of the dressings $D_c$;
the operators $G_c$ themselves lie in $\mathcal C^{(\max(1,\ell))}$.
\end{definition}

\begin{exampleu}[Levels]
The toric code is $X$-monomial of level 1 with every $D_c=1$. The code with a $CZ$ wall of Section~\ref{sec:gausslaw} has level 1 with each $D_c$ a
single $Z$; both are Pauli stabilizer codes. The type-III dressed code, with $G_{1,e}=X_1(\delta e)(-1)^{\int e\cup a_2\cup a_3\cup B}$, has $D_c$ a product of
$CZ$ gates and level 2, and its dimension 22 on the slab $T^2\times[0,1]$ shows that it is not equivalent to a code of level 1 (Corollary~\ref{cor:cert}). Twisting one
3D toric code by its gauged-SPT symmetry $(-1)^{\int a\cup a\cup a}$ also gives level 2 (Section~\ref{sec:threegroup}). The twisted $\Z_2^N$ codes
of~\cite{KZH2025chs} have level $N-1$ (Example~\ref{ex:kzh}).
\end{exampleu}

\begin{proposition}[Level rule for Clifford-hierarchy parents]\label{prop:chslevel}
Let $\mathcal C$ be $X$-monomial of level $\ell$, and let $U=D\,L$, where $L$ is a circuit of $CNOT$, $X$ and qubit-permutation gates, so that
$L|a\rangle=|Aa+b\rangle$ with $A$ invertible over $\FF_2$, and $D$ is diagonal in $\mathcal C^{(m)}$, $m\ge1$.
\begin{enumerate}[label=(\alph*)]
\item $U\mathcal C$ is $X$-monomial of level at most $\max(\ell,m-1)$. Its $Z$-checks are $\pm Z(A^{-T}v)$, and its other generators are
\[
UG_cU^\dagger=X(Ac)\,E_c\,D'_c,\qquad D'_c=LD_cL^\dagger\in\mathcal C^{(\ell)},\qquad E_c=X(Ac)^\dagger D\,X(Ac)\,D^\dagger\in\mathcal C^{(m-1)},
\]
with $D'_c$ and $E_c$ diagonal.
\item Let a group $\Gamma$ act freely by qubit permutations and permute the $G_c$ and the $Z(v)$. Every descended operator of
Definition~\ref{def:descent} has its diagonal part in $\mathcal C^{(\ell)}$, and the descended $Z$-checks are Pauli. So a dressed quotient of $\mathcal C$,
whenever the descended operators commute on the descended $P_V$, is $X$-monomial of level at most $\ell$.
\item Consequently, conjugating $\mathcal C$ by $U$ on a region (by the product $U_R=D_RL_R$ of the gates of $U$ in the region $R$) and then
descending along a free permutation action gives $X$-type checks of level at most $\max(\ell,m-1)$ and Pauli $Z$-type checks. The dressed
quotients $\mathcal C_0/\Gamma'$ of Theorem~\ref{thm:dressedquot}, with $U(\tilde\eta)$ of level $m$, are the case $L=1$ and $\ell=1$.
\end{enumerate}
In particular, twisting by $U=DL$ with $D\in\mathcal C^{(\ell+1)}$ never raises the level, and Proposition~\ref{prop:clifflevel}(b) is the case $\ell\le1$, $L=1$.
\end{proposition}
\idea A $CNOT$ circuit only relabels the computational basis, so it keeps every level. The diagonal part $D$ contributes, as in
Proposition~\ref{prop:clifflevel}, only its discrete derivative, one level lower. Descent identifies variables, which can merge monomials but
never creates one of higher degree.
\movedproof{prop:chslevel}

Three remarks delimit the statement. Only $X$-type checks carry the higher level: the $Z$-type checks stay Pauli under every twist of this
form. A twist containing Hadamard gates, such as the electromagnetic dualities of Section~\ref{sec:general}, leaves the class altogether, because
$HD_cH$ is not diagonal when $D_c$ is not Pauli; so duality quotients need Pauli parents. And the bound concerns checks only: the logical
$X$-type operators of a code of level $\ell\ge2$ are dressed as well, which is why Proposition~\ref{prop:level} does not apply
(Section~\ref{sec:autdressed}).

\begin{example}[The codes of Kobayashi, Zhu and Hsin as twisted compactifications]\label{ex:kzh}
Take $N$ one-form $\Z_2$ fields and $\omega=a_1\cup\dots\cup a_N$.
\begin{enumerate}[label=(\roman*)]
\item On a closed $(N-1)$-dimensional complex, $|\omega|=D+1$, the case $q=-1$, $B=1$ of Section~\ref{sec:higher}. The Gauss laws of
Definition~\ref{def:gauss} are $G_{i,v}=X_i(\delta v)\,(-1)^{\int a_1\cup\dots\cup v\cup\dots\cup a_N}$, with $v$ in slot $i$, and
Lemma~\ref{lem:cochain}(ii)--(iv) and Proposition~\ref{prop:twisted-code}(a),(b) hold (Appendix~\ref{app:tech-lattice}). With the
cubical cup~\eqref{eq:cubicalcup} on the square lattice, $N=3$ gives exactly the $X$-checks of the $D_4$ code of~\cite{KZH2025chs}, and the flux
checks are its $Z$-checks. Its level is $N-1$.
\item The same phases arise as twisted compactifications in the sense of this paper. Take $N$ toric codes in $N$ dimensions and a wall ($q=0$)
carrying the pump $U_\omega=(-1)^{\int a_1\cup\dots\cup a_N}$, a constant-depth circuit of $C^{N-1}Z$ gates, of level $N$. On $X\times S^1$ the
dressed code is the dressed quotient $\mathcal C_0/\langle t\circ U_\omega\rangle$ (Theorem~\ref{thm:dressedquot}(c)). On the thin slab $X\times[0,1]$
with relative $B$ (Corollary~\ref{cor:boundary}) it is a code on $X$ with the same ground-space dimension, 22 on $T^2$ for $N=3$, which is
the dimension of $D(D_4)$ (Theorem~\ref{thm:pointcount}). It is the code of (i) up to a constant-depth circuit, which removes 9 of its 15
qubits per site (Corollary~\ref{cor:thinslab}). Propositions~\ref{prop:clifflevel} and~\ref{prop:chslevel} give
level $\max(1,N-1)=N-1$, as for the codes of (i).
\item Kobayashi, Zhu and Hsin's automorphism gates have the form $U=W\,V$ of Proposition~\ref{prop:chslevel}, with $V$ a transversal $CNOT$
circuit and $W=e^{i\int\alpha}$ a diagonal circuit whose phase polynomial contains $\frac14\tilde a_2\cdots\tilde a_N$~\cite[eq.~(47)]{KZH2025chs}; so
$W\in\mathcal C_N$ and $m=N=\ell+1$. Twisting a code of (i) or (ii) by its own automorphism gate, for example by $U$ restricted to a region, keeps every
check at level $N-1$. The logical action of the gate, by contrast, reaches level $N$ with suitable boundaries~\cite{KZH2025chs}; its bulk part is
the descent of $V$ described in Section~\ref{sec:autdressed}.
\end{enumerate}
\end{example}

\subsection{Lens spaces and $\mathbb{RP}^2$ as internal spaces}
\label{sec:lens}

On a lens space or on $\mathbb{RP}^2$ every background class is torsion, and the universal cover, a sphere, trivializes it. This subsection
carries out steps 1--3 of the method for these internal spaces.

For $F$ a lens space $L(m;r)$ or $\mathbb{RP}^2$, a connected finite cover trivializes a class $B_F\in H^j(F;\FF_p)$ exactly when its degree is
divisible by $p$, except for $j=2$ on a lens space, where it must be divisible by the largest power of $p$ that divides $m$. For $m=p$ the
universal cover, a sphere, trivializes every class: the generator $x$ of $H^1$ becomes the coboundary of the sheet index, and a deck
transformation shifts the sheet index by one, so the twisted deck operator for $B_F=x$ is the deck transformation composed with a global
symmetry. The counts follow from the K\"unneth formula and the cohomology rings of these spaces. Proposition~\ref{prop:lens} (Appendix~\ref{app:tech-lattice})
states all of this precisely, with a table of the counts.

\begin{example}[Two toric codes on a lens space]\label{ex:lens}
For two 3D toric codes ($n_1=n_2=1$) with $\omega=a_1\cup a_2$ on $L(p;r)$ and $B=\beta x$, Proposition~\ref{prop:lens}(d) and Theorem~\ref{thm:hfcount}
give $k'=0$, against $k=2$ without the twist, since $x\cup\beta x$ generates $H^3$. By Proposition~\ref{prop:lens}(c), $\Gamma'$ and the bare deck group act identically
on the parent code space on $S^3$, which has no logical qudits; the two quotients, with $k'=0$ and $k=2$, differ only in how they identify
stabilizers. On $S^1\times\mathbb{RP}^2$ with $B=x^2$ the same pair keeps $k'=2$ of its 4 logical qubits: the class $x$ survives in each copy, and the
class $t$ of the circle, with $t\cup x^2\ne0$, is removed.
\end{example}

\subsection{Distance of dressed codes}
\label{sec:dresseddist}

For quadratic $\omega$ the dressed code is a Pauli stabilizer code, and Section~\ref{sec:general} bounds its distance
(Remark~\ref{rem:carry}). For cubic and higher $\omega$ there is no Pauli stabilizer group, and the counts above give only the dimension. The
following theorem gives the distance in that case. It says that a diagonal dressing of the $X$-type checks never creates a light logical
operator: a code is protected as soon as the undressed toric codes underneath it are, on regions of the relevant size. It applies to
operators that are not Pauli, and it needs no cover.

Call the code with the same $Z$-checks and the undressed $X$-type checks $X(h_v)$ the \emph{CSS shadow} of an $X$-monomial code.

\begin{theorem}[Distance of dressed codes]\label{thm:dresseddist}
Let $\tilde{\mathcal C}\ne0$ be the common $+1$ eigenspace of Pauli operators $Z(g)$, $g\in G$, and of unitaries $\tilde X_v=X(h_v)\,D_v$, $v\in V$, on
$\Z_p$ qudits, where each $D_v$ is diagonal and $\langle g,h_v\rangle=0$ for all $g,v$. Let $P$ be its projector, and let $H_Z$ and $H_X$ be the matrices with rows
$g$ and $h_v$.
\begin{enumerate}[label=(\alph*)]
\item (Local criterion.) Let $O$ act on a set $A$ of qudits. Suppose that every $x\in\ker H_Z$ supported in $A$ can be written $x=\sum_vt_vh_v$ with
$t_v\ne0$ only for $v$ in a set $V_x$, such that every $z\in\ker H_X$ supported in $A_x=A\cup\bigcup_{v\in V_x}\mathrm{supp}\,D_v$ lies in the row space of $H_Z$.
Then $POP=c(O)\,P$ for a scalar $c(O)$.
\item (Cubical complexes.) Let $M=(\Z^d\times F)/\Gamma$, with $F$ a point or an interval $[0,h]$ and $\Gamma$ acting freely by cellular
automorphisms, and suppose that the quotient map is injective on every box $[a,a+s]^d\times F$. Let the qudits be fields of degrees
$1\le n_i\le d-1$ on $M$, let the $Z$-checks be the flux checks, and let the $X$-type checks be $X_i(\delta e)\,D_{i,e}$, with $D_{i,e}$ diagonal and supported on
the faces of the $D$-cells that contain $e$. Then $POP=c(O)P$ for every operator $O$ supported on at most $(s-2)/10$ qudits, so the distance of
$\tilde{\mathcal C}$ exceeds $(s-2)/10$.
\end{enumerate}
\end{theorem}
\idea Expand $O$ in $X$-strings times diagonal operators. An $X$-string that preserves the flux-free subspace is, on a small region, a product
of undressed Gauss-law strings $X(h_v)$, and each of these equals a code stabilizer $\tilde X_v$ times the inverse of its dressing. So on the
code space the $X$-string acts as a diagonal operator, supported a little beyond $A$. Expand that operator in $Z$-strings: each either
anticommutes with some $\tilde X_v$, and so vanishes between code states, or is a cycle of the shadow, and on a small region a cycle is a product
of flux checks, which act as the identity. In (b) the two fillings are local because a box has no cohomology relative to its sides and no
homology, and a light operator splits into pieces that each fit in a box.
\movedproof{thm:dresseddist}

Part (b) is the analogue of Proposition~\ref{prop:injdist} for codes that are not Pauli stabilizer codes; that proposition handles any Pauli
parent, and this theorem handles any diagonal dressing of a CSS shadow. The constant $10$ is not optimized. Distances of twisted codes on product and sheaf complexes are
studied in~\cite{ZKH2026nonabelian,ZTKH2026sheaf}; Theorem~\ref{thm:dresseddist} is a local statement, for all operators, that covers the
geometrically local dressed codes of this paper.

\begin{corollary}[Distances of the dressed codes]\label{cor:dresseddist}
Let $\mathcal C_B$ be a dressed code (Definition~\ref{def:gauss}) on the torus $T^d_L$, or on the slab $T^d_L\times[0,h]$ with relative $B$
(Corollary~\ref{cor:boundary}), with fields of degrees $1\le n_i\le d-1$, any cup polynomial $\omega$ of any order, and any background $B$, including
$B=1$ and classes with no trivializing cover, and assume $\mathcal C_B\neq0$, which can fail (Proposition~\ref{prop:twisted-code}). Then the distance of $\mathcal C_B$ exceeds $(L-3)/10$. By Theorem~\ref{thm:dressedquot}(b) the same
holds for the dressed quotients $\mathcal C_0/\Gamma'$ of Theorem~\ref{thm:dressedquot}. In particular, a $Z$-loop of length $L$ gives the matching
upper bound in the following cases, where $\code{n,\log_2K,d}$ denotes a code of dimension $K$ on $n$ qubits, $K$ not necessarily a
power of two:
\begin{enumerate}[label=(\roman*)]
\item type III on the slab $T^2_L\times[0,1]$, the code of $D(D_4)$: $\code{15L^2,\log_222,d}$ with $(L-3)/10<d\le L$;
\item $b\cup a_1\cup a_2$ on $T^3_L\times[0,1]$, the 3D $D_4$ gauge theory: $\code{23L^3,\log_292,d}$ with $(L-3)/10<d\le L$;
\item type IV on $T^3_L\times[0,1]$: $\code{28L^3,\log_21576,d}$ with $(L-3)/10<d\le L$;
\item the twisted $\Z_2^N$ codes of~\cite{KZH2025chs} on $T^{N-1}_L$ with $N\ge3$ (Example~\ref{ex:kzh}(i), $B=1$): $d=\Theta(L)$.
\end{enumerate}
\end{corollary}
\idea A torus or a slab of side $L$ looks like flat space on boxes of side $L-1$, so part (b) of the theorem applies; a straight $Z$-loop
of length $L$ is a logical operator. \movedproof{cor:dresseddist}

\begin{remark}[What the theorem does not give]\label{rem:dresseddist}
The theorem shows that the dressing never lowers the protection that the shadow has on small regions; it cannot show that the twist
removes light logical operators that the shadow already has. Those arise for fields of degree $0$ or $d$, as in the circle examples, where the
shadow contains repetition codes, and for a small closed internal space $F$, where the shadow has logical operators that wrap $F$. For
type III on $T^2_L\times S^1_{L_z}$ with $B$ dual to $T^2$, for example, the $Z$-loops around the circle remain logical ($176=22\cdot2^3$), so $d\le L_z$.
The distance of dressed codes on lens spaces and $\mathbb{RP}^2$ with growing $X$ (open problem 1) therefore still needs an argument that uses
the dressing.
\end{remark}

\section{Symmetries of the quotient and logical gates}
\label{sec:sym}

The correspondence of Sections~\ref{sec:flatlat} and~\ref{sec:higher} translates backgrounds but says nothing about the symmetries of the compactified theory. In field
theory, symmetries of the internal space and of the parent that preserve the background become symmetries of the compactified theory and
act on the operators wrapped on $F$ (Section~\ref{sec:symft}). This section gives the lattice version, step 4 of the method: which
automorphisms of the parent descend to the quotient, how they act on logicals, and which of them can be non-Clifford. The general
principles are in Sections~\ref{sec:descent}--\ref{sec:twistgates}; the later subsections apply them to particular parents, and
Appendix~\ref{app:tech-sym} collects the technical complements.

\paragraph{The principle for symmetries.} The principle of Section~\ref{sec:method} applies to operators as well as to stabilizers. A parent
operator acts on the quotient when it is compatible with the identification by $\Gamma'$, and its logical action is read off on the logical
states of the quotient, which the twist has changed. Compatibility comes in three degrees, and the section follows them.
\begin{enumerate}[leftmargin=2em,label=(\alph*)]
\item \emph{The operator commutes with the twist.} A parent automorphism that normalizes $\Gamma'$ descends as the same local circuit
(Section~\ref{sec:descent}). Here the twist \emph{selects} the parent's gates that preserve the background; it does not create isometry gates,
and descended monomial gates stay Clifford (Section~\ref{sec:cliffordlevel}). Whether a symmetry of the quotient arises in this way depends on
the parent (Sections~\ref{sec:genon} and~\ref{sec:accidental}).
\item \emph{The operator is a cup-product circuit.} It is applied on the quotient by the same local formula, and it acts on each logical state
by the response $\omega$ evaluated on that state. The twist changes which logical states exist, so the same circuit can be trivial on the parent
and a non-trivial gate on the quotient: the twist \emph{creates} the gate (Sections~\ref{sec:twistgates}, \ref{sec:threegroup}
and~\ref{sec:toritwist}).
\item \emph{The operator commutes with the twist only up to a coboundary.} It descends after a dressing supported on the wall
(Section~\ref{sec:autdressed}), and on manifolds with boundary after boundary terms (Section~\ref{sec:bdrytrans}).
\end{enumerate}
Non-invertible parent symmetries are treated in Section~\ref{sec:noninv}. The simplest instance of (a) is the example after
Proposition~\ref{prop:descent}, of (b) the logical $CCZ$ on $\mathbb{RP}^3$ after Proposition~\ref{prop:twisted}, and of (c)
Example~\ref{ex:kzhgate}.

\paragraph{Setup.} $C$ is a parent code with Pauli stabilizer group $S$ on a qubit set $Q$. A monomial Clifford is a qubit permutation composed with
single-qubit Cliffords, and $\mathrm{Aut}(C)$ is the group of monomial Cliffords \emph{modulo Pauli operators} whose action on $\FF_2^{2n}$ preserves $S$.
Counting modulo Paulis is harmless, since two sign choices on $S$ differ by a character of $S$, which is conjugation by a Pauli; so every element
of $\mathrm{Aut}(C)$ lifts to a unitary preserving the code space, unique up to a Pauli. For qubits $\mathrm{Aut}(C)$ is a subgroup of $S_3\wr S_n$, and every
order quoted below (8 for the genon code, 80 and 360 in the running example) counts it this way, not physical unitaries. $\tilde\Gamma\subset\mathrm{Aut}(C)$ is the free group of
deck-times-symmetry operators of Sections~\ref{sec:setup}--\ref{sec:flatlat}, and $C/\tilde\Gamma$ is the quotient. Appendix~\ref{app:tech-sym} says which parent and which group we use for higher-form and cup-product twists.

\subsection{Descent of symmetries}
\label{sec:descent}

\begin{proposition}[Descent]\label{prop:descent}
Every $\varphi\in\mathrm{Aut}(C)$ with $\varphi\tilde\Gamma\varphi^{-1}=\tilde\Gamma$ induces an automorphism $\bar\varphi$ of $C/\tilde\Gamma$, acting on orbits by
$[q]\mapsto[\varphi(q)]$ with its local Cliffords transported along orbit representatives. This gives a homomorphism
$N_{\mathrm{Aut}(C)}(\tilde\Gamma)/\tilde\Gamma\to\mathrm{Aut}(C/\tilde\Gamma)$.
\end{proposition}
\idea A symmetry that carries each identification to another identification maps orbits to orbits, and so acts on the
quotient.
\begin{proof}
Quotient qubits are $\tilde\Gamma$-orbits, and a quotient check is the push-forward of a parent check along the identifications $P_q\sim g(P)_{g(q)}$. If
$\varphi g\varphi^{-1}=g'\in\tilde\Gamma$, then $\varphi$ maps orbits to orbits and intertwines the identification by $g$ with the identification by $g'$, so it maps
pushed-forward checks to pushed-forward checks. Elements of $\tilde\Gamma$ act trivially on orbits.
\end{proof}

The map need not be onto: a quotient can have accidental symmetries. In the worked example below it is onto, and
Section~\ref{sec:accidental} gives a criterion for the geometric part.

\begin{exampleu}[What descends to the Klein bottle]
For the reflection quotient of the toric code (the example after Definition~\ref{def:quot}), $\tilde\Gamma=\{1,g\}$ with $g(x,y)=(x+L,-y)$. A
translation $T_{(s,0)}$ along the glide commutes with $g$, and so does the reflection $r(x,y)=(x,-y)$, so both descend to the Klein bottle. A
translation $T_{(0,s)}$ across the glide satisfies $T_{(0,s)}\,g\,T_{(0,s)}^{-1}=T_{(0,2s)}\,g$, which lies in $\tilde\Gamma$ only if $2s\equiv0\pmod{L'}$. So
the twist removes the small translations in the $y$-direction, which the parent has and the Klein bottle does not, and keeps the others. For
the translation quotient, by contrast, every parent translation commutes with $t$ and descends.
\end{exampleu}

\paragraph{Three layers.} Descended gates come in three layers: isometries of $(F,\rho)$ combined with internal symmetries, which are qudit
permutations; mapping classes, which act on the logical space as logical Cliffords through the homology of $F$ with twisted coefficients
(Proposition~\ref{prop:logaction}); and the phases of their lifts, fixed by a quadratic refinement. Appendix~\ref{app:tech-sym} discusses the
cost of permutation gates on hardware, the action of mapping classes, and the role of translations.

\begin{corollary}[Reflection-type duality quotients]\label{cor:reflection}
Let $C$ be a two-block group-algebra code over an abelian group $G$, quotiented by $D:g\mapsto t-g$ (block swap and Hadamard). A translation
$T_u$ descends if and only if $2u=0$.
\end{corollary}
\begin{proof}
$T_uDT_{-u}=T_{2u}D$.
\end{proof}

So genon orbifolds keep only the 2-torsion translations: the four genons sit at the fixed points of the $\pi$-rotation, and $G[2]$ permutes
them.

\begin{proposition}[Cup-product gates on every quotient]\label{prop:cupgate}
Let $r$ toric codes with fields $a_i\in C^{n_i}$ live on a closed cubical complex $M$ with a cup structure and $\int\delta(\cdot)=0$, or on a
simplicial complex with a branching, and let $\omega$ be a cup polynomial with $|\omega|=D$. The diagonal circuit $U_\omega=\varpi^{\int\omega(a)}$, a product of gates on the qudits of single cubes, preserves the
code space and acts on the logical state $|\alpha\rangle$, $\alpha\in\bigoplus_iH^{n_i}(M)$, by $\varpi^{\int\omega(\alpha)}$. If $\tilde M\to M$ is a finite regular cover
with deck group $\Gamma$ and $\tilde M$ carries the pulled-back cup structure (equivalently, a $\Gamma$-invariant one), the circuits on $\tilde M$ and on $M$
are given by the same local formula: the circuit on $\tilde M$ is $\Gamma$-invariant, and taking one factor for each $\Gamma$-orbit of cubes gives the
circuit on $M$.
\end{proposition}
\idea A Gauss law shifts a field by a coboundary, and by the Leibniz rule this changes $\int\omega(a)$ only by fluxes, so $U_\omega$
commutes with the checks on flux-free states. On a code state the phase depends only on the cohomology class, which gives the logical action.
On a cover the same local formula is invariant under the deck group, so the circuit downstairs is the same formula.
\movedproof{prop:cupgate}

This is the precise sense in which a cup-product gate ``descends'': not by Proposition~\ref{prop:descent}, which concerns monomial-Clifford
automorphisms, and not as the push-forward $\varpi^{h(\pi^*a)}$ of Definition~\ref{def:descent}, which would take all $|\Gamma|$ factors of each orbit,
but as the same local formula applied on the quotient, with its own proof of code preservation. The two differ by the cover degree:

\begin{corollary}[Anomaly matching under covers]\label{cor:covers}
For a finite cover $\pi:\tilde X\to X$ of closed manifolds, $\langle\pi^*(a\cup b\cup c),[\tilde X]\rangle=\deg\pi\,\langle a\cup b\cup c,[X]\rangle$. So the parent circuit, evaluated on pulled-back
classes, gives $\deg\pi$ times the cubic form of the quotient, which vanishes modulo 2 when $\deg\pi$ is even.
\end{corollary}
\begin{proof}
$\pi_*[\tilde X]=\deg\pi\,[X]$ and naturality of the cup product.
\end{proof}

So a cubic gate pulled back along a cover survives only when the degree of the cover is odd. This constrains the lifting of cup-product gates
along covers, as in the constructions of transversal non-Clifford gates on good and almost-good codes~\cite{Lin2024sheaves,LLL2026transversal,LSWLL2026invariants}.

In twisted compactification the logical Paulis are the wrapped higher-form symmetry operators of the parent. Their commutator is a
mixed 't~Hooft anomaly of the compactified symmetries, and the code space is the smallest representation of this anomaly, so $k$ is
half the $\log_p$ of the order of the anomalous group (Corollary~\ref{cor:anomaly} in Appendix~\ref{app:tech-sym}).

\subsection{Higher groups and the Clifford level}
\label{sec:cliffordlevel}

\begin{proposition}[Logical level at most physical level]\label{prop:level}
Let $C$ be a Pauli stabilizer code and $U$ a physical unitary at level $\ell$ of the Clifford hierarchy with $UC=C$. Then the logical action of $U$ is at
level at most $\ell$ of the logical Clifford hierarchy.
\end{proposition}
\idea Conjugating a gate of level $\ell$ by a Pauli gives a gate of level $\ell-1$, and on the code space Paulis act as logical Paulis;
induct on $\ell$.
\begin{proof}
Induction on $\ell$. A Pauli that preserves $C$ commutes with every stabilizer, since one it anticommutes with would map $C$ to its orthogonal
complement; so it is a logical Pauli. For $\ell\ge2$, every logical Pauli $\bar P$ has a physical Pauli representative $P$, and $UPU^\dagger$ is a
physical level-$(\ell-1)$ operator preserving $C$. By induction its logical action is at level at most $\ell-1$, so $U$'s is at level at most $\ell$.
\end{proof}

The statement is folklore, and the same induction underlies the bounds of~\cite{BravyiKoenig2013,PastawskiYoshida2015}; we include it because
the dressed codes below need its hypothesis spelled out.

The proof needs logical Paulis with physical Pauli representatives, so it does not cover dressed codes, whose logical $X$-type operators
are dressed as well (Section~\ref{sec:chsparent}); Section~\ref{sec:autdressed} discusses the gates of such codes.

\begin{corollary}[Clifford twists stay Clifford]\label{cor:cliffordstay}
Every quotient by a monomial-Clifford $\tilde\Gamma$ is a Pauli stabilizer code, and every automorphism gate of Proposition~\ref{prop:descent} is physically
Clifford. So every descended isometry, mapping-class or duality gate is a logical Clifford, whatever higher group it sits in.
\end{corollary}
\begin{proof}
Proposition~\ref{prop:valid} and Proposition~\ref{prop:level} with $\ell=2$.
\end{proof}

In the higher groups that isometries and large diffeomorphisms produce~\cite{Gukov:2020btk,GukovHsinPei2025}, an isometry junction emits a
wrapped higher-form operator on a lower-dimensional cycle, that is, a logical Pauli or a stabilizer. On the code, $U_gU_h=U_{gh}P(g,h)$ and
$U_gPU_g^{-1}=P'$ with $P,P'$ logical Paulis: the Postnikov data is the extension class of the descended gate group by the Pauli group, not a
higher Clifford level. For 2D-local quotients that are Pauli stabilizer codes, Bravyi and K\"onig give the same conclusion for any
locality-preserving gate~\cite{BravyiKoenig2013}; the dressed 2D codes of Example~\ref{ex:kzh} evade it~\cite{KZH2025chs,HsinKobayashi2025aut}
(Section~\ref{sec:autdressed}).

\subsection{Gates created by the twist}
\label{sec:twistgates}

Descent (Proposition~\ref{prop:descent}) carries a parent symmetry to the quotient as the same local circuit, but the logical gate it gives need
not resemble the parent's. The twist changes which logical states exist, and the descended circuit acts on whatever logical states the
quotient has. A parent symmetry that acts trivially on the parent code space, or a parent with no logical qubits at all, can therefore give a
non-trivial logical gate on the quotient. This is how the method produces gates that the parent does not have.

\paragraph{The principle for gates.} A twist creates a logical gate by creating the logical states on which a parent symmetry acts
non-trivially. For a cup-product symmetry $U_\omega$ these states are labeled by flat backgrounds $\chi$ on the quotient $M$, and the gate is the
response $\langle\chi^*\omega,[M]\rangle$ evaluated on them. The twist enters only through which $\chi$ exist and through the class $[M]$. A twist can also create the
gate itself: by identifying the inputs of an $r$-copy gate, it allows the phase to be lifted to finer roots of unity, which no circuit on
$r$ independent copies can do (Section~\ref{sec:toritwist}).

Table~\ref{tab:create} lists the mechanisms in this paper, with the statement behind each row. In each row except the one on lifted
coefficients, the circuit is the same before and after the twist; only the set of states it acts on changes. This subsection contains the
first three rows and the first and last limits.

\begin{table}[ht]
\centering\small
\begin{tabularx}{\textwidth}{@{}>{\raggedright}p{3.4cm}>{\raggedright}p{4.0cm}>{\raggedright}X>{\raggedright\arraybackslash}p{2.4cm}@{}}
\toprule
Mechanism & Parent, and the gate there & Gate on the quotient & Where \\
\midrule
twisted sectors of the deck group & three 3D toric codes on $S^3$: no logical qubits & logical $CCZ$ on $\mathbb{RP}^3=S^3/\Z_2$ & Prop.~\ref{prop:twisted}, check T6 \\
the same, fermionic pump & $\Z_8$ pump on $S^3$: no logical qubits & logical $T$ on $\mathbb{RP}^3$~\cite{Barkeshli:2023bta}; $CS$ on $S^3/Q_8$ from~\cite[eq.~(26)]{Barkeshli:2023bta}; no pump known on either & Prop.~\ref{prop:twisted}, Sec.~\ref{sec:fermionic} \\
a twist that also permutes the copies & three toric codes on $T^3$; on the tricosm by a screw alone the gate is the identity & $CZ\,CZ\,CCZ$ on the tricosm by the screw composed with the cyclic permutation & Prop.~\ref{prop:internal}, check T7 \\
lifted coefficients after a copy-permuting twist & $r$ independent toric codes: multilinear phases only ($C^{r-1}Z$); the lifted circuit is not defined & one copy: qutrit $T$ on $T^3$ ($r=3$); qubit $T$, $CS$, $CCZ$ on $T^4$ ($r=4$)~\cite{BDEHT2026cupsII} & Prop.~\ref{prop:tori}, checks T9, T10 \\
a deep-hole holonomy & one qutrit toric code on $S^1\times T^2_{\rm hex}$ with the gate of the previous row: no $T$ & $T_1T_2T_3\,CCZ^{-1}$ on the cubic $T^3$, its quotient & Ex.~\ref{ex:deephole}, checks T9, T10 \\
an orientation-reversing holonomy & 3D toric code on $T^3$: the gauged-SPT symmetry acts trivially & logical $CZ$ on $K\times S^1$ & Prop.~\ref{prop:threegroup}, Ex.~\ref{ex:klein}, check T8 \\
a rotation holonomy, fermionic pump & fermionic toric code on $T^3$: $V=CCZ$, $V^2=1$ & $CCZ\,CS^\dagger$ on $T^2\rtimes_{C_2}S^1$~\cite{Barkeshli:2023bta}; $T^{\pm1}CZ$ on $T^2\rtimes_{C_4}S^1$, distance linear in the size & Props.~\ref{prop:gaussdescent}, \ref{prop:injdist}, Ex.~\ref{ex:dicosm}, Prop.~\ref{prop:quarter} \\
compactifying a non-invertible symmetry & not a unitary: its fusion has several outcomes & a logical unitary when the outcomes act as scalars & Lemma~\ref{lem:noninv} \\
\midrule
\multicolumn{4}{@{}l}{\emph{Limits}} \\
sector map through a space with $H_3=0$ & any & the identity for a bosonic response; for example the cup-product response with SWAP holonomies on $T^3$ & Cor.~\ref{cor:rank} \\
monomial twist and monomial symmetry & any Pauli stabilizer parent & a logical Clifford & Cor.~\ref{cor:cliffordstay} \\
small internal space & any & controlled by logicals no longer than the shortest essential fibre cycle & Prop.~\ref{prop:kk} \\
\bottomrule
\end{tabularx}
\caption{How a twist creates logical gates, and what limits it. Except in the row on lifted coefficients, the circuit is the same local
circuit on the parent and on the quotient; for the fermionic pump it is a finite-time quasi-local evolution.}
\label{tab:create}
\end{table}

\paragraph{What does give non-Clifford gates.} In the constructions of this paper, a parent 0-form symmetry $U_\omega$ that pumps a response $\omega$ of
degree at least 3, applied on the quotient by Proposition~\ref{prop:cupgate}: the cup
product $a_1\cup a_2\cup a_3$ for three toric codes, or the fermionic $\Z_8$ pump~\cite{Barkeshli:2022edm,Barkeshli:2023bta}. On a quotient $M$ it acts
on the logical state labeled by a flat background $\chi$ by $\exp2\pi i\langle\chi^*\omega,[M]\rangle$, so its level is the degree of
$\chi\mapsto\langle\chi^*\omega,[M]\rangle$ on $H^1$. Three statements say how the twist controls it.

\begin{proposition}[Twisted sectors]\label{prop:twisted}
Let $M$ be a closed 3-manifold carrying the codes of Proposition~\ref{prop:cupgate} with a cubic $\omega$, read as a class in
$H^3(BG;\Z_p)$, and let $[M]\in H_3(M;\Z_p)$ be its fundamental class, which exists for $p=2$ and, for odd $p$, when $M$ is oriented. Let
$c:M\to B\Gamma$ classify the cover $\tilde M\to M$, and let $\chi=\phi\circ c$ for a homomorphism $\phi:\Gamma\to G$. Then
\[
U_\omega|\chi\rangle=\varpi^{\langle(B\phi)^*\omega,\ c_*[M]\rangle}\,|\chi\rangle,\qquad c_*[M]\in H_3(B\Gamma;\Z_p).
\]
For a fermionic pump on a spin manifold the same holds with $c_*[M]\in\Omega_3^{\rm Spin}(B\Gamma)$.
\end{proposition}
\begin{proof}
$\chi^*\omega=c^*(B\phi)^*\omega$, and $\langle c^*x,[M]\rangle=\langle x,c_*[M]\rangle$.
\end{proof}

So the gate on twisted-sector logicals is a bordism invariant of the deck group, independent of the parent, which may have no logical qubits at
all. For $S^3\to\mathbb{RP}^3$ with the cup product this gives a logical $CCZ$ across three copies (check T6 of Appendix~\ref{app:checks}). With the fermionic pump,
$\Omega_3^{\rm Spin}(B\Z_2)=\Z_8$ is generated by $\mathbb{RP}^3$, which explains the logical $T$ gate of~\cite{Barkeshli:2023bta} there.
Rotation twists of the same pump are treated in Section~\ref{sec:fermionic}.

The same formula limits the gates: the phase depends on the sector only through the class it pushes forward, which vanishes whenever the
sector factors through a space of lower dimension.

\begin{corollary}[Rank obstruction]\label{cor:rank}
In the setting of Proposition~\ref{prop:twisted}, with $\omega$ bosonic, suppose the sector map $\chi:M\to BG$ is homotopic to a composite $M\to W\to BG$
with $H_3(W;\Z_p)=0$. Then $U_\omega|\chi\rangle=|\chi\rangle$. In particular, on $M=T^3$ every sector whose holonomies generate a subgroup of $G$ with at
most two generators is fixed, and if all sectors are of this kind, $U_\omega$ acts as the logical identity.
\end{corollary}
\begin{proof}
$\chi_*[M]$ is the image of a class in $H_3(W;\Z_p)=0$. On $T^3$ the holonomies commute, so they generate an abelian group $A$, a quotient of $\Z^2$
when it has at most two generators. The homomorphism $\Z^3\to A$ lifts to $\Z^3\to\Z^2$, whose kernel contains a primitive vector $v$. After a change
of basis of $\Z^3$, realized by a homeomorphism of $T^3$, the homomorphism factors through $\Z^3/\langle v\rangle\cong\Z^2$. Since $T^3$ and $T^2$ are
Eilenberg--MacLane spaces, $\chi$ is homotopic to the projection $T^3\to T^2$ followed by a map $T^2\to BA\to BG$, and $H_3(T^2;\Z_p)=0$.
\end{proof}

The argument applies whenever the logical states are labeled by flat $G$-bundles and the gate acts on them through a class in
$H^3(BG;U(1))$. For example, with a SWAP holonomy between two toric codes the logical states are labeled by flat bundles for
$D_4=(\Z_2\times\Z_2)\rtimes\Z_2$. Every abelian subgroup of $D_4$ has at most two generators, so on $T^3$ such a gate is trivial for every
arrangement of SWAP walls. This concerns the example, whose commuting holonomy images have at most two generators; it is not a general
need for non-abelian holonomy, since an abelian image with three generators can carry a cubic response, as for the transversal $CCZ$ of
three toric codes on $T^3$. Escaping the obstruction takes a different sector topology, such as a non-abelian $\pi_1(M)$, as for the
quarter-turn space $T^3/\Z_4$, or a non-abelian orbifold fundamental group, as for a copy permutation combined with an isometry that has
fixed points (Proposition~\ref{prop:internal}; for $T^3/C_3$ it is $\Z^3\rtimes C_3$), a larger holonomy image, or boundaries. 

\begin{proposition}[Internal twists]\label{prop:internal}
Let $r$ copies of the $\Z_p$ toric code with $n$-form fields live on a closed cubical complex $\tilde M$ of dimension $rn$ with a cup structure and
$\int\delta(\cdot)=0$, or on a simplicial complex with a branching. Let $R$ be a cellular automorphism of $\tilde M$ that preserves the cup
structure, such that $\langle R^r\rangle$ acts freely on cells (and preserves the orientation if $p$ is odd). Let $\tilde\Gamma$ be generated by the
permutation $(i,c)\mapsto(i-1,Rc)$ of qudits, which moves copy $i$ at the cell $c$ to copy $i-1$ at $Rc$, indices modulo $r$. Put
$\hat M=\tilde M/\langle R^r\rangle$, with projection $\pi$, let $\hat R$ be the automorphism of $\hat M$ induced by $R$, and let $M=\hat M/\langle\hat R\rangle$,
an orbifold if $\hat R$ has fixed points. $R$ itself need not act freely.
\begin{enumerate}[label=(\alph*)]
\item $\tilde\Gamma$ acts freely on the qudits, and the quotient code is one toric code on $\hat M$. A configuration $a$ on $\hat M$ corresponds to the
invariant parent configuration $a_{1+j}=\pi^*(\hat R^j)^*a$, $j=0,\dots,r-1$, so the logical classes $H^n(\hat M;\FF_p)$ of the quotient correspond to
the classes $(\alpha,\hat R^*\alpha,\dots,(\hat R^{r-1})^*\alpha)$ of $r$ copies on $\hat M$.
\item Let $V=\varpi^{T(a)}$ with $T(a)=\int_{\hat M}a\cup\hat R^*a\cup\dots\cup(\hat R^{r-1})^*a$. It is the parent circuit $\varpi^{\int a_1\cup\dots\cup a_r}$ evaluated on the
configurations of (a), with one factor for each $\langle R^r\rangle$-orbit of top cells. Each factor acts on the qudits of one top cell of $\hat M$
and of its images under $\hat R$, which lie over one cell of $M$. $V$ preserves the code space and acts on $|\alpha\rangle$ by $\varpi^{T(\alpha)}$, which
is the action of the $r$-copy circuit on the classes of (a).
\end{enumerate}
\end{proposition}
\idea A copy-permuting twist glues copy $i$ to copy $i-1$ moved by $R$, so an invariant configuration is determined by one copy:
the $r$ copies become one code on $\hat M$, and copy $j$ reads the field through $\hat R^j$. The $r$-copy circuit evaluated on such
configurations is the twisted cup product $\int a\cup\hat R^*a\cup\cdots$, and it commutes with the checks by the Leibniz rule, as in
Proposition~\ref{prop:cupgate}.
\movedproof{prop:internal}

So on the same manifold and with the same qubit count, the twist can create the gate or kill it (check T7 of Appendix~\ref{app:checks}).

Proposition~\ref{prop:internal} contains the twisted cup products of~\cite{BDEHT2026cupsII} as instances (Appendix~\ref{app:tech-sym}).

\begin{proposition}[Kaluza--Klein bound for any twist]\label{prop:kk}
Let $F\to M\xrightarrow{\pi}X$ be a fibre bundle of cell complexes with connected fibre and arbitrary monodromy, with $X$ of dimension at most 2,
carrying toric codes with qubits in degree 1, and let $\ell_F$ be the least length of a cycle in a fibre that is non-trivial in
$H_1(M;\FF_p)$. If the cubic form $T$ of a cup-product gate is non-zero on $(a,b,c)$, then one of $a,b,c$ restricts non-trivially to the fibre;
in particular $\ell_F$ is finite, and the $Z$-distance of the code is at most $\ell_F$.
\end{proposition}
\idea A non-zero cubic form needs three classes with non-zero product, and classes pulled back from a base of dimension at most two
multiply to zero. So one of the three is felt along the fibre, and a short cycle in the fibre is then a short logical.
\begin{proof}
$\ker(H^1(M)\to H^1(F))=\pi^*H^1(X)$ by the five-term exact sequence of the Leray--Serre spectral sequence, and $\pi^*x\cup\pi^*y\cup\pi^*z=\pi^*(x\cup y\cup
z)=0$ since $H^3(X)=0$. A class $a$ with $a|_F\neq0$ pairs with some cycle $\gamma\subset F$, which is therefore non-trivial in $H_1(M;\FF_p)$, and the
Wilson line on any fibre cycle that is non-trivial in $H_1(M;\FF_p)$ is a non-trivial $Z$-logical.
\end{proof}

When the internal space is small, the control of every non-Clifford gate sits in its Kaluza--Klein modes, and the distance is capped by
the size of the internal space. So a family with growing distance and a cubic cup-product gate needs a growing fibre or a base of dimension
at least 3.

\subsection{Worked example: the $\code{144,6}$ genon code}
\label{sec:genon}

The parent is the bivariate-bicycle code $\code{288,12,18}$ on $G=\Z_{12}^2$, and the quotient is by the ZX-duality $D_{(0,1)}:g\mapsto(0,1)-g$ of
Corollary~\ref{cor:reflection}, which is free because $2g=(0,1)$ has no solution. The quotient is a $\code{144,6}$ code, the \emph{genon code}; a
separate exhaustive search gives distance 15, which we do not use, while Appendix~\ref{app:checks} proves $d\ge7$. Appendix~\ref{app:tech-sym} describes how the automorphism groups below were computed.

\begin{center}\small
\begin{adjustbox}{max width=\textwidth}\begin{tabular}{@{}lll@{}}
\toprule
& Parent $\code{288,12,18}$ & Quotient $\code{144,6}$ \\
\midrule
affine automorphism group & $1728=144$ translations $\times$ 12 linear parts & 8 $=$ (centralizer of $D$, order 16)$/\langle D\rangle$ \\
monomial-Clifford automorphism group & & 8 (complete), so descent accounts for all of it \\
logical group generated & order 432 & order 2 (a triple SWAP) \\
\bottomrule
\end{tabular}\end{adjustbox}
\end{center}

The survivors are the half-period translations $(6,0)$, $(0,6)$, $(6,6)$, which are logically trivial (Corollary~\ref{cor:reflection}), and the
orientation-reversing fold $\beta=\left(\begin{smallmatrix}3&10\\10&9\end{smallmatrix}\right)$ with $\det\beta\equiv-1$, composed with Hadamard, the same
mechanism as fold-transversal gates~\cite{BreuckmannBurton2024,EberhardtSteffan2024}. By Corollary~\ref{cor:cliffordstay} no descended gate can be
non-Clifford.

\begin{example}[Running example, part 4: accidental symmetries]\label{ex:run4}
The stabilizers of minimum weight of the parent $\code{20,2,4}$ of Example~\ref{ex:run1} are its 20 plaquettes, and they generate $S$, so the
graph method of Appendix~\ref{app:checks} computes its whole monomial-Clifford automorphism group: it has order 80, the translations and the rotations by
multiples of $90^\circ$ about a site, each composed with $H^{\otimes20}$ when it exchanges the two colours. All of them normalize $\Gamma$, and
$N(\Gamma)/\Gamma$, of order 20, descends injectively onto the affine group $j\mapsto2^aj+b$ of $\Z_5$ acting on the five-qubit code; its logical image
has order 2, the identity and a gate exchanging $\bar X$ and $\bar Z$ up to Paulis. The five-qubit code itself has 360 monomial-Clifford
automorphisms (its 15 stabilizers of weight 4 generate it), and their logical image is all of $\mathrm{Sp}(2,\FF_2)\cong S_3$, including the
familiar order-3 transversal gate. So 340 are accidental, and descent is not onto: these additional automorphisms do not descend from this
specified parent action. For the order-3 gate one sees this directly: all automorphisms of this parent, and the transport by $H$, preserve the
letters $\{X,Z\}$ up to exchange, while the order-3 gate maps $X$ to $Y$. Whether some other parent and action produce them is a different
question, which the example does not address.
\end{example}

\begin{table}[ht]
\centering\small
\begin{adjustbox}{max width=\textwidth}\begin{tabular}{@{}lllll@{}}
\toprule
Internal space & Code & Symmetry that descends & Logical group & Check \\
\midrule
square torus (orbifold point $\tau=i$) & $\code{10,2,3}$, $\code{20,2}$, $\code{26,2}$ & $90^\circ$ rotation & order 2 (CNOT or SWAP) & T2 \\
triangular torus ($\tau=e^{i\pi/3}$) & $\code{21,2,3}$, $\code{39,2,4}$ & $60^\circ$ rotation & order 3, entangling & T2 \\
triangular rhombic torus & $\code{27,2}$, $\code{48,2}$ & point group $D_6$ & $GL(2,\FF_2)$, order 6 & T2 \\
Klein quartic, genus 3 & $\code{84,6,4}$ & $PSL(2,7)=\Delta(2,3,7)/K$ & order 168 in $\mathrm{Sp}(6,\FF_2)$ & T3 \\
Hantzsche--Wendt $T^3/\Z_2^2$ & $\code{48,2}$, $\code{162,2}$, $\code{384,2}$ & signed axis permutations & $S_3$ if $4\mid L$, $\Z_3$ if $L=6$ & T4 \\
\bottomrule
\end{tabular}\end{adjustbox}
\caption{Isometry gates of toric codes on quotients (Appendix~\ref{app:checks}). The Klein quartic is a case of
Proposition~\ref{prop:descent} with a non-abelian deck group: the parent is the toric code on the $\{7,3\}$ tiling, its symmetry group is the
triangle group $\Delta(2,3,7)$, and the deck group is a normal surface group $K$. These examples illustrate the dictionary and are not claimed as
new.}
\label{tab:isometry}
\end{table}

\paragraph{Relation to the literature.} Section~\ref{sec:sym} overlaps with the higher-group and refined-Clifford-hierarchy picture
of~\cite{Barkeshli:2022edm}, the gates on $T^3$, $T^2\rtimes_{C_2}S^1$ and $\mathbb{RP}^3$ of~\cite{Barkeshli:2023bta}, cohomology operations and
symmetry~\cite{HKZ2024classifying,BDET2024cups}, and triple intersections on 3-manifolds and mapping tori~\cite{ZhuSikanderPortnoyCrossBrown2023}. Our
contribution is the quotient-code reading: Proposition~\ref{prop:level} as the reason Clifford twists stay Clifford, Proposition~\ref{prop:twisted}
as the deck-group formula, Proposition~\ref{prop:internal} with T7 (a twist creating the gate), and Proposition~\ref{prop:kk}. T6 is the bosonic
analogue of the $\mathbb{RP}^3$ example of~\cite{Barkeshli:2023bta}. We do not claim the gates of T6 and T7 as new: logical $CCZ$ gates from triple
intersections on 3-manifolds and mapping tori are studied in~\cite{ZhuSikanderPortnoyCrossBrown2023,Barkeshli:2023bta}, and fold-transversal gates
in~\cite{BreuckmannBurton2024,EberhardtSteffan2024}; what T6 and T7 add is a check of Propositions~\ref{prop:twisted} and~\ref{prop:internal}.
The twisted cup products and twisted Pontryagin powers of~\cite{BDEHT2026cupsII} are, in this language, copy-cup gates descended through
copy-permuting twists (Proposition~\ref{prop:internal}); the logical $CCZ$ on $T^3$ is also in~\cite{SLK2026chain}. What we add there is the reading
as a quotient, that freeness is not needed, and Proposition~\ref{prop:tori}, which shows that the choice of torus, and so the twisted
compactification, decides whether the lifted gate contains a $T$.
Related genon gates are constructed in~\cite{BurtonGenons2024}; Corollary~\ref{cor:reflection} is stated here as a consequence of
Proposition~\ref{prop:descent}, without a claim of priority.

\subsection{Accidental symmetries and the choice of cover}
\label{sec:accidental}

Example~\ref{ex:run4} shows that descent need not be onto. For homological codes the geometric part of the question has a complete answer,
and the answer shows that whether a symmetry is accidental depends on the parent, not only on the quotient. Here a \emph{homological code} on
a cell complex is the CSS code of its cellular chain complex with qudits in a fixed degree, as for toric codes. A cellular automorphism
induces a permutation automorphism of it, and the quotient of the code on a regular cover by the deck group is the code on the base
(Theorem~\ref{thm:balanced}(b) with a point as fibre).

\begin{proposition}[Lifting criterion]\label{prop:lift}
Let $\pi:\tilde M\to M$ be a finite regular cover of connected cell complexes, with $\Pi=\pi_*\pi_1(\tilde M)$ normal in $\pi_1(M)$ and deck group
$\Gamma\cong\pi_1(M)/\Pi$, and consider homological codes on $\tilde M$ and $M$. For a cellular automorphism $\bar f$ of $M$ write $\bar f_*$ for the
automorphism of $\pi_1(M)$ that it induces after a change of base point, which is defined up to an inner automorphism.
\begin{enumerate}[label=(\alph*)]
\item $\bar f$ is covered by a cellular automorphism $f$ of $\tilde M$ if and only if $\bar f_*(\Pi)=\Pi$. Every such $f$ normalizes $\Gamma$, and the automorphism
of the parent that it induces descends, by Proposition~\ref{prop:descent}, to the automorphism of the quotient induced by $\bar f$.
\item If $\Pi$ is characteristic in $\pi_1(M)$, for instance $\Pi=\ker\big(\pi_1(M)\to H_1(M;\Z_m)\big)$, the cover of mod-$m$ homology, then every cellular
automorphism of $M$ is covered, and every automorphism of the quotient induced by one descends.
\item For a twisted compactification with fibre $Y$ and holonomy $\rho:\Gamma\to G\subseteq\mathrm{Aut}(Y)$, as in Theorem~\ref{thm:balanced}, let $\bar f$ be covered
by $f$ and let $s\in\mathrm{Aut}(Y)$. Then $(f,s)$ normalizes $\Gamma_{\rm diag}=\{(g,\rho(g))\}$, and so descends, if and only if
$\rho(fgf^{-1})=s\,\rho(g)\,s^{-1}$ for all $g\in\Gamma$.
\end{enumerate}
\end{proposition}
\idea This is covering-space theory: a map of the base lifts to the cover exactly when it preserves the subgroup of $\pi_1$ that
defines the cover. A lift normalizes the deck group, so it descends, and when the subgroup is characteristic every map lifts.
\movedproof{prop:lift}

Relative to a characteristic cover, then, every geometric automorphism of the quotient descends, and an accidental symmetry cannot be
geometric. Relative to other covers geometric automorphisms can be accidental too.

\begin{example}[The rotation of a square torus]\label{ex:accidental}
Let $Q$ be the toric code on $\Z_3\times\Z_3$, with $n=18$ and $k=2$. Its stabilizers of weight 4 are exactly its stars and plaquettes, so the
graph method of Appendix~\ref{app:checks} computes its whole monomial-Clifford automorphism group: it has order 144, made of the translations,
the point group of the square and the electromagnetic duality composed with a half-diagonal shift. As the quotient of the toric code on
$\Z_6\times\Z_3$ by the translation by $(3,0)$, only 72 of them descend. Here $\Pi=2\Z\oplus\Z$, which the rotations by $\pm90^\circ$ and the diagonal
reflections do not preserve, and indeed the parent has 144 automorphisms, all normalizing $\Gamma$, whose image is the subgroup of index 2
without these four linear parts. As the quotient of the toric code on $\Z_6\times\Z_6$ by $\Gamma=\Z_2^2$, the cover of mod-2 homology, the parent has
576 automorphisms and all 144 automorphisms of $Q$ descend. The rotation of $Q$ is therefore accidental for the first parent and not for the
second (Appendix~\ref{app:checks}).

For the running example, $\Pi=2\Lambda$ is characteristic in $\Lambda\cong\Z^2$. The cellular automorphisms of the five-site twisted torus $\Z^2/\Lambda$
are the 20 affine maps $j\mapsto2^aj+b$, the translations and the rotations by multiples of $90^\circ$ (no reflection preserves $\Lambda=(1+2i)$), and
they preserve the duality holonomy, which is the parity of the coordinate sum. So by (b) and (c) all 20 descend, as computed in
Example~\ref{ex:run4}. Each of the 340 accidental automorphisms of the five-qubit code either permutes the qubits by a non-affine permutation or
acts on the letters differently from the descended automorphism with the same permutation, as the order-3 gate does.
\end{example}

\subsection{Automorphism gates of dressed codes}
\label{sec:autdressed}

Proposition~\ref{prop:descent} descends symmetries that preserve the background. A parent symmetry that changes $\omega$ by a coboundary
does not commute with $U(\tilde\eta)$, but it still descends, with a correction supported on the wall. The principle is Stokes once more. If
$V|a\rangle=|Aa\rangle$ and $\frac1p\,\omega(Aa)-\frac1p\,\omega(a)=\delta\lambda(a)$ modulo 1 on flat configurations, then on these configurations
$U(\tilde\eta)VU(\tilde\eta)^\dagger=V\,e^{2\pi i\int\delta\lambda\cup\tilde\eta}=V\,e^{\pm2\pi i\int\lambda\cup B}$, and the last factor is a circuit on the wall alone.
So $V$ composed with that circuit preserves the dressed code, which is defined with $B$ and no primitive; Proposition~\ref{prop:transgress}
gives the precise statement. The transversal logical $T$ gate of
Kobayashi, Zhu and Hsin~\cite{KZH2025chs} has this form. In field theory this is the statement that an automorphism $\rho$ with $\rho^*\omega=\omega+d\alpha$ is a symmetry only together with the gauged SPT operator $e^{i\int\alpha}$~\cite{HsinKobayashi2025aut}. Further transversal gates of this kind are the swap gate $\mathrm{SWAP}_{12}\,CCZ$ of a $5{+}1$D code~\cite{Hsin:2024pdi}, a logical $CS$ of two 3D toric codes with a $\mathrm{SWAP}$ wall, and a gate at level 4 of the $\Z_N$ Clifford hierarchy, $N\ge3$, in two dimensions~\cite{HsinKobayashi2025aut}. Throughout, $\tilde a$ denotes the $0/1$ integer lift of a $\Z_2$ cochain, and $\omega$
applied to lifts is the integer cup polynomial, whose reduction mod 2 is $\omega$.

\begin{proposition}[Descent with a transgression dressing]\label{prop:transgress}
Let $p=2$, and let $\mathcal C_B$ be the dressed code of Section~\ref{sec:higher} on a closed cubical manifold $M$, for 1-form or higher-form
fields $a=(a_1,\dots,a_r)$ and a cup polynomial $\omega$ with background $B$ of degree $q+1$. Let $A$ be an invertible $\FF_2$-linear map of the
fields, $(Aa)_i=\sum_jA_{ij}a_j$ with $A_{ij}=0$ unless $n_i=n_j$, and $V$ the $CNOT$ circuit $V|a\rangle=|Aa\rangle$. Assume:
\begin{enumerate}[label=(\roman*)]
\item (transgression) there is a real cochain $\lambda(a)$ of degree $|\omega|-1$, whose value on each cell is a polynomial in the lifts $\tilde a$
of the qubits on faces of that cell, such that for every flat $a$, cell by cell,
\[
\tfrac12\,\omega(\widetilde{Aa})-\tfrac12\,\omega(\tilde a)\equiv\delta\lambda(a)\pmod 1;
\]
\item (lift) $B$ is the reduction of an integer cocycle $B_\Z$, for instance the Poincar\'e dual of a co-oriented wall;
\item (locality) for every cell $e$ carrying a Gauss law, $B_\Z=\delta\eta_e$ on the closed star $\overline{\mathrm{st}}(e)$ of $e$ (the cells of $\mathrm{st}(e)$ and their
faces) for an integer cochain $\eta_e$ on $\overline{\mathrm{st}}(e)$; this holds when $\overline{\mathrm{st}}(e)$ is contractible, for instance on cubical tori with all sides at least 3.
\end{enumerate}
Let $W|a\rangle=\exp\big(2\pi i\,s\int\lambda(a)\cup B_\Z\big)|a\rangle$ with $s=(-1)^{|\omega|}$, and $\tilde U=V\,W$. Then on flat configurations
\[
\tilde U\,G^B_{i,e}\,\tilde U^\dagger=\prod_j\big(G^B_{j,e}\big)^{A_{ji}},
\]
and $\tilde U$ maps flux checks to products of flux checks. So $\tilde U$ preserves $\mathcal C_B$. $W$ is a product of commuting diagonal gates, one for
each $D$-cell, of the level given by the phase polynomial of $\lambda$ (Section~\ref{sec:chsparent}).
\end{proposition}
\idea Conjugating $V$ by the twist operator $U_\eta$ multiplies it by the phase of $\int(\omega(Aa)-\omega(a))\cup\eta$. By (i) the
integrand is $\delta\lambda$, so Stokes moves the $\delta$ onto $\eta$ and leaves $\int\lambda\cup B$, a phase supported on the wall: this is $W$. When $B$ is
exact on all of $M$ this is the whole proof. In general the same computation is done on the closed star of each Gauss law, where (iii) makes
$B$ exact.
\movedproof{prop:transgress}

\begin{example}[The gate of Kobayashi, Zhu and Hsin]\label{ex:kzhgate}
Take three toric codes with fields $a_r,a_g,a_b$, $\omega=a_r\cup a_g\cup a_b$, and $A:a_g\mapsto a_r+a_g+a_b$. Then
$\omega(Aa)-\omega(a)=a_r\cup a_r\cup a_b+a_r\cup a_b\cup a_b$. For a flat 1-cochain, $\delta\tilde a/2\equiv a\cup a\pmod2$ on every square (check the
eight flat patterns), so by the Leibniz rule over $\Z$ condition (i) holds with $\lambda=\frac14\,\tilde a_r\cup\tilde a_b$; this is the identity
$a_r^2a_b+a_ra_b^2=d(\tilde a_r\tilde a_b)/2$ of~\cite[Supplemental Material, Section~III.A]{KZH2025chs}. With a wall $B=\theta_z$ at $z=0$ on $T^2\times S^1$ and $B_\Z$ its integer
lift, $s=-1$ and
\[
W=\prod_p CS^\dagger_{e_{01},e_{13}}\;CS_{e_{02},e_{23}},
\]
with $CS$ acting on the $r$-qubit of the first edge and the $b$-qubit of the second, over the squares $p$ of the layer $z=0$ with vertices
$0=x$, $1=x+e_1$, $2=x+e_2$, $3=x+e_1+e_2$. This is the circuit of~\cite[eq.~(6)]{KZH2025chs}, placed on the wall, up to inversion; the inverse also works,
because $W^2=(-1)^{\int a_r\cup a_b\cup B}$ commutes with every Gauss law on flat configurations by Stokes. So the 2D automorphism gate $U=WV$ of the
$D_4$ code is the descent, through the twisted compactification of Example~\ref{ex:kzh}(ii), of the transversal $CNOT$ symmetry $V$ of three
toric codes, dressed by the transgression of $\omega$ along $A$ on the wall. On a closed surface with $B=1$ (Example~\ref{ex:kzh}(i)) the same identity
is~\cite[Theorem~1]{KZH2025chs}.

A finite check (Appendix~\ref{app:checks}) confirms the identity of Proposition~\ref{prop:transgress} for this gate on $T^2$ with $B=1$, on closed
$T^3$ with a wall, including sizes that violate (iii), and on thin slabs, for both signs of $W$; dropping $W$ fails.
\end{example}

\paragraph{What this adds, and what it does not.} Three points. First, in two dimensions Bravyi and K\"onig restrict constant-depth logical gates
of Pauli stabilizer codes to the Clifford group~\cite{BravyiKoenig2013}, and the dressed codes of Example~\ref{ex:kzh} evade this: with three
gapped boundaries and boundary $T^\dagger$ terms, $U$ acts as a logical $T$~\cite{KZH2025chs}. For $\Z_N$ qudits with $N\ge3$, the swap automorphism of the twisted $\Z_N^3$ code on $T^2$ acts at level 4, on the logical subspace with one holonomy fixed~\cite{HsinKobayashi2025aut}; Propositions~\ref{prop:chslevel} and~\ref{prop:transgress} are stated for $p=2$ only. Proposition~\ref{prop:level} does not apply to these
codes, as noted after it, and the physical gate $W$ is itself at level 3 ($CS$). Second, Proposition~\ref{prop:transgress} explains where the
dressing comes from: $V$ is a symmetry of the parent that fails to commute with $U(\tilde\eta)$, the failure is a coboundary on flat
configurations, and Stokes moves it onto the wall. It plays the role, for $CNOT$-type automorphisms, of the identification of slots
in Proposition~\ref{prop:internal}: such an automorphism fixes $\omega$ only up to the coboundary of a half-integral cochain, and the half is what
makes $W$ a $CS$ circuit rather than a $CZ$ one. Third, boundaries are treated in Section~\ref{sec:bdrytrans}. On the slabs of
Corollary~\ref{cor:boundary}, where (iii) fails, the dressing flows onto the boundary (Lemma~\ref{lem:inflow}, Proposition~\ref{prop:slabgate}), and
the thin slab is the code of~\cite{KZH2025chs} up to a constant-depth circuit (Corollary~\ref{cor:thinslab}). On a manifold with gapped
boundaries and corners, the terms $\beta$ with $\alpha|_\partial=d\beta$ and $\gamma$ on hinges that~\cite{KZH2025chs} add are relative transgressions
(Proposition~\ref{prop:reltrans}, Example~\ref{ex:kzhbdry}).

\subsection{Twisted compactification by the 3-group of the 3D toric code}
\label{sec:threegroup}

Barkeshli, Chen, Hsin and Kobayashi~\cite{Barkeshli:2022edm} describe a 3-group symmetry of the $\Z_2$ toric code in three dimensions built from
three symmetries: a 0-form symmetry $U$ whose domain wall carries a $\Z_2$ Dijkgraaf--Witten theory, which on flux-free states is
$U=(-1)^{\int a\cup a\cup a}$ (Section~\ref{sec:higher}); the magnetic 1-form symmetry $V(\Sigma)=\prod_{e\cap\Sigma\neq\emptyset}X_e$; and the electric 2-form
symmetry $W(\gamma)=\prod_{e\in\gamma}Z_e$. Their backgrounds satisfy $dC_3=B_2\cup d\tilde C_1/2$, and on the lattice $[U(R,\sigma),V(\Sigma)]=W(\gamma)$, where
$\gamma$ is the intersection of $\Sigma$ with the orientation-reversing walls of the orientation data $\sigma$ of $U$~\cite[eqs.~(37) and~(44)]{Barkeshli:2022edm}.
We use these symmetries as twists.

\paragraph{Twisting by $U$.} $U$ is the cup-product symmetry of $\omega=a\cup a\cup a$ (type I) with $q=0$, so twisting by it is a case of
Section~\ref{sec:higher}. The Gauss laws are $CZ$-dressed, of level 2 (Proposition~\ref{prop:clifflevel}), and on $X\times S^1$ the code is the
dressed quotient $\mathcal C_0/\langle t\circ U\rangle$ of the toric code by a translation composed with $U$
(Theorem~\ref{thm:dressedquot}(c)). In field theory, compactifying the twist $\pi\int a^3\cup B$ with $B=d\theta$ gives the
Dijkgraaf--Witten twist $\pi\int a^3$ of a $2+1$D $\Z_2$ gauge theory, the double-semion theory, together with the $\Z_2$ sector of holonomies
around the circle. On $T^2$ this gives dimension $4\cdot2=8$, the dimension of the code on $T^2_2\times S^1$ (Appendix~\ref{app:checks}); the dimension alone does not distinguish this from the untwisted theory, but the dressing does. So twisting the 3D toric code by its gauged-SPT symmetry gives a $CZ$-dressed quotient code whose field theory is
the double-semion theory times a $\Z_2$ sector.

\paragraph{Twisting by $V$ or $W$.} These are Pauli operators, so by Corollary~\ref{cor:paulinogo} they change only the phases of the quotient.

\paragraph{The 3-group on quotients.} The twists that matter are those under which $U$ and $V$ fail to commute. On a closed quotient the
commutator, and the logical action of $U$, are as follows.

\begin{proposition}[The 3-group on closed quotients]\label{prop:threegroup}
Let $M$ be a closed 3-manifold, given as a cubical complex with a cup structure or as a simplicial complex with a branching, carrying the
$\Z_2$ toric code with qubits on edges, and let $U=(-1)^{\int a\cup a\cup a}$.
\begin{enumerate}[label=(\alph*)]
\item $U$ preserves the code space and acts on the logical state $|\alpha\rangle$, $\alpha\in H^1(M;\Z_2)$, by $(-1)^{q(\alpha)}$, $q(\alpha)=\langle\alpha^3,[M]\rangle$.
\item $q(\alpha+\lambda)=q(\alpha)+q(\lambda)+b(\alpha,\lambda)$ with $b(\alpha,\lambda)=\langle w_1(M)\cup\alpha\cup\lambda,[M]\rangle$. So $U$ acts as a product of logical $CZ$
gates, one for each pair of basis classes with $b=1$, and logical $Z$ gates, one for each basis class with $q=1$; it is a logical Pauli
operator if and only if $b=0$, in particular whenever $M$ is orientable.
\item For $\lambda\in H^1(M;\Z_2)$ let $\bar X(\lambda)=V(\Sigma)$ with $\Sigma$ Poincar\'e dual to $\lambda$. On the code space
\[
U\,\bar X(\lambda)\,U^\dagger\bar X(\lambda)^\dagger=(-1)^{q(\lambda)}\,W(\gamma),\qquad\gamma\ \text{Poincar\'e dual to}\ w_1(M)\cup\lambda,
\]
that is, $\gamma$ is the intersection of $\Sigma$ with a surface dual to $w_1(M)$, along which the orientation reverses.
\item If $M=\tilde M/\Gamma$ for a free action on an orientable $\tilde M$, then $w_1(M)=c^*\epsilon$, where $c:M\to B\Gamma$ classifies the cover and $\epsilon:\Gamma\to\Z_2$
records which deck transformations reverse the orientation. So $U$ is a non-Pauli logical gate on the quotient only if some deck
transformation reverses the orientation.
\end{enumerate}
\end{proposition}
\idea $U$ acts on a logical state by the cubic phase $\int\alpha^3$, so its logical gate is read off from the polarization of
this cubic function. Since $(\alpha+\lambda)^3-\alpha^3-\lambda^3=\mathrm{Sq}^1(\alpha\lambda)$, Wu's formula turns the polarization into $\int w_1\cup\alpha\cup\lambda$, which
vanishes on orientable manifolds: only an orientation-reversing twist makes the gate non-Pauli.
\movedproof{prop:threegroup}

Part (c) is the 3-group relation of~\cite{Barkeshli:2022edm} in logical form, including their sign $(-1)^{q(\lambda)}$, the parity of the triple
self-intersection of $\Sigma$. Part (d) says what the twist contributes: on any quotient by an orientation-preserving action $U$ is a Pauli
operator, as on $\mathbb{RP}^3=S^3/\Z_2$, where it is the logical $Z$, and an orientation-reversing holonomy is what makes it a $CZ$. This is the row of Table~\ref{tab:create} on orientation-reversing
holonomies. The condition is bosonic: the fermionic pump needs no orientation reversal (Section~\ref{sec:fermionic}).

\begin{example}[The logical $CZ$ of~\cite{Barkeshli:2022edm} as a quotient code]\label{ex:klein}
Let $g(x,y,z)=(x+L,-y,z)$ on the cubical torus $\tilde M=\Z_{2L}\times\Z_L\times\Z_L$. It is a free glide reflection, so $M=\tilde M/\langle g\rangle=K\times S^1$ is the
twisted compactification of the toric code on $T^2_{y,z}$ along the $x$-circle with the reflection $y\mapsto-y$ as holonomy, and by
Theorem~\ref{thm:balanced} its code is the quotient of the toric code on $\tilde M$ by a free permutation action, with $k=3$. The cubical cup
structure is not invariant under $g$, so, as for T6, we put the qubits on the edges of the barycentric subdivision, whose canonical branching
every cellular map preserves. Since $g$ reverses the orientation, $w_1(M)$ is the class of the double cover, and in the basis $u,v,z$ of $H^1(K\times S^1;\Z_2)$, with
$u^2=v^2$ the generator of $H^2(K)$, $uv=0$, $z^2=0$ and $w_1=u+v$, one finds $q(\alpha_1u+\alpha_2v+cz)=c(\alpha_1+\alpha_2)$: $U$ is a logical $CZ$ on the quotient~\cite[Section~5.1]{Barkeshli:2022edm}, while on the parent
$\tilde M$ it acts trivially. Check T8 (Appendix~\ref{app:checks}) confirms this on lattices of 448 and 1512 qubits.
\end{example}

\paragraph{Mixed backgrounds.} Twisting by $U$ and $V$ at once, along two commuting deck generators, needs a dressed membrane operator in
place of $V$; Appendix~\ref{app:tech-sym} constructs it, and the resulting code is open.

\subsection{Non-invertible parent symmetries}
\label{sec:noninv}

A non-invertible symmetry $N$ of the parent fuses with its orientation reversal as $\bar N\times N=\sum_cU_c$, a sum of operators $U_c$ supported
on cycles of its support; for a condensation defect these are the condensed operators. Gukov, Hsin and Pei observe that compactification can
make such a symmetry invertible: if the internal space has no cycles that support the $U_c$, except for the identity, the fusion becomes
invertible~\cite[Section~3.3.3]{GukovHsinPei2025}. For example, the Kramers--Wannier defect of Maxwell theory at $\tau=iN$ becomes invertible on
$S^3$, which has no 2-cycles. On a code the corresponding statement is elementary.

\begin{lemma}[Non-invertible symmetries that become gates]\label{lem:noninv}
Let $\mathcal C$ be a code space and $N$ an operator with $N\mathcal C\subseteq\mathcal C$, and suppose that on $\mathcal C$, $N^\dagger N=\sum_cU_c$, where each $U_c$ maps $\mathcal C$
to itself and acts on it as a scalar $\lambda_c$. If $\Lambda=\sum_c\lambda_c\neq0$, then $\Lambda>0$ and $\Lambda^{-1/2}N$ restricts to a unitary on $\mathcal C$, a
logical gate.
\end{lemma}
\begin{proof}
On $\mathcal C$, $N^\dagger N=\Lambda\cdot1$ is positive semidefinite and non-zero, so $\Lambda>0$. Then $V=\Lambda^{-1/2}N$ maps the finite-dimensional space $\mathcal C$ to itself
isometrically, so it is unitary there.
\end{proof}

On a quotient the scalars have two sources. A condensed operator on a cycle that becomes a boundary in the quotient is a product of
stabilizers, which is the mechanism of~\cite{GukovHsinPei2025}. And by Theorem~\ref{thm:hfcount} a higher-form twist turns into stabilizers
exactly the logicals that pair non-trivially with $B$; in the code with a $CZ$ wall of Section~\ref{sec:gausslaw} these are the two electric loops
along the wall. So a non-invertible parent symmetry, applied on the quotient by the same local formula, gives a logical gate there when its
fusion outcomes are among these operators. Appendix~\ref{app:tech-sym} states the general criterion. An example in which the gate is not already induced by an invertible parent symmetry is
open (Section~\ref{sec:discussion}).

\begin{exampleu}[A non-invertible operator that becomes a qutrit gate]
On the circle with the cubic twist $\omega=a_1\cup a_2\cup b$ and $B$ dual to a point (the example after Proposition~\ref{prop:clifflevel}),
let $n$ be the two-qubit operator
\[
n|00\rangle=|01\rangle,\qquad n|01\rangle=|10\rangle,\qquad n|10\rangle=|00\rangle,\qquad n|11\rangle=0,
\]
let $n_v$ be its copy on the vertex qubits $a_{1,v},a_{2,v}$, and let $N=\prod_vn_v$. On the undressed code $\mathcal C_0$, on a circle of any length, $N$ maps each logical product state to another one
and annihilates $|11\rangle$, so $N\mathcal C_0\subseteq\mathcal C_0$ and $N$ is not invertible there. Its fusion is
\[
N^\dagger N=\prod_v\tfrac12\big(1+CZ(a_{1,v},a_{2,v})\big)=2^{-L}\sum_{S}\,\prod_{v\in S}CZ(a_{1,v},a_{2,v}),
\]
summed over the sets $S$ of vertices. The outcomes are products of the local operators $(-1)^{a_1a_2}$, and on $\mathcal C_0$ each of them acts
as the logical $CZ$ of the two repetition codes, so $N^\dagger N$ is the projector that removes $|11\rangle$. The cubic wall makes all of them
scalars. The product of the Gauss laws of $b$ is $CZ(a_{1,0},a_{2,0})$ (the example after Theorem~\ref{thm:pointcount}), and the flux checks
identify every $CZ(a_{1,v},a_{2,v})$ with it, so each outcome acts on $\mathcal C_B$ as $+1$ and $\Lambda=1$. By Lemma~\ref{lem:noninv}, $N$ is a
logical unitary on $\mathcal C_B$. It permutes the three surviving states of the repetition codes cyclically,
$|00\rangle\to|01\rangle\to|10\rangle\to|00\rangle$, and acts trivially on $b$, so on $\mathcal C_B\cong\mathbb C^3\otimes\mathbb C^2$ it is the shift gate
of a logical qutrit. In the quotient description, $n_v$ commutes with $CCZ(a_{1,v},a_{2,v},b_{e_v})$, because it annihilates $|11\rangle$ and has
no $|11\rangle$ in its image. So $N$ on the doubled circle commutes with $t\circ\prod_vCCZ(a_{1,v},a_{2,v},b_{e_v})$, and on the quotient it is
given by the same on-site formula. Direct computation at $L=3,4$, and on the doubled circle with six vertices, confirms these statements.

Two remarks place the example. First, the outcome that becomes a scalar is not a Pauli operator. Making it $+1$ while keeping the other three
states is something no Pauli stabilizer code does, since the dimension 6 is not a power of 2 (Corollary~\ref{cor:cert}); with a quadratic wall the
outcomes that become scalars are Pauli operators, such as the two electric loops along the wall $W$ of Section~\ref{sec:gausslaw}. Second, the gate is also induced by an invertible
parent symmetry. One has $n=\sigma\cdot\frac12(1+CZ)$, where $\sigma=X_2\,CNOT_{1\to2}\,\mathrm{SWAP}_{12}$ is the permutation
$(x_1,x_2)\mapsto(x_2,x_1+x_2+1)$ of the computational basis, which acts on $|00\rangle,|01\rangle,|10\rangle$ as $n$ does and fixes $|11\rangle$. It
preserves $x_1x_2$, so it commutes with $CCZ$, and the transversal $\prod_v\sigma_v$ descends and agrees with $N$ on $\mathcal C_B$. The example
therefore illustrates the lemma, but it is not of the kind left open above.
\end{exampleu}

\subsection{Twisted cup products and the choice of torus}
\label{sec:toritwist}

With $R$ not free, Proposition~\ref{prop:internal} is the twisted cup product of~\cite{BDEHT2026cupsII}, and there the twist does one more thing.
After it identifies the $r$ inputs of the copy-cup gate, the phase can be lifted to finer roots of unity, because a power depends less on the
integral lift than a product: $(c+3e)^3\equiv c^3\pmod 9$ and $(c+2e)^4\equiv c^4\pmod 8$, while $c_1c_2c_3$ changes by $3ec_2c_3$. Corrections by
higher cup products turn this into a circuit~\cite{BDEHT2026cupsII}; their twisted Pontryagin powers give a qutrit $T$ gate on one 3D qutrit
toric code on $T^3$, and a qubit $T$, $CS$ and $CCZ$ on one 4D toric code on $T^4$~\cite[Section~7]{BDEHT2026cupsII}. On $r$ independent copies
the lifted phase is not defined (check T9), so these gates are created by the twist without being descended parent symmetries. On a torus
they are an index, and the choice of torus, which a twisted compactification can change, decides their level.

\begin{proposition}[Twisted gates on tori]\label{prop:tori}
Let $X=\mathbb R^r/\Lambda$ carry an action of a finite abelian group $G$ of order $r$ by linear maps that preserve $\Lambda$, with a
$G$-invariant triangulation with a $G$-invariant branching, that is, an order of the vertices of each simplex, compatible with faces, that $G$
preserves. Let $\Phi$ be a twisted cohomology operation of~\cite[Section~4]{BDEHT2026cupsII}
with $r$ inputs permuted regularly by $G$ and values in $\Z/p^k$, normalized so that on integrally liftable classes it is
$\int c\cup g_2^*c\cup\dots\cup g_r^*c$ (in general this is multiplied by a constant~\cite[Proposition~4.10]{BDEHT2026cupsII}). For
$x\in H^1(X;\FF_p)$ with an integral lift $c\in H^1(X;\Z)=\Lambda^*$,
\[
\Phi(x)=\det(c,g_2^*c,\dots,g_r^*c)=\pm\,[\Lambda^*:\Z[G]c]\pmod{p^k},
\]
with the determinant taken in a basis of $\Lambda^*$; it is $0$ when $\Z[G]c$ has rank less than $r$.
\begin{enumerate}[label=(\alph*)]
\item $\Phi$ takes a value prime to $p$, a primitive $p^k$-th root of unity such as the qubit or qutrit $T$, if and only if $H^1(X;\FF_p)$ is a
free $\FF_p[G]$-module; for $G$ cyclic of order $p^m$, if and only if a generator acts on it as a single Jordan block.
\item For a $G$-equivariant covering $\pi:\tilde X\to X$ of degree $\deg\pi$ with the lifted triangulation ($X$ any closed triangulated manifold
with such an action, not only a torus), $\Phi_{\tilde X}(\pi^*x)=\deg\pi\,\Phi_X(x)$.
\end{enumerate}
\end{proposition}
\idea On a torus every degree-one class lifts to an integral class $c$, and the twisted operation is the cup product of the $r$ classes
$c,g_2^*c,\dots$, which on an $r$-torus is their determinant. The determinant is prime to $p$ exactly when these classes span the lattice mod $p$,
that is, when $c$ generates $H^1$ as a module over the group.
\movedproof{prop:tori}

Part (b) is Corollary~\ref{cor:covers} with finer coefficients: a quotient can take a $(\deg\pi)$-th root of the gate. Part (a) says
what the twist has to change. Every $r$-torus carries $r$ logical qudits per code; what changes is how the residual symmetry $G$ acts on them,
and the $T$ gate detects exactly that. Two $G$-tori $\mathbb R^r/\Lambda_1$ and $\mathbb R^r/\Lambda_2$ with commensurable lattices are both
quotients of the $G$-torus $\mathbb R^r/(\Lambda_1\cap\Lambda_2)$, so they are twisted compactifications of a common parent.

\begin{example}[A deep-hole holonomy creates the qutrit $T$]\label{ex:deephole}
Let $P$ be the cyclic permutation of the coordinates of $\mathbb R^3$, with the Kuhn triangulation of the unit cubes, which $P$ preserves together
with its branching, the order of the vertices of each simplex. On $\tilde X=\mathbb R^3/L\big(\Z(1,1,1)\oplus A_2\big)=S^1\times T^2_{\rm hex}$, with $A_2$ the vectors of $\Z^3$ with zero
coordinate sum, $P$ fixes the circle and rotates the hexagonal torus, and $H^1(\tilde X;\FF_3)=J_1\oplus J_2$. The translation by
$Le_1$ has order three on $\tilde X$, commutes with $P$, and moves the circle by a third of its length and the hexagonal torus by a deep hole;
the quotient is the cubic torus $T^3_L=\mathbb R^3/L\Z^3$, with $H^1(T^3_L;\FF_3)=J_3$. For one qutrit toric code, the operation
of~\cite[eq.~(38)]{BDEHT2026cupsII} takes only the values $0,3,6$ on $\tilde X$, where it is the controlled phase $\omega^{x_0(b_1-b_2)^2}$ in a
basis $x_0$ (circle), $b_1,b_2$ (hexagonal torus) of $H^1$, and on $T^3_L$ it is $T_1T_2T_3\,CCZ^{-1}$~\cite[eq.~(54)]{BDEHT2026cupsII} (check
T9). For qubits, $\FF_2[C_3]$ is semisimple and both tori have free $H^1(\,\cdot\,;\FF_2)$; the gate is $CCZ$-type on both,
$Z_1Z_2Z_3\,CCZ$~\cite[eq.~(48)]{BDEHT2026cupsII} on $T^3_L$ and the gate of T7b on $\tilde X$. In four dimensions the same happens in steps. For
the coordinate $4$-cycle $\rho$ and $\Z/2\to\Z/8$, the six $\rho$-tori between $S^1\times S^1\times T^2_{\rm sq}$ ($\rho$ the identity, a
reflection and a quarter turn) and the cubic $T^4$ give the identity on the first, which covers $T^4$ with degree 8; $Z$ and $CZ$ on the
three of degree 4; $S$ and $CZ$ on the one of degree 2; and $T$, $CS$ and $CCZ$ on the cubic $T^4$~\cite[eq.~(53)]{BDEHT2026cupsII} (check
T10). On the tricosm the deep-hole translation descends, but on its 3-torus cover the lifted translation acts trivially on $H^1$, where the
gate is therefore $\det(c,c,c)=0$; by (b) the tricosm has no qutrit $T$ from this operation.
\end{example}

\subsection{Transgression gates on manifolds with boundary}
\label{sec:bdrytrans}

Proposition~\ref{prop:transgress} is stated on closed manifolds, and its proof needs $B$ to be locally exact. Two kinds of boundary
occur. On the slabs of Corollary~\ref{cor:boundary} the boundary is smooth and $B$ is relative; such a $B$ is never locally exact relative
to the boundary (on a vertical edge of $X\times[0,1]$ both endpoints lie in $\partial M$), and the dressing flows onto the boundary. At the
gapped boundaries of~\cite{KZH2025chs} the fields obey linear conditions that trivialize $\omega$, and the boundary terms are relative
transgressions. Throughout, $p=2$, the notation is that of Proposition~\ref{prop:transgress}, and $\partial M$ carries the orientation for which
$\int_M\delta y=\int_{\partial M}y$ for every real cochain $y$; the cubical cup product on $M$ restricts to the cubical cup product on each
boundary piece.

\paragraph{Slabs.} On a slab with relative $B$, Stokes' theorem leaves a boundary term, and the dressing flows onto the boundary: the code
is the undressed code in the bulk with the dressed code of $\partial M$, one degree lower, on the boundary (Lemma~\ref{lem:inflow}). A
transgression gate then descends to the slab with an extra boundary circuit, provided that circuit is a symmetry of the boundary code
(Proposition~\ref{prop:slabgate}); for $q=0$ the boundary code has $B=1$, and the needed identity follows from the closed case on $X\times S^1$
(Lemma~\ref{lem:bone}). These three statements are in Appendix~\ref{app:tech-sym}. For one layer of cubes they identify the slab code with a known one.

\begin{corollary}[The thin slab is the code of Example~\ref{ex:kzh}(i)]\label{cor:thinslab}
Let $M=X\times[0,1]$ have one layer of cubes, with $r$ fields of one degree $n$, relative $B$ dual to $X\times\{\frac12\}$, and $\eta$ the
indicator of the vertices of $X\times\{1\}$. Then $U_\eta=(-1)^{\int_M\omega(a)}$ is a product of commuting phase gates, one per top cell, and
\[
\mathcal C_B=U_\eta\,K\big(\mathcal C^{(1)}_X\otimes|0\rangle^{\otimes r|X_n|}\otimes|+\rangle^{\otimes r|X_{n-1}|}\big),
\]
where $\mathcal C^{(1)}_X$ is the code of Example~\ref{ex:kzh}(i) on $X\times\{1\}$ (with $B=1$), $|X_k|$ is the number of $k$-cells of $X$, and $K$ is a
$CNOT$ circuit of two rounds, of depth bounded in terms of the local geometry. For three $1$-form fields on a square lattice, $\mathcal C^{(1)}_X$ is the $D_4$ code of~\cite{KZH2025chs}.
\end{corollary}
\idea On one layer of cubes the twist operator is a single layer of phase gates. After it, by Lemma~\ref{lem:inflow}, only the top
Gauss laws are dressed, and a $CNOT$ circuit moves the bottom and vertical qubits into a product state, leaving the code of
Example~\ref{ex:kzh}(i) on the top face.
\movedproof{cor:thinslab}

So the thin-slab code of Example~\ref{ex:kzh}(ii), whose field theory is $D(D_4)$, is on the lattice the 6-qubit-per-site code
of~\cite{KZH2025chs} up to a constant-depth circuit: the slab code has 15 qubits per site, and a circuit of phase gates ($CCZ$ for
$\omega=a_r\cup a_g\cup a_b$) and $CNOT$ gates removes 9 of them. Proposition~\ref{prop:slabgate} then proves the thin-slab check of Example~\ref{ex:kzhgate}:
there $W_\partial=\exp(2\pi i\int_{X\times\{1\}}\tfrac14\tilde a_r\cup\tilde a_b)$ is the 2D circuit of~\cite{KZH2025chs} on the top face, and the
hypothesis on $\partial M$ is Lemma~\ref{lem:bone}.

\paragraph{Gapped boundaries.} At the gapped boundaries of~\cite{KZH2025chs} the fields obey linear conditions on which $\omega$ vanishes. The
bulk computation of Proposition~\ref{prop:transgress} then leaves boundary terms, and since the transgression $\lambda$ is itself a coboundary
on each boundary piece, Stokes' theorem applies again, piece by piece and then hinge by hinge, each step adding one correction to the
gate (Proposition~\ref{prop:reltrans} in Appendix~\ref{app:tech-sym}). For the triangle of~\cite{KZH2025chs} the only correction is the string
$\int\frac18\tilde a_r$ along one side, so the boundary $T$-string of~\cite{KZH2025chs} is the relative transgression of the bulk $CS$ circuit
along the boundary condition $a_r=a_b$, and the gate acts as the logical $T$ (Example~\ref{ex:kzhbdry}, check T13).

Proposition~\ref{prop:slabgate} and Proposition~\ref{prop:reltrans} are the two ways a boundary enters. At a smooth boundary with relative
$B$ the transgression flows onto the boundary and becomes the gate of a dressed code there, one degree lower; at a gapped boundary that
trivializes $\omega$ it is absorbed by relative transgressions, as in~\cite[eqs.~(53)--(56)]{KZH2025chs}.

For $N\ge4$ the $\lambda$ of~\cite[eq.~(47)]{KZH2025chs} contains $\cup_1$ products, so (i) and (bc3) need a cubical $\cup_1$ that satisfies the
Steenrod coboundary formula, which we have not implemented or checked (Section~\ref{sec:discussion}).
Warman and Sch\"afer-Nameki obtain the logical $T$ of Example~\ref{ex:kzhbdry} in the $D(D_4)$ presentation of the same code, by stacking the
non-trivial class of $H^2(D_4;U(1))$ with a counterterm on the $\langle r\rangle$ boundary~\cite{WSN2025}. Under their dictionary, their boundaries
$\langle rs\rangle$, $\langle s\rangle$ and $\langle r\rangle$ are $L_r$, $L_b$ and $L_{rb}$, and their counterterm is $\beta_{rb}$. Their circuit is diagonal in the $D_4$
group basis, while the gate $U=WV$ of Example~\ref{ex:kzhgate} is a $CNOT$ circuit followed by a diagonal one in the $\Z_2^3$ basis. Their
gates on the doubles of the larger dihedral groups lie beyond the twists of Proposition~\ref{claim:reach} (Section~\ref{sec:gauging}).

\subsection{Fermionic pumps under rotation twists}
\label{sec:fermionic}

The principle for gates also applies when the parent symmetry is not a strictly local circuit. The 3D fermionic toric code is the $\Z_2$
gauge theory whose charge is a fermion. On a closed oriented 3-manifold $M$ its logical states are labeled by the spin structures of $M$, a
torsor over $H^1(M;\Z_2)$, so $k=b_1(M;\Z_2)$~\cite{Barkeshli:2023bta}. Its $\Z_8$ symmetry is generated by the pump $V$ of a $p+ip$ layer through
space: pairs of $p+ip$ and $p-ip$ layers are nucleated between layers $2j-1$ and $2j$ and annihilated between layers $2j$ and $2j+1$, which
carries one layer once around~\cite{FidkowskiHastings2024,Barkeshli:2023bta}. The pump is a free-fermion unitary with an exponentially decaying kernel,
built from finite-time evolutions or from a few quasi-local Gaussian unitaries, so it is a finite-time quasi-local evolution rather than an
exact circuit of bounded depth and bounded range, and no such circuit can pump a chiral phase. In the field theory
of~\cite{Barkeshli:2023bta}, modulo logical Cliffords, $V$ is the product of $CCZ$, $CS^\dagger$ and $T$ gates with exponents the triple cup products $\int\sigma_j\sigma_k\sigma_l$, $\int\sigma_j\sigma_k^2$ and
$\int\sigma_j^3$ of a basis of $H^1(M;\Z_2)$~\cite[eq.~(26)]{Barkeshli:2023bta}. On a quotient $M=\tilde M/\Gamma$ the pump is built by the same rule, with
the layers glued across a seam by the deck transformation. This is the same local formula as in Proposition~\ref{prop:cupgate}, and the proof
of code preservation extends to free-fermion pumps without error.

\begin{proposition}[Descent of free-fermion pumps]\label{prop:gaussdescent}
Let a finite group $\Gamma$ act freely on a cover $p:\tilde M\to M$ by cellular maps. Fix a spin structure $\tilde s$ on $\tilde M$, given on the lattice
by the signs of the fermion hoppings, that is a pull-back $p^*s_0$ of a spin structure $s_0$ on $M$. Let $\hat G_g$ be the lift of $g$ to the
single-particle space of $(\tilde M,\tilde s)$ determined by $s_0$: the cellular map composed with parity-even on-site unitaries, with on-site part
$\rho(g)$, and with the sign changes that restore the hopping signs of $\tilde s$; the $\hat G_g$ form a unitary representation of $\Gamma$. The spin
structures $s$ on $M$ with $p^*s=\tilde s$ are $s=s_0+\varepsilon_s$, with $\varepsilon_s\in\Hom(\Gamma,\{\pm1\})=H^1(\Gamma;\Z_2)$, whose image in $H^1(M;\Z_2)$ is $\ker p^*$. For
such $s$ let $\mathcal H_s=\{\psi:\hat G_g\psi=\varepsilon_s(g)\psi\}$ be its single-particle space. Let $\tilde u$ be a single-particle unitary on
$(\tilde M,\tilde s)$, number-conserving or of Bogoliubov--de~Gennes form, that commutes with every $\hat G_g$ and whose kernel decays
exponentially, $\|\tilde u(\tilde x,\tilde y)\|\le Ce^{-{\rm dist}(\tilde x,\tilde y)/\xi}$, and let $u_s=\tilde u|_{\mathcal H_s}$.
\begin{enumerate}[label=(\alph*)]
\item $u_s$ is a single-particle unitary on $M$ in sector $s$, with kernel
\[
u_s(x,y)=\sum_{g\in\Gamma}\varepsilon_s(g)\,\tilde u(\tilde x,g\tilde y)\,\rho(g)
\]
for lifts $\tilde x,\tilde y$, with $\rho(g)$ the on-site part of $\hat G_g$, signs included. Replacing $\tilde y$ by $h\tilde y$ multiplies the kernel on the
right by the inverse of the transport factor $\varepsilon_s(h)\rho(h)$. In a local trivialization, with $\tilde x$ and $\tilde y$ in one ball of $\tilde M$ that
injects into $M$, only the term $g=1$ is not exponentially small when $x$ and $y$ are closer than the injectivity radius, so $u_s$ is the same
local formula as $\tilde u$; for lifts in a fixed fundamental domain, a pair across its boundary keeps instead the term of the $g$ with $g\tilde y$
near $\tilde x$, with its transport factor. If $\tilde u=\mathcal T\exp(-i\int_0^T\tilde k(t)\,dt)$ for a quasi-local single-particle Hamiltonian
that commutes with every $\hat G_g$, then $u_s$ is the evolution of $k_s(t)=\tilde k(t)|_{\mathcal H_s}$, whose kernel is given by the same formula.
\item If $\tilde u$ preserves the occupied subspace of a $\Gamma$-invariant atomic insulator on $\tilde M$, then $u_s$ preserves the occupied subspace of
the atomic insulator $|\Omega_s\rangle$ of sector $s$ on $M$ exactly, so every Fock-space implementation of $u_s$ maps $|\Omega_s\rangle$ to a multiple of
itself of modulus one.
\item Suppose that for each spin structure $\tilde s$ in the image of $p^*$ such a $\tilde u_{\tilde s}=\mathcal T\exp(-i\int_0^T\tilde k_{\tilde s})$ is given, with
all $\tilde k_{\tilde s}$ built by one local rule. Fix the many-body implementation by the evolution
\[
U_s(t)=\mathcal T\exp\Big(-i\int_0^tH_s(t')\,dt'\Big),\quad U_s(0)=1,\qquad H_s(t)=\tfrac12\Psi^\dagger k_s(t)\Psi,
\]
with $k_s$ the restriction of $\tilde k_{p^*s}$ and $H_s$ in the symmetric Nambu form, $\Psi=(c,c^\dagger)$, which fixes its scalar part (in the
number-conserving case one may take $c^\dagger k_sc$, whose evolution is the canonical second quantization of $u_s$), and let
$\theta(s)=\langle\Omega_s|U_s(T)|\Omega_s\rangle$, a phase by (b). After gauging fermion parity, let $V$ be the evolution of the gauged Hamiltonian,
in which the term of each pair of sites of $M$ and homotopy class of paths between them is multiplied by the Wilson line along a path in
that class. Then $V$ preserves the code space of the fermionic toric code on $M$ exactly, for every system size, and acts on it as the
diagonal logical gate $|s\rangle\mapsto\theta(s)|s\rangle$.
\end{enumerate}
\end{proposition}
\idea An operator that commutes with a representation of $\Gamma$ preserves its eigenspaces. So the pump of each twisted sector is the
restriction of the cover pump with the compatible spin structure, exactly, and the atomic insulator of each sector, which is the logical
state $|s\rangle$ after gauging, is multiplied by a phase. The phase is that of the evolution of the restricted Hamiltonian, built by one rule
for all sectors, and the gauged evolution acts in every sector by that evolution. Locality enters only through the image sum in (a).
\movedproof{prop:gaussdescent}

Appendix~\ref{app:tech-sym} explains which spin structures of $M$ one deck representation reaches, and why the convention in (c) matters
only through scalars.

\begin{example}[The controlled-$S^\dagger$ of~\cite{Barkeshli:2023bta} as a quotient]\label{ex:dicosm}
Let $\tilde M=T^2\times S^1$ with $4N$ layers along the circle, and let $g$ be the half turn $C_2$ of the square torus $T^2$ composed with the
translation by $2N$ layers. It acts freely, and $M=\tilde M/\langle g\rangle=T^2\rtimes_{C_2}S^1$, with $k=3$. The pump of~\cite[eq.~(40)]{Barkeshli:2023bta}, which
nucleates and annihilates pairs on $2N$ layers and glues layer $2N+1$ to layer 1 by the rotation, is the parent pump with one factor for each
$g$-orbit, as in Proposition~\ref{prop:gaussdescent}(a). On the parent, which is $T^3$, $V=CCZ$ and $V^2=1$; on the quotient $V=CCZ\,CS^\dagger$ and
$V^2=CZ$~\cite[eq.~(48)]{Barkeshli:2023bta}. So the controlled-$S^\dagger$ is created by the twist. Check T18 (Appendix~\ref{app:checks}) confirms this
with an explicit equivariant Chern pump: in each twisted sector the restricted cover pump is the pump built on the quotient, the code
space is preserved, and $V^2=CZ$.
\end{example}

\begin{figure}[tbp]
\centering
\begin{tikzpicture}[x={(0.95cm,0cm)},y={(0.45cm,0.40cm)},z={(0cm,0.95cm)},>=Stealth,line join=round]
\def\H{2.4}
\newcommand{\qtbox}{%
\draw[gray!60,densely dashed] (-1,1,0) -- (1,1,0) (-1,1,0) -- (-1,-1,0) (-1,1,0) -- (-1,1,\H);
\draw[gray!70!black] (-1,-1,0) -- (1,-1,0) -- (1,1,0) -- (1,1,\H) -- (-1,1,\H) -- (-1,-1,\H) -- cycle
  (1,-1,0) -- (1,-1,\H) -- (-1,-1,\H) (1,-1,\H) -- (1,1,\H);}
\begin{scope}
\fill[gray!12] (-1,-1,0) -- (1,-1,0) -- (1,1,0) -- (-1,1,0) -- cycle;
\fill[gray!12] (-1,-1,\H) -- (1,-1,\H) -- (1,1,\H) -- (-1,1,\H) -- cycle;
\qtbox
\draw[->,very thick,red!75!black] (-0.55,0,0) -- (0.55,0,0);
\draw[->,very thick,red!75!black] (0,-0.55,\H) -- (0,0.55,\H);
\draw[->,red!75!black] (-1.25,-0.2,0.15) to[out=150,in=210] node[left,font=\scriptsize,align=right]{quarter\\turn} (-1.25,-0.2,\H-0.1);
\draw[->,gray!60!black] (1.25,1,0.2) -- node[right,font=\scriptsize,align=left]{circle,\\$J$ layers} (1.25,1,\H-0.2);
\node[font=\scriptsize,below=1pt] at (0,-1,0) {fibre $T^2_L$};
\node[font=\small] at ($(0,-1,0)+(0.45cm,-0.8cm)$) {(a) gluing};
\end{scope}
\begin{scope}[xshift=5.5cm]
\qtbox
\fill[orange!55,opacity=0.40] (0,-1,0) -- (0,1,0) -- (0,1,\H) -- (0,-1,\H) -- cycle;
\fill[orange!55,opacity=0.40] (-1,0,0) -- (1,0,0) -- (1,0,\H) -- (-1,0,\H) -- cycle;
\draw[orange!70!black] (0,-1,0) -- (0,-1,\H) (1,0,0) -- (1,0,\H);
\draw[orange!70!black,very thick] (0,-1,\H) -- (0,1,\H) (-1,0,\H) -- (1,0,\H);
\fill[blue!45,opacity=0.40] (-1,-1,0.5*\H) -- (1,-1,0.5*\H) -- (1,1,0.5*\H) -- (-1,1,0.5*\H) -- cycle;
\draw[blue!60!black] (-1,-1,0.5*\H) -- (1,-1,0.5*\H) -- (1,1,0.5*\H) -- (-1,1,0.5*\H) -- cycle;
\node[font=\scriptsize,blue!60!black,anchor=west] at (1.1,1,0.5*\H) {fibre};
\node[font=\scriptsize,orange!60!black,anchor=west,align=left] at (1.1,1,0.92*\H) {cross${}\times{}$circle};
\node[font=\small] at ($(0,-1,0)+(0.45cm,-0.8cm)$) {(b) membranes};
\end{scope}
\begin{scope}[xshift=11.0cm]
\qtbox
\draw[line width=1.6pt,green!45!black] (0,0,0) -- (0,0,\H);
\fill[green!45!black] (0,0,0) circle (1.4pt) (0,0,\H) circle (1.4pt);
\draw[line width=1.6pt,violet] (-1,-0.5,0.3*\H) -- (1,-0.5,0.3*\H);
\node[font=\scriptsize,green!40!black,anchor=west,align=left] at (1.1,1,0.85*\H) {rotation axis:\\winds the circle};
\node[font=\scriptsize,violet,anchor=west,align=left] at (1.1,1,0.3*\H) {fibre cycle:\\torsion, order 2};
\node[font=\small] at ($(0,-1,0)+(0.45cm,-0.8cm)$) {(c) logical strings};
\end{scope}
\end{tikzpicture}
\caption{The quarter-turn space $M=T^2_L\rtimes_{C_4}S^1$ of Proposition~\ref{prop:quarter} (schematic). (a) A fundamental domain: the fibre
square, with opposite sides identified, times the $J$ layers of the circle. The top face is glued to the bottom face after a quarter turn,
$(x,y,0)\sim(-y,x,J)$, so the two arrows are identified. (b) The two membranes of Table~\ref{tab:quarter}: a fibre, and the cross
$\{x=0\}\cup\{y=0\}$ swept around the circle, which closes up because the quarter turn maps the cross to itself. (c) Two kinds of logical
strings. The rotation axis is a closed loop of length $J$ that winds once around the circle; the lightest logical operators of
Table~\ref{tab:quarter} are strings of this kind. A fibre cycle is the torsion class of $H_1(M;\Z)=\Z\oplus\Z_2$: twice it bounds, and the
linking form on this torsion gives the cubic term $\int\tau^3$ behind the logical $T$.}
\label{fig:quarter}
\end{figure}
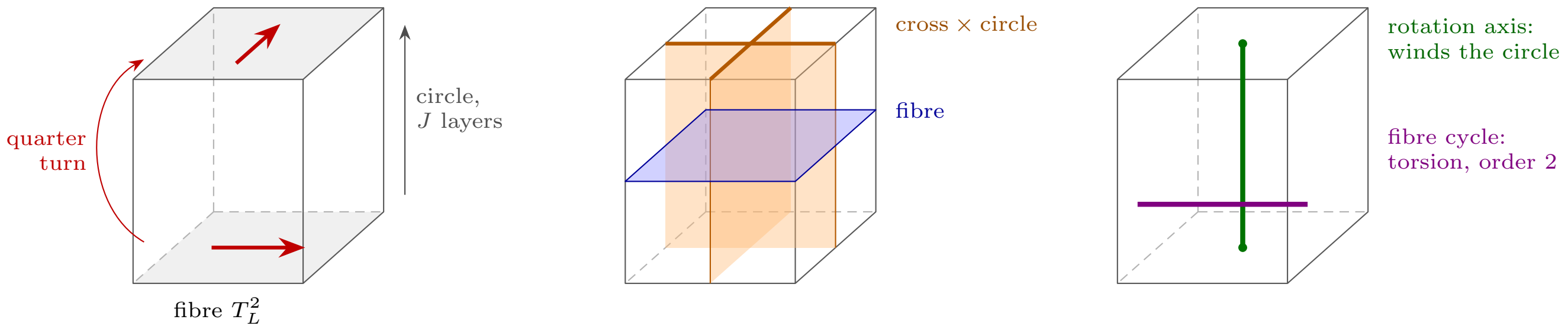

The quarter turn needs a pump of a $p+ip$ layer that commutes with the rotation. Appendix~\ref{app:tech-sym} constructs one
(Lemma~\ref{lem:pippump}). On each pair of layers a unitary $W$, built from the occupied projector of the $p+ip$
Hamiltonian, nucleates a $p+ip$ and a $p-ip$ layer from the vacuum, and $WQ=e^{iK}$, with $Q$ the parity of one layer of the pair, has a
quasi-local generator $K$ that commutes with the quarter turn. Running $K$ first on the pairs $(2j-1,2j)$ and then on the pairs $(2j+1,2j)$
pumps one layer once around the circle. The layer has Chern number $\pm1$, so it is the generator $\nu=\pm1$ of the sixteenfold way, and
the pump is the sweep of the $p+ip$ defect of~\cite{Barkeshli:2023bta}, whose logical action is their eq.~(26).
Figure~\ref{fig:quarter} shows the space on which it acts.

\begin{proposition}[A logical $T$ from a quarter turn]\label{prop:quarter}
Let $\tilde M=T^2_L\times S^1$, with the square fibre $T^2_L$ of even side $L$ and $4J$ layers along the circle, $J$ even, and let $g$ be the
quarter turn $C_4$ of the fibre composed with the translation by $J$ layers. It acts freely, and $M=\tilde M/\langle g\rangle=T^2_L\rtimes_{C_4}S^1$ is
the quarter-turn space, with $H_1(M;\Z)=\Z\oplus\Z_2$ and circle length $L_z=J$. Take the 3D fermionic toric code on $M$, the gauged atomic
insulator of the fermions on the $J$ layers glued at the seam by the quarter turn~\cite{Barkeshli:2023bta}. It has $k=2$. Fix a
reference spin structure $s_0$ on $M$, let $z\in H^1(M;\Z_2)$ be the class pulled back from the circle, and let $\tau\in H^1(M;\Z_2)$ be a class that
restricts non-trivially to the fibre, dual to the torsion of $H_1(M;\Z)$ and defined up to adding $z$. The spin structures are
$s=s_0+a\tau+bz$, $a,b\in\Z_2$, and the logical basis is $|a,b\rangle=|s_0+a\tau+bz\rangle$, with qubits $\tau$ and $z$. Let $V$ be the gauged
pump of Lemma~\ref{lem:pippump} (Appendix~\ref{app:tech-sym}) on the $J$ layers, with $\mu=\pm1$ (either chirality) and the convention of
Proposition~\ref{prop:gaussdescent}(c).
\begin{enumerate}[label=(\alph*)]
\item (Exact descent.) $V$ preserves the code space exactly, for every $L$ and $J$, and acts on it as a diagonal gate $|s\rangle\mapsto\theta(s)|s\rangle$.
\item (Locality.) With the Wilson lines routed as in the proof, $V$ is the evolution, in two unit-time steps, of quasi-local Hamiltonians
on the qubits of the code, whose terms decay exponentially with a length $\xi$ independent of $L$ and $J$; on the subspace of flat gauge
fields, which contains the code space, $V$ does not depend on the routing. For every $\varepsilon>0$ it is within $\varepsilon$ in operator norm of
a circuit of constant depth whose gates act on regions of diameter $O(\xi\log(n/\varepsilon))$, with $n$ the number of qubits.
\item (Gate.) $V$ sweeps one layer of the invertible phase $\nu=\pm1$ once through $M$, which is the $p+ip$ defect
of~\cite{Barkeshli:2023bta}. Taking as input the logical action of that sweep,~\cite[eq.~(26)]{Barkeshli:2023bta}, derived there in field
theory, $V$ is $T_\tau^{\pm1}$ modulo logical Clifford gates, with the sign set by the chirality, for every choice of $s_0$ and $\tau$.
\item (Distance.) $d=\Theta(\min(L,L_z))$.
\end{enumerate}
\end{proposition}
\idea The pump commutes with the deck generator, so it descends exactly by Proposition~\ref{prop:gaussdescent}. Its generator is
quasi-local, which gives (b). The gate is read off from the triple cup products of $H^1(M;\Z_2)$, in which only $\int\tau^3$ is non-zero, and the
distance follows from the injectivity radius. \movedproof{prop:quarter}

\paragraph{Logical basis.} By Proposition~\ref{prop:gaussdescent}(c) the phases $\theta(s)$ are fixed sector by sector, up to one
global phase, without reference to $s_0$ or $\tau$. Changing $s_0$ by $a'\tau+b'z$ relabels the basis $|a,b\rangle$ by the logical Pauli
$X_\tau^{a'}X_z^{b'}$, and replacing $\tau$ by $\tau+z$ relabels it by the logical $CNOT$ with control $\tau$. The first multiplies a diagonal
gate $D$ of level 3 by the diagonal Clifford $XDXD^\dagger$, and the second commutes with $T_\tau$, so (c) holds for every choice.

\paragraph{Why the layer determines the gate.} Item (c) uses the pump only through the invertible phase it sweeps; the input
from~\cite{Barkeshli:2023bta} is that, modulo logical Cliffords, the logical action depends on nothing else. In the field theory, two sweeps
of the same $2+1$D invertible phase differ by an $s$-independent phase and by sweeps of lower-dimensional invertible phases, Kitaev
chains on surfaces and fermion worldlines on curves, whose signs are quadratic and linear in $(a,b)$, that is, logical Cliffords; the
many-body convention of Proposition~\ref{prop:gaussdescent}(c) affects only scalars, as noted after that proposition. Since $T^2=S$, the
action modulo Cliffords then depends only on $\nu$ mod 2, and $\nu=\pm1$ by Lemma~\ref{lem:pippump}(v). The global phase and Clifford
factors, such as the $CZ_{\tau z}$ and $Z_\tau^\epsilon$ found by check T21, are what this identification cannot see and what (c) leaves open;
T21 tests the field-theory step on this pump without using it.

\paragraph{Range and depth.} No exact circuit of bounded depth and bounded range can replace the approximation in (b): restricted to a
half space, it would create a $p+ip$ layer from a product state. A Hamiltonian of finite range is a different notion, since its
finite-time evolution generally has exponential tails. The circuit in (b) has constant depth, but its gates act on regions of diameter
$O(\log(n/\varepsilon))$, not of fixed size. This is the sense in which the logical $CCZ$ of~\cite{FidkowskiHastings2024} and the $T$
of~\cite{Barkeshli:2023bta} are constant depth, and the quarter-turn $T$ meets the same standard.

 The gate is the value of~\cite[eq.~(26)]{Barkeshli:2023bta} on these spaces;
the explicit equivariant pump, its identification with the $p+ip$ defect, its exact descent, its locality and the distance are what the
proposition adds. Check T16 confirms the cup products exactly, and checks T17, T18, T20 and T21 confirm the other statements numerically,
T21 without using eq.~(26) (Appendix~\ref{app:tech-sym}); the proof uses none of them.

A short logical operator of a quotient breaks into pieces that each fit in a ball, and on a ball the quotient looks like the parent. This
gives a lower bound on the distance from the injectivity radius, for any quotient and in particular for the family of
Proposition~\ref{prop:quarter}.

\begin{proposition}[Distance from the injectivity radius]\label{prop:injdist}
Let a finite group $\Gamma$ act freely by monomial-Clifford symmetries on a Pauli stabilizer parent $S$, generated by a set of checks, with quotient code $S/\Gamma$ (Definition~\ref{def:quot}). Measure distances in the \emph{check graph}, in which two qudits are
adjacent when some check acts on both, and on the quotient in the image of this graph. Suppose that the quotient map is injective on every
ball of radius $R$ of the parent, and let $\tilde d_R$ be the least weight of an element of $S^\perp\setminus S$ supported in such a ball,
with $\tilde d_R=\infty$ if there is none; so $\tilde d_R\ge d$. Then every non-trivial logical Pauli operator of $S/\Gamma$ has weight at
least $\min(\tilde d_R,R)$.
\end{proposition}
\idea Split a light logical operator into clusters of qudits linked by checks. Each check meets at most one cluster, so each cluster
commutes with every check by itself, and a light cluster fits in a ball on which the quotient is a copy of the parent. Its lift is then
a Pauli operator of the parent that commutes with every check and is lighter than $\tilde d_R$, hence a stabilizer, and its image is a
stabilizer of the quotient.
\movedproof{prop:injdist}

The proof uses $\Gamma$ only through $\pi$ and the formula $\langle\pi a,\pi b\rangle'=\sum_g\langle a,gb\rangle$, so it applies verbatim to the
universal cover, with $\Gamma=\pi_1(M)$ acting with finitely many orbits of qudits and checks. If the parent on the universal cover has no
finitely supported logical operator, then $\tilde d_R=\infty$ for every $R$, and $d\ge R$ for every $R$ up to the injectivity radius of $M$
in the check-graph metric. For the 3D toric code this holds because a finitely supported 1-cycle of $\mathbb R^3$ bounds a finite 2-chain and a
finitely supported 1-cocycle is the coboundary of a finitely supported 0-cochain, since $H^1_c(\mathbb R^3)=0$. For the fermionic toric code it holds by Lemma~\ref{lem:ftclocal} of Appendix~\ref{app:tech-sym}, which shows that a parent of the form
$W_c=Z(\partial c)$, $G_e=X^{\delta e}Z^{\lambda_e}$ on $\mathbb R^3$ has no finitely supported logical operators. For the flat quarter-turn space the injectivity radius is half the systole,
$\min(L,L_z)/2$; a ball of check-graph radius $R$ lies in a Euclidean ball of radius $r_0R$, with $r_0$ the largest Euclidean diameter of a
check, so $d\ge\min(L,L_z)/(2r_0)$.

\paragraph{Small members of the family.} Table~\ref{tab:quarter} gives the code of Proposition~\ref{prop:quarter} on an explicit
triangulation. Each fibre square is cut into four triangles by its centre, the lattice edges are oriented by a rule invariant under the
quarter turn and the edges to a centre point to it, and the prism over a fibre triangle $(a<b<c)$ is cut into the tetrahedra
$(a_0b_0c_0c_1)$, $(a_0b_0b_1c_1)$, $(a_0a_1b_1c_1)$. This branched triangulation is $g$-invariant and descends to $M$, with a qubit on each
of the $n=24L^2J$ triangles and the checks $W_t=Z(\partial t)$ and $G_e=X(\delta e)Z(\lambda_e)$, $\lambda_e(f)=\int\delta e\cup_1f$, of the exact
bosonization~\cite{ChenKapustin2019,Chen2020bosonization}. On each instance the checks commute, $k=2$, and the T-junction
test~\cite{LevinWen2003} finds a fermion. A logical operator that anticommutes with the membrane $Z(\Sigma)$, for $\Sigma$ the fibre or the
cross $\{x=0\}\cup\{y=0\}$ swept around the circle (Figure~\ref{fig:quarter}(b)), has as $X$-part a non-trivial dual cycle; we find the least weight of such operators
exactly by enumerating these cycles up to that weight and minimizing the $Z$-part near each one. Every other non-trivial logical operator
anticommutes with all logical strings of two of the three non-trivial classes, so its weight is at least the number of pairwise disjoint
such strings, which a greedy search bounds from below. In each instance this bound is at least the first minimum, so $d$ is that minimum, exactly. In these instances $d=4J+1$, carried by a
string that winds once around the circle, independently of $L$; by (d), the fibre strings take over once $J$ is large compared
with $L$.

\begin{table}[ht]
\centering
\caption{Small members of the family of Proposition~\ref{prop:quarter}, on the triangulation of the text, all with $k=2$: the
number of qubits, the lengths of the shortest non-trivial dual cycles that wind around the circle and that lie in the torsion class of
$H_1(M;\Z)$, the exact distance, and the lower bound for the logical operators that commute with both membranes (check T25).}
\label{tab:quarter}
\begin{tabular}{@{}ccrcccc@{}}
\toprule
$L$ & $J$ & $n=24L^2J$ & circle cycle & torsion cycle & $d$ & other operators \\
\midrule
4 & 2 & 768 & 8 & 15 & 9 & $\ge12$ \\
6 & 2 & 1728 & 8 & 15 & 9 & $\ge17$ \\
8 & 2 & 3072 & 8 & 15 & 9 & $\ge29$ \\
4 & 4 & 1536 & 14 & 16 & 17 & $\ge20$ \\
6 & 4 & 3456 & 14 & 24 & 17 & $\ge27$ \\
8 & 4 & 6144 & 14 & 27 & 17 & $\ge38$ \\
\bottomrule
\end{tabular}
\end{table}

\paragraph{Why no orientation reversal is needed.} By Proposition~\ref{prop:threegroup}(d), the bosonic gauged SPT $(-1)^{\int a\cup a\cup a}$ is a
non-Pauli gate on a quotient only if some deck transformation reverses the orientation. The fermionic pump needs no reversal, because its
cubic term $\int\sigma^3=\int\sigma\cup\beta\sigma$, with $\beta$ the Bockstein, is the linking form on the 2-torsion of $H_1(M;\Z)$, which the spin structure
refines to $\Z_8$. For a mapping torus $T^2\rtimes_RS^1$ the torsion of $H_1$ is $\mathrm{coker}(R-1)$ on $\Z^2$. For $R=1$ there is none, and the gate on
$T^3$ is the $CCZ$ from $b_1=3$. For the half turn it is $\Z_2^2$ with a hyperbolic linking form, which gives the $CS^\dagger$ of
Example~\ref{ex:dicosm}. For the third and sixth turns it is $\Z_3$ and $0$, with no 2-torsion, and the gate is Pauli. For the quarter turn it is
$\Z_2$ with linking form $\pm\frac12$, which the refinement makes a $T$.

\paragraph{Non-abelian deck groups.} On the quaternionic space $S^3/Q_8$, $k=2$, and on the barycentric subdivision $\int x^2y=\int xy^2=1$ and
$\int x^3=\int y^3=0$ (check T16), so modulo Cliffords the formula~\cite[eq.~(26)]{Barkeshli:2023bta} gives a logical $CS$. The parent $S^3$ has no
logical qubits, so the twist creates the logical action, as for the $CCZ$ on $\mathbb{RP}^3$. On $\mathbb{RP}^3$ itself the cup product gives the $T$ of Proposition~\ref{prop:twisted}, and on $L(4,1)=S^3/\Z_4$ only a Clifford.
For these spaces no pump is known, because the pumps we know need a layered space and $S^3$ is not layered. Barkeshli, Hsin and
Kobayashi realize $\mathbb{RP}^3$ inside $T^3\#\mathbb{RP}^3$ by a partial rotation, where the pump projects onto the code space as a non-unitary
operator proportional to $T$~\cite{Barkeshli:2023bta}.

\paragraph{Internal twists.} Twisting the fermionic toric code by its internal symmetries is expected to give no gate
beyond Paulis, or no code: a holonomy $V^\nu$ around a circle gives the chiral theory $\nu$ of Kitaev's sixteenfold way, which is not
expected to have a commuting-projector realization with bounded-range checks, and magnetic, fermion-line and Kitaev-chain backgrounds
leave no state or lift some sectors. We leave these cases open.

\section{Discussion and open problems}
\label{sec:discussion}

\paragraph{Summary.} Quotient codes and twisted compactification are one construction. For codes, this makes the design of a code the
choice of a parent, a symmetry and an internal space, and the transfer theorem and its relatives compute the result. For field theory, it
gives each compactification a lattice realization with explicit checks, logical operators and distance. The gluing need not be a
permutation: entangling twists give codes beyond the Pauli stabilizer framework, and twists create logical gates, such as the $CCZ$ on
$\mathbb{RP}^3$ and the pump on the quarter-turn space that acts as a logical $T$ up to Clifford factors.

\paragraph{Design rules.} The results also say in advance when a search cannot succeed. A Clifford twist, of degree at most 2, never gives
non-abelian order (Proposition~\ref{prop:clifflevel}). Monomial-Clifford automorphisms that descend through a monomial quotient of a Pauli
parent induce only logical Clifford gates (Corollary~\ref{cor:cliffordstay}); other code-preserving circuits on the same quotient can induce
non-Clifford gates, such as the cup-product $CCZ$ on $\mathbb{RP}^3$ (Proposition~\ref{prop:twisted}). For a bosonic response, a twisted-sector gate is the
identity on every sector whose classifying map factors through a space with $H_3=0$; for the SWAP example of
Section~\ref{sec:twistgates}, the bosonic cup-product response therefore acts trivially on every flat sector on $T^3$
(Corollary~\ref{cor:rank}). Within the twists of Proposition~\ref{claim:reach}, the higher-level gates of~\cite{WSN2025} on the doubles of larger
dihedral groups are out of reach; they need gauging. Table~\ref{tab:create} lists these limits next to the ways in which a twist does create
gates.

\paragraph{Open problems.} The correspondence turns questions about codes into questions about twisted compactification and back. There are
several future directions we would like to explore:

\begin{enumerate}[leftmargin=1.5em]
\item \emph{Transfer and distance for dressed quotients.} Theorem~\ref{thm:general} gives the rate, the logical operators and distance bounds
of a quotient, but it needs Pauli stabilizers. For cubic and higher cup defects the dressed quotient is an $X$-monomial code
(Definition~\ref{def:xmono}, Section~\ref{sec:chsparent}); its dimension is a point count (Theorem~\ref{thm:pointcount}), and
Theorem~\ref{thm:dresseddist} bounds its distance from below by local data of its CSS shadow, but no formula yet gives its logical operators.
Find the analogue of Theorem~\ref{thm:general} for these codes, with a Tate term and a distance bound that uses the dressing. Distances of
families with growing $X$ are open even for quadratic twists on lens spaces and $\mathbb{RP}^2$ (Section~\ref{sec:lens},
Remark~\ref{rem:dresseddist}), where the split of the quotient logicals into winding sectors and
parent classes also lacks a formula: the lattice shows that it depends on whether $p=2$.
\item \emph{A cohomological formula for cubic counts.} For quadratic twists $k'=\dim\ker\cup[B]$ (Theorem~\ref{thm:hfcount}), and
Proposition~\ref{prop:cone} realizes this count by a two-step filtration with $d_1=\cup[B]$. For cubic twists the dimension is a point count on
the zero fibre of $Q_B$ (Theorem~\ref{thm:pointcount}), such as $22$ for type III on the slab over $T^2$, which no linear count of logical qubits
gives. Find a filtration of the zero fibre, for instance by the rank of the linearization, whose pieces are counted by cohomological data.
\item \emph{Codes beyond the quotient construction.} A class that survives every finite cover, such as a non-zero $B\in H^2(S^2;\Z_p)$, still defines a
dressed Gauss-law code, which Theorem~\ref{thm:dressedquot} does not produce (Remark~\ref{rem:cover}). Is there a quotient presentation with a
natural parent, for instance a parent with a gauged higher-form symmetry? Non-invertible walls beyond condensation have no quotient description at all. Such a defect becomes invertible when the
internal space has no cycles supporting its fusion outcomes~\cite[Section~3.3.3]{GukovHsinPei2025}; does the twist then give an honest
quotient? In the other direction, Lemma~\ref{lem:noninv} turns a non-invertible parent symmetry into a logical gate on any quotient on which
its fusion outcomes act as scalars. Find an example in which this gate is not induced by an invertible parent symmetry; a natural candidate is
the swap of the twisted $\Z_2^3$ gauge theory, which does not preserve $[\omega]$ and is realized as a non-invertible sandwich of condensation
walls~\cite[Section~2.3.1]{HsinKobayashi2025aut}.
\item \emph{Twists for $T$-type gates on growing families.} Section~\ref{sec:sym} computes the gates that a given twist creates. The converse,
choosing the twist so that a logical $T$ appears on a family with growing distance, is open along three routes. Which free actions on
3-manifolds give classes of order 8 in $\Omega_3^{\rm Spin}(B\Gamma)$ (Proposition~\ref{prop:twisted})? Which $\Z_3$ actions on 3-manifolds,
free or not, have a non-degenerate twisted cubic form $\int a\cup R^*a\cup(R^2)^*a$ (Proposition~\ref{prop:internal};~\cite{BDEHT2026cupsII}
gives instances)? Proposition~\ref{prop:tori} answers this on tori, but on other manifolds the torsion classes do not lift integrally. And
the transgression gates of Propositions~\ref{prop:transgress}, \ref{prop:slabgate} and~\ref{prop:reltrans} extend to $N\ge4$ copies only with
a cubical $\cup_1$ that satisfies the Steenrod coboundary formula, since the $\lambda$ of~\cite[eq.~(47)]{KZH2025chs} and the swap
automorphism of $\frac{2\pi}{N}\int a_1\cup a_2\cup a_3$~\cite{HsinKobayashi2025aut} contain $\cup_1$ products. For the fermionic pump the
quarter turn of Proposition~\ref{prop:quarter} gives such a family (Proposition~\ref{prop:injdist}, Lemma~\ref{lem:ftclocal}), with a
finite-time evolution, and so a constant-depth circuit of gates of logarithmic range, in place of a strictly local circuit, which no chiral
pump admits. Whether the pumps on $S^3/Q_8$ and $\mathbb{RP}^3$, which are not layered, can be realized in the same way is open
(Section~\ref{sec:fermionic}).
\item \emph{Twisting versus gauging.} Summing over $G$-bundles instead of fixing one gives quantum doubles, with ground space
$\CC[\Hom(\pi_1X,G)/G]$ for non-abelian $G$, and the non-abelian qLDPC codes of~\cite{ZKH2026nonabelian,ZTKH2026sheaf} are obtained by
gauging. Proposition~\ref{prop:nilp} draws one line between the two: the twists of Proposition~\ref{claim:reach} stay at nilpotency class 2, while a single
gauging step, of the charge conjugation of the $\Z_{4N}$ toric code, gives $D(D_{4N})$, of class above 2 for $N\ge2$. Conversely, for a
central extension $1\to Z\to G\to K\to1$ the anyons of $D(G)$ with flux in $Z$ and charge pulled back from $K$ form a Lagrangian
subcategory $\mathrm{Rep}(\hat Z\times K)$, so $D(G)\simeq D^{\omega}(\hat Z\times K)$ for some $\omega$~\cite{DGNO2007,NaiduNikshych2008}. Here
$\omega=\hat c\cup e$ with $e$ the extension class, which check T19 confirms on the anyon data of $D(D_8)$ and $D(D_{16})$. This
presents each step $D(D_4)\to D(D_8)\to D(D_{16})$ as a twist of a non-abelian parent, the $K$ gauge theory written in its own gauge field,
which is not a Pauli parent. Relate gauging to twisted compactification at the level of codes,
and find when gauging preserves distance.
\end{enumerate}

\section*{Acknowledgment}
P.-S. H. is supported by Department of Mathematics, King's College London.
Claude Opus 5.5 was used to assist with drafting and editing text and with writing and running the scripts for the computations
of Appendix~\ref{app:checks}. The author is responsible for the content of the paper.
All scripts and data used for the computations in this paper are available from the author upon request.

\appendix
\section{The dictionary, known codes and known gates}
\label{app:dict}

Table~\ref{tab:dict} lists the correspondence between twisted compactification and quotient codes entry by entry. The left column is the
field-theory side, from Section~\ref{sec:ft}, and the right column the same statement for codes, with the result where it appears. The upper part of the table is the construction of codes (steps 1--3 of the method), and the lower
part their symmetries and gates (step 4).

\begin{table}[tbp]
\centering\small
\begin{tabularx}{\textwidth}{@{}>{\raggedright}p{6.2cm}>{\raggedright\arraybackslash}X@{}}
\toprule
Field theory & Codes \\
\midrule
theory on $X\times F$ & parent on $Y\times\tilde F$ \\
0-form symmetry $G$ & monomial (Clifford) action \\
flat bundle, holonomy $\rho$ with finite image & regular cover, deck group $\Gamma=\mathrm{im}\,\rho$; balanced product, and the quotient by $\Gamma_{\rm diag}$ when the action is monomial on the bases (Thm.~\ref{thm:balanced}) \\
circle with one defect & mapping torus; Wang sequence \\
infinite-order holonomy (Anosov map) & mapping-torus code, not a quotient of a finite product \\
EM or Hodge duality wall & $\Gamma$ contains a ZX-duality; output need not be CSS (``Hadamard $+$ half shift''); the 4D 2-form toric code gives the 3D fermionic toric code \\
$q$-form background $B\in H^{q+1}(F)$ & dressed Gauss laws (Def.~\ref{def:gauss}, Lemma~\ref{lem:cochain}); on a trivializing cover, the dressed quotient by deck $\circ$ $U(\zeta_g)$ (Thm.~\ref{thm:dressedquot}) \\
$B$ with no finite trivializing cover ($H^2(S^2)$) & dressed Gauss-law code only \\
reduced action $M^B_{rs}$ & logicals pairing with $B$ become stabilizers, $k'=\dim\ker\cup[B]$ (Thm.~\ref{thm:hfcount}, Cor.~\ref{cor:boundary}); a two-step spectral sequence with $d_1=\cup[B]$ (Prop.~\ref{prop:cone}) \\
Clifford level of the defect & level of the Gauss-law dressing, at most one lower (Prop.~\ref{prop:clifflevel}) \\
defect of level $m$ on a Clifford-hierarchy parent of level $\ell$ & checks of level at most $\max(\ell,m-1)$; $Z$-checks stay Pauli (Prop.~\ref{prop:chslevel}) \\
cubic defect $\int a_1a_2a_3B$ & $CZ$-dressed Gauss laws; point count (Thm.~\ref{thm:pointcount}); not a Pauli stabilizer code when the count is not a power of 2 (Cor.~\ref{cor:cert}); distance $\Theta(L)$ (Cor.~\ref{cor:dresseddist}); $D(D_4)$ in field theory \\
condensation wall & extra stabilizers, not a quotient \\
winding sectors & Tate group $\hat H^0(\Gamma;S)$, extending the logical classes with invariant representatives (Thm.~\ref{thm:general}) \\
gauging & dynamical gauge qubits; non-abelian $G$ gives quantum doubles; those with $[G,G]$ not central have class above 2 and lie beyond the twists of Prop.~\ref{claim:reach} (Prop.~\ref{prop:nilp}) \\
\midrule
\multicolumn{2}{@{}l}{\emph{Symmetries} (Section~\ref{sec:sym})} \\
symmetry of $X\times F$ preserving the background & element of $N(\tilde\Gamma)/\tilde\Gamma$, an automorphism of the quotient; all geometric ones from a characteristic cover (Prop.~\ref{prop:lift}) \\
isometry of $F$ (Kaluza--Klein symmetry) & permutation automorphism gate; logical action filtered, with $\bar f_*$ on $E^2=H_*(F;H(Y)_\rho)$ as associated graded \\
finite subgroup of $\mathrm{MCG}(F)$, $F$ a surface & isometries at an orbifold point of moduli (Nielsen), hence permutation gates \\
infinite-order mapping class (Dehn twist) & code-space symmetry only; code deformation \\
wrapped higher-form symmetries and their anomaly & logical Paulis and their commutation \\
projective MCG action (quadratic refinement) & Pauli and phase part of an automorphism gate \\
anomaly matching under covers & cubic forms scale by the cover degree; even covers kill pulled-back $CCZ$ actions \\
higher group from isometries or mapping classes & Clifford gates whose products pick up logical Paulis \\
0-form symmetry pumping a cubic SPT response & non-Clifford diagonal gate on every quotient (Prop.~\ref{prop:cupgate}); the source of all non-Clifford gates in this paper \\
fermionic $\Z_8$ pump under a free rotation & quasi-local pump restricted to each twisted sector, exactly code-preserving (Prop.~\ref{prop:gaussdescent}); logical $T$ on the quarter-turn space (Prop.~\ref{prop:quarter}) \\
response on holonomy sectors of the deck group & gate on twisted-sector logicals $=\langle\omega,c_*[M]\rangle$, a bordism invariant \\
twist by a symmetry commuting with the pump & pump on twisted cohomology; can create or kill the gate \\
3-group of the 3D toric code~\cite{Barkeshli:2022edm} & gauged-SPT gate is a quadratic form with polar form $\langle w_1\alpha\lambda\rangle$; logical $CZ$ from an orientation-reversing holonomy (Prop.~\ref{prop:threegroup}) \\
non-invertible symmetry whose fusion outcomes are trivial on $F$ & logical gate on the quotient (Lemma~\ref{lem:noninv}) \\
automorphism preserving $\omega$ up to $\delta\lambda$~\cite{HsinKobayashi2025aut} & descends with the wall dressing $e^{2\pi i\int\lambda\cup B_\Z}$ (Prop.~\ref{prop:transgress}); the 2D logical $T$ of~\cite{KZH2025chs} \\
gate controlled by Kaluza--Klein modes of a small $F$ & distance at most the shortest essential fibre cycle \\
\bottomrule
\end{tabularx}
\caption{The dictionary between twisted compactification and quotient codes. Each code-side entry holds under the hypotheses of the cited
statement.}
\label{tab:dict}
\end{table}

Table~\ref{tab:known} writes known code families as a parent, a symmetry, an internal space and a twist. It also contains phases that are
not usually described as quotients: twisted quantum doubles such as $D(D_4)$~\cite{deWildPropitius1995} and the Clifford-hierarchy stabilizer
codes of~\cite{KZH2025chs} are compactifications of copies of toric codes with a wall that carries a gauged SPT phase
(Sections~\ref{sec:hfcountsec} and~\ref{sec:chsparent}). These codes and phases are known; what we add is their place in one construction.

\begin{table}[tbp]
\centering\small
\begin{tabularx}{\textwidth}{@{}>{\raggedright}p{3.7cm}>{\raggedright}p{4.1cm}>{\raggedright}p{3.1cm}>{\raggedright\arraybackslash}X@{}}
\toprule
Code family & Fibre $Y$ with symmetry & Internal $F$ & Twist \\
\midrule
translation-twisted torus & slice code & $S^1$ & translation \\
balanced, fibre-bundle products; Haah's code & Tanner or fractal code & $S^1$, general $F$ & permutation \\
BB / two-block, abelian $G$ & $\FF_2[G]\xrightarrow{\cdot\sum B}\FF_2[G]$ & rose $\FF_2[G]\xrightarrow{\sum A\cdot}\FF_2[G]$ & $a\in A$, abelian \\
non-abelian two-block & same, $G=D_m\times\Z_2$ & same & non-abelian \\
lifted product~\cite{PanteleevKalachev2022good} & Tanner complex of $\Cay(G,B)$ & Tanner rose, local code $C_A$ & $A\subset G$, expanding \\
quantum Tanner~\cite{LeverrierZemor2022} & Tanner sheaf on $\hCay(G,B)$ & dipole $\Theta_\Delta=\hCay(G,A)/G$ & $A$; total space the quadripartite complex \\
five-qubit code, $\code{t^2+(t+1)^2,1,2t+1}$ & 2D toric code & $S^1$, $T^2$ & duality \\
genon codes & BB code & $S^1$ of length $\frac12$ & ZX-duality \\
triality codes, $\code{32,2,7}$ & colour code & $S^1$ & $X\to Y\to Z$ \\
higher-form dressed codes & $n$-form toric code & $T^{q+1}$, $D^{q+1}$ & $B\in H^{q+1}$ \\
gauged-SPT codes, $D(D_4)$ & copies of toric codes & $D^{q+1}$, $S^1$ & cubic, quartic cup defects \\
Clifford-hierarchy stabilizer codes~\cite{KZH2025chs} & $r$ toric codes in $r$ dimensions & $D^1$, $S^1$ & wall carrying $(-1)^{\int a_1\cdots a_r}$ \\
\bottomrule
\end{tabularx}
\caption{Known codes as twisted compactifications. $\hCay$ is the bipartite double cover; the dipole $\Theta_\Delta$ has two vertices joined by
$\Delta$ edges.}
\label{tab:known}
\end{table}

\paragraph{Five cases in detail.} Gap~1 of Section~\ref{sec:intronew} names five cases that show the reach of the construction. Here they
are in detail.
\begin{itemize}[leftmargin=1.5em]
\item \emph{Duality twists.} The optimal twisted toric family $\code{t^2+(t+1)^2,1,2t+1}$ of Kovalev and Pryadko and of Bomb\'in and
Martin-Delgado~\cite{KovalevPryadko2012,BombinMartinDelgado2007} is the 2D toric code on the torus with lattice
$\langle(1,2t+1),(0,2(t^2+(t+1)^2))\rangle$ and one $e\leftrightarrow m$ wall. For $t=1$ it is the five-qubit code, the running example of this
paper (Examples~\ref{ex:run1}--\ref{ex:run4}).
\item \emph{Beyond dualities.} If $C_X=C_Z$, every transversal single-qubit Clifford is a symmetry. The 2D colour code has a triality symmetry
$X\to Y\to Z$, whose twist defects are studied in~\cite{KesselringColorTwist2018}. The colour code on a $12\times12$ torus modulo the translation $(4,4)$
composed with $X\to Y\to Z$ gives a $\code{32,2,7}$ code (exact distance, Appendix~\ref{app:checks}) with weight-6 checks that uses all three
Pauli types.
\item \emph{Haah's code.} The balanced product $C_1\otimes_GC_2=(C_1\otimes C_2)/G_{\rm diag}$ is the case of Theorem~\ref{thm:balanced} in
which $F$ is the classifying complex of the action. Let $C_A$ be the cyclic code of $1+x+y+z$ and $C_B$ that of $1+xy+yz+zx$ on $\Z_L^3$.
Their tensor product, in the convention of~\cite{Zhang:2026onf}, modulo the diagonal $\Z_L^3$ is Haah's code~\cite{PhysRevA.83.042330}, with
$(n,k)=(16,6),(54,2),(128,14)$ for $L=2,3,4$.
\item \emph{Non-abelian two-block codes.} For $G=D_m\times\Z_2$, $m=3,6$, the quotient of the hypergraph-type product by the diagonal $G$
is the non-abelian two-block group-algebra code of~\cite{LinPryadko2024} (Appendix~\ref{app:checks}). The internal space is a rose with
non-abelian holonomy. This is the simplest example of what gap~3 of Section~\ref{sec:intronew} calls type (N1): the twist group is non-abelian, but the code
is still a Pauli stabilizer code.
\item \emph{Good codes.} The good quantum LDPC codes~\cite{PanteleevKalachev2022good,LeverrierZemor2022} are twisted compactifications on a
rose or a dipole of $\Delta$ circles, whose holonomies act on a fibre that is itself a classical expander code. The untwisted version is a
hypergraph-type product, with distance $\Theta(\sqrt n)$; the expanding non-abelian twist makes the distance linear.
\end{itemize}

\paragraph{Sources of the example gates.} The example gates of Section~\ref{sec:sym} are known, except for the two fermionic evaluations
named in gap~4 of Section~\ref{sec:intronew}. The logical $CCZ$ on $\mathbb{RP}^3$ is the bosonic analogue of the $T$ gate
of~\cite{Barkeshli:2023bta}, and triple-intersection gates on 3-manifolds and mapping tori are studied
in~\cite{ZhuSikanderPortnoyCrossBrown2023,Barkeshli:2023bta}. The 3-group of the 3D toric code, its lattice commutator and the logical $CZ$ on
$K\times S^1$ are due to Barkeshli, Chen, Hsin and Kobayashi~\cite{Barkeshli:2022edm}. The 2D automorphism gate with logical $T$ and its
boundary and hinge terms are due to Kobayashi, Zhu and Hsin~\cite{KZH2025chs}. Warman and Sch\"afer-Nameki obtain the same $T$ in the
$D(D_4)$ presentation and extend it to the doubles of all dihedral groups of order $8N$~\cite{WSN2025}; by Proposition~\ref{prop:nilp}, the
extension to $N\ge2$ is beyond the twists of Proposition~\ref{claim:reach}. The fermionic $\Z_8$ pump and its logical $CCZ$ on $T^3$ are due to
Fidkowski and Hastings~\cite{FidkowskiHastings2024}, and the controlled-$S^\dagger$ on the half-turn space and the $T$ on $\mathbb{RP}^3$ to
Barkeshli, Hsin and Kobayashi~\cite{Barkeshli:2023bta}. The twisted cup products and the qutrit and qubit $T$ gates on tori are
in~\cite{BDEHT2026cupsII}. That compactification can make a non-invertible symmetry invertible is the observation of Gukov, Hsin and
Pei~\cite{GukovHsinPei2025}, of which Lemma~\ref{lem:noninv} is the lattice form. Whether a symmetry of a quotient is accidental depends on
the parent: the rotation of the $3\times3$ toric code descends from the $6\times6$ parent but not from the $6\times3$ parent
(Proposition~\ref{prop:lift}, Example~\ref{ex:accidental}).

\section{Technical complements}
\label{app:tech}

This appendix collects the conventions, auxiliary statements and side remarks that Sections~\ref{sec:ft}--\ref{sec:sym} use, in the order
of the sections they support. Three of them carry the proofs of main results: the local cochain lemma (Lemma~\ref{lem:cochain}), on which the
dressed Gauss laws rest; the equivariant $p+ip$ pump (Lemma~\ref{lem:pippump}), on which Proposition~\ref{prop:quarter} rests; and
Lemma~\ref{lem:ftclocal}, which gives its distance bound. Appendix~\ref{app:worked} works out the cochain lemma by hand in three cases.
Proofs not given here are in Appendix~\ref{app:proofs}.

\subsection{Complements to Section~\texorpdfstring{\ref{sec:ft}}{2}}
\label{app:tech-ft}

\paragraph{Spin theories and non-invertible walls.} For a spin TQFT the exchange of the two circles exchanges their spin structures. The
left channel counts states only when $S^1_t$ is antiperiodic, so the formula then holds with $S^1_t$ antiperiodic in $\mathcal H(Y\times S^1_t)$, and
with $(-1)^F\hat W$ in place of $\hat W$ when $S^1_W$ is periodic; the codes of this paper are bosonic. For a non-invertible wall the two
channels give $\dim\mathcal H_W(Y)=\TrW\hat W$ with $\hat W$ the operator of the TQFT, which has no phase ambiguity; for a condensation wall it is
the sum of the condensed operators over the cycles of $Y\times S^1$ times a positive normalization fixed by higher
gauging~\cite{Roumpedakis:2022aik,GukovHsinPei2025}. The codes of this paper use only invertible walls.

When $\hat W$ acts on the code space as a Clifford unitary, write it as $D_v\,\mu(g)$, with $g$ its symplectic action on the logical Paulis,
$\mu$ the Weil representation and $D_v$ a logical Pauli. For odd prime $N$, $|\TrW D_v\mu(g)|^2=|\ker(g-1)|$ if $v$ lies in the image of $g-1$,
and the trace vanishes otherwise~\cite{Howe1973Weil,Gerardin1977,ThomasWeil2008,Gukov:2020btk}; for instance a logical Pauli other than the
identity, $g=1$ and $v\ne0$, has trace zero. When the trace is non-zero, $k'=\frac12\dim\ker(g-1)$. On the lattice we do not use this formula: the counts for permutation and duality twists are given by
Section~\ref{sec:general} and Theorem~\ref{thm:ss}, and the running example tests the trace formula for a duality wall at $N=2$
(Example~\ref{ex:run3}).

\subsection{Complements to Section~\texorpdfstring{\ref{sec:lattice}}{3}}
\label{app:tech-lattice}

\paragraph{Composite dimensions.} All results are stated for prime $p$, and most examples have $p=2$. For composite $N$ the Pauli group
modulo phases is the $\Z_N$-module $\Z_N^{2n}$, and a Pauli stabilizer group is a submodule $S$ with code-space dimension $N^n/|S|$, which need not be a
power of $N$: on one four-level qudit, the Pauli stabilizer group generated by $Z^2$ has a two-dimensional code space. The proofs of
Proposition~\ref{prop:valid} and Theorem~\ref{thm:general} use only orders of finite groups and go through with dimensions replaced by
$\log_N$ of orders; Proposition~\ref{prop:count} and the rank arguments of Section~\ref{sec:higher} and Appendix~\ref{sec:cert} use vector spaces over
$\FF_p$ and do not.

\paragraph{Factorization in practice.} We checked that one-step and two-step quotients agree on small instances ($\code{12,1}$ and
$\code{24,4}$, both not CSS), as Theorem~\ref{thm:factor} requires.

\begin{lemma}[Checks meeting each orbit once]\label{lem:once}
If the support of every check meets every $\Gamma$-orbit of qubits in at most one point, then each quotient check has the same weight as its
parent, and its letter on each quotient qubit is the parent letter transported by the prescribed single-qubit Clifford $c_h^{-1}$; no further
letters arise from collisions within an orbit. In particular, if every transport is the identity or a Hadamard and the parent is CSS, the
quotient checks have no $Y$ letters.
\end{lemma}
\begin{proof}
$\pi$ restricted to the support of such a check is injective, so letters are never multiplied together; each letter is only transported.
Identity and Hadamard map $X$ and $Z$ to $X$ or $Z$.
\end{proof}

The letters themselves are not preserved in general: in the running example the pure $X$ and $Z$ plaquettes become the mixed checks
$XZZXI,\dots$, because the transport on half of the sites is $H$.

\begin{remark}[Invariant classes and invariant representatives]\label{rem:invclasses}
The map $(S^\perp)^\Gamma/S^\Gamma\to\mathcal L^\Gamma$, $\mathcal L=S^\perp/S$, is injective but need not be onto. Taking invariants of $0\to S\to S^\perp\to\mathcal L\to0$
gives the exact sequence $0\to S^\Gamma\to(S^\perp)^\Gamma\to\mathcal L^\Gamma\to H^1(\Gamma;S)$, so an invariant logical class need not contain an invariant
Pauli operator, and the obstruction lies in $H^1(\Gamma;S)$. If $p\nmid|\Gamma|$, then $H^1(\Gamma;S)=0$ and the two spaces agree. In the modular case, which is
where winding sectors occur, they can differ: in the running example $\mathcal L^\Gamma$ is two-dimensional but only one dimension has invariant
representatives (Example~\ref{ex:run2}). So the transfer theorem describes the quotient logicals as an extension of the parent logical classes
\emph{with invariant representatives} by the Tate group, not as an extension of all invariant logical classes.
\end{remark}

\paragraph{Integration.} We assume $\int\delta x=0$ for every $(D-1)$-cochain $x$ that occurs in Sections~\ref{sec:higher}--\ref{sec:sym}. This holds when $[M]$ is the fundamental
cycle of a closed cubical manifold, orientable if $p$ is odd. It also holds when $[M]$ is the relative fundamental chain of a cubical manifold
with boundary and the background $B$ and the cochains $\eta,\mu$ of Section~\ref{sec:higher} are relative cochains, that is, vanish on every cell contained in $\partial M$:
every such $x$ has the form $y\cup B$, $y\cup\eta$ or $y\cup\mu$, which vanishes on $\partial M$ because $(y\cup B)(\sigma)$ uses $B$ only on faces of $\sigma$, so
$\int\delta x=\langle x,\partial[M]\rangle=0$. No collar is needed. No other topological hypothesis is used until Section~\ref{sec:hfcountsec}; in particular,
stars of cells need not be contractible. Only the Leibniz rule and locality of the cup product are used, so everything in Sections~\ref{sec:higher}--\ref{sec:chsparent} also
holds for the simplicial cup product of a simplicial complex with a branching (a local ordering of vertices); Section~\ref{sec:sym} uses this
on barycentric subdivisions, whose canonical branching is preserved by every cellular automorphism.

\paragraph{The case $q=-1$.} We also allow $|\omega|=D+1$ on a closed $M$, with $B=1\in Z^0(M;\Z_p)$ the constant 0-cochain, so $q=-1$. Then
$y\cup B=y$, and every statement of Sections~\ref{sec:higher}--\ref{sec:chsparent} whose proof uses only Stokes for $y\cup B$, the Leibniz rule and locality holds verbatim; this
covers Lemma~\ref{lem:cochain}(ii)--(iv) and Proposition~\ref{prop:twisted-code}(a),(b), which are what Example~\ref{ex:kzh}(i) and
Section~\ref{sec:bdrytrans} use.

\begin{lemma}[Local cochain lemma]\label{lem:cochain}
Let $e$ be an $(n_i-1)$-cell and $e'$ an $(n_j-1)$-cell. A \emph{curvature term} is an integral $\int y\cup\eta$ or $\int y\cup B$ in which $y$ is a
cup product with $\delta a_k$, for some field $k$, in one of its slots.
\begin{enumerate}[label=(\roman*)]
\item (Conjugation up to curvature.) Let $\eta$ be a cochain with $\delta\eta=B$ on the faces of the cells of $\mathrm{st}(e)$, and
$U_\eta=\varpi^{F_\eta(a)}$ with $F_\eta(a)=\int\omega(a)\cup\eta$. Then $U_\eta X_i(\delta e)U_\eta^\dagger=G_{i,e}\,\varpi^{R^\eta_{i,e}(a)}$, where $R^\eta_{i,e}$ is a
combination of curvature terms with $e$ in another slot. In particular $R^\eta_{i,e}=0$ on the range of $P_{V_e}$.
\item (Fluxes.) $G_{i,e}$ commutes with every flux check.
\item (Gauss laws.) $G_{i,e}G_{j,e'}=\varpi^{\Delta(a)}G_{j,e'}G_{i,e}$, where $\Delta$ is a combination of curvature terms $\int y\cup B$ with $e$ and $e'$ in two
further slots of $y$. If $\omega$ has order at most 2, $\Delta=0$ identically; in general $\Delta=0$ on the range of $P_{V_e}P_{V_{e'}}$.
\item (Order.) $G_{i,e}^p=\varpi^{\Phi(a)}$, where $\Phi$ is a combination of curvature terms $\int y\cup B$ with $e$ in two further slots of $y$. If $\omega$
has order at most 2, $G_{i,e}^p=1$; in general $G_{i,e}^p=1$ on the range of $P_{V_e}$.
\end{enumerate}
\end{lemma}
\idea All four statements come from Stokes' theorem, $\int\delta(\cdot)=0$, applied to cup products. Conjugating $X_i(\delta e)$ by $U_\eta$
multiplies it by $\varpi^{F_\eta(a+\delta e)-F_\eta(a)}$, a sum of cup products containing $\delta e$. Write each as a coboundary and move the $\delta$ onto
the other factors: on $\eta$ it gives $\delta\eta=B$, which is the dressing, and on a field it gives a flux $\delta a_k$, a curvature term that vanishes on
flux-free states. Commutation and order are the same manipulation applied to two Gauss laws, or to one Gauss law $p$ times.
\movedproof{lem:cochain}

\paragraph{Bounded depth.}\label{rem:depth} A product of commuting diagonal gates, one for each $D$-cell, is not a depth-one circuit when the gates overlap. Suppose the geometry is
uniformly bounded: $D$ and the number $r$ of fields are fixed, and every cell is a face of at most $K$ cells of dimension $D$. A gate acts on
the qudits on the faces of one $D$-cell, at most $r\,3^D$ of them, and it shares a qudit with at most $3^DK$ other gates. Colouring the overlap
graph with $3^DK+1$ colours schedules the circuit in depth at most $3^DK+1$, independent of the size of $M$. The same holds for $U_\eta$, $U(\tilde\eta)$
and $U(\zeta_g)$ of Section~\ref{sec:dressedquot}. If the gates are Clifford ($\omega$ of order at most 2), conjugation maps a single-qudit Pauli operator to one supported on at most
$1+r\,3^DK$ qudits, so it changes the weight of any Pauli operator, and hence the distance of a Pauli stabilizer code, by at most that factor.
Every statement about constant depth or bounded weight change in this paper assumes this bounded geometry.

\paragraph{Collapse on cochains and on logical states.} For a 1-form field the cubic polynomial does not collapse on cochains: on a cube,
$(a\cup a\cup a)(x,\{1,2,3\})$ is a sum of products of three different edge variables, so $U=(-1)^{\int a\cup a\cup a}$ is a circuit of $CCZ$
gates. On flux-free configurations it equals the circuit $\prod_cCS^{\pm1}$ of~\cite[eq.~(40)]{Barkeshli:2022edm}, because
$a\cup a\equiv\delta\tilde a/2\pmod2$ on every flat square (Example~\ref{ex:kzhgate}). Its action on logical states can still collapse. By
Proposition~\ref{prop:cupgate}, $U$ acts on $|\alpha\rangle$, $\alpha\in H^1(M;\Z_2)$, by $(-1)^{q(\alpha)}$ with $q(\alpha)=\int\alpha\cup\alpha\cup\alpha$, and $q$
is a quadratic rather than a cubic function of $\alpha$, with polarization $\langle w_1(M)\cup\alpha\cup\lambda,[M]\rangle$
(Proposition~\ref{prop:threegroup}). So this level-3 circuit acts on the code space as a circuit of $CZ$ and $Z$ gates, and as a Pauli operator
when $M$ is orientable. On $T^3$ it is trivial, and on $\mathbb{RP}^3$ it is the logical $Z$, since $q(x)=x$. On $K\times S^1$, with $K$ the Klein
bottle, $H^1=\langle u,v,z\rangle$, $u^2=v^2$ the generator of $H^2(K)$, $uv=0$, $z^2=0$ and $w_1=u+v$, one finds
$q(\alpha_1u+\alpha_2v+cz)=c(\alpha_1+\alpha_2)$: the logical gate is $CZ_{13}CZ_{23}$ in this basis, a single $CZ$ after a logical change of basis,
which is the logical $CZ$ that Barkeshli, Chen, Hsin and Kobayashi find on $K\times S^1$ from the 3-group of the 3D toric
code~\cite[Section~5.1]{Barkeshli:2022edm}. Since $K\times S^1$ is $T^3$ twisted by a reflection around one circle, this $CZ$ is created by an
orientation-reversing holonomy (Section~\ref{sec:threegroup}, and check T8 in Appendix~\ref{app:checks}).

\paragraph{Checks of the cochain lemma.} Finite checks of Lemma~\ref{lem:cochain} at $p=2,3,5$, including the signs of~\eqref{eq:gauss}, are listed in Appendices~\ref{app:worked}
and~\ref{app:checks}; dropping the sign $(-1)^{\mathrm{pos}(\min S)}$ makes the conjugation and gauge identities fail.

\paragraph{Descent and phases.} For Pauli operators ($h_\alpha$ affine) the descent agrees with the quotient map $\pi$ of Definition~\ref{def:quot} for a permutation action
after forgetting phases, but not always with them: on two qubits exchanged by $\Z_2$ the commuting stabilizers $X_1Z_2$ and $Z_1X_2$ are
permuted, and both descend to $XZ$, with $(XZ)^2=-1$. As for the stabilizers of a monomial quotient, a consistent choice of phases may then be
needed, here $Y$. In general the descent is not a homomorphism of operator algebras, which is why it is defined on generators. So the dressed
quotient is a construction on a code together with a generating family, not on the code alone, and in general it can depend on the family,
since the descent of a product need not be the product of the descents; for Pauli generators the quotient up to phases does not, since
descent is then the linear map $\pi$.

\paragraph{Without the hypotheses of Theorem~\ref{thm:dressedquot}.} Without either hypothesis of the theorem, the descended Gauss laws can differ from those of $M$ by terms in which a cube contains two
lifts of the same cell. In the type-I case on a circle of length 1, which satisfies neither hypothesis, they agree in every case computed
(Appendix~\ref{app:checks}).

\begin{remark}[Existence of trivializing covers]\label{rem:cover}
For $q=0$ the group $\Gamma'$ consists of deck transformations composed with powers of the global symmetry $U_\omega$
(Theorem~\ref{thm:dressedquot}(c)). For $q\ge1$, $U(\zeta_g)$ is a $CZ$-type circuit on a
codimension-$q$ slab, so $\Gamma'$ is free on positions but not monomial, and Definition~\ref{def:quot} does not apply to $(S,\Gamma')$ directly; it applies
to $(S_{\tilde B},\Gamma)$. If $[B]\neq0$, every trivializing cover $\pi:\tilde F\to F$ has degree divisible by $p$: its transfer $\mathrm{tr}$ satisfies
$\mathrm{tr}(\pi^*[B])=\deg\pi\,[B]$, so $\pi^*[B]=0$ forces $\deg\pi\,[B]=0$, and $[B]$ has order $p$. So $p\mid|\Gamma|$ and the Tate term of
Section~\ref{sec:general} can be non-zero. A trivializing cover need not exist: a non-zero $B\in H^2(S^2;\Z_p)$ survives every finite cover, since
$S^2$ is simply connected. Then Theorem~\ref{thm:dressedquot} does not apply, and the Gauss-law description of Section~\ref{sec:gausslaw}
defines the code. This does not show that the code has no quotient presentation of another kind, and without restrictions on the parent
and the action no such statement could hold: every code $C$ is the quotient of $C\otimes C$ by the exchange of the two copies.
\end{remark}

\begin{remark}[What carries over]\label{rem:carry}
For $\omega$ of order at most 2, Theorem~\ref{thm:dressedquot}(d) makes $S_B$ a monomial quotient of the dressed parent $S_{\tilde B}$ by a free
permutation action. Section~\ref{sec:general} applies to the pair $(S_{\tilde B},\Gamma)$ verbatim, and Section~\ref{sec:sym} applies with parent automorphism
group $\mathrm{Aut}(S_{\tilde B})$. The dressed parent has the same $k$ as the undressed one, and under the bounded geometry of
Appendix~\ref{rem:depth} its distance differs from the undressed one by at most the factor $1+r\,3^DK$, because $U(\tilde\eta)$ is a bounded-depth
Clifford circuit. For order at least 3 the construction need not produce a Pauli stabilizer code, and the results of Section~\ref{sec:general} do not
apply without a separate identification of a Pauli stabilizer group; the dimension is given by Theorems~\ref{thm:dressdim} and~\ref{thm:pointcount}, and the distance by Theorem~\ref{thm:dresseddist}; and
permutation automorphisms of $\mathcal C_{\tilde B}$ that normalize $\Gamma$ descend by the proof of Proposition~\ref{prop:descent}.
\end{remark}

\begin{proposition}[The quadratic count as a spectral sequence]\label{prop:cone}
In the setting of Theorem~\ref{thm:hfcount}, write the dressed Gauss laws as $G_{i,e}=X_i(\delta e)\,Z(\Phi e)$ up to a phase, with $\Phi$ linear in $e$,
identify $Z$-type vectors on $n$-cells with $C^n$ through the cell basis, so that $\partial=\delta^\top$, and let $K$ be the Pauli stabilizer complex of
$\mathcal C_B$,
\[
K^0=\bigoplus_i\big(C^{n_i-1}\oplus C^{n_i+1}\big)\xrightarrow{D^0}K^1=\bigoplus_i\big(C^{n_i}\oplus C^{n_i}\big)\xrightarrow{D^1}K^2=\bigoplus_i\big(C^{n_i+1}\oplus C^{n_i-1}\big),
\]
with $D^0(e,f)=(\delta e,\ \delta^\top f+\Phi e)$, which sends Gauss laws and flux checks to their $X$- and $Z$-parts, and $D^1(x,z)=(\delta x,\ \Phi^\top x-\delta^\top z)$, the
syndrome map.
\begin{enumerate}[label=(\alph*)]
\item $D^1D^0=0$, $\operatorname{im}D^0=S_B$ and $\ker D^1=S_B^\perp$, so $H^1(K)$ is the logical group and $\dim H^1(K)=2k'$.
\item The $Z$-parts form a subcomplex $K_Z$ whose quotient is the complex $K_X$ of $X$-parts, and both are undressed:
$H^1(K_X)=\bigoplus_iH^{n_i}(M)$ and $H^1(K_Z)=\bigoplus_iH_{n_i}(M)$ are the electric and magnetic logicals of the undressed code. The spectral sequence
of the filtration $K_Z\subset K$ has $E_1=H(K_X)\oplus H(K_Z)$; its only differential is $d_1$, the connecting map of $0\to K_Z\to K\to K_X\to0$; and
$E_2=E_\infty$.
\item Under Poincar\'e duality $d_1$ is $\cup[B]$. For $\omega=a_1\cup a_2$, $K$ is the sum of the subcomplexes of $(X_1,Z_2)$ and $(X_2,Z_1)$. On the
first, $d_1:H^0(K_X)=Z^{n_1-1}\to H_{n_2}(M)$ is $\lambda\mapsto[\Phi\lambda]$, which factors through $H^{n_1-1}(M)$ and has the rank of $\cup[B]$ on
$H^{n_2}$, and $d_1:H^{n_1}(M)\to H^2(K_Z)=C_{n_2-1}/B_{n_2-1}$ is $[x]\mapsto[\Phi^\top x]$, with kernel $\ker(\cup[B]|_{H^{n_1}})$. So this summand of $E_2=E_\infty$ in
degree 1 has dimension $\dim\ker(\cup[B]|_{H^{n_2}})+\dim\ker(\cup[B]|_{H^{n_1}})$, the other summand has the same dimension, and $k'=\frac12\dim H^1(K)$ is
Theorem~\ref{thm:hfcount}(i). For one field the pairing is multiplied by $\kappa(1+(-1)^n)$, which gives (ii).
\end{enumerate}
\end{proposition}
\idea Write the Pauli stabilizer code as a chain complex: generators, Pauli vectors, syndromes. Splitting every vector into its $X$- and
$Z$-parts filters this complex in two steps. The two graded pieces are the undressed complexes, whose homology is the undressed logical
space, and the twist enters only through the connecting map between them, which removes logicals in pairs, as in the proof of
Theorem~\ref{thm:hfcount}.
\movedproof{prop:cone}

In this form the dressing is a perturbation of the undressed code: $E_1$ is the undressed logical space, and $d_1$ removes the electric classes
$\alpha$ with $\alpha\cup[B]\ne0$ together with the magnetic classes dual to them, which become stabilizers. The same complex with relative $B$ gives
Corollary~\ref{cor:boundary}. Theorem~\ref{thm:ss} does not apply here, since the dressed code is not CSS in general.

\paragraph{Repeated fields.} Theorem~\ref{thm:pointcount}
covers repeated fields (types I and II, and the $Q_8$ theories $b\,(a_1^2+a_1a_2+a_2^2)$): there the dressing of $G_{i,e}$ involves $a_i$, and Step 1 of
the proof (Appendix~\ref{app:proofs3}) is what makes the slot sum correct. Check T11 (Appendix~\ref{app:checks}) confirms Step 2 at the cochain level, on random flat
configurations that are not class representatives.

\begin{proposition}[Lens spaces and $\mathbb{RP}^2$]\label{prop:lens}
Let $F$ be a lens space $L(m;r)$, or $F=\mathbb{RP}^2$ with $m=2$, so that $\pi_1F=\Z_m$ and the universal cover is $S^{D_F}$ with $D_F=\dim F$. Let
$p\mid m$ be prime and $0\ne B_F\in H^j(F;\FF_p)$, $1\le j\le D_F$; every such class comes from torsion, since $H^i(F;\Z)$ is finite for $0<i<D_F$.
Let $x$ generate $H^1(F;\FF_p)$.
\begin{enumerate}[label=(\alph*)]
\item A connected finite cover of $F$ of degree $d$ trivializes $B_F$ iff $p\mid d$, except for $j=2$ on a lens space, where it does iff
$p^v\mid d$, with $p^v$ the largest power of $p$ dividing $m$. So for $m=p$ the universal cover, of degree $p$, trivializes every class, and $p$
is the least degree that Remark~\ref{rem:cover} allows; for $L(4;1)$ and $p=2$ the class in $H^2$ needs the universal cover, of degree 4.
\item Let $m=p$ and let $g$ generate the deck group. Since $S^{D_F}$ is simply connected, $\pi^*x=\delta h$ for a $\Z_p$-valued 0-cochain $h$, and
$g^*h-h$ is a non-zero constant, which we normalize to 1. For $B_F=x$ ($q=0$) take $\tilde\eta=h$: then $\zeta_g=-1$ and
$\mathcal C_B=\mathcal C_0/\langle g\circ U_\omega^{-1}\rangle$, the deck transformation of $S^{D_F}\to F$ composed with a global symmetry, as in
Theorem~\ref{thm:dressedquot}(c). For $B_F=\beta x=\delta\hat x/p$ ($q=1$), where hats denote lifts to $\{0,\dots,p-1\}$, the carry cochain
$\tilde\eta=(\pi^*\hat x-\delta\hat h)/p$, with values in $\{0,1\}$, is a primitive of $\pi^*B_F$, and $\zeta_g=-\delta\chi$, where $\chi$ is the indicator
of the sheet $\{\hat h=p-1\}$.
\item If $m=p$ and $1\le q\le D_F-1$, every $\zeta_g$ is exact, so $U(\zeta_g)$ is the identity on flux-free configurations and $\Gamma'$ acts on the
parent code space $\mathcal C_0$ exactly as the bare deck group does. The quotients still differ (Example~\ref{ex:lens}): a quotient identifies
stabilizers, not states.
\item Let $M=X\times F$ with $X$ a closed manifold, orientable if $p$ is odd, and $B=\pi_F^*B_F$. Then
$\dim\ker(\cup[B]|_{H^s(M)})=\sum_ib_{s-i}(X)\,\nu_i$, where $b_i=\dim H^i(X;\FF_p)$ and $\nu_i=\dim\ker(\cup B_F|_{H^i(F;\FF_p)})$ is listed below,
with $y$ a generator of $H^2$ and $xy$ of $H^3$, and Theorem~\ref{thm:hfcount} gives $k'$ for every quadratic twist.
\end{enumerate}
\end{proposition}

\begin{center}\small
\begin{tabular}{@{}lll@{}}
\toprule
$F$, $p$ & $B_F$ & $(\nu_0,\nu_1,\dots,\nu_{D_F})$ \\
\midrule
$L(m;r)$, $p$ odd or $4\mid m$ & $x$; $y$; $xy$ & $(0,1,0,1)$; $(0,0,1,1)$; $(0,1,1,1)$ \\
$L(m;r)$, $p=2$, $m\equiv2\bmod4$ (e.g.\ $\mathbb{RP}^3$) & $x$; $x^2$; $x^3$ & $(0,0,0,1)$; $(0,0,1,1)$; $(0,1,1,1)$ \\
$\mathbb{RP}^2$, $p=2$ & $x$; $x^2$ & $(0,0,1)$; $(0,1,1)$ \\
untwisted & $0$ & $(1,\dots,1)$ \\
\bottomrule
\end{tabular}
\end{center}

\idea Everything is controlled by $\pi_1F=\Z_m$. A class on $F$ dies on a cover exactly when its restriction to the corresponding
subgroup vanishes, a computation in the cohomology of cyclic groups. On the universal cover the generator $x$ is the coboundary of the sheet
index $h$, and a deck transformation shifts $h$ by one, so $\zeta_g$ is a constant. The counts follow from the K\"unneth formula and the cohomology
rings of lens spaces.
\movedproof{prop:lens}

\paragraph{Lens spaces: checks and winding sectors.} Check T15 (Appendix~\ref{app:checks}) confirms Proposition~\ref{prop:lens}(d) and the quotient description on lens spaces and
$\mathbb{RP}^2$. The split of Theorem~\ref{thm:general}(ii) into winding sectors and
parent classes depends on $p$: for fields of degrees $(0,1)$ on $L(p;r)$ with $B_F=xy$, both quotient logical operators are winding sectors when $p$ is
odd, while for $p=2$ one of them lifts to the parent logical $Z_1$, because the norm $\sum_{t\in\Z_p}g^{t*}h=p(p-1)/2$ of the sheet index is non-zero
only for $p=2$.

\subsection{Worked cochain calculations}
\label{app:worked}

This subsection carries out Lemma~\ref{lem:cochain} by hand in three cases and records the odd-prime check of its signs. Throughout, $\int\delta(\cdot)=0$
and the Leibniz rule $\delta(\alpha\cup\beta)=\delta\alpha\cup\beta+(-1)^{|\alpha|}\alpha\cup\delta\beta$ are the only inputs, and $G=X(\delta e)\varpi^{f(a)}$ acts by
$G|a\rangle=\varpi^{f(a)}|a+\delta e\rangle$. Two Gauss laws therefore commute up to $\varpi^{\Delta(a)}$ with
$\Delta(a)=[f_{i,e}(a+\delta e')-f_{i,e}(a)]-[f_{j,e'}(a+\delta e)-f_{j,e'}(a)]$, and $G^p_{i,e}=\varpi^{\Phi(a)}$ with $\Phi(a)=\sum_{t=0}^{p-1}f_{i,e}(a+t\,\delta e)$.

\paragraph{Mutual quadratic, $\omega=a_1\cup a_2$.} Here $|\omega|=n_1+n_2$, and~\eqref{eq:gauss} gives, with $\mathrm{pos}=0$ for slot 1 and $\mathrm{pos}=n_1$ for slot 2,
\[
f_{1,e}(a)=(-1)^{n_1+n_2}\!\int e\cup a_2\cup B,\qquad f_{2,e'}(a)=(-1)^{n_2}\!\int a_1\cup e'\cup B,
\]
which is~\eqref{eq:mutualgauss}. \emph{Commutator.} Only the field shifted by the other Gauss law enters, so
$\Delta=(-1)^{n_1+n_2}\int e\cup\delta e'\cup B-(-1)^{n_2}\int\delta e\cup e'\cup B$. Integrating
$\delta(e\cup e'\cup B)=\delta e\cup e'\cup B+(-1)^{n_1-1}e\cup\delta e'\cup B$ (using $\delta B=0$) gives $\int\delta e\cup e'\cup B=(-1)^{n_1}\int e\cup\delta e'\cup B$, so
$\Delta=0$ for every $a$: the Gauss laws commute on the whole Hilbert space. \emph{Order.} $f_{1,e}$ does not involve $a_1$, so $\Phi=p\,f_{1,e}(a)\equiv0$.
\emph{Conjugation.} For $B=\delta\eta$ and $F_\eta(a)=\int a_1\cup a_2\cup\eta$, $F_\eta(a+\delta e)-F_\eta(a)=\int\delta e\cup a_2\cup\eta$, and integrating
$\delta(e\cup a_2\cup\eta)$ gives
\[
\int\delta e\cup a_2\cup\eta=(-1)^{n_1+n_2}\!\int e\cup a_2\cup B+(-1)^{n_1}\!\int e\cup\delta a_2\cup\eta=f_{1,e}(a)+R^\eta_{1,e}(a).
\]
The curvature term $R^\eta_{1,e}$ is linear in $\delta a_2$, so $\varpi^{R}$ is a product of flux checks, and it vanishes on the flux-free subspace.

\paragraph{Multilinear cubic, $\omega=a_1\cup a_2\cup a_3$ with 1-form fields.} Here $|\omega|=3$, $e,e'$ are 0-cochains, and~\eqref{eq:gauss} gives
\[
f_{1,e}(a)=-\int e\cup a_2\cup a_3\cup B,\qquad f_{2,e'}(a)=\int a_1\cup e'\cup a_3\cup B,
\]
diagonal phases of degree 2, products of $CZ$ and $Z$ gates for $p=2$. \emph{Commutator.} $\Delta=-\int e\cup\delta e'\cup a_3\cup B-\int\delta e\cup e'\cup a_3\cup B$.
Integrating $\delta(e\cup e'\cup a_3\cup B)=\delta e\cup e'\cup a_3\cup B+e\cup\delta e'\cup a_3\cup B+e\cup e'\cup\delta a_3\cup B$ gives
\[
\Delta(a)=\int e\cup e'\cup\delta a_3\cup B,
\]
a single curvature term with $e$ and $e'$ in two slots. It vanishes when $\delta a_3=0$ on the cells having both $e$ and $e'$ as faces, as
Lemma~\ref{lem:cochain}(iii) states, and it is non-zero for suitable $a$ with flux there: the $CZ$-dressed Gauss laws do not commute off the
flux-free subspace, which is why Proposition~\ref{prop:twisted-code} multiplies them by $P_{V_e}$. \emph{Order.} $f_{1,e}$ does not involve $a_1$, so
$\Phi\equiv0$ identically, as for every multilinear $\omega$.

\paragraph{Repeated field: the order identity.} For one $n$-form field and $\omega=\kappa\,a\cup a$, the three subsets $S=\{1\},\{2\},\{1,2\}$ give
\[
f_e(a)=\kappa\Big[\int e\cup a\cup B+(-1)^n\!\int a\cup e\cup B+\int e\cup\delta e\cup B\Big],
\]
the last term an $a$-independent phase. Then $\Phi(a)=p\,f_e(a)+\kappa\big(\sum_{t=0}^{p-1}t\big)\big[\int e\cup\delta e\cup B+(-1)^n\int\delta e\cup e\cup B\big]$. For odd $p$, $\sum_tt=p(p-1)/2\equiv0$.
For $p=2$, $\sum_tt=1$, and integrating $\delta(e\cup e\cup B)$ shows that the bracket is even. So $G_e^p=1$ on the whole Hilbert space, as
Lemma~\ref{lem:cochain}(iv) states for order 2; in the language of its proof, $u=2$ and $\sum_{0<j<2,\,(p-1)\mid j}\binom2j\equiv0$. With a third field the
identity can hold only on the flux-free subspace. For odd $p$ the factor $\sum_tt$ again kills the correction, but for $\omega=a\cup a\cup b$ with
1-forms and $p=2$ the same computation gives
\[
\Phi(a,b)\equiv\int e\cup\delta e\cup b\cup B+\int\delta e\cup e\cup b\cup B\equiv\int e\cup e\cup\delta b\cup B\pmod2,
\]
a curvature term with $e$ in two slots, so $G_e^2=(-1)^{\int e\cup e\cup\delta b\cup B}$, which is 1 when $b$ has no flux on the cells containing $e$ but not in general.
This is the type-II failure of $G^2=1$ on random configurations seen in the $\Z_2$ check.

\paragraph{Odd-prime check of the signs.} The signs $(-1)^{\mathrm{pos}(\min S)+|\omega|}$ in~\eqref{eq:gauss} are invisible at $p=2$. We
implemented~\eqref{eq:cubicalcup} and~\eqref{eq:gauss} over $\Z_p$ on the cubical torus $(\Z_3)^D$ and
tested, on configurations with no flux, the conjugation identity (with $B=\delta\eta$), the commutator, the order identity and the gauge identity
(with $B$ a non-exact cocycle made random by a gauge transformation), 40 random trials per case. With the signs of~\eqref{eq:gauss} there are no
failures, and the Leibniz rule holds on random cochains of all degrees. With the sign $(-1)^{\mathrm{pos}(\min S)}$ dropped, the identities fail:
\begin{center}\small
\begin{tabular}{@{}llcc@{}}
\toprule
$\omega$ & $p$, $D$ & failures, signs of~\eqref{eq:gauss} & failures (conjugation, gauge), sign dropped \\
\midrule
$a_1\cup a_2$, $n_1=n_2=1$ & 3, 3 & 0 & 12, 13 \\
$a_1\cup a_2$, $n_1=n_2=1$ & 5, 3 & 0 & 16, 15 \\
$a_1\cup a_2$, $n_1=1$, $n_2=2$ & 3, 3 & 0 & 13, 12 \\
$2\,a\cup a$, $n=1$ & 3, 3 & 0 & 24, 31 \\
$a\cup a$, $n=2$ & 3, 4 & 0 & 0, 0 (the dropped sign is $+1$ here) \\
$a_1\cup a_2\cup a_3$ & 3, 3 & 0 & 12, 8 \\
$a_1\cup a_2\cup a_3$ & 5, 3 & 0 & 10, 10 \\
$a_1\cup a_1\cup a_2$ & 3, 3 & 0 & 11, 8 \\
$a\cup a\cup a$ & 3, 3 & 0 & 24, 20 \\
$2\,a_1a_2a_3+a_2a_1a_1$ & 3, 3 & 0 & 14, 13 \\
\bottomrule
\end{tabular}
\end{center}
For the quadratic cases the commutator and order identities also hold on random configurations with flux, as Lemma~\ref{lem:cochain} states;
for the cubic cases the commutator fails on a few random configurations with flux, which are the curvature terms computed above.

\subsection{Complements to Section~\texorpdfstring{\ref{sec:sym}}{4}}
\label{app:tech-sym}

\paragraph{Parents for higher-form twists.} For quadratic $q$-form twists we
take as parent the dressed code $S_{\tilde B}$ on the trivializing cover, with $\tilde\Gamma$ the deck group acting by permutations; then
Definition~\ref{def:quot} applies to $(S_{\tilde B},\tilde\Gamma)$ directly and $S_B=S_{\tilde B}/\tilde\Gamma$ (Theorem~\ref{thm:dressedquot}(d)). The undressed parent
with $\Gamma'=\{g\circ U(\zeta_g)\}$ is conjugate to this pair by $U(\tilde\eta)$, which maps monomial automorphisms to Clifford automorphisms that are in
general not monomial, so statements in that frame are obtained by conjugation and not by applying Section~\ref{sec:sym} to $(S,\Gamma')$. For cubic
backgrounds the parent is the commuting-projector code $\mathcal C_{\tilde B}$, and only permutation automorphisms are considered.

\begin{proposition}[Logical action on a product parent]\label{prop:logaction}
In the setting of Theorem~\ref{thm:balanced}(b), suppose $G$ acts faithfully on $C(Y)$, and let $\tilde\Gamma=\Gamma_{\rm diag}$ with $\Gamma=\rho(\pi_1F)$
acting on $\tilde F$ by deck transformations. Let $f$ be a cellular automorphism of $\tilde F$ that normalizes the deck group,
$f\gamma f^{-1}=\theta(\gamma)$ with $\theta\in\mathrm{Aut}(\Gamma)$, and covers $\bar f$ on $F$. Let $s$ be a chain automorphism of $C(Y)$, monomial on the
bases, with $sgs^{-1}=\alpha(g)$ for an automorphism $\alpha$ of $G$ that preserves $\Gamma$.
\begin{enumerate}[label=(\alph*)]
\item $f\otimes s$ normalizes $\tilde\Gamma$, and so descends, if and only if $\alpha|_\Gamma=\theta$. In terms of the holonomy this reads
$\rho\circ\bar f_*=\alpha\circ\rho$, up to an inner automorphism of $\Gamma$ that depends on the choice of base points.
\item In that case the descended automorphism acts on the $Z$-type logical operators $H_n\big(C(Y\times_\rho F)\big)$ by the map that
$f\otimes s$ induces on the balanced product. This map preserves the filtration of $H_n$ by $F$-degree, and on the associated graded
$\bigoplus_pE^\infty_{p,n-p}$ it is the map induced by the morphism of spectral sequences whose $E^2$ term is the map
$(\bar f,s_*)_*$ on $H_p(F;H_q(Y)_\rho)$.
\end{enumerate}
\end{proposition}
\idea A pair of a map of $F$ and a symmetry of the fibre descends exactly when it carries the holonomy to itself. Its logical
action is then the map it induces on the balanced product, and since it respects the fibration, it respects the spectral sequence of
Theorem~\ref{thm:ss}, acting on $H_p(F;H_q(Y)_\rho)$ in the evident way.
\movedproof{prop:logaction}

On the abutment the induced map preserves the filtration, so $E^2$ determines it only up to block-triangular corrections.

\paragraph{Three layers.} (1) \emph{Isometries} of $(F,\rho)$ that preserve the cell structure, combined with internal symmetries, are
permutation automorphisms. Three properties of a qudit permutation should be kept apart. As a map of wires it sends each qudit to one
qudit, so it does not spread errors, and it is fault-tolerant in that sense. Where qudits can be relabeled or moved, as in architectures
with movable qubits, it costs no gates. On hardware with fixed geometrically local connectivity it must be implemented by SWAP gates, in
depth at least proportional to the distance a qudit is moved, which for an isometry of $F$ is up to the diameter of $F$; it is constant depth
only when $F$ has bounded size or the permutation is local. (2) \emph{Mapping classes.} The stabilizer of $[\rho]$ in
$\mathrm{MCG}(F)\times\mathrm{Out}(G)$ acts on the logical space $H_n\big(C(Y\times_\rho F)\big)$ preserving its symplectic form, which on the code side is the
logical Clifford group modulo Paulis; this is the map $\mathrm{MCG}(M)\to\mathrm{Aut}(H^*(M;D),\text{pairing})$ of~\cite{Gukov:2020btk}, with twisted
coefficients. The logical space is the abutment of the spectral sequence of Theorem~\ref{thm:ss}, not $\bigoplus_{p+q=n}H_p(F;H_q(Y)_\rho)$ itself: the action
preserves the filtration, its associated graded is the action on $E^\infty$, and the extension data between the graded pieces can carry
additional off-diagonal terms. A mapping class is a permutation gate only where it is isometric for some lattice structure. For a surface $F$,
Nielsen realization~\cite{Kerckhoff1983} puts every finite subgroup of $\mathrm{MCG}(F)$ at an orbifold point of moduli; we do not use it in higher
dimensions, where a finite subgroup of the mapping class group need not be realized by isometries. Infinite-order classes such as Dehn
twists act on the code space but not on the stabilizer group, so they need code deformation. (3) \emph{Phases.} The physical gate is a projective lift of the symplectic map, fixed by a
quadratic refinement. The lift is invisible for CSS permutation gates; it matters for non-CSS quotients with $Y$-type checks and for $\Z_N$
with $N$ even. The identity component of $\mathrm{Iso}(F)$ acts trivially on $H_*(F;D)$, so translations are logically trivial for topological
parents; they act non-trivially on bivariate-bicycle codes, a sign that their logicals carry momentum along $F$ and have no TQFT limit.

\begin{corollary}[Logical Paulis as an anomaly]\label{cor:anomaly}
Let $C$ be a Pauli stabilizer code on prime-dimensional qudits with $k$ logical qudits, for example a quotient code. The logical Pauli operators,
restricted to the code space, generate a group $\mathcal H$ whose centre $Z$ consists of scalars, with $\mathcal H/Z\cong L=\mathsf N(S)/S\cong\FF_p^{2k}$ and
commutator given by the symplectic form of $L$. The code space is an irreducible representation of $\mathcal H$, and every irreducible
representation of $\mathcal H$ with the same central character has dimension $p^k=|L|^{1/2}$.
\end{corollary}
\idea The logical Paulis form a Heisenberg group. Every non-central element has trace zero in any representation with the given
centre, and this forces the dimension of an irreducible representation to be $p^k$.
\begin{proof}
The symplectic form is non-degenerate on $\mathsf N(S)/S$. Representatives of a symplectic basis act on the code space as the generalized
Paulis of $k$ qudits under an encoding isomorphism, and these span all operators on $(\CC^p)^{\otimes k}$, so the representation is irreducible. In
any representation $V$ with the given central character, a non-central $h\in\mathcal H$ has some $h'$ with $h'hh'^{-1}=\varpi h$ for a root of unity $\varpi\ne1$,
so $\Tr_Vh=0$. The character of an irreducible $V$ then has norm $1=|\mathcal H|^{-1}|Z|(\dim V)^2$, so $(\dim V)^2=|L|$.
\end{proof}

In twisted compactification the logical Paulis are the wrapped higher-form symmetry operators of the parent, which become 0-form
symmetries of the compactified theory. Their commutator, braiding times intersection on $F$, is a mixed 't~Hooft anomaly, and the corollary
says that the code space is its smallest representation, so $k=\frac12\log_p$ of the order of the anomalous group. For a polarization the
remaining symmetry is $L^\vee$ and the anomaly is the extension class of $L\to\mathcal H\to L^\vee$~\cite{Gukov:2020btk}.

\paragraph{The rank obstruction and twisted operations.} The same obstruction appears
in~\cite[Section~4]{BDEHT2026cupsII}: a natural operation vanishes on a degree-one class that lifts integrally, because the class factors
through $S^1$. A twisted operation replaces $x$ by its orbit $(x,g^*x,\dots)$, a map to $(S^1)^r$, and on liftable classes with $r=\dim M$ the
gate is the degree of that map; on $T^3$ it is $\det(x,Ax,A^2x)$, with $A$ the action on $H^1(T^3;\Z)$, non-zero exactly on cyclic vectors.
For qubits on the closed $T^3$ an order-two twist gives nothing in this framework: its operations in degree three contain a Bockstein and
vanish on liftable classes, which are all classes of $T^3$. The $CS$ and $T$ gates of~\cite[Section~8.3]{BDEHT2026cupsII} on $T^3$ use folds and
relative cohomology, that is, boundaries.

\paragraph{Internal twists and twisted cup products.} When $\hat R$ acts freely, (a) is Shapiro's lemma~\cite[Section~III.6]{Brown1982}: $r$ copies on $M$ whose copies are cycled by the
holonomy of $\hat M\to M$ have $H^n(M;\FF_p[C_r])\cong H^n(\hat M;\FF_p)$. The circuit $V$ is not the parent circuit $\varpi^{\int a_1\cup\dots\cup a_r}$ on $r$
independent copies, which the copy permutation does not preserve at the level of cochains; it is the local formula of that circuit evaluated
after the identification of (a), and only its action on the twisted classes is inherited from the parent.

Proposition~\ref{prop:internal} is the twisted cup product of~\cite{BDEHT2026cupsII}: their $CCZ$ on $T^3$, with $R$ the coordinate $3$-cycle,
which fixes the diagonal circle~\cite[Section~6.1]{BDEHT2026cupsII}, and their gates on the Bolza surface times a circle and on the codes of
Golowich, Tamo and Zhu~\cite[Sections~6.2 and~6.4]{BDEHT2026cupsII} are instances with non-free $R$, and check T7b is a free instance.
When $R$ has fixed points the quotient reads as $r$ copies on the orbifold $\tilde M/\langle R\rangle$, joined along the fixed locus by a line
that cycles them; the holonomies then generate a subgroup of $G\wr C_r$ through $\pi_1^{\rm orb}$, which for $T^3/C_3$ is $\Z^3\rtimes C_3$.

\paragraph{Automorphism groups of the genon code.} We
enumerated all affine monomial symmetries of the parent: maps $g\mapsto Mg+v_b$ on each block with $M\in GL(2,\Z_{12})$, block swaps and Hadamards.
For the quotient we computed the whole monomial-Clifford automorphism group, not only its affine part. An exhaustive search shows that
its only stabilizers of weight at most 6 are its 144 checks, so every such automorphism permutes the checks; the automorphisms of a
coloured graph with vertices for checks, qubits and the three Pauli letters of each qubit are exactly these permutations
(Appendix~\ref{app:checks}).

\paragraph{Mixed backgrounds.} Twisting by $U$ and $V$ at once, along two commuting deck generators $g_1,g_2$ of a cover, uses the twisted
deck operators $g_1\circ U$ and $g_2\circ V(\Sigma)$. By Proposition~\ref{prop:threegroup}(c), applied on the cover, their commutator acts on the
parent code space as $(-1)^{q(\lambda)}W(\gamma)$ with $\gamma$ dual to $w_1(\tilde M)\cup\lambda$. On an orientable cover this is the sign
$(-1)^{q(\lambda)}$, an 't~Hooft anomaly of the 3-group~\cite{Barkeshli:2022edm} that obstructs the background when it is $-1$; on a non-orientable
cover it is a Wilson line, which the 2-form background $C_3$ must absorb, as the relation $dC_3=B_2\cup d\tilde C_1/2$ requires. We find the
twisted deck operator for $g_2$. On the code twisted by $U$ the bare membrane $V(\Sigma)=X(\sigma)$, $\sigma$ dual to $\Sigma$, is no longer a symmetry
where $\Sigma$ meets the wall. For one toric code on $T^3_L$ with the wall $B_1=\theta_x$ and $\Sigma$ the plane $y=0$, the dressed membrane
\[
V'(\Sigma)=X(\sigma)\;i^{\int(\tilde\sigma\cup\tilde a)\cup B_{1,\Z}}\;\prod_zCZ\big(e^y_{(0,0,z)},e^z_{(0,0,z-1)}\big),
\]
a string of $S$ gates on the $z$-edges at $(0,1,z)$ and a chain of $CZ$ gates along the line $\Sigma\cap\{x=0\}$, commutes with every Gauss law
(check T14), while no product of $CZ$ and $Z$ gates near the line does. The phase $i$ is forced, because the local primitive of
$\mathrm{Sq}^1x=x^2$ is the half-integral cochain $\tilde x/2$; in field theory the wall carries the double-semion theory and the line
is a semion string. By Proposition~\ref{prop:chslevel}(c), any code twisted by $g_2\circ V'(\Sigma)$ has $X$-checks of level at most 2 and Pauli
$Z$-checks. Constructing it and counting its dimension is open; Theorem~\ref{thm:pointcount} does not apply, since the
twisted deck operator is Clifford but not diagonal.

\paragraph{Non-invertible symmetries.} The general criterion is $N^\dagger N|_{\mathcal C}=\Lambda\,1_{\mathcal C}$ with $\Lambda>0$, which is necessary
and sufficient for $\Lambda^{-1/2}N$ to be unitary on $\mathcal C$; Lemma~\ref{lem:noninv} gives a sufficient condition in terms of the fusion outcomes. A converse
needs a condition that rules out cancellation, since non-scalar outcomes can sum to a scalar. For instance, if the restriction of each $U_c$
to $\mathcal C$ is a multiple of a logical Pauli operator, then, since distinct logical Paulis are linearly independent, $N^\dagger N$ is a scalar on
$\mathcal C$ exactly when, for each logical Pauli other than the identity, the coefficients of the outcomes that act as multiples of it sum to zero. This answers, as a
criterion, whether a non-invertible symmetry before compactification gives a logical gate on the quotient code; an example in which the gate is not already induced by an invertible parent symmetry is open (Section~\ref{sec:discussion}).

\paragraph{Slabs.} Throughout, $p=2$, and the notation is that of Proposition~\ref{prop:transgress} and Section~\ref{sec:bdrytrans}.

\begin{lemma}[Inflow on slabs]\label{lem:inflow}
Let $M$ be a compact cubical manifold with absolute fields and a relative background $B$ as in Corollary~\ref{cor:boundary}, and suppose
that $B$ has an integer lift $B_\Z=\delta\eta_\Z$ with $\eta_\Z$ an absolute integer cochain; this holds for $M=X\times D^{q+1}$ and $B$ the
pull-back of the generator of $H^{q+1}(D^{q+1},\partial D^{q+1})$, since that class vanishes in $H^{q+1}(D^{q+1})$. Let $\eta=\eta_\Z\bmod2$
and $z=\eta|_{\partial M}$, a $q$-cocycle on the closed manifold $\partial M$ because $\delta\eta=B$ vanishes on $\partial M$. For a cell
$e\subset\partial M$ let $f^{\partial,z}_{i,e}$ be the dressing of Definition~\ref{def:gauss} on $\partial M$ with background $z$, and let
$f^{\partial,z}_{i,e}=0$ for $e\not\subset\partial M$; for $q=0$ this is the case $B=1$ of Example~\ref{ex:kzh}(i). Then on flat configurations
\[
U_\eta\,X_i(\delta e)\,U_\eta^\dagger=G^B_{i,e}\,D_{i,e},\qquad D_{i,e}|a\rangle=(-1)^{f^{\partial,z}_{i,e}(a|_{\partial M})}|a\rangle .
\]
Consequently $\mathcal C_B=U_\eta\,\mathcal C_0^{\partial z}$, where $\mathcal C_0^{\partial z}$ has the flux checks of the undressed code, the Gauss
laws $X_i(\delta e)$ for $e\not\subset\partial M$, and the Gauss laws $X_i(\delta e)D_{i,e}$ for $e\subset\partial M$: the undressed code on $M$
whose boundary carries the dressed code of $\partial M$ with background $z$, one degree lower.
\end{lemma}
\idea This is the computation of Lemma~\ref{lem:cochain}(i), but Stokes on a manifold with boundary leaves a boundary term. That
term is a cup product with the restriction $z$ of $\eta$ to the boundary, which is exactly the dressing of a code on $\partial M$ with background
$z$: the bulk dressing flows onto the boundary, one degree lower.
\movedproof{lem:inflow}

\begin{proposition}[Transgression descent on slabs]\label{prop:slabgate}
In the setting of Lemma~\ref{lem:inflow}, let $A$, $V$ and $\lambda$ satisfy (i) of Proposition~\ref{prop:transgress}, and let
$W_\partial|a\rangle=\exp\big(2\pi i\int_{\partial M}\lambda(a)\cup\eta_\Z\big)|a\rangle$, which depends only on $a|_{\partial M}$. Then on flat configurations
\[
U_\eta VU_\eta^\dagger=V\,W\,W_\partial,\qquad W|a\rangle=\exp\Big(2\pi i\,s\int_M\lambda(a)\cup B_\Z\Big)|a\rangle,\quad s=(-1)^{|\omega|}.
\]
Suppose that, on flat configurations, $VW_\partial^{-1}$ maps each Gauss law $X_i(\delta_{\partial}e)D_{i,e}$ of the dressed code on $\partial M$
with background $z$ to $\prod_j(X_j(\delta_\partial e)D_{j,e})^{A_{ji}}$. Then $\tilde U=VW$ maps $G^B_{i,e}$ to $\prod_j(G^B_{j,e})^{A_{ji}}$ on flat
configurations, maps flux checks to products of flux checks, and preserves $\mathcal C_B$. The hypothesis on $\partial M$ is
Proposition~\ref{prop:transgress} on $\partial M$ when $q\ge1$, and Lemma~\ref{lem:bone} when $q=0$.
\end{proposition}
\idea Conjugate $V$ by $U_\eta$ as in the exact case of Proposition~\ref{prop:transgress}. Stokes now produces, besides the wall term
$W$, a boundary term $W_\partial$, and the bulk statement follows once $VW_\partial^{-1}$ is a symmetry of the boundary code, which is the
hypothesis.
\movedproof{prop:slabgate}

\begin{lemma}[The background $B=1$ from a wall]\label{lem:bone}
Let $X$ be a closed cubical manifold carrying the code of Example~\ref{ex:kzh}(i) with $B=1$, let $A$, $V$, $\lambda$ satisfy (i), and suppose
that $X\times S^1_m$, $m\ge3$, satisfies (iii) of Proposition~\ref{prop:transgress} (for instance, $X$ a cubical torus with all sides at least
3). Then for either sign, $VW_X^{\pm1}$ with $W_X=\exp(2\pi i\,s\int_X\lambda)$ maps each Gauss law $G_{i,v}$ to $\prod_jG_{j,v}^{A_{ji}}$ on flat
configurations.
\end{lemma}
\idea The background $B=1$ on $X$ is what a wall $X\times\{0\}$ looks like from inside the wall. So the identity on $X$ is the identity
of Proposition~\ref{prop:transgress} on $X\times S^1$, restricted to configurations pulled back from $X$.
\movedproof{lem:bone}

\paragraph{Gapped boundaries.} Let $M$ be compact, with $\partial M$ a union of subcomplexes $\partial_kM$, and let all $r$ fields have one
degree $n$. Each piece carries a subspace $L_k\subseteq\FF_2^r$: for every $n$-cell $c\subset\partial_kM$ the vector $a(c)=(a_i(c))_i$ lies in $L_k$,
enforced by $Z$-type checks on $c$. The Gauss laws at an $(n-1)$-cell $e$ are the products $G^B_{u,e}=\prod_i(G^B_{i,e})^{u_i}$, in a fixed
order, for the $u\in\FF_2^r$ that lie in $L_k$ for every $k$ with $e\subset\partial_kM$; the flux checks are as before. A configuration is
\emph{admissible} if it satisfies the boundary checks. For KZH's triangle, $L_r=\{u_r=0\}$, $L_b=\{u_b=0\}$ and $L_{rb}=\{u_g=0,\ u_r=u_b\}$.

\begin{proposition}[Relative transgressions at gapped boundaries]\label{prop:reltrans}
Let $A$, $V$, $\lambda$ satisfy (i) of Proposition~\ref{prop:transgress} on admissible flat configurations, let $B$ satisfy (ii) and (iii) (closed
stars taken in $M$; $B$ need not vanish on $\partial M$), and assume:
\begin{enumerate}[label=(bc\arabic*)]
\item $AL_k\subseteq L_k$ for every $k$;
\item the multilinear form $\hat\omega$ of $\omega$, with $\omega(a)(\sigma)=\sum\pm\hat\omega(a(f_1),\dots,a(f_{|\omega|}))$ over the cup terms of a cell $\sigma$,
vanishes on $L_k\times\dots\times L_k$ for every $k$;
\item there are real cochains $\beta_k$ on $\partial_kM$, cellwise polynomials in the lifts, with $\lambda(a)|_{\partial_kM}\equiv\delta\beta_k(a)\pmod1$
for every admissible flat $a$;
\item on each hinge $h_{kl}=\partial_kM\cap\partial_lM$, oriented as part of $\partial(\partial_kM)$, there are $\gamma_{kl}$ with
$\beta_k-\beta_l\equiv\delta\gamma_{kl}\pmod1$, and so on on the lower strata, the last alternating sum vanishing mod 1 on the strata of
dimension 0.
\end{enumerate}
Then
\[
\tilde U=V\exp2\pi i\,s\Big(\int_M\lambda\cup B_\Z-\sum_k\int_{\partial_kM}\beta_k\cup B_\Z+\sum_{k<l}\int_{h_{kl}}\gamma_{kl}\cup B_\Z-\cdots\Big)
\]
maps each Gauss law $G^B_{u,e}$ to $G^B_{Au,e}$ on admissible flat configurations, maps $Z$-type checks to products of $Z$-type checks, and
preserves the code.
\end{proposition}
\idea At a gapped boundary the fields obey linear conditions on which $\omega$ vanishes, so the bulk computation of
Proposition~\ref{prop:transgress} leaves boundary terms $\int_{\partial_kM}\lambda\cup\eta$. On each boundary piece $\lambda$ is itself a coboundary, so Stokes
applies again, piece by piece and then hinge by hinge, until only terms on the wall remain; each step contributes one correction to $\tilde U$.
\movedproof{prop:reltrans}

\begin{example}[The boundary terms of Kobayashi, Zhu and Hsin]\label{ex:kzhbdry}
Take the fields, $\omega=a_r\cup a_g\cup a_b$, $A$ and $\lambda=\frac14\tilde a_r\cup\tilde a_b$ of Example~\ref{ex:kzhgate}, and KZH's boundary pieces
$L_r$, $L_b$, $L_{rb}$. Condition (bc1) holds because $A(u_r,u_g,u_b)=(u_r,u_g+u_r+u_b,u_b)$, and (bc2) because $\hat\omega(v_1,v_2,v_3)=(v_1)_r(v_2)_g(v_3)_b$
vanishes when $(v_1)_r=0$, when $(v_3)_b=0$ and when $(v_2)_g=0$. On $L_r$ and $L_b$, $\lambda=0$ and $\beta=0$. On $L_{rb}$, with $a=a_r=a_b$ and the
signed cubical cup $(x\cup y)(\square)=x_{01}y_{13}-x_{02}y_{23}$ on a square with vertices $0,1,2,3$ as in Example~\ref{ex:kzhgate},
\[
\tfrac14\,\tilde a\cup\tilde a=\delta\big(\tfrac18\tilde a\big)
\]
exactly, on each of the eight flat patterns of a square; so $\beta_{rb}=\frac18\tilde a_r$ satisfies (bc3). (With the unsigned product
$x_{01}y_{13}+x_{02}y_{23}$ the identity fails by $\frac12$.) At every hinge the fields in $\beta$ vanish ($a_r=0$ on $L_r\cap L_{rb}$, $a_r=a_b=0$ on
$L_b\cap L_{rb}$), so $\gamma=0$. Proposition~\ref{prop:reltrans} on $X\times S^1_m$ with a wall, followed by the reduction of
Lemma~\ref{lem:bone}, whose proof uses only the layer $X\times\{0\}$ and holds with these boundary conditions, gives on the triangle $X$ with $B=1$
\[
W=\exp2\pi i\Big(-\int_X\tfrac14\tilde a_r\cup\tilde a_b+\int_{\partial X\cap L_{rb}}\tfrac18\tilde a_r\Big),
\]
the complex conjugate of~\cite[eq.~(57)]{KZH2025chs}, with the same relative sign of the bulk and boundary terms. So the boundary
$T$-string of~\cite{KZH2025chs} is the relative transgression of the bulk $CS$ circuit along the boundary condition $a_r=a_b$. Check T13
(Appendix~\ref{app:checks}) confirms this on KZH's square of sides 2, 3 and 4, with $L_r$ on the left, $L_b$ on top and $L_{rb}$ on the bottom and
right: $k=1$, the gate maps every Gauss law correctly, and its logical action is $\mathrm{diag}(1,e^{2\pi i/8})=T$, and $T^\dagger$
for~\cite[eq.~(57)]{KZH2025chs}, as in~\cite[Theorem~2]{KZH2025chs}; dropping the boundary term fails.
\end{example}

\paragraph{Spin structures and deck lifts.} One deck representation reaches only the spin structures of $M$ over $\tilde s$, which differ by
classes in $\ker p^*$, the image of $H^1(\Gamma;\Z_2)$ in the exact sequence $0\to H^1(\Gamma;\Z_2)\to H^1(M;\Z_2)\to H^1(\tilde M;\Z_2)^\Gamma\to H^2(\Gamma;\Z_2)$
of the cover; letting $\tilde s$ run over the image of $p^*$, as in (c), reaches all of them. For a $\Gamma$-invariant $\tilde s$ that is not a
pull-back the lifts satisfy only $\hat G_g\hat G_h=\pm\hat G_{gh}$, a representation of an extension of $\Gamma$ by $\Z_2^F$ whose class in $H^2(\Gamma;\Z_2)$
obstructs descending $\tilde s$; in Bogoliubov--de~Gennes form particle--hole symmetry allows only sign changes of the lifts, which cannot
remove it, and such $\tilde s$ contribute no sector. For the quarter turn of Proposition~\ref{prop:quarter}, $\ker p^*$ is spanned by the class
$z$ along the circle while $p^*\tau\ne0$, so the four sectors come from the two $C_4$-invariant spin structures of the fibre, each with both
characters of $\Z_4$; for the half turn of Example~\ref{ex:dicosm} the eight sectors come from the four fibre spin structures. Checks T17 and
T18 build the sectors in this way.

The convention in (c) matters only through scalars. Another choice of the scalar part, such as $c^\dagger hc$ in place of the symmetric form,
changes $\theta(s)$ by a factor that depends on $s$ only through terms of $\tilde k$ that join a site to one of its deck images; the pumps of Section~\ref{sec:fermionic}
have none, since their deck elements move every site by at least two layers and their terms join neighbouring layers. What the convention
excludes is an implementation of the endpoint $u_s$ chosen separately in each sector, which would leave the relative phases, and so the
logical gate, undetermined.

\paragraph{The equivariant $p+ip$ pump.} The quarter turn of Proposition~\ref{prop:quarter} needs a $p+ip$ layer, and its pump is constructed as follows.

\begin{lemma}[The equivariant $p+ip$ pump]\label{lem:pippump}
On a square lattice take spinless fermions with Nambu spinor $\Psi_k=(c_k,c^\dagger_{-k})$ and the $p+ip$ Hamiltonian
\[
h(k)=\begin{pmatrix}\xi&\Delta\\ \bar\Delta&-\xi\end{pmatrix},\qquad\xi=-2(\cos k_x+\cos k_y)-\mu,\qquad\Delta=\sin k_x+i\sin k_y,
\]
with $E=(\xi^2+|\Delta|^2)^{1/2}$. For $0<|\mu|<4$ it is gapped, $E\ge E_0(\mu)>0$, in the weak-pairing phase.
Let $P=\frac12(1-h/E)$ project on its occupied vector, and $x=P_{22}=\frac12(1+\xi/E)\in[0,1]$, the weight of that
vector on the hole component, which spans the occupied space of the vacuum. On a pair of layers, the first carrying $h$ and the second $h^*$
(a $p-ip$ layer), in the Nambu basis $((1,p),(1,h),(2,p),(2,h))$, let
\[
W=\begin{pmatrix}-P_{22}&-P_{12}&-P_{12}&P_{11}\\ P_{21}&-P_{22}&P_{11}&P_{21}\\ -P_{21}&P_{11}&P_{22}&P_{21}\\ P_{11}&P_{12}&-P_{12}&P_{22}\end{pmatrix},\qquad
Q=\mathrm{diag}(-1,-1,1,1),
\]
so that $Q$ is the fermion parity of the first layer, and let $D_\eta=\mathrm{diag}(\eta,\bar\eta)$, $\eta=e^{i\pi/4}$, $D=D_\eta\oplus\bar D_\eta$ and $R(k_x,k_y)=(-k_y,k_x)$.
\begin{enumerate}[label=(\roman*)]
\item $W(k)$ is unitary and maps the occupied space of the vacuum, spanned by $(1,h)$ and $(2,h)$, onto the occupied space of $h(k)\oplus h(k)^*$.
\item $\tau_xW(-k)^*\tau_x=W(k)$, with $\tau_x$ exchanging $p$ and $h$ in each layer, and $DW(k)D^\dagger=W(Rk)$; $Q$ commutes with $\tau_x$ and $D$.
\item $WQ+(WQ)^\dagger=2x$, and $WQ=e^{iK}$ with $K=-i\,g(x)\,(WQ-x)$, $g(x)=\arccos x/\sqrt{1-x^2}$. $K$ is Hermitian, $\|K\|\le\pi/2$,
$\tau_xK(-k)^*\tau_x=-K(k)$ and $DK(k)D^\dagger=K(Rk)$, and $K$ is analytic and bounded on a complex strip $|{\rm Im}\,k_x|,|{\rm Im}\,k_y|<a$, $a=a(\mu)>0$. So its kernel on the
infinite lattice satisfies $|K(r)|\le Ce^{-|r|/\ell}$, with $C$ and $\ell$ depending only on $\mu$, and on a torus of any side, with either
boundary condition, the kernel is the signed sum of the images of this one and obeys the same bound in the distance on the torus, with a
constant independent of the side.
\item On layers $1,\dots,N$, $N$ even, with layer $N+1$ identified with layer 1 by the boundary condition, let $K_1$ be $K$ on the pairs
$(2j-1,2j)$ and $K_2$ be $K$ on the pairs $(2j+1,2j)$, odd layer first. The pump is the evolution for $0\le t\le2$ of
$H(t)=-\frac12\Psi^\dagger K_1\Psi$ for $t\le1$ and $H(t)=\frac12\Psi^\dagger K_2\Psi$ for $t>1$. Its single-particle unitary is
$e^{-iK_2}e^{iK_1}=Q_{\rm odd}\,u_W\,Q_{\rm odd}$, with $u_W$ the pump built in the same way from $W$ and $Q_{\rm odd}$ the parity of the odd layers; it
preserves the occupied space of the vacuum, and it commutes with the quarter turn acting by $D_\eta$ on odd and by $\bar D_\eta$ on even layers.
\item (The pumped phase.) The occupied band of $h$ has Chern number $\pm1$, of opposite signs for $0<\mu<4$ and $-4<\mu<0$. So $h$ is the
invertible phase $\nu=\pm1$ of Kitaev's sixteenfold way~\cite{Kitaev:2005hzj}, of chiral central charge $\pm\frac12$, the $p+ip$ superconductor
of~\cite{Barkeshli:2023bta}. After the first step of (iv) the odd layers carry $\nu$ and the even layers $-\nu$, and the second step removes
them in the pairs $(2j+1,2j)$. On an open stack of layers one layer of $\nu$ would be left at one end and one of $-\nu$ at the other, so on
the circle the pump sweeps one layer of $\nu$ once around it.
\end{enumerate}
\end{lemma}
\begin{proof}
Write $\psi=(a,b)$ for a unit occupied vector of $h(k)$ and $\chi=(-\bar b,\bar a)$. The columns of $W$ are $(b\chi,\bar a\bar\chi)$, $(-\bar b\psi,a\bar\psi)$,
$(a\chi,-\bar b\bar\chi)$ and $(\bar a\psi,b\bar\psi)$, which do not change when $\psi$ is multiplied by a phase and give the displayed entries, since
$P=\psi\psi^\dagger$. They are orthonormal, and the second and fourth span $(\psi,0)$ and $(0,\bar\psi)$, the occupied space of $h\oplus h^*$; this is (i).
For (ii), $\xi$ is even and $R$-invariant, $\Delta(-k)=-\Delta(k)$ and $\Delta(Rk)=i\Delta(k)$, so $\sigma_xh(-k)^*\sigma_x=-h(k)$ and $D_\eta h(k)D_\eta^\dagger=h(Rk)$.
Hence $P_{11}$ and $P_{22}$ are even and $R$-invariant, $P_{12}(-k)=-P_{12}(k)$ and $P_{12}(Rk)=iP_{12}(k)$, and both identities follow by substitution.
For (iii), every diagonal entry of $WQ$ is $P_{22}=x$, and its off-diagonal part $A=WQ-x$ is anti-Hermitian, because $P$ is Hermitian with
real diagonal; so $WQ+(WQ)^\dagger=2x$. Unitarity gives $A^2=x^2-1$, so for $x<1$ we have $A=i\sin\vartheta\,N$ with $\vartheta=\arccos x\in[0,\pi/2]$, $N$
Hermitian and $N^2=1$, and $WQ=e^{i\vartheta N}$ with $\vartheta N=-ig(x)A=K$; for $x=1$, $A=0=K$. The symmetries pass to $K$, a fixed function of $WQ$
and of the even, $R$-invariant $x$. Since $E^2\ge E_0^2$ on the real torus, by compactness there is $a>0$ such that on the strip $|{\rm Im}\,k_x|,|{\rm Im}\,k_y|<a$ the
analytic continuation of $E^2$ has positive real part and that of $x$, which lies in $[0,1]$ on the real torus, stays away from $(-\infty,-1]$.
There $E$, $P$ and $x$ are analytic, and so is $g(x)=\vartheta/\sin\vartheta$, which is analytic off $(-\infty,-1]$; so $K$ is analytic and
bounded on a slightly smaller strip. Shifting the contour of its Fourier integral gives $|K(r)|\le Ce^{-a'(|r_x|+|r_y|)}$ for every $a'<a$. On
a torus of side $L$ the kernel is $\sum_{n\in\Z^2}\pm K(r+nL)$ by Poisson summation, with signs fixed by the boundary condition, and the sum
obeys the same bound with a constant independent of $L$. For (iv), $e^{iK_1}$ is $WQ$ on each pair of the first
step and $e^{-iK_2}=QW^\dagger$ on each pair of the second, with $Q$ on the odd layer in both, which gives $Q_{\rm odd}u_WQ_{\rm odd}$. By (i) the first step
maps the occupied space of the vacuum onto that of $h\oplus h^*$ on each pair, the second maps it back, and $Q_{\rm odd}$ preserves it. The
symmetry under the quarter turn is (ii) and (iii) on each pair. For (v), $h=d\cdot\sigma$ with $d=(\sin k_x,-\sin k_y,\xi)$, and up to a
sign fixed by convention the Chern number of the occupied band is the degree of $d/|d|:T^2\to S^2$. The preimages of the north pole are the
points with $\sin k_x=\sin k_y=0$ and $\xi>0$, where the local degree is the sign of $-\cos k_x\cos k_y$. For $0<\mu<4$ the only one is
$(\pi,\pi)$, with local degree $-1$; for $-4<\mu<0$ they are $(0,\pi)$, $(\pi,0)$ and $(\pi,\pi)$, with local degrees $1$, $1$ and $-1$. So
the degree is $\mp1$ for $\pm\mu\in(0,4)$, and the $p-ip$ layer $h^*$ has the opposite Chern number. The statement on the layers follows
from (i) and (iv).
\end{proof}
So $W$ and $WQ$ nucleate the same state, a $p+ip$ and a $p-ip$ layer on each pair, and the two pumps differ by conjugation with an on-site
parity that fixes the vacuum: they pump the same invertible phase. $WQ$ is used because $W$ has eigenvalues at both $\pm1$, while those of
$WQ$ have real part $x\ge0$. On the cover $\tilde M$ of Proposition~\ref{prop:quarter} the deck generator is the quarter turn, by $D_\eta$ on odd and
$\bar D_\eta$ on even layers, composed with the translation by $J$ layers. Since $D_\eta^4=-1$, its fourth power is 1 exactly when the fermions are
antiperiodic along the circle of $\tilde M$, which is the circle spin structure of every pull-back $p^*s$, as a fermion going once around that
circle crosses the seam four times; with the two $C_4$-invariant fibre spin structures, doubly periodic and doubly antiperiodic, these are
the two spin structures of the cover in the image of $p^*$.

\paragraph{Numerical evidence.} The proof does not use the following computations, which are double-precision checks on finitely many
instances (Appendix~\ref{app:checks}). Check T21 computes $\theta(s)$ without eq.~(26) of~\cite{Barkeshli:2023bta}: it evolves the atomic insulator of
each sector in Fock space under the Hamiltonians of the equivariant pump above, with the convention of Proposition~\ref{prop:gaussdescent}(c), block by block
over the $C_4$-orbits of fibre momenta. For $L=4,6,8$ with $J=2$ and $L=4,6$ with $J=4$, both signs of $\mu$, and rotations about a site and
about a plaquette centre, it finds
\[
V=T_\tau^{\mp1}\,CZ_{\tau z}\,Z_\tau^{\epsilon}\qquad(\mu=\pm1)
\]
up to a global phase, within $2\cdot10^{-15}$ rad, with $\epsilon\in\{0,1\}$ depending on the centre, the sign of $\mu$ and $J$. Check T17 finds the
same pattern from the rotation eigenvalues at the seam~\cite[eq.~(49)]{Barkeshli:2023bta} on fibres $L=4,6,8$, in agreement with the Ising-theory
reading, in which the quarter turn acts on the doubly periodic sector as the modular $S$, with eigenvalue $e^{7\pi i/4}$ on
$|\sigma;\psi\rangle$~\cite[Section~5.3]{Barkeshli:2023bta}. Check T18 builds an explicit $C_4$-equivariant Chern pump on fibres $L=4,6$ and confirms
Proposition~\ref{prop:gaussdescent} sector by sector, with $V^2=S^\dagger_\tau$; in the instances $L=6,8$, truncating the nucleation kernel to range 1
and re-unitarizing gave the exact logical phase and leakage at most $4\cdot10^{-14}$, an observation on these instances for which we have no
proof. Check T20 confirms the identity for $WQ$, proved above by hand, to $6\cdot10^{-14}$ on a $160\times160$ momentum grid, and finds a decay
length of $K$ of 1.3 to 1.4 lattice spacings.

\begin{lemma}[No finitely supported logical operators]\label{lem:ftclocal}
Take a locally finite cell structure on $\mathbb R^3$ with one qubit on each 2-cell. Suppose the checks are $W_c=\prod_{f\subset\partial c}Z_f$
for the 3-cells $c$ and $G_e=X^{\delta e}Z^{\lambda_e}$ for the 1-cells $e$, where $X^{\delta e}=\prod_{f\supset e}X_f$ is the product over the
2-cells whose boundary contains $e$, and $\lambda_e$ is any finite set of 2-cells for which the checks commute. Then every finitely supported
Pauli operator that commutes with every check is a product of checks, up to a phase.
\end{lemma}
\begin{proof}
Write the operator as $X^uZ^v$ with $u,v$ finitely supported 2-cochains. It commutes with $W_c$ if and only if $(\delta u)(c)=0$, so $\delta u=0$.
Since $H^2_c(\mathbb R^3;\Z_2)=0$, there is a finitely supported 1-cochain $w$ with $u=\delta w$, and multiplying by $\prod_eG_e^{w(e)}$ leaves
$Z^{v'}$, up to a phase. It commutes with $G_e$ if and only if $(\partial v')(e)=0$, reading $v'$ as a finite 2-chain, so $\partial v'=0$. Since
$H_2(\mathbb R^3;\Z_2)=0$, there is a finite 3-chain $C$ with $v'=\partial C$, and $Z^{v'}=\prod_{c\in C}W_c$.
\end{proof}

The dressing $\lambda_e$ plays no role, so the lemma covers the dual form of the 3D toric code, with $\lambda_e=\emptyset$, and the fermionic
toric code. In the exact bosonization of~\cite{ChenKapustin2019}, and of~\cite{Chen2020bosonization} on triangulations such as the barycentric
subdivisions used here, qubits sit on 2-cells, the fermion parity of a 3-cell is $W_c$, a fermion hopping across a 2-cell $f$ is $X_f$ times
$Z$'s, and the gauge constraint $G_e$ is the product of the hoppings across the 2-cells around $e$, times parities; so $G_e=X^{\delta e}Z^{\lambda_e}$.
The fermionic toric code is the gauged atomic insulator, whose checks are the $W_c$, which fix the insulator, and the $G_e$. So the bound
$d\ge\min(L,L_z)/(2r_0)$ for the quarter-turn family needs no further input.

\clearpage
\section{Proofs}
\label{app:proofs}

This appendix contains the longer proofs. Each statement is in the main text or in Appendix~\ref{app:tech}, followed by the idea of its
proof; the proofs below are grouped by the section that uses the statement.

\subsection{Proofs for Section~\texorpdfstring{\ref{sec:ft}}{2}}
\label{app:proofs2}

\phantomsection\label{pf:prop:nilp}
\begin{proof}[Proof of Proposition~\ref{prop:nilp}]
(a) The simple objects of $D^\omega(A)$ are pairs $(a,\chi)$ of a flux $a\in A$ and an irreducible $\theta_a$-projective representation $\chi$ of $A$, and
fusion adds fluxes. The dual of $(a,\chi)$ has flux $-a$, so $x\otimes\bar x$ has flux 0. For normalized $\omega$, $\theta_0$ is trivial, so the flux-0 simple
objects are the characters of $A$, which are invertible. So $D^\omega(A)_{\rm ad}$ is pointed, and the adjoint subcategory of a pointed category is
$\mathrm{Vec}$. (b) $(\mathcal B\boxtimes\mathcal B')_{\rm ad}=\mathcal B_{\rm ad}\boxtimes\mathcal B'_{\rm ad}$, and $\mathcal D_{\rm ad}\subseteq\mathcal B_{\rm ad}$ for a fusion subcategory $\mathcal D\subseteq\mathcal B$. For a
condensable algebra $E$ the free-module functor $F:\mathcal B\to\mathcal B_E$ is a surjective tensor functor: each simple $E$-module $y$ is a summand of some
$F(x)$, so $\bar y$ is a summand of $F(\bar x)$ and $y\otimes_E\bar y$ of $F(x\otimes\bar x)$. If $\mathcal B_{\rm ad}$ is pointed, $F(x\otimes\bar x)$ is a sum of invertible objects,
since tensor functors preserve invertibility, so $(\mathcal B_E)_{\rm ad}$ is pointed. The local modules form a fusion subcategory of $\mathcal B_E$.
(c) $\mathrm{Rep}(G)$ is a fusion subcategory of $D(G)$, and $\mathrm{Rep}(G)_{\rm ad}=\mathrm{Rep}(G/Z(G))$~\cite{GelakiNikshych2008}, which is pointed only if $G/Z(G)$ is
abelian. So if $D(G)$ has class at most 2, then $[G,G]\subseteq Z(G)$. Conversely, let $G/Z(G)$ be abelian, and consider the anyons $(z,\chi)$ of $D(G)$
with central flux $z$ and a character $\chi$ of $G$ that is trivial on $Z(G)$. There are $|Z(G)|\,|G/Z(G)|=|G|$ of them, all invertible. They fuse by
$(z,\chi)\otimes(z',\chi')=(zz',\chi\chi')$, and their mutual braiding $\chi(z')\chi'(z)$ and their spins $\chi(z)$ are trivial. So they form a Lagrangian
subcategory equivalent to $\mathrm{Rep}(A)$ for an abelian group $A$ of order $|G|$, and a non-degenerate braided fusion category with such a
subcategory is a twisted double $D^\omega(A)$~\cite{DGNO2007,NaiduNikshych2008}. By (a) it has class at most 2.
\end{proof}

\subsection{Proofs for Section~\texorpdfstring{\ref{sec:lattice}}{3}}
\label{app:proofs3}

\phantomsection\label{pf:thm:general}
\begin{proof}[Proof of Theorem~\ref{thm:general}]
Freeness makes the lift an isomorphism onto $\calP^\Gamma$, with $\langle\pi a,P'\rangle'=\langle a,P'^\uparrow\rangle$ and lift$\circ\pi=\mathsf N$. So $P'$ is a
quotient logical iff $P'^\uparrow\in S^\perp$, and $\pi(S)\cong\mathsf N(S)$. (i) follows from $\mathsf N(gs)=\mathsf N(s)$ and $n=|\Gamma|\,n'$, and (ii) from $\mathsf N(S)\subseteq S^\Gamma\subseteq
(S^\perp)^\Gamma$. For (iii), a non-trivial quotient logical lifts to an element of $(S^\perp)^\Gamma\setminus\mathsf N(S)$; either it is not in $S$, and
then its weight is at least $d$, or it lies in $S^\Gamma\setminus\mathsf N(S)$. For (iv), the descent of $\mathsf N(\ell)$ is a non-trivial logical of weight
at most $|\ell|$.
\end{proof}

\phantomsection\label{pf:prop:count}
\begin{proof}[Proof of Proposition~\ref{prop:count}]
Because $\Gamma$ acts freely, the Pauli module is a free $\FF_p[\Gamma]$-module, induced from the orbit representatives. For this free module
$\mathsf N$ factors as $\bar{\mathsf N}\circ\pi$ with $\bar{\mathsf N}$ an isomorphism from coinvariants to invariants, and $\langle\pi a,\pi b\rangle'=\langle a,\mathsf Nb\rangle$. So
$\dim\pi(S)^\perp=\dim(S^\perp)^\Gamma$ and $\dim\pi(S)=\dim\mathsf N(S)$, which gives the first formula. If $p\nmid|\Gamma|$, $\mathsf N/|\Gamma|$ projects onto
invariants, so $\mathsf N(S)=S^\Gamma$ and taking invariants is exact. For $\Gamma=\Z_2$ and $p=2$, $\dim\mathsf N(M)=\dim M-\dim M^\Gamma$ and $\dim
M^\Gamma=\frac12(\dim M+\hat h(M))$ on any module, so $2k'=\frac12(\dim S^\perp+\hat h(S^\perp))-\frac12(\dim S-\hat h(S))$. Finally $\hat h(S^\perp)=\hat h(S)$:
$S^\perp\cong(\calP/S)^*$, Tate cohomology of $\Z_2$ is unchanged by duality and shifted by one by the free module $\calP$, and it is periodic.
\end{proof}

\phantomsection\label{pf:thm:balanced}
\begin{proof}[Proof of Theorem~\ref{thm:balanced}]
(a) Each $C_j(\tilde F)$ is a free $\FF_p[\Gamma]$-module with basis the lifted $j$-cells, so $C_j(\tilde F)\otimes_\Gamma C(Y)\cong\bigoplus_{\sigma\in F_j}C(Y)$. A
boundary term $\partial\tilde\sigma=\sum_i\gamma_i\tilde\tau_i$ becomes $\sum_i\rho(\gamma_i)\otimes\tau_i$, the local-coefficient differential. The deck group acts
freely on the cells of $\tilde F$, hence on product cells.

(b) A chain map acts on chains by $g$ and on cochains by the contragredient $(g^{-1})^\top$, so for $g\,e_c=\lambda e_{gc}$ it maps $Z$ on $c$ to $Z^\lambda$ on $gc$ and $X$ on $c$
to $X^{\lambda^{-1}}$ on $gc$; this is conjugation by the permutation composed with the multiplication gate $M_{\lambda^{-1}}:|j\rangle\mapsto|\lambda^{-1}j\rangle$, a
single-qudit Clifford. Because $g$ commutes with $\partial$ and $\delta$, it maps $Z$-checks $Z(\partial f)$ to $Z$-checks and $X$-checks $X(\delta c)$ to $X$-checks. The
quotient map of Definition~\ref{def:quot} on $Z$-type operators is the coinvariant projection $C_n(\tilde F\times Y)\to C_n(\tilde F\times Y)_\Gamma$, which is the
balanced product, so the $Z$-checks of the quotient are the images $\bar\partial\bar f$. On $X$-type operators it is the norm followed by restriction to orbit
representatives; since the module is free, $\mathsf N$ maps $C^{n-1}$ onto the invariant cochains, which are the cochains of the balanced product, and
$\mathsf N\delta=\delta\mathsf N$. So the $X$-checks of the quotient are $X(\bar\delta\bar c)$, those of the CSS code of the balanced product.
\end{proof}

\phantomsection\label{pf:lem:cochain}
\begin{proof}[Proof of Lemma~\ref{lem:cochain}]
All cochains below that contain $e$ as a factor are supported on cells having $e$ as a face, by locality of the cup product, and their
values involve $\eta$ only on faces of cells of $\mathrm{st}(e)$; we extend $\eta$ arbitrarily elsewhere. We use one identity repeatedly. Let $w$
be an ordered cup product of cochains in slots $1,\dots,k$, followed by a last factor $\theta$, and write $w^{(s)}$ for $w$ with the factor in slot
$s$ replaced by its coboundary. Since $\int\delta(\,\cdot\,)=0$,
\begin{equation}
\sum_s(-1)^{\mathrm{pos}_w(s)}\int w^{(s)}+(-1)^{\deg w-\deg\theta}\int w|_{\theta\to\delta\theta}=0,
\label{eq:stokes}
\end{equation}
where $\mathrm{pos}_w(s)$ is the degree in front of slot $s$ in $w$. A slot holding $\delta e$ or $\delta e'$ contributes nothing, and a slot holding
$a_k$ contributes a curvature term.

(i) Conjugation gives $U_\eta X_i(\delta e)U_\eta^\dagger=X_i(\delta e)\varpi^{F_\eta(a+\delta e)-F_\eta(a)}$, where $a+\delta e$ shifts the field $a_i$ only. By
multilinearity $F_\eta(a+\delta e)-F_\eta(a)=\sum_mc_m\sum_{\emptyset\ne S\subseteq I_i(m)}T_S$, with $T_S=\int y_S\cup\eta$ and $y_S$ the monomial with $\delta e$ in the
slots $S$ and $a$ elsewhere. Apply~\eqref{eq:stokes} to $w=m^e_S(a)$, $\theta=\eta$. The slot $\min S$ gives $(-1)^{\mathrm{pos}(\min S)}T_S$, the last
factor gives $(-1)^{|\omega|-1}\int m^e_S(a)\cup B$, and the other slots give curvature terms. Hence
$T_S=(-1)^{\mathrm{pos}(\min S)+|\omega|}\int m^e_S(a)\cup B+(\text{curvature})$, and summing over $m$ and $S$ gives $f_{i,e}+R^\eta_{i,e}$.

(ii) $X_i(\delta e)$ does not change $\delta a$, since $\delta\delta e=0$, and the dressing is diagonal.

(iii) Applying the two operators in both orders to $|a\rangle$ gives
$\Delta(a)=[f_{i,e}(a+\delta e')-f_{i,e}(a)]-[f_{j,e'}(a+\delta e)-f_{j,e'}(a)]$, where $\delta e'$ shifts $a_j$ and $\delta e$ shifts $a_i$. Expanding,
both brackets are sums over a monomial $m$ and disjoint non-empty sets $S\subseteq I_i(m)$, $T\subseteq I_j(m)$ of the same cochain $y_{S,T}$, with
$\delta e$ in the slots $S$, $\delta e'$ in the slots $T$ and $a$ elsewhere, except that in the first bracket the slot $\min S$ holds $e$ (call this
$y^e$) and in the second the slot $\min T$ holds $e'$ ($y^{e'}$). The contribution of $(m,S,T)$ to $\Delta$ is
$c_m(-1)^{|\omega|}\big[(-1)^{\mathrm{pos}(\min S)}\int y^e\cup B-(-1)^{\mathrm{pos}(\min T)}\int y^{e'}\cup B\big]$. Suppose $\min S<\min T$ and apply~\eqref{eq:stokes} to
$w$, the product $y_{S,T}$ with $e$ in slot $\min S$ and $e'$ in slot $\min T$, and $\theta=B$. The slot $\min S$ gives $(-1)^{\mathrm{pos}(\min S)}\int y^{e'}\cup B$,
the slot $\min T$ gives $(-1)^{\mathrm{pos}(\min T)-1}\int y^e\cup B$, $\delta B=0$, and the remaining slots give curvature terms containing $e$ and $e'$. So the
bracket is a combination of curvature terms; the case $\min T<\min S$ is the same with the roles exchanged. If $\omega$ has order at most 2, the
slots $S\cup T$ are all the slots and no curvature term occurs. The curvature terms involve $\delta a$ only on faces of cells containing both $e$
and $e'$, so they vanish on the range of $P_{V_e}P_{V_{e'}}$.

(iv) $G_{i,e}^p|a\rangle=\varpi^{\Phi(a)}|a\rangle$ with $\Phi(a)=\sum_{t=0}^{p-1}f_{i,e}(a+t\,\delta e)$. Expanding $f_{i,e}(a+t\,\delta e)$, the slots of $I_i(m)$ outside $S$ that
receive $\delta e$ form a set $T$ and come with $t^{|T|}$. Since $\sum_{t\in\FF_p}t^j\equiv-1$ if $j\ge1$ and $(p-1)\mid j$, and $\equiv0$ otherwise,
$\Phi=-\sum_mc_m\sum_{U}\sum_{S\sqcup T=U}(-1)^{\mathrm{pos}(\min S)+|\omega|}\int y^{e}_{U,\min S}\cup B$, the inner sum over $S,T\ne\emptyset$ with $(p-1)\mid|T|$, where
$y^e_{U,s}$ has $e$ in slot $s$, $\delta e$ in the other slots of $U$ and $a$ elsewhere. By the argument of (iii) with $e'=e$, the value
$(-1)^{\mathrm{pos}(s)}\int y^e_{U,s}\cup B$ is the same for all $s\in U$ up to curvature terms with $e$ in two slots. So $\Phi$ equals, up to such terms, a
combination of these common values with coefficients $\sum_{0<j<u,\ (p-1)\mid j}\binom uj$, $u=|U|$. These vanish modulo $p$: evaluating
$\sum_{x\in\FF_p^\times}(1+x)^u$ once by the binomial theorem and once as $\sum_{y\in\FF_p,\,y\ne1}y^u$ gives $\sum_{0\le j\le u,\,(p-1)\mid j}\binom uj\equiv1+[(p-1)\mid u]$,
which are exactly the terms $j=0$ and $j=u$. For order at most 2 no curvature term occurs.
\end{proof}

\phantomsection\label{pf:prop:twisted-code}
\begin{proof}[Proof of Proposition~\ref{prop:twisted-code}]
(a) By Lemma~\ref{lem:cochain}(ii) $G_{i,e}$ commutes with $P_{V_e}$; on its range $G_{i,e}$ is unitary with $G_{i,e}^p=1$ by (iv), so the average
of its powers is the orthogonal projector onto its fixed space there, and $\Pi_{i,e}$ is a Hermitian idempotent. Two such operators commute:
$\Pi_{i,e}\Pi_{j,e'}=P_{V_e}P_{V_{e'}}\,\frac1{p^2}\sum_{t,t'}G_{i,e}^tG_{j,e'}^{t'}$, and on the range of $P_{V_e}P_{V_{e'}}$, which both $G$'s preserve, they commute by
(iii). Flux projectors commute with everything by (ii). A vector is fixed by all the projectors iff its fluxes are trivial and it is fixed by
every $G_{i,e}$.

(b) For order at most 2, $f_{i,e}$ is affine in $a$, so $G_{i,e}$ is an $X$-string times a $Z$-string times a phase; it commutes with all checks by
Lemma~\ref{lem:cochain}(ii)--(iii), and $G_{i,e}^p=1$ by (iv). Commuting Pauli operators of order $p$ generate a Pauli stabilizer group unless they
generate a non-trivial scalar, in which case their common $+1$ eigenspace is zero.

(c) The dressing is linear in $B$. By Lemma~\ref{lem:cochain}(i) with $B$ replaced by $\delta\mu$ and $\eta=\mu$,
\[
U_\mu G^B_{i,e}U_\mu^\dagger=U_\mu X_i(\delta e)U_\mu^\dagger\,\varpi^{f^B_{i,e}}=G^{B+\delta\mu}_{i,e}\,\varpi^{R^\mu_{i,e}},
\]
and $R^\mu_{i,e}=0$ on the range of $P_{V_e}$, which $U_\mu$
preserves; so the projectors are conjugate. $U_\mu$ is a product over $D$-cells of diagonal gates on the qudits of one cube. For order at most 2,
$R^\mu_{i,e}$ is linear in $\delta a$, so $\varpi^{R^\mu_{i,e}}$ is a product of flux checks.

(d) The formula~\eqref{eq:gauss} is built from the cup product and $\int$, both preserved by $\phi$ when $\phi$ preserves the cup structure.
\end{proof}

\phantomsection\label{pf:prop:clifflevel}
\begin{proof}[Proof of Proposition~\ref{prop:clifflevel}]
(a) Each term of~\eqref{eq:gauss} has $e$ in one slot, so at most $o-1$ slots hold fields. (b) Lemma~\ref{lem:cochain}(i): on the range of
$P_{V_e}$, $f_{i,e}(a)=F_\eta(a+\delta e)-F_\eta(a)$, and a discrete derivative lowers the degree by at least one. The bounds are not equalities in
general: the degree can drop when the phase polynomial collapses on the lattice. For a $\Z_2$ 0-form field, cup products of 0-cochains are
pointwise products of bits and $a_v^2=a_v$, so $a\cup a\cup a=a$ as functions of the configuration: the twist operator
$(-1)^{\int a\cup a\cup a\cup B}=(-1)^{\int a\cup B}$ is a Pauli $Z$-string, and the Gauss laws are Pauli, one level below the bound.
\end{proof}

\phantomsection\label{pf:thm:dressedquot}
\begin{proof}[Proof of Theorem~\ref{thm:dressedquot}]
(a) Proposition~\ref{prop:twisted-code}(c) on $\tilde M$ with $B=0$ and $\mu=\tilde\eta$. Deck transformations are cellular automorphisms preserving the
pulled-back cup structure and $[\tilde M]$, act freely on cells, and fix $\tilde B$, so Proposition~\ref{prop:twisted-code}(d) shows that they permute the Gauss
laws.

(b) The $X$-parts descend to $X(\pi_*\delta\tilde e)=X(\delta e)$, since $\pi_*$ commutes with $\delta$ for a covering. For the phases we use the
projection formula $\int_{\tilde M}y_1\cup\dots\cup\tilde x\cup\dots\cup y_k=\int_M\bar y_1\cup\dots\cup\pi_*\tilde x\cup\dots\cup\bar y_k$, valid when all factors but $\tilde x$ are
pull-backs $y_s=\pi^*\bar y_s$; it holds cube by cube, because the lifts of a cube are in bijection with the lifts of each of its faces. Each
term of $f^{\tilde B}_{i,\tilde e}(\pi^*a)$ has $\tilde e$ in one slot, $\delta\tilde e$ in $|S|-1$ slots, and pull-backs elsewhere. If $|S|=1$, which is always the
case for multilinear $\omega$, the projection formula turns it into the corresponding term of $f^B_{i,e}(a)$. If $|S|\ge2$ and the covering is
injective on closed cubes, then on every cube containing $\tilde e$ the cochain $\delta\tilde e$ agrees with $\pi^*\delta e$, because no other lift of $e$ is a
face of that cube; we replace it and apply the projection formula. Flux checks descend to flux checks. The descended operators are
therefore the dressed Gauss laws of $M$, which satisfy Lemma~\ref{lem:cochain}. They depend on $\tilde B$ but not on $\tilde\eta$, and not on the
representatives by Definition~\ref{def:descent}.

(c) $g$ acts on configurations by $g|\tilde a\rangle=|(g^{-1})^*\tilde a\rangle$, and $\int(g^{-1})^*y=\int y$. So
$U^\dagger gU|\tilde a\rangle=\varpi^{\int\omega(\tilde a)\cup\tilde\eta-\int\omega(\tilde a)\cup g^*\tilde\eta}\,g|\tilde a\rangle$, which is $g\,U(\zeta_g)|\tilde a\rangle$; and $\delta\zeta_g=\tilde B-g^*\tilde B=0$. A
0-cocycle on a connected complex is constant.

(d) For order at most 2 all operators are Pauli. After forgetting phases, descent is then the map $\pi$ of Definition~\ref{def:quot}, and
the dressed Gauss laws of (b) supply consistent phases, as in the paragraph after Proposition~\ref{prop:valid}.
\end{proof}

\phantomsection\label{pf:thm:hfcount}
\begin{proof}[Proof of Theorem~\ref{thm:hfcount}]
By Proposition~\ref{prop:twisted-code}(b) the code is a Pauli stabilizer code with $k'=n-\dim S_B$. The $X$-parts of the generators are those of the
undressed code, so $\dim S_B$ exceeds the undressed rank by the number of new independent $Z$-type elements. These are products of Gauss laws
whose $X$-parts cancel, that is, over cocycles $\lambda\in Z^{n_1-1}$ (for field 1; field 2 is the same). By~\eqref{eq:mutualgauss} such a product is
$Z_2(\Phi\lambda)$ up to a phase, with $\Phi\lambda(c)=\pm\int\lambda\cup c\cup B$. It commutes with the field-2 Gauss laws, since
$\pm\int\lambda\cup\delta e'\cup B=\pm\int\delta(\lambda\cup e'\cup B)=0$, so $\Phi\lambda$ is a cycle. If $\lambda=\delta\kappa$, then $\Phi\lambda(c)=\pm\int\kappa\cup\delta c\cup B$, so $\Phi\lambda$ is a
boundary and $Z_2(\Phi\lambda)$ is a product of flux checks. In case (i) the dressing of field 1 involves only $a_2$, so these products carry no
extra phase and the group contains no non-trivial scalar. The new $Z$-stabilizers therefore remove exactly $r_1$ logical qubits of field 2, where
$r_1$ is the rank of the pairing $H^{n_1-1}\times H^{n_2}\to\Z_p$, $([\lambda],[c])\mapsto\langle[\lambda]\cup[c]\cup[B],[M]\rangle$. By Poincar\'e duality
$H^{n_1-1}\cong(H^{n_2+q+1})^*$, so $r_1=\rk\big(\cup[B]:H^{n_2}\to H^{n_2+q+1}\big)$, and field 2 keeps $b_{n_2}-r_1=\dim\ker(\cup[B]|_{H^{n_2}})$ logical qubits. For
one field, $\Phi\lambda(c)=\kappa\big[\int\lambda\cup c\cup B+(-1)^n\int c\cup\lambda\cup B\big]$ up to signs, and graded commutativity on cohomology,
$[c]\cup[\lambda]=(-1)^{n(n-1)}[\lambda]\cup[c]=[\lambda]\cup[c]$, turns the pairing into $\kappa(1+(-1)^n)\langle[\lambda]\cup[c]\cup[B],[M]\rangle$, which is zero unless $n$ is even
and $p$ is odd.
\end{proof}

\phantomsection\label{pf:prop:cone}
\begin{proof}[Proof of Proposition~\ref{prop:cone}]
(a) The rows of $D^0$ are the generators and $D^1$ computes symplectic products with them. $D^1D^0=0$ says $\delta\delta=0$ and
$\Phi^\top\delta=\delta^\top\Phi$, which is the commutation of the Gauss laws (Proposition~\ref{prop:twisted-code}(b)).

(b) $D^0$ and $D^1$ map $Z$-parts to $Z$-parts, so $K_Z$ is a subcomplex, with differential $\delta^\top=\partial$; the induced differential on $K_X$ is
$\delta$. A filtration with two steps has $d_r=0$ for $r\ge2$, and over a field the long exact sequence gives
$\dim H^1(K)=\dim\coker\big(d_1|_{H^0(K_X)}\big)+\dim\ker\big(d_1|_{H^1(K_X)}\big)$, which is $\dim E_\infty$ in degree 1.

(c) To compute $d_1$, lift a cocycle of $K_X$ with zero $Z$-part and apply $D$. For $\lambda\in Z^{n_1-1}$, $D^0(\lambda,0)=(0,\Phi\lambda)$, and the proof of
Theorem~\ref{thm:hfcount} shows that $\Phi\lambda$ is a cycle, a boundary when $\lambda$ is exact, and pairs with $[c]\in H^{n_2}$ by
$\pm\langle[\lambda]\cup[c]\cup[B],[M]\rangle$; by Poincar\'e duality the rank of this pairing is the rank of $\cup[B]$ on $H^{n_2}$. For $x\in Z^{n_1}$,
$D^1(x,0)=(0,\Phi^\top x)$. Over a field $B_{n_2-1}=\partial C_{n_2}$ is the annihilator of $Z^{n_2-1}$, so $[\Phi^\top x]=0$ if and only if
$\langle\Phi^\top x,\lambda'\rangle=\langle x,\Phi\lambda'\rangle=\pm\int x\cup\lambda'\cup B$ vanishes for every $\lambda'\in Z^{n_2-1}$. This depends only on $[x]$ and $[\lambda']$,
and since $H^{n_2-1}\cong(H^{n_1+q+1})^*$ by Poincar\'e duality, it vanishes for all $\lambda'$ if and only if $[x]\cup[B]=0$. The one-field case is the
same computation with the pairing of the proof of Theorem~\ref{thm:hfcount}(ii).
\end{proof}

\phantomsection\label{pf:thm:pointcount}
\begin{proof}[Proof of Theorem~\ref{thm:pointcount}]
$\mathcal C_B$ is the joint $+1$ eigenspace of the dressed $X$-checks $G_{i,e}$ on the flux-free subspace, so Theorem~\ref{thm:dressdim} applies. The
$X$-check relations are the cocycles $\lambda\in Z^{n_i-1}$ of each field, and relations of different fields act on different qudits, so their
characters multiply. It remains to compute the character of one relation. Signs are irrelevant mod 2 throughout.

\emph{Step 1: products of Gauss laws.} For an $(n_i-1)$-cochain $\mu$ let $G_i(\mu)=\prod_{e\in\mu}G_{i,e}$; on flux-free configurations the factors
commute and square to 1 (Lemma~\ref{lem:cochain}(iii),(iv)), so the order does not matter. Let
\[
\Phi_i(\mu;a)=\sum_mc_m\sum_{\emptyset\ne S\subseteq I_i(m)}\int m^\mu_S(a)\cup B,
\]
where $m^\mu_S(a)$ is the monomial $m$ with $\mu$ in the slot $\min S$, $\delta\mu$ in the other slots of $S$ and $a$ elsewhere, so that
$f_{i,e}=\Phi_i(e;\cdot)$ by~\eqref{eq:gauss}. We show that $G_i(\mu)|a\rangle=(-1)^{\Phi_i(\mu;a)}|a+\delta\mu\rangle$ for every flux-free $a$, where $\delta\mu$
shifts the field $a_i$. For one cell this is Definition~\ref{def:gauss}. If it holds for $\mu$, then
$G_{i,e}G_i(\mu)|a\rangle=(-1)^{\Phi_i(\mu;a)+\Phi_i(e;a+\delta\mu)}|a+\delta\mu+\delta e\rangle$, so it suffices to show that
$\Phi_i(\mu+e;a)=\Phi_i(\mu;a)+\Phi_i(e;a+\delta\mu)$ on flux-free $a$. Expand both sides by multilinearity. In $\Phi_i(\mu+e;a)$ every slot of $S$
receives either the $e$-part or the $\mu$-part; let $S_e$ and $S_\mu$ be the slots that receive them. The terms with $S_e=\emptyset$ form $\Phi_i(\mu;a)$.
In $\Phi_i(e;a+\delta\mu)$ a term is labeled by $S_e$, which holds $e$ and $\delta e$, and by the set $S_\mu$ of slots of $I_i(m)$ outside $S_e$ that
receive $\delta\mu$. When $S_\mu=\emptyset$ or $\min S_e<\min S_\mu$, the two expansions have the same term for $(S_e,S_\mu)$. When
$\min S_\mu<\min S_e$, the term of $\Phi_i(\mu+e;a)$ has $\mu$ in the slot $\min S_\mu$ and $\delta e$ in the slot $\min S_e$, and the term of
$\Phi_i(e;a+\delta\mu)$ has $\delta\mu$ and $e$ there instead. Apply~\eqref{eq:stokes} to the product with $\mu$ in the slot $\min S_\mu$, $e$ in the slot
$\min S_e$, the other slots as in these terms, and $\theta=B$: the two slots give the two terms, slots holding $\delta e$ or $\delta\mu$ give nothing,
$\delta B=0$, and slots holding a field give curvature terms, which vanish on flux-free $a$. So the two terms are equal, and the claim follows.

\emph{Step 2: characters.} For a relation $\lambda\in Z^{n_i-1}$, $\delta\lambda=0$, so $G_i(\lambda)$ is diagonal on flux-free configurations and every term of
$\Phi_i(\lambda;a)$ with $|S|\ge2$ vanishes. Hence $G_i(\lambda)|a\rangle=(-1)^{q_\lambda(a)}|a\rangle$ with
$q_\lambda(a)=\sum_mc_m\sum_{s\in I_i(m)}\int m^\lambda_{\{s\}}(a)\cup B=\int\lambda\cup\partial_i\omega(a)\cup B$. For multilinear $\omega$ the dressing $f_{i,e}$ does not
involve $a_i$, and Step 1 is not needed; for repeated fields it is. By the Leibniz rule and $\int\delta x=0$, $q_\lambda$ depends only on the classes of
$\lambda$ and $a$: every factor is a cocycle, and replacing one or more of them by coboundaries gives an exact integrand.

\emph{Step 3: the count.} So the count of Theorem~\ref{thm:dressdim} over cocycles is $|B^n|$ times the count over classes, and
$2^{-r_X}|B^n|=1$ since $r_X=\dim B^n$ field by field. Every monomial has at least two slots, so after one slot is replaced by $\lambda$ at least one
field remains, and every $q_\lambda$ is homogeneous of degree at least 1; so the class $\alpha=0$ is always counted. A linear term $c\,a_i$ would
contribute the constant $c\int\lambda\cup B$ to $q_\lambda$, which can be non-zero and empty the code.
\end{proof}

\phantomsection\label{pf:cor:boundary}
\begin{proof}[Proof of Corollary~\ref{cor:boundary}]
For every cochain $y$, $y\cup B$ vanishes on $\partial M$, so $\int\delta(y\cup B)=0$ (Setting); the same holds with $\eta$ or $\mu$ in place of $B$. The proofs of
Lemma~\ref{lem:cochain} and Proposition~\ref{prop:twisted-code} use Stokes only in this form. In the proof of Theorem~\ref{thm:hfcount}, the relations among the
field-1 Gauss laws are the absolute cocycles $\lambda\in Z^{n_1-1}(M)$, since every $(n_1-1)$-cell carries a Gauss law. The new $Z$-type stabilizers are
$Z_2(\Phi\lambda)$ with $\Phi\lambda(c)=\pm\int\lambda\cup c\cup B$; $\Phi\lambda$ is a cycle by Stokes, and for $\lambda=\delta\kappa$, $\kappa$ absolute, $Z_2(\Phi\lambda)$ is a product of
flux checks, since every $(n_2+1)$-cell carries one. The pairing is $H^{n_1-1}(M)\times H^{n_2}(M)\to\FF_p$, $([\lambda],[c])\mapsto\langle[\lambda]\cup([c]\cup[B]),[M,\partial M]\rangle$,
with $[c]\cup[B]\in H^{n_2+q+1}(M,\partial M)$. Lefschetz duality states that $\langle\,\cdot\cup\cdot\,,[M,\partial M]\rangle:H^j(M)\times H^{D-j}(M,\partial M)\to\FF_p$ is perfect;
with $j=n_1-1$ and $D-j=n_2+q+1$, it identifies $H^{n_1-1}(M)$ with the dual of $H^{n_2+q+1}(M,\partial M)$, so the rank of the pairing is the rank of
$\cup[B]:H^{n_2}(M)\to H^{n_2+q+1}(M,\partial M)$, which gives (i). Case (ii) is the same, with graded commutativity of the absolute cup product. For (iii), $q_\lambda(a)$
depends only on $[\lambda]\in H^{n_i-1}(M)$ and $[a]\in H^{n_i}(M)$ by the same Stokes argument, and the count in the proof of Theorem~\ref{thm:pointcount} is unchanged.
For $M=X\times D^{q+1}$, the K\"unneth formula gives $H^n(M)\cong H^n(X)$ and $H^{n+q+1}(M,\partial M)\cong H^n(X)\otimes H^{q+1}(D^{q+1},\partial D^{q+1})$, and $\cup[B]$ is the
cross product with the generator, an isomorphism.
\end{proof}

\phantomsection\label{pf:prop:chslevel}
\begin{proof}[Proof of Proposition~\ref{prop:chslevel}]
(a) $L$ permutes the computational basis affinely, so $LX(c)L^\dagger=X(Ac)$ and $LZ(v)L^\dagger=\pm Z(A^{-T}v)$, and $LD_cL^\dagger$ is diagonal. It lies in
$\mathcal C^{(\ell)}$ because $L$ is a Clifford. Since diagonal operators commute,
\[
UG_cU^\dagger=D\,X(Ac)\,D'_c\,D^\dagger=X(Ac)\,\big(X(Ac)^\dagger DX(Ac)D^\dagger\big)\,D'_c .
\]
The phase polynomial of $E_c$ is $h(a+Ac)-h(a)$, a discrete derivative of that of $D$. Flipping the bits in a set $T$ sends $a^S$ to
$\prod_{j\in S\cap T}(1-a_j)\prod_{j\in S\setminus T}a_j$. In $c_S(a^S(a+t)-a^S(a))$, $t=Ac$, the monomials other than $a^S$ have lower degree and
denominator at most $2^{j_S}$, and $a^S$ has coefficient $((-1)^{|S\cap T|}-1)c_S$, which is 0 or $-2c_S$, with denominator at most $2^{j_S-1}$. So
every monomial of the derivative has $|S|+j_S-1\le m-1$, and $E_c\in\mathcal C^{(m-1)}$. Then $E_cD'_c\in\mathcal C^{(\max(\ell,m-1))}$ because diagonal gates of a given level form a group, and
the Pauli factor $X(Ac)$ does not change the level. Conjugation by $U$ preserves the commutation relations and maps $P_V$ to the $+1$
eigenspace of the new $Z$-checks, since $D$ commutes with every $Z$.

(b) The descent of $X(c)e^{2\pi ih(a)}$ is $X(\pi_*c)e^{2\pi ih(\pi^*a)}$. Each bit of $\pi^*a$ is a bit of $a$, so $h\circ\pi^*$ is obtained from $h$ by
identifying variables. A monomial $a^S$ becomes a monomial $a^{S'}$ with $|S'|\le|S|$, since $a_j^2=a_j$, and coefficients only add, which does
not increase denominators. So the level does not increase. $Z(z)$ descends to the Pauli operator $Z(\pi_*z)$.

(c) $D_R$ is a product of commuting diagonal gates of level at most $m$, hence in $\mathcal C^{(m)}$, and $L_R$ is again a $CNOT$, $X$ and permutation
circuit; apply (a) to $U_R$ and then (b).
\end{proof}

\phantomsection\label{pf:prop:lens}
\begin{proof}[Proof of Proposition~\ref{prop:lens}]
(a) Connected covers of $F$ correspond to the subgroups $\langle g^d\rangle\cong\Z_{m/d}$, and for $i<D_F$ the classifying map induces isomorphisms
$H^i(B\Z_m;\FF_p)\to H^i(F;\FF_p)$, compatibly with restriction to subgroups. In degree 1, $x$ restricts to $x(g^d)=d\,x(g)$, which vanishes iff
$p\mid d$. In degree 2 on a lens space, $H^2(F;\FF_p)$ is spanned by the reduction of a generator $c$ of $H^2(B\Z_m;\Z)=\Z_m$, which restricts to a
generator of $H^2(B\Z_{m/d};\Z)=\Z_{m/d}$; its reduction mod $p$ vanishes iff $p\nmid m/d$. In the top degree a covering of degree $d$ multiplies
$H^{D_F}(\,\cdot\,;\FF_p)\cong\FF_p$ by $d$ (mod 2 for $\mathbb{RP}^2$).

(b) $g^*h-h$ is a 0-cocycle, hence a constant $c$; if $c=0$, $h$ descends to $F$ and $x=\delta h$ would be exact. For $B_F=x$,
$\zeta_g=h-g^*h=-1$, and Theorem~\ref{thm:dressedquot}(c) applies. For $\beta x$: $\pi^*\hat x-\delta\hat h\equiv0\pmod p$, and its entries lie in
$(-p,2p)$, so $\tilde\eta$ is an integral $\{0,1\}$-valued cochain with $\delta\tilde\eta=\pi^*(\delta\hat x)/p$. Finally $g^*\hat h-\hat h=1-p\chi$, so
$\zeta_g=\tilde\eta-g^*\tilde\eta=\delta(g^*\hat h-\hat h)/p=-\delta\chi$.

(c) Take $\tilde\eta$ pulled back from $S^{D_F}$, as in (b); a different primitive changes $\zeta_g$ by $c-g^*c$ for a cocycle $c$, which is exact because
$g$ acts trivially on $H^q(X\times S^{D_F};\FF_p)$ for $q<D_F$. Then $\zeta_g$ is pulled back from $H^q(S^{D_F};\FF_p)=0$, so $\zeta_g=\delta\lambda$, and by Stokes
$\int\omega(a)\cup\delta\lambda=\pm\int\delta(\omega(a))\cup\lambda$, which vanishes when every $\delta a_i$ does, by the Leibniz rule.

(d) By the K\"unneth theorem $H^*(M;\FF_p)\cong H^*(X;\FF_p)\otimes H^*(F;\FF_p)$ as rings, and $\cup(1\otimes B_F)$ is $\pm\,\mathrm{id}\otimes(\cup B_F)$. The table
follows from the rings $H^*(L(m;r);\FF_p)=\Lambda(x)\otimes\FF_p[y]/(y^2)$ for $p$ odd or $4\mid m$, $\FF_2[x]/(x^4)$ for $m\equiv2\bmod4$, and
$H^*(\mathbb{RP}^2;\FF_2)=\FF_2[x]/(x^3)$, in which $xy\ne0$ by Poincar\'e duality.
\end{proof}

\phantomsection\label{pf:thm:dresseddist}
\begin{proof}[Proof of Theorem~\ref{thm:dresseddist}]
(a) Expand $O=\sum_xX(x)\,D_x$ over $x\in\Z_p^A$, with $D_x$ diagonal on $A$. The projector $P$ lies below the projector $P_Z$ onto the common $+1$
eigenspace of the $Z(g)$, and $X(x)$ maps that eigenspace to the eigenspace with syndrome $H_Zx$. So $PX(x)D_xP=0$ unless $x\in\ker H_Z$. For such $x$
write $x=\sum_vt_vh_v$ as in the hypothesis. Then $X(x)=\prod_{v\in V_x}X(h_v)^{t_v}=\prod_{v\in V_x}(\tilde X_vD_v^\dagger)^{t_v}$. For a diagonal $E$,
$E\tilde X_v=\tilde X_v\,D_v^\dagger\big(X(h_v)^\dagger EX(h_v)\big)D_v$, and conjugation by $X(h_v)$ keeps $E$ diagonal with the same support. Moving every diagonal
factor to the right therefore gives $X(x)D_x=\big(\prod_{v\in V_x}\tilde X_v^{t_v}\big)E_x$ with $E_x$ diagonal and supported in $A_x$. Since $P\tilde X_v=P$ (both
$\tilde X_vP=P$ and unitarity give $P\tilde X_v=(\tilde X_v^\dagger P)^\dagger=P$), $PX(x)D_xP=PE_xP$. Expand $E_x=\sum_zc_zZ(z)$ over $z$ supported in $A_x$. If
$\langle z,h_v\rangle\ne0$ for some $v$, then $Z(z)\tilde X_v=\varpi^{\langle z,h_v\rangle}\tilde X_vZ(z)$, because $D_v$ commutes with $Z(z)$, and
$PZ(z)P=PZ(z)\tilde X_vP=\varpi^{\langle z,h_v\rangle}P\tilde X_vZ(z)P=\varpi^{\langle z,h_v\rangle}PZ(z)P$, so $PZ(z)P=0$. Otherwise $z\in\ker H_X$, so by hypothesis $Z(z)$ is a product
of the $Z(g)$ up to a phase, and $PZ(z)P$ is a multiple of $P$. Summing, $POP$ is a multiple of $P$.

(b) Measure distances in $\Z^d$ with the $\ell^\infty$ metric, applied to the projections of closed cells to $\Z^d$, and on $M$ along lifts; a cell
has diameter at most $1$. Every flux check acts on the faces of one cell, a set of diameter at most $1$; every $X_i(\delta e)$ acts on the cells that contain
$e$, a set of diameter at most $2$; and $D_{i,e}$ acts within distance $1$ of $e$.

\emph{Filling in a box.} Let $Q=[a,b]^d\times F$ and $Q^+=[a-1,b+1]^d\times F$, with side boundary $\partial'Q^+=\partial[a-1,b+1]^d\times F$. (i) Let $x$ be an
$n$-cocycle of $\Z^d\times F$, $1\le n\le d-1$, supported on cells contained in $Q$. It vanishes on the cells of $\partial'Q^+$, so it is a relative cocycle
of $(Q^+,\partial'Q^+)$, and $H^n(Q^+,\partial'Q^+;\Z_p)\cong H^n([a-1,b+1]^d,\partial;\Z_p)=0$ for $n\ne d$. So $x=\delta w$ on $Q^+$, with $w$ vanishing on $\partial'Q^+$.
Extended by zero, $w$ satisfies $\delta w=x$ everywhere: a cell not contained in $Q^+$ meets $Q^+$ only in faces lying in $\partial'Q^+$, where $w=0$. (ii) An
$n$-cycle supported in $Q$, $n\ge1$, is the boundary of an $(n+1)$-chain in $Q$, since $H_n(Q;\Z_p)=0$.

Let $O$ act on a set $A$ of $m\le(s-2)/10$ qudits; we verify the hypothesis of (a). Let $x\in\ker H_Z$ be supported in $A$. Call two cells of its support
linked if a flux check acts on both, and split $x=\sum_jx_j$ along the classes of the generated equivalence relation. A flux check meets at most one
class, so each $x_j\in\ker H_Z$. If $x_j$ has $m_j$ cells, a path of links joins any two of them, and it lifts to $\Z^d\times F$; so $x_j$ lifts to a
cochain $\tilde x_j$ supported in a box $Q_j$ of side $m_j$. The quotient map is injective on $Q_j^{++}$, the box enlarged by $2$ on every side, of side
$m_j+4\le s$, so $\tilde x_j$ is a cocycle and, by (i), $\tilde x_j=\delta\tilde w_j$ with $\tilde w_j$ supported in $Q_j^+$. Projecting, $x_j=\delta w_j$, and we take for $V_x$
the union of the supports of the $w_j$. Then $A_x$ lies in the union of the $m$ cells of $A$, of diameter at most $1$ each, and of the images of the boxes
$Q_j^{++}$, of diameter at most $m_j+4$ each.

Let $z\in\ker H_X$ be supported in $A_x$. Call two cells of its support linked if some $X_i(\delta e)$ acts on both, and split $z=\sum_lz_l$ in the same
way; each $z_l\in\ker H_X$. The cells of $z_l$ lie in a union of some of the sets just listed, and consecutive cells of a path of links are at distance
at most $2$. So $z_l$ lifts into a box of side at most
\[
m+\sum_j(m_j+4)+2(m+k)\le m+5m+4m=10m\le s-2,
\]
where $k\le m$ is the number of classes $x_j$. The quotient map is injective on this box enlarged by $1$, so the lift is a cycle, and by (ii) it is
the boundary of a chain in the box. Projecting, $z_l$ is a sum of flux checks, and so is $z$. Part (a) gives $POP=c(O)P$. Since this holds for every
operator on at most $(s-2)/10$ qudits, the distance exceeds $(s-2)/10$.
\end{proof}

\phantomsection\label{pf:cor:dresseddist}
\begin{proof}[Proof of Corollary~\ref{cor:dresseddist}]
The torus is $\Z^d/L\Z^d$ and the slab is $(\Z^d\times[0,h])/L\Z^d$; the quotient map is injective on boxes of side $s=L-1$. By
Proposition~\ref{prop:twisted-code}(a), $\mathcal C_B$ is the common $+1$ eigenspace of the flux checks and the $G_{i,e}=X_i(\delta e)\varpi^{f_{i,e}}$, and $f_{i,e}$ involves only
qudits on faces of the $D$-cells that contain $e$ (Definition~\ref{def:gauss}, since the cubical cup product is local). So Theorem~\ref{thm:dresseddist}(b)
applies. For the upper bounds, let $z$ be a straight 1-cycle of length $L$ of a 1-form field $a_i$. The operator $Z_i(z)$ commutes with every check and acts
on the code space by $(-1)^{\langle\alpha_i,z\rangle}$ on the state of the class $\alpha$ (Theorem~\ref{thm:pointcount}), and in each case some allowed class has
$\langle\alpha_i,z\rangle\ne0$, so $Z_i(z)$ is a non-trivial logical operator.
\end{proof}

\subsection{Proofs for Section~\texorpdfstring{\ref{sec:sym}}{4}}
\label{app:proofs4}

\phantomsection\label{pf:prop:logaction}
\begin{proof}[Proof of Proposition~\ref{prop:logaction}]
(a) For $\gamma\in\Gamma$, $(f\otimes s)(\gamma\otimes\gamma)(f\otimes s)^{-1}=\theta(\gamma)\otimes\alpha(\gamma)$. Distinct deck transformations are distinct maps of $\tilde F$,
and by faithfulness distinct elements of $G$ are distinct maps of $C(Y)$, so this lies in $\Gamma_{\rm diag}$ if and only if $\alpha(\gamma)=\theta(\gamma)$; conjugation
by $f\otimes s$ is injective and $\Gamma_{\rm diag}$ is finite, so mapping it into itself is the same as normalizing it. For the holonomy, fix $\tilde x_0$ over
$x_0$; for a loop $c$ at $x_0$, $\rho(c)$ is the deck transformation that sends $\tilde x_0$ to the end point of the lift of $c$ from $\tilde x_0$. The map $f$
sends this lift to the lift of $\bar f\circ c$ from $f\tilde x_0$, which ends at $f\rho(c)\tilde x_0=\theta(\rho(c))f\tilde x_0$. So with base point $f\tilde x_0$ the
holonomy of $\bar f\circ c$ is $\theta(\rho(c))$, and moving the base point conjugates the holonomy by an element of $\Gamma$.

(b) By Theorem~\ref{thm:balanced}(b) the quotient code is the CSS code of $(C(\tilde F)\otimes C(Y))_\Gamma$, and the descended automorphism of
Proposition~\ref{prop:descent} is the chain automorphism that $f\otimes s$ induces on these coinvariants, which exists by (a). A monomial chain
automorphism maps $Z(z)$ to $Z$ of the image of $z$, up to the multiplication gates of Theorem~\ref{thm:balanced}(b), so it acts on $Z$-type
logicals by the induced map on $H_n$. Since $f$ is cellular, $f\otimes s$ preserves the filtration by $F$-degree, and a filtered chain map
induces a morphism of the associated spectral sequences that is compatible with the filtration of the abutment~\cite[Chapter~2]{McCleary2001}.
On $E^1_{pq}=C_p(F)\otimes H_q(Y)$ it sends the copy of $H_q(Y)$ over a $p$-cell $\sigma$ to the copy over $\bar f(\sigma)$ by $s_*$, so on $E^2$ it is the
induced map on homology with local coefficients.
\end{proof}

\phantomsection\label{pf:prop:cupgate}
\begin{proof}[Proof of Proposition~\ref{prop:cupgate}]
$U_\omega$ commutes with the flux checks. By Lemma~\ref{lem:cochain}(i) with $B=0$ and $\eta=1$, the constant 0-cochain,
$U_\omega X_i(\delta e)U_\omega^\dagger=X_i(\delta e)\varpi^{R(a)}$ with $R$ a combination of curvature terms, so on the flux-free subspace $U_\omega$ commutes with every
Gauss law and preserves the code space. On a flux-free $a$, $\int\omega(a)$ depends only on the classes $[a_i]$ by the Leibniz rule, and the code
state $|\alpha\rangle$ is the uniform superposition over the configurations in the class $\alpha$. The last statement holds because the cup product is
computed cell by cell and deck transformations preserve the pulled-back cup structure. A cubical structure that is not $\Gamma$-invariant,
such as the coordinate structure of $\partial[-m,m]^4$ under the antipodal map, does not satisfy the hypothesis; there we pass to the barycentric
subdivision, whose canonical branching (vertices ordered by the dimension of the cell they subdivide) is preserved by every cellular map.
\end{proof}

\phantomsection\label{pf:prop:internal}
\begin{proof}[Proof of Proposition~\ref{prop:internal}]
(a) The element $g^j$ of $\tilde\Gamma$, $g$ the generator, sends $(i,c)$ to $(i-j,R^jc)$. It fixes a qudit only if $r\mid j$ and $R^j$ fixes a cell,
so only if $R^j=1$ by freeness of $\langle R^r\rangle$, and then $g^j$ is the identity. The orbit of $(i,c)$ meets copy 1 exactly in
$\{(1,R^{i-1+rt}c)\}_t$, one $\langle R^r\rangle$-orbit of cells, so copy-1 representatives identify the quotient qudits with the $n$-cells of $\hat M$. A
check of copy $i$ at a cell is identified with the same check of copy 1 at its image under $R^{i-1}$, since $R$ is cellular and commutes with $\delta$
and $\partial$, and then with the check of $\hat M$ below it, because $\pi$ is a covering. So the quotient checks are the toric-code checks of
$\hat M$. A configuration is invariant if and only if $a_{i-1}(Rc)=a_i(c)$ for all $i,c$, that is $a_i=R^*a_{i-1}$, which forces $a_1=(R^r)^*a_1$, so
$a_1=\pi^*a$ and $a_{1+j}=(R^j)^*\pi^*a=\pi^*(\hat R^j)^*a$. Passing to classes gives the last statement.

(b) $\pi$ commutes with the cup product, because $\hat M$ carries the cup structure induced from $\tilde M$, so
$\int_{\tilde M}a_1\cup\dots\cup a_r=\int_{\tilde M}\pi^*\big(a\cup\hat R^*a\cup\dots\big)$, and every top cell of $\hat M$ has $|\langle R^r\rangle|$ preimages with the same
value; keeping one of them gives $V$. The factor at a top cell $c$ involves $(\hat R^j)^*a$ on the faces of $c$, that is $a$ on the faces of
$\hat R^jc$. $V$ is diagonal, so it commutes with the flux checks. On a flux-free configuration, the Gauss law $X(\delta x)$ replaces each factor
$(\hat R^j)^*a$ by $(\hat R^j)^*a+\delta(\hat R^j)^*x$. Every term of $T(a+\delta x)-T(a)$ is a cup product of cocycles and at least one coboundary, which is
a coboundary by the Leibniz rule, so its integral vanishes, and $V$ commutes with every Gauss law on the flux-free subspace. Hence $V$
preserves the code space, and the same argument shows that $T$ depends only on the class of $a$, which gives the action on $|\alpha\rangle$.
\end{proof}

\phantomsection\label{pf:prop:lift}
\begin{proof}[Proof of Proposition~\ref{prop:lift}]
(a) By the lifting criterion of covering-space theory, $\bar f\circ \pi:\tilde M\to M$ lifts through $\pi$ if and only if $(\bar f\circ\pi)_*\pi_1(\tilde M)=\bar f_*(\Pi)$ is contained
in a conjugate of $\Pi$, that is, in $\Pi$; since $\bar f_*$ is an automorphism and $\Pi$ has finite index, containment is equality. A lift $f$ is a cellular map
covering $\bar f$, and composing it with a lift of $\bar f^{-1}$ gives a map covering the identity, which is a deck transformation; so $f$ is invertible.
For $g\in\Gamma$, $fgf^{-1}$ covers the identity, so it lies in $\Gamma$. A cellular automorphism permutes cells and preserves the chain complex,
so $f$ induces a permutation automorphism of the parent normalizing $\Gamma$, and Proposition~\ref{prop:descent} applies. Conversely, a cellular
automorphism $f$ of $\tilde M$ normalizing $\Gamma$ maps fibres of $\pi$ to fibres, so it covers a cellular automorphism $\bar f$, and $\bar f_*(\Pi)=\Pi$ because
$f_*$ preserves $\pi_1(\tilde M)$.
(b) A characteristic subgroup is invariant under every automorphism, inner ones included. The kernel of $\pi_1\to H_1(\,\cdot\,;\Z_m)$ is
characteristic because every automorphism of $\pi_1$ induces one of $H_1$.
(c) $(f,s)(g,\rho(g))(f,s)^{-1}=(fgf^{-1},\,s\rho(g)s^{-1})$, which lies in $\Gamma_{\rm diag}$ if and only if $s\rho(g)s^{-1}=\rho(fgf^{-1})$.
\end{proof}

\phantomsection\label{pf:prop:transgress}
\begin{proof}[Proof of Proposition~\ref{prop:transgress}]
The Gauss laws at one cell $e$ commute on the flat range (Proposition~\ref{prop:twisted-code}), so the order of the product does not matter,
and $V$ maps $X_i(\delta e)$ to $\prod_jX_j(\delta e)^{A_{ji}}$ and flux checks to products of flux checks.

\emph{Exact case.} Suppose first that $B_\Z=\delta\eta_\Z$ on all of $M$ for an integer cochain $\eta_\Z$, and let $\eta$ be its reduction, so $B=\delta\eta$.
For a flat $a$, since $(-1)^{\int\omega(a)\cup\eta}=e^{\pi i\int\omega(\tilde a)\cup\eta_\Z}$,
\[
U_\eta VU_\eta^\dagger|a\rangle=e^{2\pi i\int(\frac12\omega(\widetilde{Aa})-\frac12\omega(\tilde a))\cup\eta_\Z}|Aa\rangle
=e^{2\pi i\int\delta\lambda(a)\cup\eta_\Z}|Aa\rangle=e^{2\pi i\,s\int\lambda(a)\cup B_\Z}|Aa\rangle=\tilde U|a\rangle .
\]
The second step uses (i): the difference is an integer cochain, and its cup with $\eta_\Z$ integrates to an integer. The third is Stokes over
$\RR$, $\int\delta\lambda\cup\eta_\Z=(-1)^{|\lambda|+1}\int\lambda\cup\delta\eta_\Z$, with $|\lambda|=|\omega|-1$. By Lemma~\ref{lem:cochain}(i), $U_\eta X_j(\delta e)U_\eta^\dagger=G^B_{j,e}$
on the flat range, which $U_\eta$, $V$ and $W$ preserve. Hence, on the flat range,
$\tilde UG^B_{i,e}\tilde U^\dagger=U_\eta VX_i(\delta e)V^\dagger U_\eta^\dagger=\prod_j(U_\eta X_j(\delta e)U_\eta^\dagger)^{A_{ji}}=\prod_j(G^B_{j,e})^{A_{ji}}$.

\emph{General case.} Fix $(i,e)$. Both sides of the identity map $|a\rangle$ to a multiple of $|a+A\,\delta e_i\rangle$, where $\delta e_i$ is $\delta e$ in field $i$;
let $\Phi(a)$ be the ratio of the two multiples. The left side is
$W(y)^{-1}\,(-1)^{f^B_{i,e}(y)}\,W(y+\delta e_i)$ with $y=A^{-1}a$. By locality of the cup product, $\lambda(y+\delta e_i)-\lambda(y)$ is supported on cells that
contain $e$, and its cup with $B_\Z$ only involves $D$-cells of $\mathrm{st}(e)$; the dressings $f^B_{j,e}$ also only involve $\mathrm{st}(e)$
(Lemma~\ref{lem:cochain}). So $\Phi(a)$ depends only on $a$, $B$ and $B_\Z$ on $\overline{\mathrm{st}}(e)$. By (iii), extend $\eta_e$ by zero to an integer cochain
$\eta'$ on $M$; then $B'_\Z=\delta\eta'$ agrees with $B_\Z$ on $\overline{\mathrm{st}}(e)$, because $\overline{\mathrm{st}}(e)$ is closed under taking faces, and $B'=B'_\Z\bmod2$ agrees with $B$ there.
The exact case for $(B',B'_\Z)$ gives $\Phi=1$ on flat configurations, and $\Phi$ is the same function for $(B,B_\Z)$.
\end{proof}

\phantomsection\label{pf:prop:threegroup}
\begin{proof}[Proof of Proposition~\ref{prop:threegroup}]
(a) is Proposition~\ref{prop:cupgate} for one 1-form field and $\omega=a\cup a\cup a$, with $|\omega|=3=D$.
(b) The cup product on $H^*(M;\Z_2)$ is commutative, so $(\alpha+\lambda)^3=\alpha^3+\alpha^2\lambda+\alpha\lambda^2+\lambda^3$. By the Cartan formula and $\mathrm{Sq}^1x=x^2$ for $|x|=1$,
$\alpha^2\lambda+\alpha\lambda^2=\mathrm{Sq}^1(\alpha\lambda)$, and by Wu's formula on a closed 3-manifold $\langle\mathrm{Sq}^1y,[M]\rangle=\langle v_1\cup y,[M]\rangle$ for $|y|=2$, with
$v_1=w_1(M)$. So the polarization of $q$ is $b$, which is bilinear. A function on $\FF_2^k$ with $q(0)=0$ and bilinear polarization is
$q(\sum_i\alpha_ie_i)=\sum_i\alpha_iq(e_i)+\sum_{i<j}\alpha_i\alpha_jb(e_i,e_j)$, which gives the gate. A diagonal gate $(-1)^{q}$ is Pauli exactly when $q$ is affine, that is,
when $b=0$.
(c) $\bar X(\lambda)$ maps $|\alpha\rangle$ to $|\alpha+\lambda\rangle$, so by (a) the left side acts by $(-1)^{q(\alpha)+q(\alpha+\lambda)}=(-1)^{q(\lambda)+b(\alpha,\lambda)}$, and
$W(\gamma)|\alpha\rangle=(-1)^{\langle\alpha,\gamma\rangle}|\alpha\rangle=(-1)^{\langle\alpha\cup w_1\cup\lambda,[M]\rangle}|\alpha\rangle$.
(d) The tangent bundle of $M$ is the quotient of that of $\tilde M$, so the orientation character of $M$ is the composite $\pi_1(M)\to\Gamma\xrightarrow{\epsilon}\Z_2$.
\end{proof}

\phantomsection\label{pf:prop:tori}
\begin{proof}[Proof of Proposition~\ref{prop:tori}]
Every class of a torus lifts integrally, so $\Phi(x)=\int_X c\cup g_2^*c\cup\dots\cup g_r^*c$~\cite[Proposition~4.10]{BDEHT2026cupsII}, and on
$T^r$ the cup product of $r$ integral degree-one classes, evaluated on $[X]$, is their determinant in a basis of $\Lambda^*$ dual to an
oriented basis of $\Lambda$. The $r$ vectors $g_i^*c$ span $\Z[G]c$, so their determinant is $\pm$ its index. (a) $p\nmid[\Lambda^*:\Z[G]c]$ if and
only if $c$ generates $\Lambda^*\otimes\Z_{(p)}$ as a $\Z_{(p)}[G]$-module, if and only if, by Nakayama's lemma, $x$ generates
$\Lambda^*/p\Lambda^*=H^1(X;\FF_p)$ as an $\FF_p[G]$-module. A cyclic module of dimension $|G|$ is free, and for $G$ cyclic of order $p^m$,
$\FF_p[G]=\FF_p[t]/(t-1)^{p^m}$, on which $t$ is a single Jordan block. (b) The cochain formula for $\Phi$ is natural for equivariant simplicial
maps, and $\pi_\#[\tilde X]=\deg\pi\,[X]$.
\end{proof}

\phantomsection\label{pf:lem:inflow}
\begin{proof}[Proof of Lemma~\ref{lem:inflow}]
The phase of $U_\eta X_i(\delta e)U_\eta^\dagger$ on $|a\rangle$ is $(-1)^{\int_M(\omega(a+\delta e_i)-\omega(a))\cup\eta}$. As in the proof of
Lemma~\ref{lem:cochain}(i), for flat $a$ the difference $\omega(a+\delta e_i)-\omega(a)$ is $\delta P_{i,e}(a)$, where $P_{i,e}$ is the sum over
the non-empty sets $S$ of slots holding field $i$ of the term with $e$ in the first slot of $S$, $\delta e$ in its other slots and $a$
elsewhere. Stokes now has a boundary term: $\int_M\delta P\cup\eta=\int_{\partial M}P\cup\eta+\int_MP\cup\delta\eta\pmod2$. The second term is
the dressing $f^B_{i,e}$. In the first, $P\cup\eta$ evaluated on a cell of $\partial M$ involves only faces of that cell, so it vanishes unless
$e\subset\partial M$, and then it is $\int_{\partial M}P_{i,e}(a|_{\partial M})\cup z$, the defining formula of $f^{\partial,z}_{i,e}$. The flux checks
commute with $U_\eta$, and $D_{i,e}$ commutes with $U_\eta$, so $U_\eta X_i(\delta e)D_{i,e}U_\eta^\dagger=G^B_{i,e}$.
\end{proof}

\phantomsection\label{pf:prop:slabgate}
\begin{proof}[Proof of Proposition~\ref{prop:slabgate}]
For flat $a$, $U_\eta VU_\eta^\dagger|a\rangle=e^{2\pi i\int_M\delta\lambda(a)\cup\eta_\Z}|Aa\rangle$ by (i), as in the exact case of
Proposition~\ref{prop:transgress}. Stokes over $\RR$ with boundary gives
$\int_M\delta\lambda\cup\eta_\Z=\int_{\partial M}\lambda\cup\eta_\Z+(-1)^{|\lambda|+1}\int_M\lambda\cup B_\Z$, and $(-1)^{|\lambda|+1}=s$. Since $W_\partial$
commutes with $U_\eta$, $VW=U_\eta(VW_\partial^{-1})U_\eta^\dagger$. By Lemma~\ref{lem:inflow} it suffices that $VW_\partial^{-1}$ maps the Gauss laws of
$\mathcal C_0^{\partial z}$ by the rule $A$. For $e\not\subset\partial M$, no cell of $\delta e$ lies in $\partial M$ (a cell of $\partial M$ has all its
faces in $\partial M$), so $X_i(\delta e)$ commutes with $W_\partial$ and $V$ maps it to $\prod_jX_j(\delta e)^{A_{ji}}$. For $e\subset\partial M$,
$X_i(\delta e)=X_i(\delta_\partial e)\,X_i(\delta e\setminus\partial M)$, the second factor commutes with $W_\partial$ and with $D_{i,e}$, and the first is the
hypothesis. Flux checks are as in Proposition~\ref{prop:transgress}.
\end{proof}

\phantomsection\label{pf:lem:bone}
\begin{proof}[Proof of Lemma~\ref{lem:bone}]
Apply Proposition~\ref{prop:transgress} on $X\times S^1_m$ with the wall $B=\theta$ dual to $X\times\{0\}$. In the cubical cup, a term of
$y\cup\theta$ on a prism $c\times[0,1]$ is $\pm y(c\times\{0\})\,\theta$, so the dressings of the Gauss laws at $v\times\{0\}$ and the dressing $W$ depend only
on the values on $X\times\{0\}$, where they are $f^{1}_{i,v}$ and $W_X$; the horizontal part of $\delta(v\times\{0\})$ is $\delta_Xv\times\{0\}$, and $V$ acts
cell by cell. So the ratio $\Phi$ of the proof of Proposition~\ref{prop:transgress} at $(i,v\times\{0\})$, evaluated on the flat pull-back of a flat
$a$ on $X$, equals the corresponding ratio on $X$ at $(i,v)$, which is therefore 1. The code is real and $V$ is a permutation, so the
complex conjugate $VW_X^{-1}$ works as well.
\end{proof}

\phantomsection\label{pf:cor:thinslab}
\begin{proof}[Proof of Corollary~\ref{cor:thinslab}]
Every cube of the slab has its last vertex in $X\times\{1\}$, so $\eta=1$ there and $U_\eta=(-1)^{\int_M\omega}$. By Lemma~\ref{lem:inflow},
$\mathcal C_B=U_\eta\mathcal C_0^{\partial z}$ with $z$ the indicator of the top face, so only top Gauss laws are dressed. Write the qubits of each
field as $t_c$ and $b_c$ on $c\times\{1\}$ and $c\times\{0\}$ ($c$ an $n$-cell of $X$) and $h_e$ on $e\times[0,1]$ ($e$ an $(n-1)$-cell). The vertical flux
checks say $t+b+\delta h=0$. After a $CNOT$ from each $t_c$ to $b_c$, the vertical checks read $b'=\delta h$, the top Gauss law at $e$ becomes
$X(t\text{ on }\delta_Xe)X(b'\text{ on }\delta_Xe)X(h_e)$ times its dressing, which depends on $t$ only, and multiplying by the Pauli bottom Gauss
law $X(b'\text{ on }\delta_Xe)X(h_e)$ leaves the Gauss law of $\mathcal C^{(1)}_X$ on $t$. The top flux checks are those of $\mathcal C^{(1)}_X$, and the
bottom ones are products of the others. After a second layer of $CNOT$s from each $h_e$ to the $b'_c$ with $c\in\delta_Xe$, the remaining
checks are $Z(b''_c)$ and $X(h_e)$ (the vertical Gauss laws, for $n\ge2$, are products of the $X(h_e)$).
\end{proof}

\phantomsection\label{pf:prop:reltrans}
\begin{proof}[Proof of Proposition~\ref{prop:reltrans}]
\emph{Exact case,} $B_\Z=\delta\eta_\Z$ on $M$ with $\eta_\Z$ any integer cochain. In the proof of Lemma~\ref{lem:inflow} the boundary term is
$\int_{\partial M}P_{u,e}\cup\eta$, and $P_{u,e}$ restricted to $\partial_kM$ is a sum of values of $\hat\omega$ on vectors in $L_k$, since $a$, $u$ and $\delta u$
are $L_k$-valued there; by (bc2) it vanishes. So $U_\eta X_u(\delta e)U_\eta^\dagger=G^B_{u,e}$, and $U_\eta\mathcal C_0^{bc}=\mathcal C_B^{bc}$, where
$\mathcal C_0^{bc}$ is the undressed code with the same boundary conditions. By (bc1), $V$ maps $X_u(\delta e)$ to $X_{Au}(\delta e)$ with $Au$ allowed, and
maps the boundary $Z$-checks, which impose the linear forms vanishing on $L_k$, to boundary $Z$-checks; so $V$ preserves $\mathcal C_0^{bc}$. By
(i), $U_\eta VU_\eta^\dagger=V\exp(2\pi i\int_M\delta\lambda\cup\eta_\Z)$ on admissible flat configurations, and Stokes on each stratum gives
\[
\int_M\delta\lambda\cup\eta_\Z=s\int_M\lambda\cup B_\Z+\sum_k\int_{\partial_kM}\lambda\cup\eta_\Z,\qquad
\int_{\partial_kM}\lambda\cup\eta_\Z\equiv\int_{\partial(\partial_kM)}\beta_k\cup\eta_\Z-s\int_{\partial_kM}\beta_k\cup B_\Z\pmod1,
\]
using (bc3) and $|\beta_k|=|\lambda|-1$; the hinge terms combine into $\int_{h_{kl}}(\beta_k-\beta_l)\cup\eta_\Z$, and (bc4) continues the recursion,
which ends on the strata of dimension 0. So $U_\eta VU_\eta^\dagger=\tilde U$, and $\tilde U$ preserves $U_\eta\mathcal C_0^{bc}=\mathcal C_B^{bc}$ with the stated
action on Gauss laws.

\emph{General case.} The proof of Proposition~\ref{prop:transgress} applies verbatim: the ratio $\Phi$ depends only on the data on the closed star
$\overline{\mathrm{st}}(e)$ in $M$, and the reference background $B'_\Z=\delta\eta'$, with $\eta'$ the extension by zero of $\eta_e$, falls under the exact case, which
places no condition on $\eta'$ at $\partial M$.
\end{proof}

\phantomsection\label{pf:prop:gaussdescent}
\begin{proof}[Proof of Proposition~\ref{prop:gaussdescent}]
The $\hat G_g$ form a unitary representation of $\Gamma$ on the single-particle space of $(\tilde M,\tilde s)$, on Nambu space in the
Bogoliubov--de~Gennes case, where they commute with particle--hole conjugation because they are parity-even. The fermion fields on $M$ with
spin structure $s=s_0+\varepsilon_s$ are the fields on $(\tilde M,\tilde s)$ that are invariant under the lift $\varepsilon_s(g)\hat G_g$ of the deck action that $s$
determines, so their single-particle space is $\mathcal H_s$, the eigenspace of the character $\varepsilon_s$.
(a) Since $\tilde u$ commutes with every $\hat G_g$, it preserves $\mathcal H_s$, and $u_s$ is unitary on $\mathcal H_s$. For $\psi\in\mathcal H_s$,
$\psi(g\tilde y)=\varepsilon_s(g)\rho(g)\psi(\tilde y)$, so $(\tilde u\psi)(\tilde x)=\sum_{y\in M}\sum_g\tilde u(\tilde x,g\tilde y)\,\varepsilon_s(g)\rho(g)\,\psi(\tilde y)$, which is the kernel in (a). The kernel
of a quasi-local $\tilde u$ is bounded by $Ce^{-{\rm dist}/\xi}$. If $\tilde x$ and $\tilde y$ lie in a ball that injects into $M$ and $x,y$ are closer than the
injectivity radius $r$, then for $g\ne1$ the distance from $\tilde x$ to $g\tilde y$ is at least $2r-{\rm dist}(x,y)>r$, since $g$ moves every point by at
least $2r$; replacing $\tilde y$ by $h\tilde y$ and using $\psi(h\tilde y)=\varepsilon_s(h)\rho(h)\psi(\tilde y)$ gives the transport factor. If $\tilde u=\mathcal T\exp\big(-i\int\tilde k(t)\,dt\big)$ with every $\tilde k(t)$ commuting with every $\hat G_g$, then restriction to
$\mathcal H_s$ is a $*$-homomorphism on the commutant of the representation, so it commutes with the norm-convergent Dyson series, and
$u_s=\mathcal T\exp\big(-i\int\tilde k(t)|_{\mathcal H_s}\,dt\big)$ is the evolution on $M$ in sector $s$ by the restricted generator. (b) The occupied projector $\tilde Q$ of the $\Gamma$-invariant atomic insulator commutes with every $\hat G_g$, and
$\tilde u\tilde Q\tilde u^\dagger=\tilde Q$. Restricting to $\mathcal H_s$ gives $u_sQ_su_s^\dagger=Q_s$, with $Q_s=\tilde Q|_{\mathcal H_s}$ the occupied projector of the atomic insulator of
sector $s$. A free-fermion state is determined by its occupied subspace up to a phase, so every Fock-space implementation of $u_s$ maps
$|\Omega_s\rangle$ to a multiple of itself; for the canonical second quantization in the number-conserving case the multiple is $\det(Q_su_sQ_s)$ on
the occupied space. (c) Since $[H_s(t),\Psi^\dagger\phi]=\Psi^\dagger k_s(t)\phi$, $U_s(t)$ implements $u_s(t)$, so $U_s(T)|\Omega_s\rangle=\theta(s)|\Omega_s\rangle$ by (b). The
gauged Hamiltonian is a sum of gauge-invariant terms, each a fermion bilinear times a Wilson line, so it commutes with the Gauss laws and
the flux checks, and on $|A\rangle\otimes|\phi\rangle$, for a gauge field $A$ in the $Z$ basis, it acts as $|A\rangle\otimes H_A|\phi\rangle$, with $H_A$ the fermion
Hamiltonian whose terms carry the signs of $A$. For a flat $A_s$ representing $s$ the Wilson line of a path depends only on its homotopy
class, and it is the transport factor of $s$ along that class: the hopping signs of $\tilde s$ for the part in $\tilde M$, and $\varepsilon_s(g)$ for a deck
element $g$. So $H_{A_s}$ is given by the image sum of (a), that is, it is $H_s$, including its scalar part. The code space of the fermionic
toric code on $M$ is spanned by the gauged atomic insulators $|s\rangle=P_G(|A_s\rangle\otimes|\Omega_s\rangle)$~\cite[Section~4]{Barkeshli:2023bta}, with $P_G$ the
projector onto gauge-invariant states, which commutes with $V$. Hence $V|s\rangle=P_G(|A_s\rangle\otimes U_s(T)|\Omega_s\rangle)=\theta(s)|s\rangle$, for every system
size.
\end{proof}

\begin{remark}[Interacting quasi-local pumps: a sketch]\label{rem:quasilocal}
Let a Pauli or commuting-projector parent on $\tilde M$ carry a free action of $\Gamma$ by local symmetries, with quotient checks $s=\pi_*(\tilde s)$, and
let $U$ be generated by a $\Gamma$-equivariant interaction $\Phi_t$ with decay length $\xi$. Let $U_M$ be generated on $M$ by the push-forward of $\Phi_t$,
with one lift for each orbit of interaction terms of diameter at most $R_0$. Assume (L) that for each parent check $\tilde s$ there is an operator
$A_R$ supported within distance $R$ of $\tilde s$ with $\|U^\dagger\tilde sU-A_R\|\le\epsilon(R)$ that fixes the product of the check projectors within distance
$R+w$ of $\tilde s$, and (I) that $R+w$ is below the injectivity radius of $M$. Then $\|(1-P_M)U_MP_M\|\le\sqrt m\,\delta$, where $P_M$ is the code projector
on $M$, $m$ the number of checks and $\delta=\epsilon(R)+Ce^{-(R_0-R)/\xi'}$. So the logical action $P_MU_MP_M$ is within $m\delta^2$ of a unitary, and two
cutoffs give logical actions within $2\sqrt m\,\delta$ of each other.
\end{remark}
\begin{proof}[Sketch]
(1) $U_M$ and the cover evolution are generated by interactions that agree on each ball that injects into $M$, so a Lieb--Robinson comparison
gives $\|U_M^\dagger sU_M-\pi_*(A_R)\|\le\delta$. (2) $\pi_*(A_R)$ fixes the push-forward of the local check projector, which is at least $P_M$ because the
pushed-forward checks are quotient checks, so $\|(1-s)U_MP_M\|\le\delta$ for every check $s$. (3) For commuting projectors
$1-\prod_s\frac{1+s}2\le\sum_s\frac{1-s}2$. For a strictly local diagonal circuit, (L) holds with $\epsilon=0$ and $R=O(1)$, which is the code-preservation step of
Proposition~\ref{prop:cupgate}. For the gauged free-fermion pump it holds with $\epsilon(R)\sim e^{-R/\xi}$, because the pump preserves the atomic occupied
subspace, so the remark also covers interacting perturbations of the pumps of Section~\ref{sec:fermionic}.
\end{proof}

\phantomsection\label{pf:prop:injdist}
\begin{proof}[Proof of Proposition~\ref{prop:injdist}]
Let $O$ be a logical Pauli operator of $S/\Gamma$ of weight $w<\min(\tilde d_R,R)$. Call two qudits of its support linked if some quotient
check acts on both, and let $O=\prod_cO_c$ be the restriction of $O$ to the classes of the equivalence relation this generates. A check
that met two classes would link them, so each check meets at most one class, and its symplectic product with $O$ equals that with the one
$O_c$ it meets. Hence every $O_c$ commutes with every quotient check. Fix a class $c$ and a qudit $q$ of it. Every qudit of $c$ is joined to
$q$ by a path of fewer than $w$ links, and every check that meets $c$ lies within distance $w$ of $q$. Paths in the quotient graph lift, so
$c$ lifts into the ball $\tilde B$ of radius $w-1$ about a lift $\tilde q$, and since $\pi$ is injective on the ball of radius $w\le R$, there is a
unique Pauli $\tilde O_c$ supported in $\tilde B$ with $\pi(\tilde O_c)=O_c$. Let $\tilde s$ be a parent check that meets $\tilde B$. It lies in the ball of
radius $w$, and no translate $g\tilde s$ with $g\ne1$ meets $\tilde B$: if it met $\tilde B$ at $\tilde q_1$, the ball of radius $w$ would contain the
distinct qudits $\tilde q_1$ and $g^{-1}\tilde q_1$ of one orbit.
By the formula in the proof of Proposition~\ref{prop:valid},
$\langle\tilde O_c,\tilde s\rangle=\sum_g\langle\tilde O_c,g\tilde s\rangle=\langle O_c,\pi(\tilde s)\rangle'=0$. A check that does not meet $\tilde B$ commutes with $\tilde O_c$
trivially. So $\tilde O_c\in S^\perp$, and it is supported in a ball of radius $R$ with weight at most $w<\tilde d_R$, so $\tilde O_c\in S$. Then
$O_c=\pi(\tilde O_c)\in\pi(S)=S/\Gamma$, and $O=\prod_cO_c\in S/\Gamma$ is not a non-trivial logical operator.
\end{proof}

\phantomsection\label{pf:prop:quarter}
\begin{proof}[Proof of Proposition~\ref{prop:quarter}]
(a) By Lemma~\ref{lem:pippump} the pump on $\tilde M$ is the evolution of quadratic Hamiltonians that commute with the deck generator and have exponentially
decaying kernels, built by one local rule, the kernel of $K$ wrapped onto $T^2_L$ with the hopping signs of the spin structure, for each of
the two spin structures of the cover described after its construction. Proposition~\ref{prop:gaussdescent} applies, and gives (a).

(b) By the construction of the pump, $V$ is the evolution for two unit-time steps of the gauged forms of $-\frac12\Psi^\dagger K_1\Psi$ and $\frac12\Psi^\dagger K_2\Psi$. Lift to
the universal cover $\mathbb R^3$ of $M$, on which $\pi_1(M)$ acts freely by isometries of the lattice. There the kernels of $K_1$ and $K_2$
are bounded by $Ce^{-r/\xi}$ between sites at distance $r$, with $\xi=\ell$ and $C$ depending only on $\mu$, by (iii) of Lemma~\ref{lem:pippump}, since they
couple only neighbouring layers. A term of the gauged Hamiltonian on $M$ is labelled by a pair of sites of $M$ and a homotopy class of
paths between them, that is, by a $\pi_1(M)$-orbit of pairs $(\tilde x,\tilde y)$ of sites of the cover. We route its Wilson line as follows:
take the representative with $\tilde x$ in a fixed fundamental domain, join $\tilde x$ to $\tilde y$ by the staircase path that changes the
three coordinates one after the other, in a fixed order, and project to $M$. The gauged terms commute with the fluxes of the $\Z_2$ gauge
field, and where all fluxes are trivial two paths in the same homotopy class give the same operator. So $V$ preserves the subspace of flat
gauge fields, which contains the code space, and its restriction there, and with it (a) and (c), does not depend on the routing. The
quarter-turn symmetry is used only for the single-particle kernels on the cover, in (a); it is not a symmetry of $M$, and the routing needs
none. A term of length $r=|\tilde x-\tilde y|$ is an operator of norm at most $Ce^{-r/\xi}$ supported within distance $O(1)$ of its path,
which lies in the box with diagonal $\tilde x\tilde y$, so the support has diameter at most $r+O(1)$. If the term acts on a qubit $q$, with a
fixed lift $\tilde q$, some image $(g\tilde x,g\tilde y)$ of its representative has $g\tilde x$ within distance $r+O(1)$ of $\tilde q$, and
distinct terms give distinct pairs. So a qubit lies in the support of at most $C'(r+1)^5$ terms of length between $r$ and $r+1$, a bound
independent of $L$ and $J$. So the interaction $\Phi$ satisfies $\sup_q\sum_{X\ni q}\|\Phi_X\|\,|X|^2e^{{\rm diam}(X)/\xi'}<\infty$ for every $\xi'>\xi$,
uniformly in $L$ and $J$, since $|X|$ grows only polynomially with ${\rm diam}(X)$. This is the decay condition of~\cite[Appendix~C,
eq.~(15)]{HHKL2018}, under which their Lieb--Robinson bound holds in the form of their Lemma~5 (their eq.~(41)); their commutator
condition, eq.~(14), holds with $\eta=1$. Their Theorem~1 is stated for terms of finite range, but its proof uses the locality only through
that bound, in their Lemma~6, which is stated for any interaction obeying it. For an evolution of time $O(1)$ that proof writes the
evolution, within $\varepsilon$ in operator norm, as a product of a constant number of layers of unitaries, each the evolution of the
terms supported in a block of diameter $O(\xi\log(n/\varepsilon))$. By Proposition~\ref{prop:gaussdescent}(c) the pump on $M$ is the evolution of
these Hamiltonians, for two unit-time steps, so it is within $\varepsilon$ of a circuit of constant depth whose gates act on regions of diameter
$O(\xi\log(n/\varepsilon))$. The error $\varepsilon$ bounds both the leakage out of the code space and the error of the logical gate.

(c) By (v) of Lemma~\ref{lem:pippump}, $V$ sweeps one layer of $\nu=\pm1$, the $p+ip$ defect of~\cite{Barkeshli:2023bta}, once through $M$. The
input we take from~\cite[eq.~(26)]{Barkeshli:2023bta} is that, modulo logical Cliffords, this sweep acts as the product of $CCZ$, $CS^\dagger$
and $T$ gates with exponents the triple cup products of a basis of $H^1(M;\Z_2)$. It is derived there in field theory and is the one step of
the proof not carried out on the lattice; check T21 tests it on this pump without using it. Here $\int\tau^3=\int\tau\cup\beta\tau$ is the linking form on the 2-torsion of $H_1(M;\Z)$. For the mapping torus of the quarter
turn $R$ this torsion is $\mathrm{coker}(R-1)=\Z_2$, generated by a fibre cycle $x$. With $y=Rx$, $(R-1)(x+y)=-2x$, so $2x$ bounds the cylinder
swept by $x+y$ once around the circle, which a parallel copy of $x$ in another fibre meets once, since $x\cdot y=\pm1$ on $T^2$. So the linking
form is $\pm\frac12$, and $\int\tau^3=1$; $z^2=0$, since $z$ is pulled back from the
circle, so $\int\tau z^2=\int z^3=0$; and $z$ is Poincar\'e dual to the fibre $T^2$, so $\int\tau^2z=\int_{T^2}(\tau|_{T^2})^2=0$, since the
square of a degree-one class on $T^2$ vanishes mod 2. Check T16, an exact computation on the barycentric subdivision, confirms these values. So $V$ is
a $T$ gate on $\tau$, with no $CS$ or $CCZ$ term.

(d) With fibre $T^2_L$ and circle length $L_z$, $\tau$ has a Wilson line of length about $L$ in the fibre and a membrane of area about $LL_z$, and
$z$ has a Wilson line of length $L_z$ and the fibre as membrane, so $d\le c\min(L,L_z)$. Proposition~\ref{prop:kk} gives only $d\le L$, because
$\tau$ restricts non-trivially to the fibre. Proposition~\ref{prop:injdist} and Lemma~\ref{lem:ftclocal} give $d\ge\min(L,L_z)/(2r_0)$. Table~\ref{tab:quarter} gives exact values on small instances.
\end{proof}

\section{Certificates}
\label{sec:cert}

This appendix collects the certificates used in the main text. Two kinds of output need them. Duality quotients are written in non-CSS form but might be CSS after a local Clifford, and
cubic-defect outputs are written as commuting projectors but might be equivalent to a Pauli stabilizer code. The five-qubit code of
Example~\ref{ex:run1} illustrates the first point trivially: no CSS code has parameters $\code{5,1,3}$, so the duality quotient is not
local-Clifford CSS, and its checks keep the weight of the parent, with letters transported by the prescribed local Clifford, by Lemma~\ref{lem:once}, because each plaquette meets each orbit
at most once.

\subsection{Not local-Clifford CSS}
\label{app:cert-css}

\paragraph{CSS in disguise.} Let $C$ be a two-block group-algebra code over an abelian group $G$ with generating sets $A$, $B$, quotiented by
the ZX-duality $D_t:g\mapsto t-g$ (block swap and Hadamard). If a homomorphism $\phi:G\to\Z_2$ vanishes on $A$ and $B$ and $\phi(t)=1$, Hadamards
on $t+\ker\phi$ make every check pure, and the quotient is the two-block code over $\ker\phi$; here $\ker\phi$ is a set of orbit representatives,
since $D_t$ exchanges $\ker\phi$ and $t+\ker\phi$. In the language of
Section~\ref{sec:flatlat} the duality twist is then gauge-equivalent to a CSS twist on a smaller group: the wall is removable.

\begin{lemma}[Rigidity]\label{lem:rigid}
For a set $T$ of qubits let $S_T=\{s\in S:\supp s\subseteq T\}$. Suppose every qubit of $T$ carries a single Pauli letter in all elements of $S_T$
that touch it, and the binary code of supports of $S_T$ is a connected matroid on $T$. If $U$ is a local Clifford with $USU^\dagger$ CSS, then $U$
maps all of $S_T$ to pure Paulis of one type. In particular, if $S_{\supp g}=\{1,g\}$ for every natural generator $g$ and the generators are not
local-Clifford CSS, then neither is the group.
\end{lemma}
\idea A CSS form splits each element into an $X$-part and a $Z$-part that again belong to the group. When every qubit carries a
single letter and the support code is connected, the supports cannot be split, so all of $S_T$ has to become one type.
\begin{proof}
Let $S'=USU^\dagger$. Every $s'\in S'$ factors as $s'=xz$ with $x,z\in S'$ of $X$ and $Z$ type and $\supp s'=\supp x\cup\supp z$, so $S'_T=US_TU^\dagger$ is
closed under these factors. It is single-letter on each qubit, like $S_T$. A qubit with letter $Y$ would give an $X$ factor with letter $X$
there, so every letter is $X$ or $Z$. If $Q_X$ and $Q_Z$ are the qubits with letters $X$ and $Z$, the factors are the restrictions to them, so
the support code is the direct sum of its restrictions, and by connectivity one of them is empty.
\end{proof}

\paragraph{Colouring tests.} A local Clifford maps different letters on a qubit to different letters, so a CSS form of a generating set with
$Y$ letters needs a 2-colouring of the generators in which, on each qubit, equal letters get equal colours; a qubit carrying all three letters
rules this out. Lemma~\ref{lem:rigid} turns this test on generators into a statement about the group.

\subsection{Not a Pauli stabilizer code}
\label{app:cert-pauli}

Let $C$ be a qubit CSS code with $X$-checks $X(h_v)$, $v\in V$, and dress them by diagonal operators: $\tilde X_v=X(h_v)D_v$, with each $D_v$ diagonal,
each $\tilde X_v$ a Hermitian involution on the $Z$-syndrome-free subspace $P_Z$, and the $\tilde X_v$ commuting on $P_Z$. The dressed Gauss laws of
Section~\ref{sec:higher} at $p=2$ are of this form, by Lemma~\ref{lem:cochain}. Let $\tilde P\subseteq P_Z$ be the joint $+1$ eigenspace.

\begin{theorem}[Dimension of dressed codes]\label{thm:dressdim}
Let $m'=\dim\ker H_X^{\top}$ be the number of independent $X$-check relations and $r_X=\rk H_X$. For a relation $u\in\ker H_X^\top$, $g_u=\prod_{v\in u}\tilde X_v$ is
diagonal on $P_Z$, $g_u|a\rangle=\chi_u(a)|a\rangle$ with $\chi_u(a)=(-1)^{q_u(a)}$ for $a\in\ker H_Z$, and $\deg q_u\le e$ if every dressing has degree at most $e$
($e=2$ for $CZ$ dressing). As a function of $a$, $\chi_u$ is not a character in general, since $q_u$ can be quadratic; for each fixed $a$, the map
$u\mapsto\chi_u(a)$ is a character of the relation group $\ker H_X^\top$. Then, with no further hypothesis,
\[
\dim\tilde P=2^{-r_X}\,\#\{a\in\ker H_Z:\ \chi_u(a)=1\ \forall u\}\ \le\ 2^k .
\]
\end{theorem}
\idea Write the projector onto the code as the average of all products of Gauss laws. A product that moves $Z$-basis states
has zero trace, so only the products over relations survive, and these act diagonally by the phases $\chi_u(a)$. For a fixed configuration $a$,
the average of $\chi_u(a)$ over the relations is 1 if all of them are $+1$ and 0 otherwise, so the trace counts configurations.
\begin{proof}
On $P_Z$ the $\tilde X_v$ commute and square to 1, so $u\mapsto g_u$ is a homomorphism from $\FF_2^V$, and
\[
\dim\tilde P=\Tr\big(P_Z\prod_v\tfrac{1+\tilde X_v}{2}\big)=2^{-|V|}\sum_u\Tr(P_Zg_u).
\]
If $H_X^\top u\neq0$, $g_u$ shifts every $Z$-basis state, so its trace vanishes.
Otherwise $\Tr(P_Zg_u)=\sum_{a\in\ker H_Z}\chi_u(a)$. Since $u\mapsto g_u$ is a homomorphism and the $g_u$ for relations are diagonal, for fixed $a$ the map
$u\mapsto\chi_u(a)$ is a character of $\ker H_X^\top$, so $\sum_{u\in\ker H_X^\top}\chi_u(a)$ is $|\ker H_X^\top|$ if $\chi_u(a)=1$ for all $u$ and $0$ otherwise. Exchanging
the two sums therefore counts the common zeros of the $q_u$ with weight $|\ker H_X^\top|$. This gives the formula, using $2^{-|V|}|\ker H_X^\top|=2^{-r_X}$. The count is at most $|\ker H_Z|$, and
$\dim\ker H_Z-r_X=k$.
\end{proof}

\begin{proposition}[Lower bound, conditional]\label{prop:warning}
In the setting of Theorem~\ref{thm:dressdim}, suppose the equations $q_u(a)=0$, for $u$ in a basis of the relations, have a common solution in
$\ker H_Z$, equivalently $\tilde P\ne0$. Then $\dim\tilde P\ge2^{k-e\,m'}$. The hypothesis holds if $q_u(0)=0$ for every $u$, and in particular for the
dressed Gauss laws of a cup polynomial with no linear terms (Theorem~\ref{thm:pointcount}), whose relation polynomials are
homogeneous of positive degree.
\end{proposition}
\begin{proof}
The $m'$ polynomials $q_u$, $u$ in a basis, have degrees summing to at most $em'$ in $\dim\ker H_Z$ variables. By Warning's second
theorem~\cite{Warning1935}, a system over $\FF_2$ with a common zero has at least $2^{\dim\ker H_Z-em'}$ of them, and $\dim\ker H_Z-r_X=k$.
\end{proof}

The hypothesis cannot be dropped. Take one qubit, no $Z$-checks, and two $X$-checks with the same support dressed as $\tilde X_1=X$ and
$\tilde X_2=-X$. They are commuting Hermitian involutions and the relation $u=\{1,2\}$ has $q_u\equiv1$. Here $k=0$, $m'=1$ and the dressings are constants,
so the bound would give $2^{k-em'}=1$, but $\tilde P=0$. The exact formula of Theorem~\ref{thm:dressdim} gives 0, as it should.

\begin{corollary}[Certificate against Pauli stabilizer codes]\label{cor:cert}
If the count in Theorem~\ref{thm:dressdim} is neither 0 nor a power of 2, then $\tilde C$ is not unitarily equivalent to any qubit Pauli stabilizer code, local or not.
Examples, in the convention of Section~\ref{sec:hfcountsec} (the thin slab $M=X\times[0,1]$ with relative $B$), are $22$ (the codes of $D(D_4)$
and $D(Q_8)$ over $X=T^2$), $92$ (the 3D $D_4$ and $Q_8$ gauge theories over $X=T^3$) and $1576$ (type IV over $X=T^3$); on the closed
$T^2\times S^1$ the type-III code has dimension 176.
\end{corollary}
\begin{proof}
Unitaries preserve the dimension of the code space, and qubit Pauli stabilizer codes have dimension $2^k$. The counts quoted are evaluations of
the exact formula by Theorem~\ref{thm:pointcount}, which covers the $Q_8$ theories, whose cup polynomials have a repeated field; direct
evaluation on the lattice gives the same numbers.
\end{proof}

That no Pauli stabilizer code realizes the phase $D(D_4)$ also follows from the structure of two-dimensional topological Pauli stabilizer
codes, whose anyons are abelian~\cite{BombinDuclosCianciPoulin2012,Bombin2014,Haah2021TI}; Pauli stabilizer models realize only abelian twisted
quantum doubles~\cite{Ellison2022}. The corollary is a statement about the code rather than the phase, and needs no locality or translation
invariance: no Pauli stabilizer code on any number of qubits has this code space, because its dimension is not a power of 2.

The characters are the only place the dressing enters the count. For the type-III construction with 1-form fields the relations are
indexed by $H^0$, and the characters are the trilinear pairing $H^1(F_1)\times H^1(F_2)\times H^0(F_3)\to\FF_2$ and its permutations, the ``two
fluxes cross and leave a charge'' rule of type-III gauge theory. For fields of degrees $n_i$ the relation spaces are $H^{n_i-1}$
(Theorem~\ref{thm:pointcount}), and the non-trivial characters decide which cohomology classes survive the count. If $m'=0$ there are no
relations, and the dressed code is diagonally conjugate to $C$. The certificate excludes Pauli stabilizer codes; it does not
exclude constant-depth equivalence to some other code that is not a Pauli stabilizer code, for which the logical commutator group is a natural candidate
invariant.

\section{Computations: inputs, methods and outputs}
\label{app:checks}

\paragraph{Software.} The computations used Python 3.13 with NumPy 2.5, networkx 3.7 for the graph method below, SciPy 1.18 for one
rank computation and for breadth-first searches, OR-Tools 9.12 (CP-SAT) for the local minimizations of check T25, and a C program, compiled with gcc 13.3, for exhaustive distance searches. Randomized checks use fixed seeds, so every
output is deterministic. Two computations are independent cross-checks. The automorphism group of $\code{144,6}$ was computed with nauty and
again with the VF2 matcher of networkx, which lists every colour-preserving automorphism of the graph described below; both find the same 8.
The distance of the triality code was certified by a separate exhaustive search on that code alone, which also verifies the weight-7 witness.
The running example has its own certificate, computed independently of the first computation (Appendix~\ref{app:runcert}).

\paragraph{What ``exact'' means.} A distance is called exact only when the output contains a logical operator of that weight, checked to
commute with every check and to lie outside the stabilizer group, together with a lower bound obtained by minimizing over \emph{all} operators,
not over a list of coset representatives. Four complete minimizations are used: (a) a listing of every element of the normalizer
$S^\perp$, keeping the least weight outside $S$ (for the running example, where $\dim S^\perp=22$); (b) a listing of every Pauli operator
of weight at most $d-1$, each tested for commuting with the checks and for lying outside $S$ (the C program for $n\le64$, and direct
searches for the certificate of the running example and for the codes of Table~\ref{tab:isometry}); (c) for the $Z$-logicals of a homological code, a
shortest non-trivial cycle, found among the fundamental cycles of breadth-first-search trees grown from every vertex, which minimizes over
every cycle in every non-trivial homology class; (d) for the quarter-turn codes (T25), the operators that anticommute with a
membrane: the $X$-part of such an operator is a dual cycle in a non-trivial class, which is simple when the target weight is less than
the shortest such cycle plus the girth of the dual graph, so every simple cycle up to the target weight is listed, and for each one the
least number of triangles that the $Z$-part needs outside it is found exactly by the CP-SAT solver~\cite{ortools} on the region within that
number of steps of the cycle, where every lighter $Z$-part must lie. A list of one representative for each of the $p^{2k}-1$ non-trivial logical cosets only
bounds the distance from above, since a coset may contain lighter elements; such lists are used here only to produce witnesses and logical
actions, never for a lower bound. A count is exact when it is a rank over $\FF_p$ or a complete enumeration of the classes involved.
Checks T17, T18, T20, T21 and T22 are double-precision floating-point computations: they report the observed deviations and are evidence on the
instances computed, not proofs. Checks labeled exact use integer or $\FF_p$ arithmetic; checks labeled finite check test identities on
finitely many configurations.

\paragraph{Light stabilizers of $\code{144,6}$, without a distance assumption.} The automorphism computation needs the set of all
stabilizers of weight at most 6. A depth-first search grows an operator from its lowest-index qubit. At each step it takes the
first check that anticommutes with the partial operator, branches over the qubits of that check not yet used (with the three letters on each),
and records the partial operator as soon as it commutes with every check; every qubit lies in at most six checks, which justifies the
pruning bound of the search. Let $P\ne1$ commute with every check and have weight at most 6. A check that anticommutes with a restriction
of $P$ but commutes with $P$ meets the rest of the support of $P$, so the branch that follows the support of $P$ stays inside it, and some
non-empty restriction $P|_A$ that commutes with every check is recorded. The output is exactly the 144 checks (136 of weight 6, 8 of
weight 5). So $P|_A$ is a check $C$, and $PC^{-1}=P|_{\mathrm{supp}P\setminus A}$ again commutes with every check and is lighter. By induction
$P$ is a product of checks with pairwise disjoint supports, and since every check has weight at least 5, $P$ is a single check. Hence the
stabilizers of weight at most 6 are exactly the 144 checks and, as a by-product, $d\ge7$. The argument uses no distance claim; the exact
value $d=15$ comes from a separate exhaustive search and is not used here.

\paragraph{What the automorphism graph encodes.} For a Pauli stabilizer group $S$ and a set $W$ of its elements, the graph has a vertex for each
element of $W$, for each qubit, and for each of the three Pauli letters at each qubit; each letter vertex is joined to its qubit, and an
element of $W$ is joined to the letter it carries at each qubit of its support. Vertex colours separate the three kinds. An automorphism of
this coloured graph is a permutation of the qubits together with a permutation of the three letters at each qubit, which is always induced
by a single-qubit Clifford up to Paulis, that maps $W$ to itself. If $W$ is mapped to itself by every monomial-Clifford automorphism of $S$ and
generates $S$, the graph automorphisms are therefore exactly the monomial-Clifford automorphisms of $S$ modulo Paulis, which is the
convention for $\mathrm{Aut}(\mathcal C)$ in Section~\ref{sec:sym}. Both conditions hold when $W$ is the set of all stabilizers of weight at most $w$
and these generate $S$: for $\code{144,6}$ with $w=6$, by the previous paragraph, and for the running example with $w=4$, where every
element of $S$ is listed (Appendix~\ref{app:runcert}). Automorphisms are found on binary vectors, that is, up to signs; each lifts to an exact
automorphism after a Pauli correction, since two sign choices on $S$ differ by a character of $S$, which is conjugation by a Pauli.

\begin{center}\small
\begin{tabularx}{\textwidth}{@{}>{\raggedright}p{4.2cm}>{\raggedright\arraybackslash}X@{}}
\toprule
Statement & Input; method; output \\
\midrule
running example (Examples~\ref{ex:run1}--\ref{ex:run4}) & rotated toric code on $\Z^2/2\Lambda$, $\Gamma=\Lambda/2\Lambda$;
exact distances by listing the normalizer; automorphisms by the graph method with $W$ the minimum-weight stabilizers; output
$\code{20,2,4}\to\code{5,1,3}$, Tate dimensions 2 and 1, $|\mathrm{Aut}|=80$ and 360, 20 descend\\
running example, certificate (Appendix~\ref{app:runcert}) & independent of the previous
computation; every element of $S$ listed; every Pauli below the distance searched; the ranks of Theorem~\ref{thm:general} and of $\mathcal L^\Gamma$\\
Lemma~\ref{lem:cochain}, $p=2$ & five $\Z_2$ defect types on $\Z_3^2\times\Z_4$, seed 2026; 150 random
nearby pairs of Gauss laws per type on random, flat and locally flat configurations (finite check); failures only off the flux-free
subspace, and for commutator and order only for cubic $\omega$; the gauge identity exact on flat configurations; a deleted $CZ$ is detected\\
Lemma~\ref{lem:cochain}, signs at $p=3,5$ (Appendix~\ref{app:worked}) & $\Z_p$ cubical tori;
Leibniz rule; conjugation, commutator, order and gauge identities, 40 random trials per case (finite check); 0 failures, and failures in
every case with the sign dropped\\
Theorem~\ref{thm:balanced}, non-abelian two-block codes; transfer with Tate term; factorization &
hypergraph-type products over $D_m\times\Z_2$, $m=3,6$; ranks over $\FF_2$ (exact); quotient equals the two-block code, $\dim\hat H^0=2$ and 0, one-
and two-step quotients agree\\
Theorem~\ref{thm:dressedquot}, quadratic: quotient identity, gauge invariance &
$\Z_p$ $n$-form toric codes on cubical tori, $p=2,3$; check matrices compared after push-forward (finite check)\\
Lemma~\ref{lem:cochain}(i) and Theorem~\ref{thm:dressedquot}, cubic & types I--III and $(1,1,2)$ in
$D=4$; dressing against conjugation in 1550 evaluations; Gauss-law polynomials pushed forward cell by cell, 0 differences (finite check)\\
Theorem~\ref{thm:hfcount} (about 50 cases); Theorem~\ref{thm:pointcount} and Corollary~\ref{cor:boundary} (30 cases) & lattice $k'$ from ranks (exact per instance) and the code-space count by complete enumeration of class
representatives, against the cohomology formulas\\
22 for type III on $T^2$, $L=2,3$ & complete enumeration\\
22, 92, 1576 from group theory & complete enumeration of $\Hom(\Z^n,G)$\\
\bottomrule
\end{tabularx}
\end{center}

\begin{center}\small
\begin{tabularx}{\textwidth}{@{}>{\raggedright}p{4.2cm}>{\raggedright\arraybackslash}X@{}}
\toprule
Statement & Input; method; output \\
\midrule
automorphism group of $\code{144,6}$ and its descent & the quotient of Section~\ref{sec:genon}; light stabilizers as above; graph method with $W$
the 144 checks, by nauty and independently by VF2; affine parent symmetries enumerated over $GL(2,\Z_{12})$; output
$|\mathrm{Aut}|=8$, centralizer of $D$ of order 16\\
isometry gates (Table~\ref{tab:isometry}) & distances $\code{10,2,3}$, $\code{21,2,3}$, $\code{39,2,4}$ by listing all Paulis of weight $\le5$; Klein
quartic and Hantzsche--Wendt by shortest non-trivial cycles; logical groups generated by the induced symplectic maps\\
logical $CCZ$ on $\mathbb{RP}^3$ (T6) & $S^3=\partial[-m,m]^4$, $m=1,2$; $d_Z$ by breadth-first search
(exact), $d_X$ an upper bound only; cubic form under four random gauge shifts\\
a copy-permuting twist creates a logical $CCZ$ (T7) & $(L,m)=(3,2),(4,2),(3,3)$; cubic form on all
of $H^1$ (finite check)\\
$\code{32,2,7}$ triality code & colour code on a $12\times12$ torus modulo
$(4,4)$ composed with $X\to Y\to Z$; every Pauli of weight $\le6$ searched, none is a logical; witness
$XXXI^{12}YI^3XI^3YI^2XI^5$ of weight 7\\
Proposition~\ref{prop:transgress} and Example~\ref{ex:kzhgate} & three toric codes,
$\omega=a_ra_ga_b$; $\tilde UG\tilde U^\dagger$ against the predicted product for every Gauss law and 8--10 random flat configurations per site, on
$T^2_L$, $L=3,4$, with $B=1$, closed $T^3$ of sides $(3,3,3)$, $(4,4,4)$, $(2,3,4)$, $(2,2,2)$, $(3,3,1)$ with $B=\theta_z$ (the last three violate (iii)), and
the thin slabs $T^2_L\times[0,1]$, $L=3,4$ (finite check); 0 failures for both signs of $W$, 7--34\% for the controls (dropping $W$, or $W^2$ in
its place); the exact-cover identity $U_\eta VU_\eta^\dagger=VW$ in 80 trials\\
Lemma~\ref{lem:inflow}, Proposition~\ref{prop:slabgate} (T12) & the same codes on the thin slabs $T^2_L\times[0,1]$, $L=3,4$ (135 and 240 qubits),
relative $B$, $\eta$ the top layer; the inflow identity for every Gauss law, $U_\eta VU_\eta^\dagger=VWW_\partial$, and $VW_\partial^{\pm1}$ on $\mathcal C_0^{\partial z}$, on 6
random flat configurations per site (finite check); 0 failures; the boundary term is non-zero in 19\% of evaluations, and dropping $W_\partial$
fails in 13--15\%\\
Proposition~\ref{prop:reltrans}, Example~\ref{ex:kzhbdry} (T13) & KZH's square with boundaries $L_r$, $L_b$, $L_{rb}$, $B=1$, sides 2, 3, 4 (36, 72, 120
qubits), and the squares of sides 2, 3 times a circle of length 3 with a wall (189, 360 qubits); $k$ by ranks; commutation of the truncated
Gauss laws; $\lambda=\delta\beta$ on the $L_{rb}$ side faces; $\tilde UG\tilde U^\dagger$ against the predicted product on 10 (square) or 6 (with circle)
configurations per Gauss law (finite check); logical action from a gauge path from $\alpha$ to $A\alpha$; output $k=1$, 0 failures, $T$ for the
derived $W$ and $T^\dagger$ for~\cite[eq.~(57)]{KZH2025chs}; controls fail in 3--21\%\\
the dressed membrane of Section~\ref{sec:threegroup} (T14) & one toric code on $T^3_L$, type-I twist with $B_1=\theta_x$, $\Sigma$ the plane $y=0$; ratios
$X(\sigma)G_vX(\sigma)^\dagger G_v^{-1}$ on 12 flat configurations per vertex, $L=3,4$; dressings solved over $\FF_2$ from 30 flat configurations per
vertex next to the line and verified on 40 fresh ones per vertex, $L=5,6$ (finite check); no $CZ$/$Z$ dressing on the edges within distance 1
($L=5$); the $S$-string with an $L$-gate $CZ$ chain: 0 failures in 5000 and 8640 evaluations\\
Proposition~\ref{prop:threegroup}, Example~\ref{ex:klein} (T8) & cubical $T^3$ with sides
$(2L,L,L)$ and its quotient $K\times S^1$ by the glide $(x+L,-y,z)$, $L=2,3$, on barycentric subdivisions; $b_1$ by ranks; $q$ on all 8 classes and after
3 random gauge shifts each; polar form against $\langle w\cup\cdot\cup\cdot\rangle$ on all 64 pairs (finite check); output $b_1=3$, $q=0$ on the cover, polar
form of rank 2 on $K\times S^1$\\
Example~\ref{ex:accidental} & toric codes on $\Z_3^2$, $\Z_6\times\Z_3$, $\Z_6^2$; graph method with $W$
the stars and plaquettes, every automorphism listed by VF2 (exact); output $|\mathrm{Aut}|=144,144,576$, images of descent 72 and 144, the $90^\circ$
rotation descends only from $\Z_6^2$\\
twisted cups as copy-permuting twists (Proposition~\ref{prop:internal}, Example~\ref{ex:deephole}) & one toric code on the Kuhn-triangulated cubic $T^3$, $L=2,3,4$, and $S^1\times T^2_{\rm hex}$, $L=2,3$; $\int a\cup P^*a\cup
P^{2*}a$ on all 8 classes under random gauge shifts, against $\det(y,Ay,A^2y)$ (finite check); output~\cite[eq.~(48)]{BDEHT2026cupsII} and the gate
of T7b\\
Example~\ref{ex:deephole}, qutrits (T9) & the same tori, $L=3$; $\int c\cup P^*c\cup P^{2*}c\bmod9$ on
all 27 classes, with 5 random changes of integral lift each; the cochain formula~\cite[eq.~(38)]{BDEHT2026cupsII} on random representatives for
four sign choices of its correction terms and without them; three independent copies (finite check)\\
Proposition~\ref{prop:tori}, Example~\ref{ex:deephole}, 4D (T10) &
every $G$-stable lattice between $\Z[G]$ and its maximal order for $C_4$ and $V_4$ on $T^4$ and $C_3$ on $T^3$, by determinants (exact);
Kuhn-triangulated cubic and split $T^4$, $N=2$, all 16 classes with 3 integral lifts each (finite check); output 6 and 17 lattices,
gates as in Example~\ref{ex:deephole}\\
Proposition~\ref{prop:lens}, Example~\ref{ex:lens} (T15) & $S^3=C_{2p}*C_{2p}$ with $\Z_p$ acting by rotations $(2,2r)$, $(p,r)=(2,1),(3,1),(5,1),(5,2)$, the
16-cell for $\mathbb{RP}^3$, the octahedron and its barycentric subdivision for $\mathbb{RP}^2$, and products with a circle of length 3; every degree of $B_F$ and every
$(n_1,n_2)$; $k'$ by ranks against $\dim\ker\cup[B]$ computed on the lattice, push-forward of the dressed parent against the Gauss laws on $X\times F$, Tate and
invariant dimensions; least trivializing cover of $L(m;1)$, $m\le9$; the carry primitive and $\zeta_g=-\delta\chi$ (exact); 0 mismatches\\
Theorem~\ref{thm:pointcount}, Step 2, repeated fields (T11) & types I and II, the $Q_8$ polynomial and $a_1a_2a_1+a_2a_2a_1$ on $\Z_2\times\Z_2\times\Z_3$ and $\Z_3\times\Z_2\times\Z_3$ with $B$ a single
slice; relation characters by applying the Gauss laws one by one, against the slot sum, on 40 random flat cochains each (finite check);
0 mismatches in 640 evaluations, 92 of them non-trivial\\
\bottomrule
\end{tabularx}
\end{center}

\begin{center}\small
\begin{tabularx}{\textwidth}{@{}>{\raggedright}p{4.2cm}>{\raggedright\arraybackslash}X@{}}
\toprule
Statement & Input; method; output \\
\midrule
cup products for the fermionic pump (T16) & barycentric subdivisions of $T^3$, $T^2\rtimes_{C_2}S^1$, $T^2\rtimes_{C_4}S^1$, the Hantzsche--Wendt space,
$\mathbb{RP}^3$, $L(4,1)$ and $S^3/Q_8$, up to 14368 cells; $H^1(\,\cdot\,;\Z_2)$ by ranks and every triple cup product (exact); the class
of~\cite[eq.~(26)]{Barkeshli:2023bta} modulo Cliffords under random changes of basis; output the gates of Section~\ref{sec:fermionic}\\
rotation phases at the seam (T17) & Chern insulator and $p+ip$ superconductor on the $L\times L$ fibre, $L=4,6,8$, rotations about a site and a
plaquette centre; free-fermion expectation values of $C_n(-1)^{Fa_z}$ in each sector, $n=2,4$, as in~\cite[eq.~(49)]{Barkeshli:2023bta}; output the
phases of Example~\ref{ex:dicosm} and Proposition~\ref{prop:quarter}, and the $C_2$ values of~\cite{Barkeshli:2023bta} as a control\\
Proposition~\ref{prop:gaussdescent}, Example~\ref{ex:dicosm} and Proposition~\ref{prop:quarter} (T18) & explicit $C_4$-equivariant Chern pump on fibres $L=4,6$, two layers on
the quotient and 2 ($C_2$) or 4 ($C_4$) times as many on the cover; the sectors from each $C_n$-invariant spin structure of the fibre, with both
characters of the deck group; commutation with the deck generator, spectra of the restricted and of the
directly built pump in each sector, $|\langle\mathrm{atomic}|U_M|\mathrm{atomic}\rangle|$ (finite check); $p+ip$ nucleation tested for equivariance and
particle--hole compatibility; range-$r$ truncation with polar re-unitarization at $L=6,8$; output deviations at most $1.3\cdot10^{-15}$, equal
spectra, overlaps 1 to 12 digits, $V^2=1$, $CZ$, $S^\dagger_\tau$ on $T^3$, $T^2\rtimes_{C_2}S^1$, $T^2\rtimes_{C_4}S^1$, leakage at most $4\cdot10^{-14}$, and the
exact phase from $r=1$\\
Proposition~\ref{prop:nilp}, open problem 5 (T19) & $D(G)$ for $G=\Z_2^3$, $D_4$, $Q_8$, $D_6$, $D_8$, $D_{12}$, $D_{16}$, with $D_n$ dihedral of order $2n$; anyons
from projective irreducible representations of centralizers, $S$-matrix, Verlinde fusion and the chain of adjoint subcategories; for
$D^{\hat c\cup e}(\Z_2\times D_4)$ and $D^{\hat c\cup e'}(\Z_2\times D_8)$ the cocycle conditions, anyon count, quantum dimensions and spins; output
class 1, 2, 2, 3, 4 for $\Z_2^3$, $D_4$, $Q_8$, $D_8$, $D_{16}$, no nilpotency for $D_6$, $D_{12}$, and agreement with $D(D_8)$ (46 anyons) and
$D(D_{16})$ (142 anyons); $D(D_4)\simeq D^{\omega_{\rm III}}(\Z_2^3)$ as a control\\
Proposition~\ref{prop:quarter}, generator of the $p+ip$ pump (T20) & the $C_4$-equivariant $p+ip$ nucleation unitary $W(k)$ of T18 for $\mu=\pm1$,
composed with the parity $Q$ of the first layer, on a $160\times160$ momentum grid with the high-symmetry points; $WQ+(WQ)^\dagger-2x$, the
eigenphases of $WQ$, Hermiticity of $K$ and $e^{iK}=WQ$, particle--hole and $C_4$ symmetry of $K$, the occupied-space map, and the real-space
decay of $K$ (finite check; the identity is also proved by hand); output deviations at most $6\cdot10^{-14}$, eigenphases and $\|K\|$ at most
$\pi/2$, decay length 1.3 and 1.4 lattice spacings\\
Proposition~\ref{prop:quarter}(c), many-body phases of the $p+ip$ pump (T21) & the pump of Lemma~\ref{lem:pippump} on $T^2_L\rtimes_{C_4}S^1$,
$L=4,6,8$ with $J=2$ and $L=4,6$ with $J=4$, $\mu=\pm1$, rotations about a site and a plaquette centre; the closed formula for $W$ against its
frames, the Chern numbers, $\hat G_g^4=1$ with the antiperiodic circle, $[K_i,\hat G_g]=0$; in each sector and each $C_4$-orbit of momenta the
restricted Nambu Hamiltonians in a Fock space of at most 16 modes, $\langle\Omega_s|e^{-iH_2}e^{-iH_1}|\Omega_s\rangle$, and the Bogoliubov action of the
evolution (finite check); output orbit factors of modulus 1 within $3\cdot10^{-15}$, and $V=T_\tau^{\mp1}CZ_{\tau z}Z_\tau^\epsilon$ for $\mu=\pm1$ within $2\cdot10^{-15}$ rad\\
Theorem~\ref{thm:dresseddist}, Knill--Laflamme on non-Pauli dressed codes (T22) & dressed codes with $B=1$ on $T^2$: type I ($4\times4$, 32 qubits), type II ($3\times3$, 36 qubits) and type III ($2\times3$, 36 qubits), and the same codes without the dressing; the code space in the flux-free configuration basis, Knill--Laflamme matrices by Walsh--Hadamard transforms (tolerance $10^{-9}$), the least weight of a Pauli operator that violates Knill--Laflamme, over all Pauli operators up to that weight, and every Pauli operator on a $2\times2$ patch of the type I code (12 qubits, 8192 operators) and a $1\times1$ patch of the type II code (8 qubits, 256 operators); output dimensions 4, 16, 22 against 4, 16, 64 without the dressing, distances 4, 3, 2 with and without it, and no violation on the patches\\
Theorem~\ref{thm:dresseddist}(b), filling in a box (T23) & 1- and 2-form fields on $T^2_9$, $T^3_6$, $T^2_9\times[0,1]$ and $T^3_6\times[0,1]$, every box side that fits; ranks over $\FF_2$ of the cocycles and cycles supported in a box against the coboundaries and boundaries from the box (exact); output every cocycle and every cycle in a box filled inside the box, and, as a negative control, 2-form fields on $T^2_9\times[0,1]$, with $n=d$, not filled\\
the \hyperref[ex:emfour]{example at the end of Section~\ref*{sec:general}}, 4D electromagnetic duality (T24) & the quotient of the 4D 2-form toric code by the half shift composed with $H$,
for the shift $(\frac12,\frac12,\frac12,M+\frac12)$ with $M=0$ ($L=2,\dots,5$) and $M=1$ ($L=2,3$); ranks over $\FF_2$; the checks compared one by one with
$W_c$ and $G_e=X(\delta e)Z_{e+h}Z_{e-h}$, and with $\int(e\cup f+f\cup e)$; the Levin--Wen T-junction with three Pauli strings along the axes,
$L=8$ to $14$, legs of length 3 to 5, with the undressed checks as control; the all-$Y$ staircase for $L\le6$; exact distances by a
syndrome-guided depth-first search over all operators of weight at most $2L-1$, and a witness of weight $2L$, for $L\le7$ (exact); output
$k'=3$ in every case, equal checks, a fermion (a boson for the control), and $d=2L$ for $2\le L\le7$\\
Proposition~\ref{prop:quarter}(d), Table~\ref{tab:quarter} (T25) & the codes of Table~\ref{tab:quarter}; $k$ by ranks over $\FF_2$;
the Levin--Wen T-junction for ten random triples of strings; shortest dual cycles by breadth-first search; the string minimum by the
minimization (d); disjoint strings found greedily (exact); output $k=2$, a fermion, and Table~\ref{tab:quarter}\\
\bottomrule
\end{tabularx}
\end{center}

\begin{table}[ht]
\centering\small
\begin{adjustbox}{max width=\textwidth}\begin{tabular}{@{}llll@{}}
\toprule
Check & Parent and quotient & $k$ & Descended gate acts as \\
\midrule
T6 & three toric codes on $S^3=\partial[-m,m]^4\to\mathbb{RP}^3$ (antipodal map), $m=1,2$, barycentric subdivision & parent 0; quotient 1 per copy & logical $CCZ$ ($\int x^3=1$) \\
T7a & three toric codes on $T^3\to$ tricosm, deck map $R$ a $120^\circ$ screw & 1 per copy & identity ($\int w^3=0$) \\
T7b & same parent, deck map $R\sigma$, $\sigma$ cycling the copies & 3 & $CZ\cdot CZ\cdot CCZ$ on $(c,a_1,a_2)$ \\
T7c & same, $R$ a pure shift & 3 & identity \\
T9 & one qutrit toric code, $S^1\times T^2_{\rm hex}\to T^3$ (deep-hole holonomy), $P$ the coordinate $3$-cycle & 3 & $\omega^{x_0(b_1-b_2)^2}$ on the cover; $T_1T_2T_3\,CCZ^{-1}$ on $T^3$ \\
T10 & one 4D toric code, $C_4$-tori from $S^1\times S^1\times T^2_{\rm sq}$ to the cubic $T^4$ & 4 & identity; $Z$, $CZ$; $S$, $CZ$; $T$, $CS$, $CCZ$ \\
\bottomrule
\end{tabular}\end{adjustbox}
\caption{Checks T6--T10: Propositions~\ref{prop:twisted}, \ref{prop:internal} and~\ref{prop:tori} and Example~\ref{ex:deephole}. In T6 one copy is $\code{1808,1}$ with $d_Z=6$ ($m=1$) and
$\code{14368,1}$ with $d_Z=12$ ($m=2$); the value was 1 under four random gauge shifts and 0 for a trivial class, and on $S^3$ it is $2\times1=0$
(Corollary~\ref{cor:covers}). T7 matched for $(L,m)=(3,2),(4,2),(3,3)$, independently of gauge; T7a and T7b have the same qubit count.
T9 and T10 use Kuhn triangulations, whose simplices and the order of the vertices of each simplex the coordinate permutations preserve. T9 ($L=3$, 567 and 189 qutrits)
evaluates $\int c\cup P^*c\cup P^{2*}c\bmod9$ on integral lifts, independently of the lift, and the cochain formula~\cite[eq.~(38)]{BDEHT2026cupsII}
on random representatives; that formula was gauge invariant and equal to the former only with $\mu_1(a,b,c)=+a\cup_1(b\cup c)$, $\cup_1$ as
in~\cite[eq.~(8)]{BDEHT2026cupsII}. On three independent copies $\int c_1\cup c_2\cup c_3\bmod9$ changes from 1 to 4 under $c_1\mapsto4c_1$. T10
evaluates the two end tori on lattices (1920 and 240 qubits, $N=2$) and all six tori by the determinant of Proposition~\ref{prop:tori}; the cubic
$T^4$ gives~\cite[eq.~(53)]{BDEHT2026cupsII} up to the sign set by $\rho$ versus $\rho^{-1}$.}
\label{tab:t67}
\end{table}

\subsection{Certificate for the running example}
\label{app:runcert}
This subsection lists every number of Examples~\ref{ex:run1}--\ref{ex:run2} with its witness, as printed by a computation
that uses only NumPy and shares no code with the first computation of the running example.

\paragraph{Parent.} Qubit $i=10x+y$ sits at the site $(x,y)$, $0\le x\le1$, $0\le y\le9$, of $\Z^2/2\Lambda$ with $2\Lambda=\langle(2,4),(0,10)\rangle$; a
point $(x,y)$ reduces to $(x\bmod2,\ y-4\lfloor x/2\rfloor\bmod10)$. The rows below, in order of $i$ and four to a line, are the plaquettes with lower-left corner at site $i$, of type $X$
for $x+y$ even and $Z$ for $x+y$ odd:
\begin{center}\footnotesize\ttfamily
\begin{tabular}{@{}llll@{}}
XXIIIIIIIIXXIIIIIIII & IZZIIIIIIIIZZIIIIIII & IIXXIIIIIIIIXXIIIIII & IIIZZIIIIIIIIZZIIIII\\
IIIIXXIIIIIIIIXXIIII & IIIIIZZIIIIIIIIZZIII & IIIIIIXXIIIIIIIIXXII & IIIIIIIZZIIIIIIIIZZI\\
IIIIIIIIXXIIIIIIIIXX & ZIIIIIIIIZZIIIIIIIIZ & IIIIIIZZIIZZIIIIIIII & IIIIIIIXXIIXXIIIIIII\\
IIIIIIIIZZIIZZIIIIII & XIIIIIIIIXIIIXXIIIII & ZZIIIIIIIIIIIIZZIIII & IXXIIIIIIIIIIIIXXIII\\
IIZZIIIIIIIIIIIIZZII & IIIXXIIIIIIIIIIIIXXI & IIIIZZIIIIIIIIIIIIZZ & IIIIIXXIIIXIIIIIIIIX
\end{tabular}
\end{center}
The rank is 18, so $k=2$, and $\dim S^\perp=22$. Listing all $2^{18}=262144$ elements of $S$ shows that the minimum weight is 4, attained
exactly by the 20 plaquettes; this is the set $W$ used for the automorphism group. The generators $g_v=T_vH^{\otimes20}$ and
$g_{iv}=T_{iv}H^{\otimes20}$, $v=(1,2)$, $iv=(-2,1)$, permute the qubits by
\begin{center}\footnotesize
$g_v:\ (12,13,14,15,16,17,18,19,10,11,8,9,0,1,2,3,4,5,6,7)$,\qquad $g_{iv}:\ (5,6,7,8,9,0,1,2,3,4,15,16,17,18,19,10,11,12,13,14)$,
\end{center}
listing the image of qubits $0,\dots,19$; the third element is $g_{v+iv}=T_{(-1,3)}$. We checked that each $g$ maps every plaquette into $S$
and moves every qubit. A basis of logical representatives is
\begin{center}\footnotesize\ttfamily
XIIIIIIIIIIIIIIXXXXX,\quad IXIIIIIIIIIIXXXIIIII,\quad ZIIIIIIIIIIZZZIIIIII,\quad IIIIIIIZIIIIZZZZZIII
\end{center}
(weights 6, 4, 4, 6). All 32550 Pauli operators of weight 1 to 3 were tested and none is a non-trivial logical; $ZZZI^{10}ZI^{6}$ is a
logical of weight 4, so $d=4$.

\paragraph{Transfer quantities.} As ranks over $\FF_2$,
\[
\dim S^\Gamma=5,\quad \dim\mathsf N(S)=4,\quad \dim\hat H^0(\Gamma;S)=1,\quad \dim(S^\perp)^\Gamma=6,\quad
\dim\frac{(S^\perp)^\Gamma}{S^\Gamma}=1,\quad \dim\mathcal L^\Gamma=2,
\]
so $k'=5-4=1$, one invariant logical class has an invariant representative, and the other does not (Remark~\ref{rem:invclasses}). The least
weight of an element of $S^\Gamma\setminus\mathsf N(S)$ is 12, attained for instance by $IIZYZIIXYXYZIIXYXIIZ$, a winding stabilizer.

\paragraph{Quotient.} Quotient qubit $j$ is the orbit class $y-2x\bmod5$, represented by its site of odd coordinate sum. The images of the 20
plaquettes are the five checks $IXZZX$, $XIXZZ$, $XZZXI$, $ZXIXZ$, $ZZXIX$, of rank 4, so $k'=1$. All 105 Pauli operators of weight 1 or 2 were
tested and none is a non-trivial logical; $XYXII$ is a logical of weight 3, so $d'=3$. The operators $XXXXX$ and $ZZZZZ$ commute with the checks,
are not stabilizers, and anticommute, so they form a logical basis. The winding stabilizer above descends to $IIXYX$, a non-trivial logical
of weight 3: this is the Tate contribution of Theorem~\ref{thm:general}(ii), and it attains $d_{\rm wind}=3$.

\bibliographystyle{utphys}
\bibliography{biblio}
\end{document}